\documentclass[longauth]{aa} 
\usepackage{graphicx}
\usepackage{txfonts}
\usepackage[breaklinks=true]{hyperref} 
\usepackage{natbib,twoopt,rerunfilecheck}
\bibpunct{(}{)}{;}{a}{}{,}             
\makeatletter
  \newcommandtwoopt{\citeads}[3][][]{\href{http://adsabs.harvard.edu/abs/#3}%
    {\def\hyper@linkstart##1##2{}%
     \let\hyper@linkend\@empty\citealp[#1][#2]{#3}}}
  \newcommandtwoopt{\citepads}[3][][]{\href{http://adsabs.harvard.edu/abs/#3}%
    {\def\hyper@linkstart##1##2{}%
     \let\hyper@linkend\@empty\citep[#1][#2]{#3}}}
  \newcommandtwoopt{\citetads}[3][][]{\href{http://adsabs.harvard.edu/abs/#3}%
    {\def\hyper@linkstart##1##2{}%
     \let\hyper@linkend\@empty\citet[#1][#2]{#3}}}
  \newcommandtwoopt{\citeyearads}[3][][]%
    {\href{http://adsabs.harvard.edu/abs/#3}
    {\def\hyper@linkstart##1##2{}%
     \let\hyper@linkend\@empty\citeyear[#1][#2]{#3}}}
\makeatother
\begin{document} 

\title{Gaia-ESO Survey: The analysis of high-resolution UVES spectra of FGK-type stars\thanks{Based on 
observations made with the ESO/VLT, at Paranal Observatory, under program 188.B-3002 (The Gaia-ESO Public Spectroscopic Survey, PIs Gilmore and Randich). }}

   \author{R. Smiljanic\inst{1,2}
          \and
          A.~J. Korn\inst{3}
          \and
%
%
          M. Bergemann\inst{4,5}
          \and
          A. Frasca\inst{6}
          \and
          L. Magrini\inst{7}
          \and
          T. Masseron\inst{8}
          \and
          E. Pancino\inst{9,10}
          \and
          G. Ruchti\inst{11}
          \and
          I. San Roman\inst{12}
          \and           
          L. Sbordone\inst{13,14,15}
% ...
          }
          
   \institute{   
   European Southern Observatory, Karl-Schwarzschild-Str. 2, 85748 Garching bei M\"unchen, Germany
              \and 
              Department for Astrophysics, Nicolaus Copernicus Astronomical Center, 
          ul. Rabia\'nska 8, 87-100 Toru\'n, Poland \\
                        \email{rsmiljanic@ncac.torun.pl}
         \and
             Department of Physics and Astronomy, Division of Astronomy and Space Physics, 
             Uppsala University, Box 516, 75120 Uppsala, Sweden \\
             \email{andreas.korn@physics.uu.se}
         \and
         Max Planck Institute for Astrophysics, Karl-Schwarzschild Str. 1 85741 Garching, Germany
         \and
         Institute of Astronomy, University of Cambridge, Madingley Road, Cambridge, CB3 0HA, United Kingdom
         \and         
         INAF - Osservatorio Astrofisico di Catania, via S. Sofia 78, I-95123 Catania, Italy 
         \and
         INAF - Osservatorio Astrofisico di Arcetri, Largo Enrico Fermi 5, 50125 Florence, Italy          
         \and
         Universit\'e Libre de Bruxelles, Campus Plaine, CP 226, Boulevard du Triomphe, 1050 Bruxelles, Belgium
         \and
         INAF - Osservatorio Astronomico di Bologna, Via Ranzani 1, I-40127 Bologna, Italy
         \and
           \pagebreak
         ASI Science Data Center, via del Politecnico SNC, I-00133, Roma, Italy
         \and
         Lund Observatory, Department of Astronomy and Theoretical Physics, Box 43, 221 00, Lund, Sweden
         \and
         Departamento de Astronom\'{\i}a, Casilla 160, Universidad de Concepci\'{o}n, Concepci\'{o}n, Chile
         \and 
         Zentrum f\"ur Astronomie der Universit\"at Heidelberg, Landessternwarte, K\"onigstuhl 12, D-69117 Heidelberg, Germany
         \and
        Millennium Institute of Astrophysics, Av. Vicu\~na Mackenna 4860, 782-0436 Macul, Santiago, Chile
        \and
        Pontificia Universidad Cat\'olica de Chile, Av. Vicu\~na Mackenna 4860, 782-0436 Macul, Santiago, Chile
% ...
}

   \date{Received 2014; accepted 2014}

\titlerunning{Gaia-ESO analysis of UVES spectra of FGK-type stars}
\authorrunning{Smiljanic et al.}

 
  \abstract
   {The Gaia-ESO Public Spectroscopic Survey is obtaining high-quality spectroscopic data for about 10$^{5}$ stars using FLAMES at the VLT. With the FLAMES-UVES link, high-resolution spectra are being collected for about 5\,000 FGK-type stars.}
   {These UVES spectra are analyzed in parallel by several state-of-the-art methodologies. Our aim is to present how these analyses were implemented, to discuss their results, and to describe how a final recommended parameter scale is defined. We also discuss the precision (method-to-method dispersion) and accuracy (biases with respect to the reference values) of the final parameters. These results are part of the Gaia-ESO second internal release and will be part of its first public release of advanced data products.}
   {The final parameter scale is tied to the one defined by the Gaia benchmark stars, a set of stars with fundamental atmospheric parameters. In addition, a set of open and globular clusters is used to evaluate the physical soundness of the results. Each of the implemented methodologies is judged against the benchmark stars to define weights in three different regions of the parameter space. The final recommended results are the weighted-medians of those from the individual methods.}
  {The recommended results successfully reproduce the benchmark stars atmospheric parameters and the expected $T_{\rm eff}$-$\log~g$ relation of the calibrating clusters. Atmospheric parameters and abundances have been determined for 1301 FGK-type stars observed with UVES. The median of the method-to-method dispersion of the atmospheric parameters is 55\,K for $T_{\rm eff}$, 0.13\,dex for $\log~g$, and 0.07\,dex for [Fe/H]. Systematic biases are estimated to be between 50-100 K for $T_{\rm eff}$, 0.10-0.25 dex for $\log~g$, and 0.05-0.10 dex for [Fe/H]. Abundances for 24 elements were derived: C, N, O, Na, Mg, Al, Si, Ca, Sc, Ti, V, Cr, Mn, Fe, Co, Ni, Cu, Zn, Y, Zr, Mo, Ba, Nd, and Eu. The typical method-to-method dispersion of the abundances varies between 0.10 and 0.20 dex.}
   {The Gaia-ESO sample of high-resolution spectra of FGK-type stars will be among the largest of its kind analyzed in a homogeneous way. The extensive list of elemental abundances derived in these stars will enable significant advances in the areas of stellar evolution and Milky-Way formation and evolution.}

                              
   \maketitle
%
\section{Introduction}

Following the seminal paper of \citetads{1957RvMP...29..547B}, it is now well established that the vast majority of the chemical elements are produced inside stars. These elements and their isotopes are synthesized by various processes in stars of different masses and of different generations \citepads[][for a review]{1997RvMP...69..995W}. 

Modern astrophysics strives to trace the processes of synthesis and dispersion of the chemical elements, and use them to decode the history of formation and evolution of planets, of stars, and of the Galaxy. Multi-element abundance information is a key requirement in this context, as the abundances of distinct elements are shaped by different physical processes. The investigation of large samples of long-lived stars, formed in different places and times in the Galaxy, is needed to put together a complete picture of Galactic and stellar evolution.

Obtaining the spectroscopic data to achieve this goal is demanding. For determining accurate detailed elemental abundances we need high-resolution, high signal-to-noise (S/N) spectra with a broad wavelength coverage. For robust statistics and to cover all Galactic populations, the observation of large stellar samples, including faint stars beyond the solar neighborhood, is needed. To achieve this, a number of spectroscopic surveys are now being conducted and/or planned as, for example, the APO Galactic Evolution Experiment \citepads[APOGEE,][]{2014ApJS..211...17A}, the GALactic Archaeology with HERMES \citepads[GALAH,][]{2012ASPC..458..421Z}, the LAMOST Experiment for Galactic Understanding and Exploration \citepads[LEGUE,][]{2012RAA....12..735D}, the RAdial Velocity Experiment \citepads[RAVE,][]{2006AJ....132.1645S,2013AJ....146..134K}, the Sloan Extension for Galactic Understanding and Exploration \citepads[SEGUE,][]{2009AJ....137.4377Y}, and the Gaia-ESO Survey \citepads[][]{2012Msngr.147...25G,2013Msngr.154...47R}.

The Gaia-ESO Survey\footnote{\url{http://www.gaia-eso.eu}} is an ambitious public spectroscopic survey that is obtaining medium- and high-resolution spectra of more than 10$^{5}$ stars. The observations started on December 31, 2011 and are carried out at the Very Large Telescope (VLT), at the Paranal Observatory, Chile. All the data collected by the Survey is homogeneously reduced and analyzed by the Gaia-ESO consortium. Catalogs with astrophysical parameters are being produced to be made available to the community.

The Survey targets represent all major Galactic components (halo, bulge, thin and thick disks) and include a large number of open clusters, selected to cover the parameter space of age, total stellar mass, distance, and metallicity. The targets include early- and late-type stars (from O- to M-type), giants, dwarfs, and pre-main-sequence stars.

Observations are conducted with the FLAMES (Fiber Large Array Multi-Element Spectrograph) multi-fiber facility \citepads{2002Msngr.110....1P}. Medium-resolution spectra ($R$ $\sim$ 20\,000) of about $\sim$ 10${^5}$ stars are being obtained with Giraffe and high-resolution spectra ($R$ $\sim$ 47\,000) of about $\sim$ 5\,000 stars are being obtained with UVES \citepads[Ultraviolet and Visual Echelle Spectrograph,][]{2000SPIE.4008..534D}.

\subsection{The Gaia-ESO release papers}

This paper is part of a series that presents a complete description of the Gaia-ESO Survey, in preparation for its first public release of advanced data products. The Survey is organized in different working groups (WGs) that deal with all the relevant tasks, from target selection and observations, to data analysis and data archiving. While it is beyond the goals of this paper to describe the Survey's internal organization, we provide an overview of the release papers for clarity and completeness. 

Two papers will provide the Survey overview describing science goals, observation plan, team organization, target selection strategy, and data release schedules. For the Milky Way part of the Survey this description will be presented in Gilmore et al. (2014, in prep.), while for the open clusters part of the Survey it will be presented in Randich et al. (2014, in prep.). The data and procedures used to select probable member stars to be observed in each selected open cluster will be presented in Bragaglia et al. (2014, in prep.). Description of the data reduction aspects will be presented in Lewis et al. (2014, in prep.) for the Giraffe spectra, and are described in \citetads{2014A&A...565A.113S} for the UVES spectra.

The analysis of different types of stars is performed by different WGs. The analysis of the Giraffe spectra of FGK-type stars will be described in Recio-Blanco et al. (2014, in prep.). The analysis of pre-main-sequence stars will be described in Lanzafame et al. (2014, submitted). The analysis of OBA-type stars, which are all observed in young open clusters, will be described in Blomme et al. (2014, in prep.). The analysis of non-standard objects and outliers will be part of Gilmore et al. (2014, in prep.). The analysis of the UVES spectra of FGK-type stars is the topic of the present paper. 

A considerable effort is dedicated to the observation of a comprehensive set of targets for internal and external calibration of the Survey parameter scale. Calibration targets include open and globular cluster stars, the Gaia\footnote{See \url{http://sci.esa.int/gaia/} for more details on the European Space Agency (ESA) Gaia space mission.} benchmark stars, and stars from the CoRoT fields \citepads[Convection Rotation and planetary Transits,][]{2006cosp...36.3749B}. The selection and observation of these targets will be described in Pancino et al. (2014, in prep.). Because the Survey includes the analysis of different types of stars, additional steps are needed to homogenize the final results, correcting systematic effects where needed. This additional step is taken to ensure that the results for early- and late-type stars, for dwarfs, giants, and pre-main-sequence stars are all on a single consistent scale. This Survey-wide homogenization process will be discussed in Fran\c cois et al. (2014, in prep.)\footnote{This final Survey-wide homogenization uses as anchoring point the results of the analysis of UVES spectra of FGK-type stars that is discussed in this paper. Therefore, our results are currently not changed by the final homogenization.}.

\subsection{The UVES analysis}

This paper describes the analysis of the UVES spectra of FGK-type stars in the Gaia-ESO Survey conducted within Working Group 11 (WG11) and as implemented for the first release of advanced data products. The products resulting from this analysis include: equivalent widths (EWs) of spectral lines, stellar atmospheric parameters, and elemental abundances. 

The analysis process in the Survey is performed in cycles, following the data reduction of newly observed spectra. Each new analysis cycle improves upon the last one, as some of the input data is updated (e.g. atomic and molecular data), as the teams improve their analysis methods, and as the method used to define the final recommended set of atmospheric parameters and abundances evolves.

We have now completed the analysis of two internal data releases (hereafter iDR). An iDR consists of reduced data that are ready to be analyzed and is initially available only within the Gaia-ESO consortium. New iDRs happen roughly every 6 months, after which a new analysis cycle is started. 

The second internal release (iDR2) included a revision in data formats of all the observations done by the Survey, superseding iDR1. There were also significant differences between the analysis strategy applied to the iDR1 and iDR2 data sets. The discussion presented here will concentrate on the analysis of iDR2. These are the results that will be part of the first Gaia-ESO public release, together with the results of iDR3. 

The analysis of the iDR3 data set, which is an incremental release, is currently ongoing. The iDR3 is incremental because it only includes new observations, completed after iDR2 was made available. Exactly the same analysis strategy applied to iDR2 is being applied to iDR3.

For completeness, we also present the analysis of iDR1 in an Appendix. There, we discuss the main differences between the analysis implementation for these two iDRs. The first Gaia-ESO science verification papers were based on iDR1, and it is therefore important to document how this analysis was conducted.

We stress again that the Gaia-ESO spectrum analysis is under continuos development. With improvements in the analysis, the complete Survey data set will be re-analyzed. Therefore, future releases of data products will supersede previous ones. Below, a description of iDR2 is given:

\begin{itemize}

\item[$\bullet$] {\bf Internal Data Release 2 (iDR2):} This data release consisted of all spectra obtained from the beginning of the Survey up to the end of June 2013, and some additional archival data included for calibration purposes. For the WG11 analysis, it included a total of 1708 spectra of 1447 FGK-type stars (multiple exposures of benchmark stars were analyzed separately, see Section \ref{sec:bench}). From these stars, 1412 were observed by Gaia-ESO, 35 of them were obtained from data archives, and 22 of them had both Gaia-ESO and archival spectra. \emph{The astrophysical results obtained from the analysis of the iDR2 data set will be part of the first Gaia-ESO public release of advanced data products.} The public release will be available through a dedicated Gaia-ESO Survey science archive\footnote{\url{http://ges.roe.ac.uk/index.html}} hosted by the Wide Field Astronomy Unit (WFAU) of the Institute for Astronomy, Royal Observatory, Edinburgh, UK. The results of iDR2 supersede the science verification results of iDR1 presented in the Appendix.
\end{itemize}

For the analysis of the UVES spectra of FGK-type stars we implemented multiple parallel methodologies as opposed to adopting one single analysis pipeline. The main advantage of a multiple analysis strategy in a broad survey like ours is that we can identify the different pipelines that perform better in different regions of the parameter space. We are therefore not constrained by the limitations of a single pipeline, that would introduce different systematics in different regions of the parameter space. With multiple analyses we can in addition quantify the precision of the spectroscopic analyses, by reviewing how well the multiple pipelines agree in all and each star of the sample. Nevertheless, this strategy also adds a level of complexity to the understanding of the results. A single pipeline would be internally more homogeneous and provide results that are easier to reproduce and correct when (and if) needed.

In this paper, we present a comparison of these multiple pipelines applied to iDR2. Our final parameter scale is built by implementing a homogenization process that ties it to the fundamental scale defined by the Gaia benchmark stars. Different pipelines give better results in different regions of the parameter space. Homogeneity is ensured by guaranteeing that the final results reproduce well the "real" parameters of the reference stars in each of the parameter space regions. We discuss how we use the multiple analyses to define the precision of our results, how the benchmarks are used to define the accuracy of the results, and present the limitations of the final catalog. This is a technical paper describing the spectrum analysis and its results. The scientific implications of the results will be discussed elsewhere.

The paper is structured as follows. Section \ref{sec:targets} presents a summary of the sample of FGK-type stars that is analyzed within WG11. In Sect.\ \ref{sec:data}, we present the general characteristics of the spectroscopic data used in the analysis. Section \ref{sec:analysis} describes the general properties of our multiple-analyses strategy and our homogenization procedure. It is followed by Sect.\ \ref{sec:tools}, where we describe the common tools that have been defined for the analysis. The subsequent sections present each of the data products determined in our analysis, discussing the method-to-method comparisons and, when applicable, comparing the final recommended results to the reference parameters of calibrators. Equivalent widths are discussed in Sect.\ \ref{sec:eqw}, stellar atmospheric parameters in Sect.\ \ref{sec:atm}, and elemental abundances in Sect.\ \ref{sec:abun}. Section \ref{sec:end} summarizes the analysis and highlights the scientific value of the data produced here. Two appendixes complete the paper. The first one, Appendix \ref{sec:nodes}, contains the details of the individual methodologies employed by each of the Nodes\footnote{Following the adopted Gaia-ESO Survey terminology, each of the independent analysis groups is referred to as a different analysis ``Node''.} involved in the data analysis. The second one, Appendix \ref{sec:idr1}, presents the science verification analysis of iDR1 and discusses the differences between that and the one implemented for iDR2.

\section{The FGK-type stars observed with UVES}\label{sec:targets}

\begin{figure*}[t]
\centering
\includegraphics[height = 11cm]{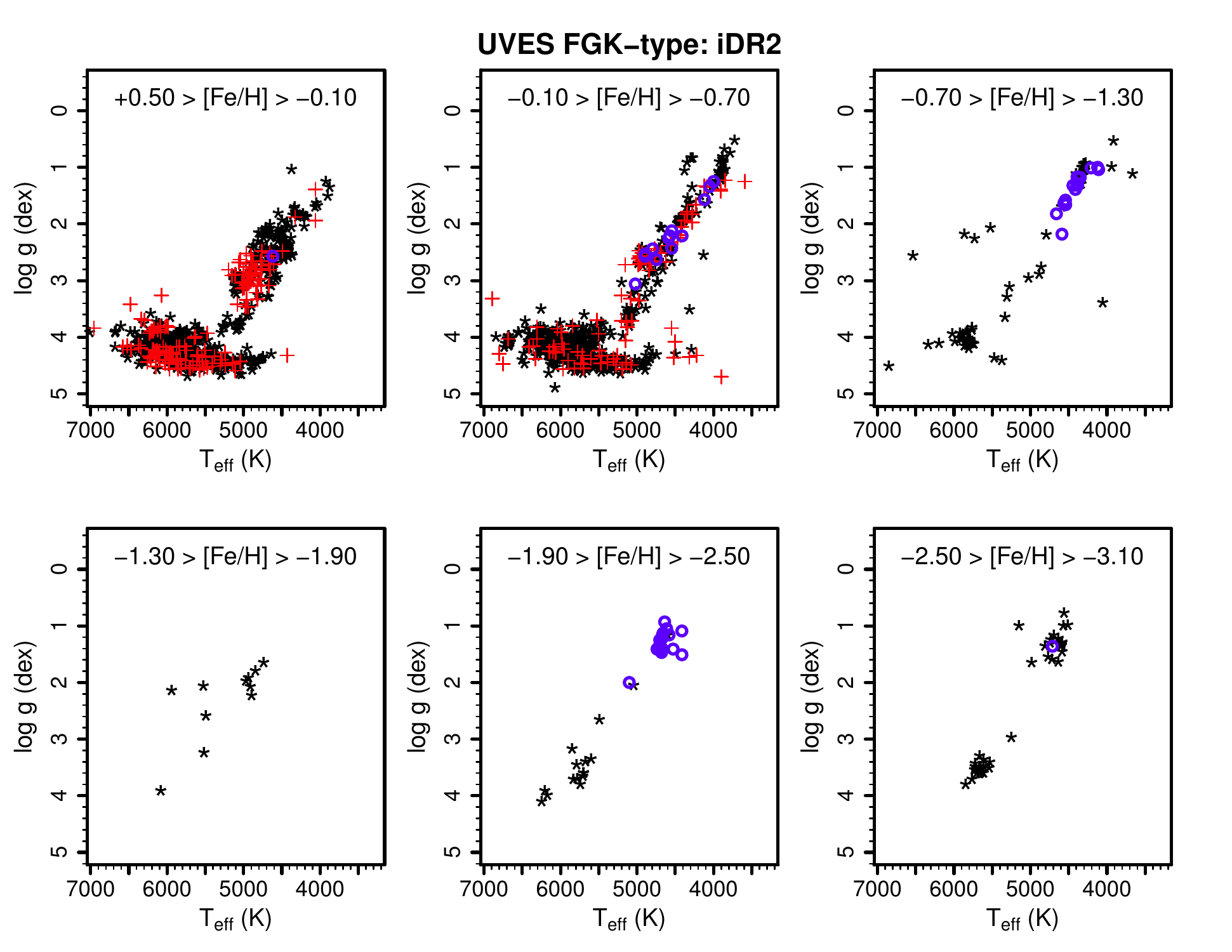}
 \caption{Distribution of FGK-type stars from iDR2 in the $T_{\rm eff}$-$\log g$ plane. The panels are divided according to metallicity. Black stars are field stars, red ``plus'' signs are stars observed in open cluster fields, and blue circles are stars observed in globular cluster fields.}\label{fig:hridr2}%
\end{figure*}

The main late-type targets observed with UVES in the Gaia-ESO Survey are FG-type dwarfs in the solar neighborhood and clump giants in old (age $>$ 1 Gyr) and intermediate-age (0.1 Gyr $<$ age $<$ 1 Gyr ) open clusters. In addition, the following targets are also present: 1) candidate clump giants in the inner disk and bulge; 2) K-type giants in the outer regions of the disk; 3) main-sequence and PMS stars in young clusters and in close-by intermediate-age clusters; 4) field stars in the line of sight of open clusters; 5) giants in a few globular clusters observed for calibration purposes; 6) giants and dwarfs in fields observed by the CoRoT satellite, used here for calibration purposes. Figure \ref{fig:hridr2} shows how the stars that are part of iDR2 are distributed in the $T_{\rm eff}$-$\log g$ plane (computed as described in Section \ref{sec:atm}).

The observations of the Milky Way solar neighborhood targets with UVES are made in parallel with the Giraffe observations. This means that the exposure times are planned according to the observations being executed with the Giraffe fibers (targets down to $V$ = 19 mag). These UVES targets are chosen according to their near-infrared colors to be FG-dwarfs/turn-off stars with magnitudes down to $J$ = 14 mag. The goal is to observe a sample of $\sim$ 5\,000 FG-type stars within 2\,kpc of the Sun to derive the detailed kinematic-multi-element distribution function of the solar neighborhood. This sample includes mainly thin and thick disk stars, of all ages and metallicities, but also a small fraction of local halo stars.

The target selection is based on 2MASS \citepads{2006AJ....131.1163S} photometry (point sources with quality flags "AAA"). A box is defined in a color-magnitude diagram with limits 12 $< J <$ 14 and 0.23 $<$ ($J-K$) $<$ 0.45 + 0.5 $\times$ E($B-V$). The \citetads{1998ApJ...500..525S} maps are used to determine the extinction E($B-V$). The targets selected before April 2012 had a brightest cut on $J$ = 11 instead of 12 . When there were not enough targets, the red edge was extended. When there were too many potential targets an algorithm selected roughly the same number of stars per magnitude bin with the rest being marked as lower priority. A complete discussion of field target selection will be given in Gilmore et al. (2014, in prep.).

In open clusters, while Giraffe is used to target complete samples of members down to $V$ = 19 mag, with the UVES fibers key brighter objects (down to $V$ = 16.5 mag) are observed. The spectra are used for accurate atmospheric parameters and abundances determination. For old and intermediate-age clusters, the UVES fibers are allocated mostly to red-clump giants. Main-sequence stars are also observed in close-by intermediate-age clusters. In young clusters, the UVES fibers are also used to observe selected main-sequence and PMS stars. These objects are first analyzed by the PMS ana\-lysis WG  (see Lanzafame et al. 2014, submitted). Those stars that are considered to be normal FGK-type stars are later added to our analysis sample (i.e., PMS stars without veiling, non-cluster members, and main-sequence stars). In the clusters, the exposure times are planned for the observations being executed with the UVES fibers. Close to $\sim$ 1\,000 FGK-type stars should be observed with UVES in clusters by the end of the Survey. The information obtained with the UVES spectra will enable the robust chemical characterization of the clusters, the study of possible star-to-star chemical variations, and will be critical inputs for studies of stellar evolution.

\section{The data}\label{sec:data}

\begin{table*}[t]
 \caption[]{\label{tab:release} Number of FGK-type stars observed with UVES and part of the iDR2 data set.}
\centering
\begin{tabular}{lcl}
\hline
\hline
Gaia-ESO Type & Stars & Comments \\
\hline
Total & 1447 & Gaia-ESO and archival data. \\
Gaia-ESO  & 1412 & Gaia-ESO only, no archival data. \\
GES\_MW  & 941 & Stars from Milky Way fields. \\
GES\_CL  & 314 & Stars from open cluster fields. \\
GES\_SD  & 157 & Calibration targets. \\
AR\_SD  & 55 & Calibrators from archival data (M 67 and the Blanco-Cuaresma et al. library). \\
\hline
GES\_SD\_BM  & 20 & Benchmark stars with Gaia-ESO spectra. \\
GES\_SD\_PC  &  2 & Peculiar stars templates. \\
GES\_SD\_GC & 51 & Stars from calibration globular clusters. \\
GES\_SD\_OC  & 23 & Stars from calibration open clusters. \\
GES\_SD\_CR  & 55 & Stars from the CoRoT fields. \\
\hline
\end{tabular}
\end{table*}
\begin{figure*}
\centering
\includegraphics[height = 6cm]{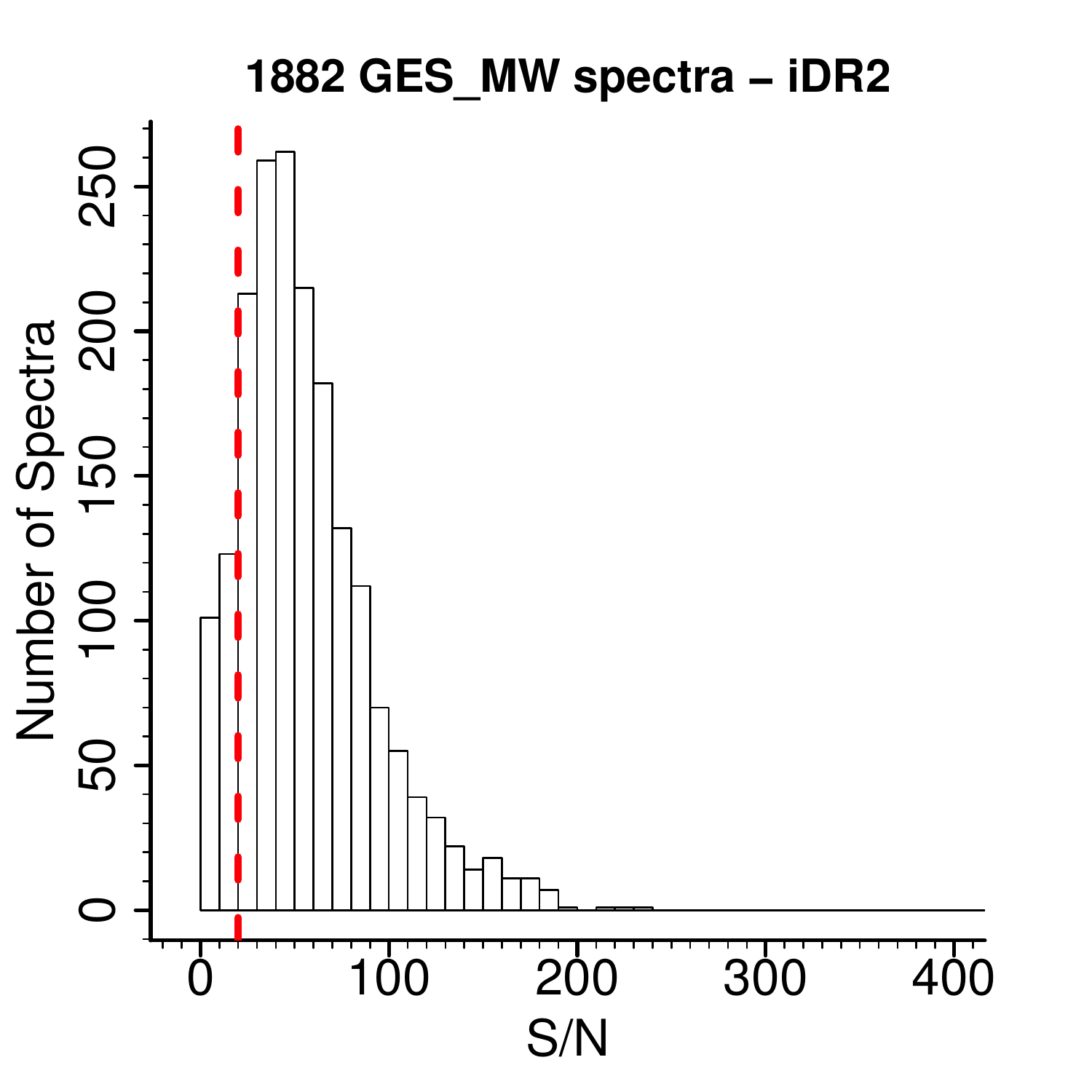}
\includegraphics[height = 6cm]{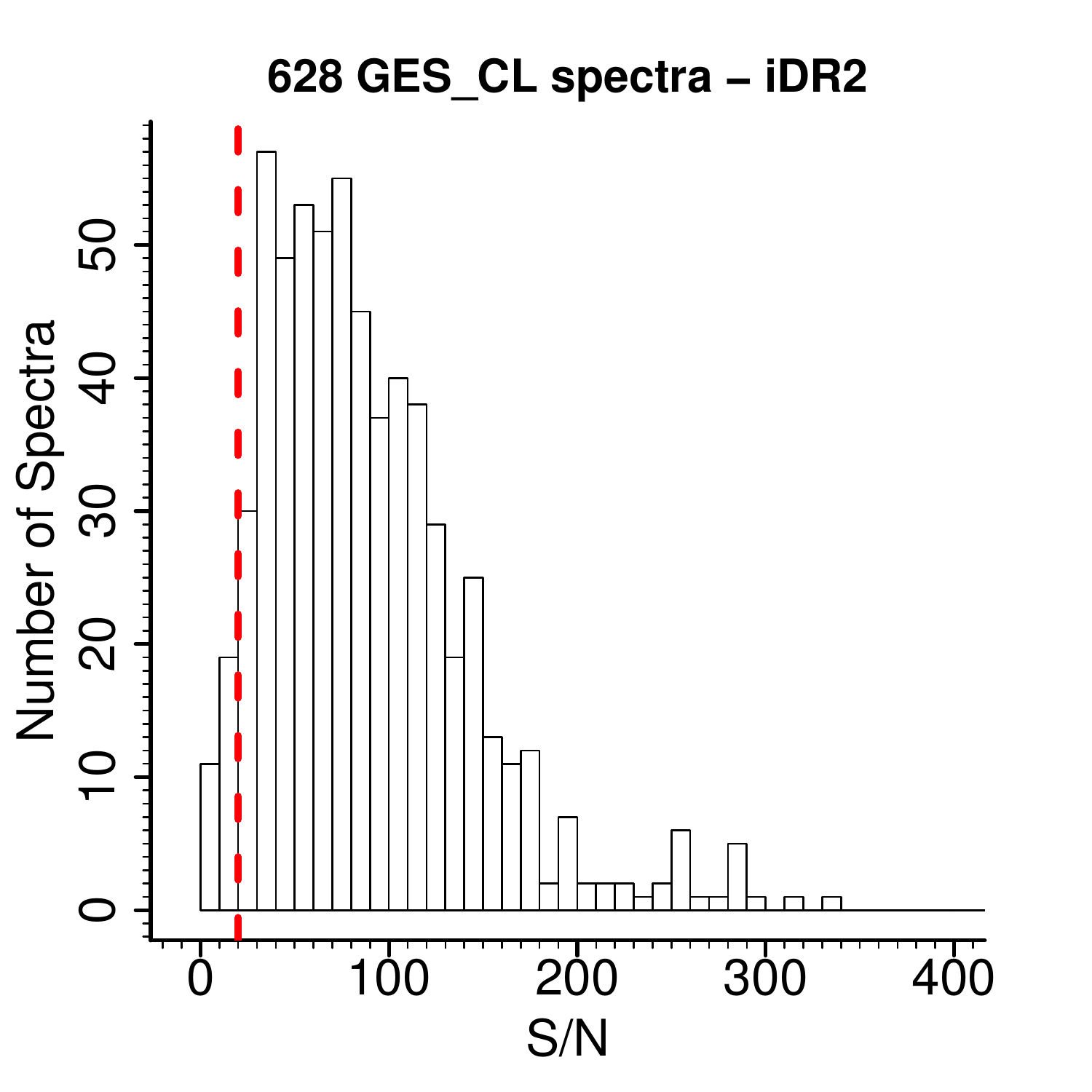}
\includegraphics[height = 6cm]{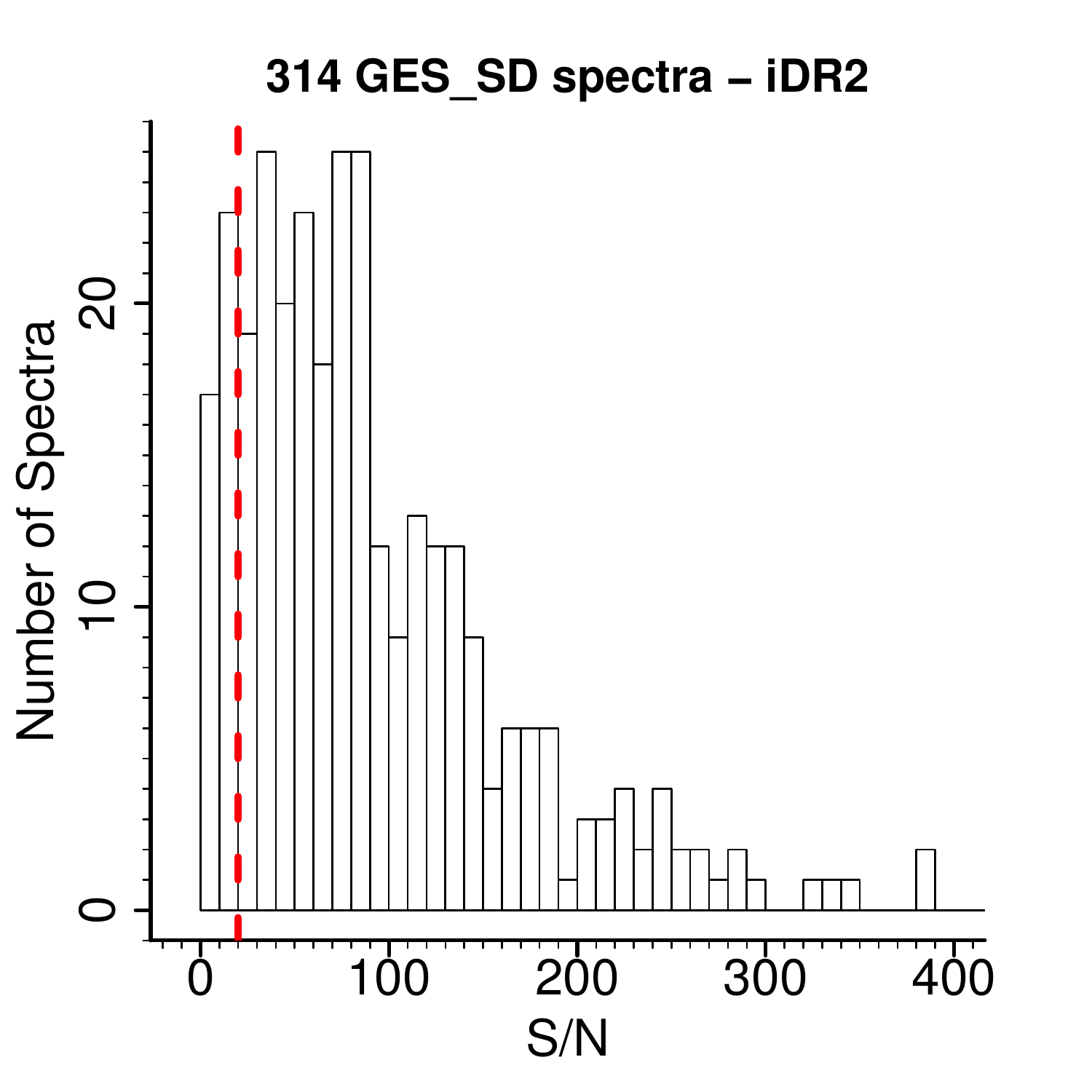}
\caption{Distribution of the median S/N per pixel of the spectra observed with UVES and that are part of iDR2 (1412 FGK-type stars). Each of the two UVES spectrum parts (from each CCD) is counted separately (thus, two spectra per star). The red dashed line indicates S/N = 20. The samples of the solar neighborhood (GES\_MW), open clusters (GES\_CL), and calibration targets (GES\_SD) are shown separately. Only final stacked spectra were included in this plot; the single exposure spectra, even when analyzed, are not counted.}\label{fig:sndist}
\end{figure*}

Late-type stars are observed with UVES in the setup centered at 580\,nm. The spectrum is exposed onto two CCDs, resulting in a wavelength coverage of 470--684 nm with a gap of $\sim$ 5 nm in the center. The FLAMES-UVES fibers have an aperture on the sky of 1$\arcsec$, resulting in a resolving power of $R$ = 47\,000.  

The UVES data are reduced with the ESO UVES pipeline and dedicated scripts described in \citetads{2014A&A...565A.113S}. Some data products are already constrained at this stage: the radial velocity ($v_{\rm rad}$) and its potential variation, and a first guess of the projected rotational velocity ($v \sin i$). The spectra are delivered to the analysis groups in a multiple-extensions {\sf FITS} format. The data are made available through an operational database hosted by the Cambridge Astronomical Survey Unit (CASU) of the Institute of Astronomy at the University of Cambridge, UK.

Different versions of the spectra are available: 1) wavelength calibrated, sky subtracted, and heliocentric corrected merged spectra; 2) continuum normalized version of the previous spectra; and 3) individual single orders, wavelength calibrated, sky subtracted, and heliocentric corrected. The inverse variance of the spectra listed before are also available. The auxiliary data collected during the sample selection phase (such as photometry and proper motions) or derived during the data reduction phase (such as $v_{\rm rad}$ and $v \sin i$) are also provided. Correction of telluric features is not implemented yet.

The distribution of the S/N per pixel of the iDR2 data is shown in Fig. \ref{fig:sndist}. The stars from the solar neighborhood sample, from open cluster fields, and the calibration targets are shown separately. The use of the calibration targets in the analysis is discussed in Sect. \ref{sec:atm}.

In addition to the Gaia-ESO sample, iDR2 also includes the library of high-resolution, high S/N observed spectra compiled by  \citetads{2014A&A...566A..98B}. We analyzed spectra of 30 Gaia benchmark stars taken from this library. The Gaia benchmark stars are defined as well-known bright stars for which well-determined $T_{\rm eff}$ and $\log g$ values are available from direct methods, independent from spectroscopy (Heiter et al. 2014a, in prep). Their metallicities are well constrained from a careful spectroscopic study \citepads{2014A&A...564A.133J}, applying some of the same analysis methods used in the Gaia-ESO Survey. 

As described in Sect. \ref{sec:atm}, the analysis of these benchmark stars is used to test the internal accuracy of the Gaia-ESO analysis and as an anchor for the scale of the Gaia-ESO parameters. In addition, these stars will be used as a first-level calibration for the Gaia results \citepads{2013A&A...559A..74B}. Their inclusion in our sample is a step towards guaranteeing a high degree of homogeneity between the results of Gaia-ESO and Gaia. Similarly, other large spectroscopic surveys can use these stars (and other stars of the Gaia-ESO calibration sample) to compare their astrophysical parameters scale with ours \citepads[see][and Pancino et al. 2014, in prep.]{2012arXiv1206.6291P}. This effort can eventually lead to a global scale of astrophysical parameters across different large spectroscopic surveys. Table \ref{tab:release} summarizes the number of stars included in the iDR2 data set.

\section{The analysis strategy}\label{sec:analysis}

Due to their high-resolution and large wavelength coverage, the UVES spectra allow for the determination of a large number of quantities. The list includes the stellar atmospheric parameters: effective temperature ($T_{\rm eff}$), surface gravity ($\log g$), microturbulence ($\xi$); the stellar metallicity [Fe/H]\footnote{The metallicity as an atmospheric parameter refers to the global content of metals in the stellar photosphere. Usually the Fe abundance is used as a proxy of the metallicity. That is true for some of the analysis methodologies employed here, but for others a global metallicity value is determined (See each method in Appendix \ref{sec:nodes}).}; elemental abundances for as many elements as the S/N and astrophysical parameters permit; and chromospheric activity indicators\footnote{Chromospheric activity indicators have not been derived yet from the spectra discussed here. The calculation of these quantities is planned and the methods used will be discussed in papers describing future data releases.}, where relevant.

\begin{figure*}[t]
\centering
\includegraphics[height = 10cm]{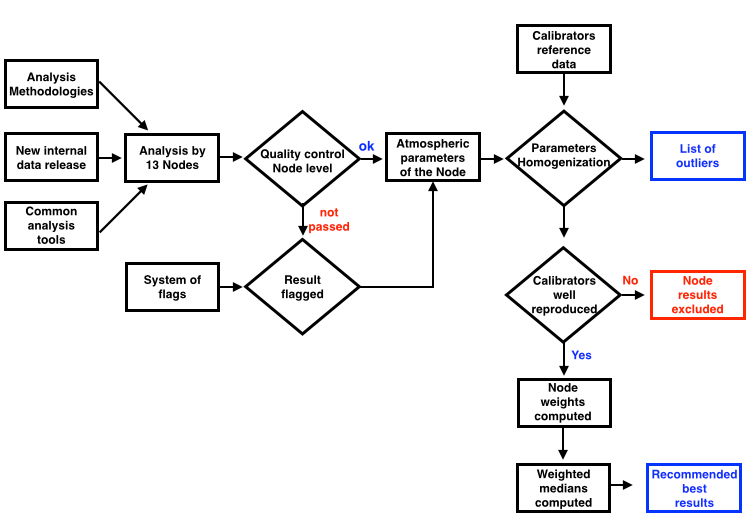}
\caption{Example of the analysis flow for the atmospheric parameters with the main tasks and decisions indicated.}\label{fig:flow}
\end{figure*}

In this section, we summarize the general strategy of our spectroscopic analysis, including the steps of quality control and homogenization. We present only the analysis strategy adopted during the iDR2 analysis, used to compute the quantities that will be included in the first public release. The strategy applied during the iDR1 analysis for science verification differed from the current one in several respects. These differences are discussed in Appendix \ref{sec:idr1}.

\subsection{Multiple pipelines strategy}

The Gaia-ESO Survey consortium includes specialists in many major state-of-the-art standard and special-purpose spectrum analysis methodologies currently employed in the literature. This gives the unique opportunity of applying multiple parallel pipelines to the same large data set. This possibility has two main advantages:

\begin{enumerate}

\item No single pipeline is optimal to analyze all stellar types that are included in our sample (e.g., giants vs. dwarfs; metal-poor vs. metal-rich stars). With multiple pipelines we can identify and use those that give best results in different regions of the parameter space. All types of objects can thus be properly analyzed, even if they require special treatment.

\item We can investigate and quantify different sources of errors, including method-dependent effects. This gives a robust measurement of what is the precision of spectroscopic analyses. This is an opportunity to give, on a star-by-star basis, what is the degree to which their absolute parameters can be trusted. 

\end{enumerate}
 
Both are invaluable advantages in a survey with targets spanning a broad range in atmospheric parameters. We mention in passing that a system of multiple analyses has been implemented to deal with the Gaia data \citepads{2013A&A...559A..74B}. Obviously, the multiple pipelines strategy also adds some complexity to the analysis process and the understanding of the results. In addition, the comparison between the analyses can not capture all the systematic uncertainties in spectroscopic analyses, caused by the limitations of the model atmospheres or by, e.g., ignoring non-LTE effects in the atmospheric parameters (as done by most of the methods implemented here apart from the LUMBA Node -- Sect.\ \ref{sec:lumba}).

To guarantee that we can deliver the best possible results, with well quantified uncertainties, we have established a series of critical tests to evaluate the results, and to bring them to a single parameter scale. Figure \ref{fig:flow} depicts the flow of our analysis strategy, as applied to the atmospheric parameters. We discuss now the general properties of these steps, and specific results for each data product are presented in Sect.\ \ref{sec:eqw} for EWs, in Sect.\ \ref{sec:atm} for atmospheric parameters, and in Sect.\ \ref{sec:abun} for the abundances.

\subsection{Node analyses}\label{sec:detailnodes}

The spectroscopic analysis is performed by 13 different ana\-lysis Nodes. The methodologies and codes used by each Node are described in detail in Appendix \ref{sec:nodes} and summarized in Table \ref{tab:nodsum}. The implementation and limitations of each code are discussed in detail elsewhere (references are given in the Appendix).

\begin{table*}[t]
\caption{Summary of the analysis methodologies used by each Node involved in WG11.}
\label{tab:nodsum} 
\centering
\begin{tabular}{llll}
\hline\hline
Node &  Contact & Codes & Type of method \\
\hline 
Bologna       & E. Pancino & {\sf DAOSPEC} and {\sf GALA} & Equivalent widths \\
Catania        & A. Frasca & {\sf ROTFIT} & Library of observed spectra \\
CAUP           & S. Sousa & {\sf ARES} and {\sf MOOG} & Equivalent widths \\
Concepcion  & S. Villanova & {\sf DAOSPEC} and {\sf GALA} & Equivalent widths \\
EPInArBo      & L. Magrini & {\sf DAOSPEC} and {\sf FAMA} & Equivalent widths \\
IAC-AIP         & C. Allende Prieto & {\sf FERRE} & Library of synthetic spectra \\
Li\`ege         & T. Morel & {\sf GAUFRE} & Equivalent widths \\
LUMBA        & S. Feltzing & {\sf SGU} based on {\sf SME} & Synthetic spectra computed on the fly\\
Nice              & V. Hill & {\sf MATISSE} & Library of synthetic spectra \\
Paris-Heidelberg & L. Sbordone & {\sf MyGIsFOS} & Library of synthetic spectra \\
UCM            & D. Montes  & {\sf ARES} and {\sf StePar} & Equivalent widths \\
ULB              & S. Van Eck & {\sf BACCHUS} & Synthetic spectra computed on the fly\\
Vilnius          & G. Tautvai{\v s}ien{\.e} & {\sf DAOSPEC} and {\sf MOOG} & Equivalent widths \\
\hline
\end{tabular}
\end{table*}

We stress that all Nodes analyze the same data, as data reduction is a step completed independently from the spectroscopic analysis. In addition, a number of ``common tools'' have been defined to guarantee some degree of homogeneity in the end results. These tools include: the use of a common line list (of atomic and molecular lines), the use of one single set of model atmospheres, and the analysis of common calibration targets. These constraints are shared with other Gaia-ESO working groups, particularly with the one responsible for the analysis of Giraffe spectra of FGK-type stars. These steps are also taken to facilitate the job of putting the full Gaia-ESO results into a single homogeneous scale (see Fran\c cois et al. 2014, in prep.). In addition, for use when needed by some analysis methodologies, a microturbulence calibration is recommended and a synthetic spectrum library computed with the same line list is available. More details on these tools are given in Sect.\ \ref{sec:tools}, or will be discussed in forthcoming publications.

Each Node performs a first quality control of their own results. They identify objects where the analysis has failed, and investigate the limits to which their results can be trusted. A dictionary of flags is used at this stage. The flags include the possibility to identify: i) phenomenological peculiarities (e.g., emission lines, multiplicity, or fast rotation); ii) stellar classification remarks, indicating for example a particular evolutionary stage (e.g. white dwarfs, post-AGBs) or properties like strong lines caused by carbon enhancement, and iii) technical issues, as for example problems with data reduction, with signal-to-noise, or analysis convergence issues. The flags will be part of the released products, and the complete dictionary will be described elsewhere (Gilmore et al. 2014, in prep. and in the release documentation).

\subsection{Parameters homogenization}

By parameter homogenization we mean the procedure of checking the performance of the Node analyses and establishing the final recommended values. At this homogenization step, stars for which only few Nodes (three or less) have provided parameters have the spectra individually checked. For the vast majority of these cases, the reasons for the analysis failure is easily detected (e.g. fast rotation, emission lines, data reduction issues). A list of outliers is produced, including the appropriate flags. Atmospheric parameters for these stars are not provided, and the list is forwarded to the WG responsible for outlier objects for further investigation.

For critically evaluating the performance of the Nodes we use a series of calibrators. The Gaia benchmark stars, a set of $\sim$ 30 stars with well defined fundamental parameters (Heiter et al. 2014a, in prep.), are the first level of calibration. They are also used as an anchor to define the final scale of the Gaia-ESO parameters. A second level of calibration is given by a series of open and globular clusters, where the consistency of the $T_{\rm eff}$ vs. $\log~g$ values can be checked. Another level of calibration will be possible with the stars observed by the CoRoT satellite, for which asteroseismic $\log~g$ values are being computed. This third check will be implemented in future releases.

The performance with respect to the benchmarks is judged separately in three regions of the parameter space: 1) \emph{metal-rich dwarfs}: stars with [Fe/H] $>$ $-$1.00 and $\log~g$ $>$ 3.5; 2) \emph{metal-rich giants}: stars with [Fe/H] $>$ $-$1.00 and $\log~g$ $\leq$ 3.5; and 3) \emph{metal-poor stars}: stars with [Fe/H] $\leq$ $-$1.00. Node results that fail the tests with the calibrators are excluded. For the remaining results, we define weights according to how well they can reproduce the reference parameters of these stars.

These weights are used to compute a weighted-median value for each atmospheric parameter. The weighted medians are adopted as the recommended best value of the atmospheric parameters. Medians are used as they are robust against outliers, minimizing the influence of less consistent results. The weights help to select the best methods in each region of the parameter space, and to force the scale to reproduce the real parameters of the benchmark stars. This is a significant advantage of our approach with respect to the usual one of using the Sun as sole reference star.

A weighted-median approach is also used for the abundances. The difference is that, apart from the Sun, there are no fundamental references of stellar abundances. We thus combined the individual Node values using the same weights defined for the atmospheric parameters. Weighted medians were computed on a line-by-line basis. The final abundance of an element is the median of the line values. In the following Sections \ref{sec:eqw}, \ref{sec:atm}, and \ref{sec:abun}, we discuss in detail the approach used to define the final recommended values of EWs, atmospheric parameters, and abundances, respectively.

When using the Gaia-ESO results, the final recommended values should be preferred, together with their uncertainty values in terms of accuracy and precision. These are the values that have been critically evaluated and calibrated to the system defined by the Gaia benchmarks.

\section{Common analysis tools}\label{sec:tools}

\subsection{Line list}\label{sec:linelist}

The Gaia-ESO Survey line list is a compilation of experimental and theoretical atomic and molecular data. As with the analysis strategy, the line list will keep evolving, being updated and improved before new analysis cycles. The details of this compilation, and the full line list will be given in a separated publication (Heiter et al. 2014b, in prep.).

Version 4.0 of the line list was used to analyze the iDR2 data. The list of molecules includes: C$_2$ ($^{12,13}$C$^{12,13}$C), CaH, $^{12,13}$CH, $^{12,13}$CN, FeH, MgH, NH, OH, SiH, $^{46,47,48,49,50}$TiO, VO, and $^{90,91,92,94,96}$ZrO. Atomic transitions needed for both spectrum synthesis and equivalent width analysis are included. Where needed, isotopic shifts and hyperfine structure (HFS) were included (for \ion{Sc}{i}, \ion{V}{i}, \ion{Mn}{i}, \ion{Co}{i}, \ion{Cu}{i}, \ion{Ba}{ii}, \ion{Eu}{ii}, \ion{La}{ii}, \ion{Pr}{ii}, \ion{Nd}{ii}, \ion{Sm}{ii}). Some atomic oscillator strengths have been newly calculated for the Survey \citepads{2014MNRAS.441.3127R}. Collisional broadening by hydrogen is considered following the theory developed by \citetads{1991MNRAS.253..549A} and \citetads{1998MNRAS.300..863B}, where available, including some new broadening computations still unpublished that will be discussed in Heiter et al. (2014b, in prep.).

The lines used for the EWs analyses have been critically reviewed by the line-list group. A system of flags has been designed and is made available together with the line list (also to be published). The flags indicate the quality of the transition probability and the blending properties of the line, as evaluated in the spectra of the Sun and of \object{Arcturus}.

It is perhaps necessary to stress here that while all Nodes have access to the same Gaia-ESO ``master'' line list, this does not mean that the methods make use of the same selected subsample of spectral lines. The choice of lines used to constrain the parameters and abundances is made by each Node according to the details of their methodology. As is common, some groups using EWs will prefer to select a restricted set of the best lines, while others will prefer to rely on the statistical properties of many lines. Other groups prefer to use strong lines such as H$\alpha$ to assist in constraining the parameters. In addition to that, there are the methods that rely on fitting large portions of the observed spectra in comparison with synthetic ones. These methods need more extensive line lists, not only the ones useful for an EW analysis. This is to emphasize that, even though a common line list is adopted, there is still considerable freedom as to how this line list is finally employed by each Node.

\subsection{Model atmosphere}

For model atmospheres we adopted the MARCS grid of \citetads{2008A&A...486..951G}. The grid consists of spherically-symmetric models complemented by plane-parallel models for stars of high surface gravity (between $\log g$ = 3.0 and 5.0, or 5.5 for the cooler models). It assumes hydrostatic equilibrium, local thermodynamic equilibrium (LTE), and uses the mixing-length theory of convection. The MARCS models assume the solar abundances of \citetads{2007SSRv..130..105G} and are $\alpha$-enhanced at low metallicities. 

We remark here that the coverage of the MARCS grid in the metal-poor regime is sometimes incomplete. Some of the analysis methods need to be able to interpolate among a grid of models on the fly. For metal-poor stars, it often happens that some of the models needed for this interpolation are not available. These methods will then fail when the border of the grid is reached. This aspect introduces one additional complication to the analysis of metal-poor stars.

Within Gaia-ESO we decided to list the abundances in the ``$\log \epsilon$'' format\footnote{$\log \epsilon$(X) = $\log$ [N(X)/N(H)] + 12, i.e. a logarithmic abundance by number on a scale where the number of hydrogen atoms is 10$^{12}$.}, without assuming a solar composition. Nevertheless, when metallicities as an atmospheric parameter in the format [Fe/H]\footnote{[A/B] = $\log$ [N(A)/N(B)]$_{\rm \star}$ $-$ $\log$ [N(A)/N(B)]$_{\rm\odot}$} are quoted in this work, we adopt the solar Fe abundance of \citetads{2007SSRv..130..105G}, $\log \epsilon$(Fe)$_{\odot}$ = 7.45, unless otherwise noted.

\subsection{Spectrum library}\label{sec:grid}

For analysis methodologies that make use of pre-computed synthetic spectra, a Gaia-ESO library of synthetic spectra is provided. Here we provide only a short description of the library, a complete discussion is given in Recio-Blanco et al. (2014, in prep.). 

The synthetic spectra were calculated using the same software used to compute the AMBRE grid of synthetic spectra \citepads{2012A&A...544A.126D}. The spectra have $R$ $\sim$ 300\,000 and cover the whole wavelength region of the UVES setup with a sampling of 0.004\,\AA. Spectra with different degrees of alpha-enhancement were computed, to account for the feedback of important $\alpha$-element electron donors on the atmospheric structure. This grid was computed using the complete Gaia-ESO line list (atoms + molecules). With each update of the line list, a new grid is computed.

\begin{figure*}[t]
\centering
\includegraphics[height = 4.5cm]{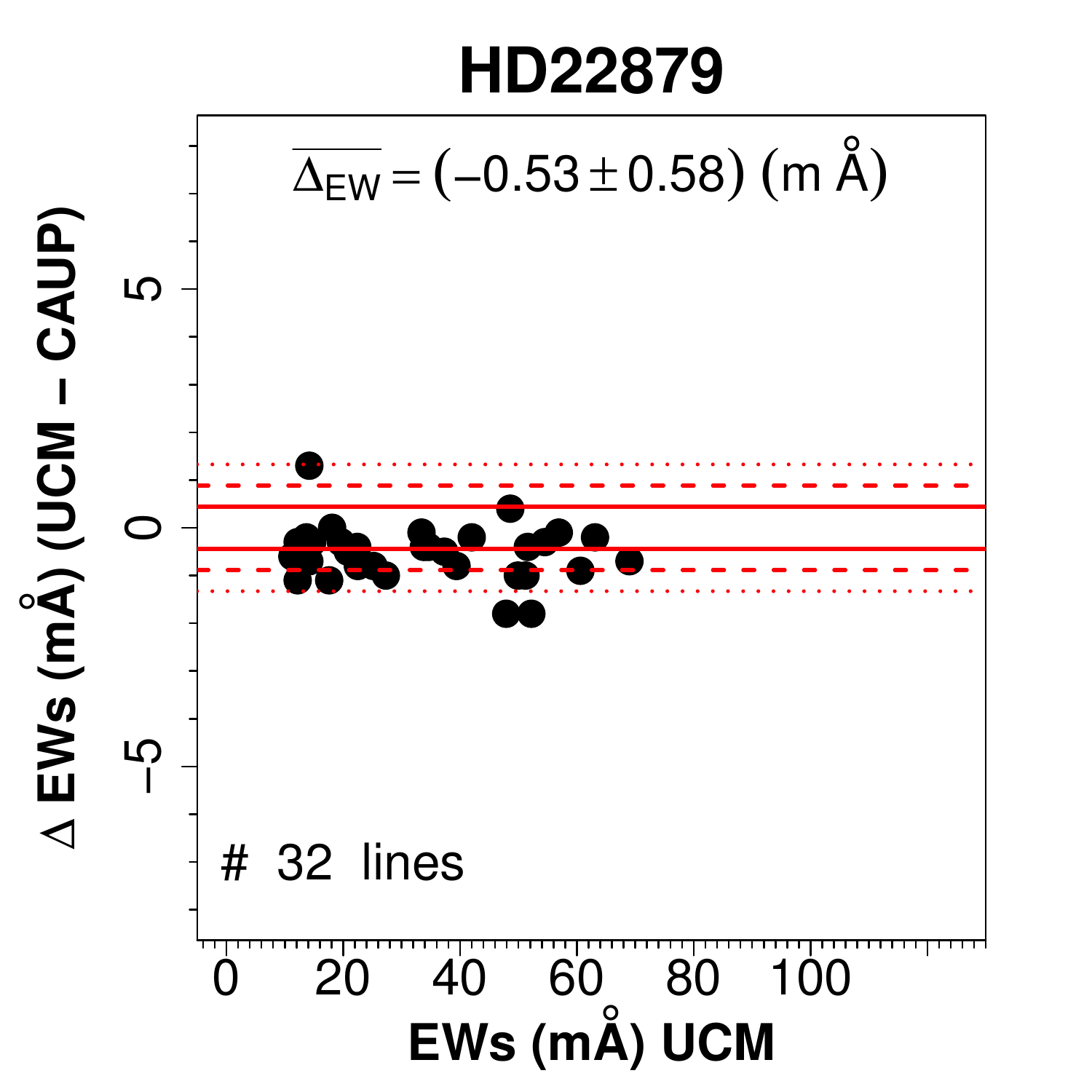}
\includegraphics[height = 4.5cm]{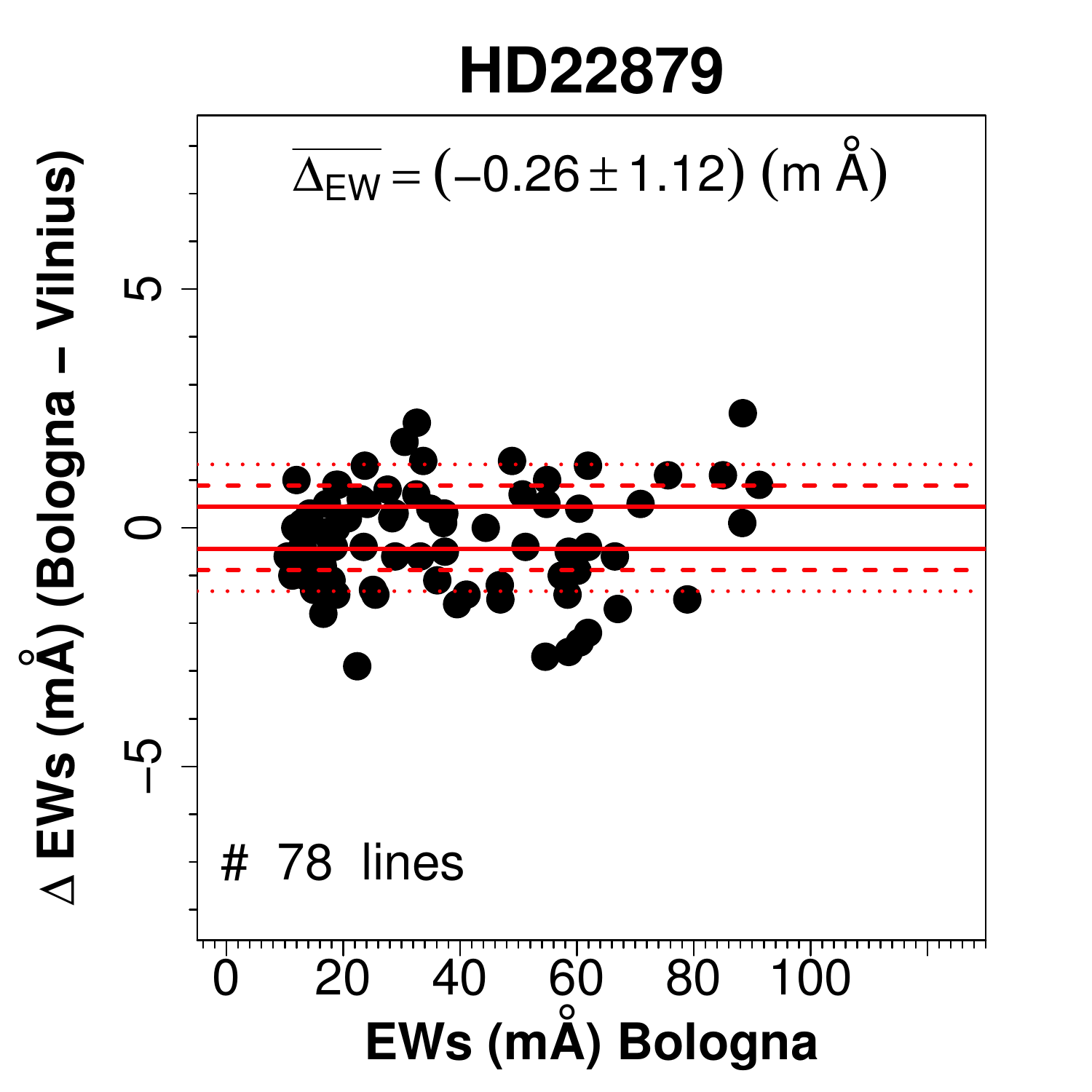}
\includegraphics[height = 4.5cm]{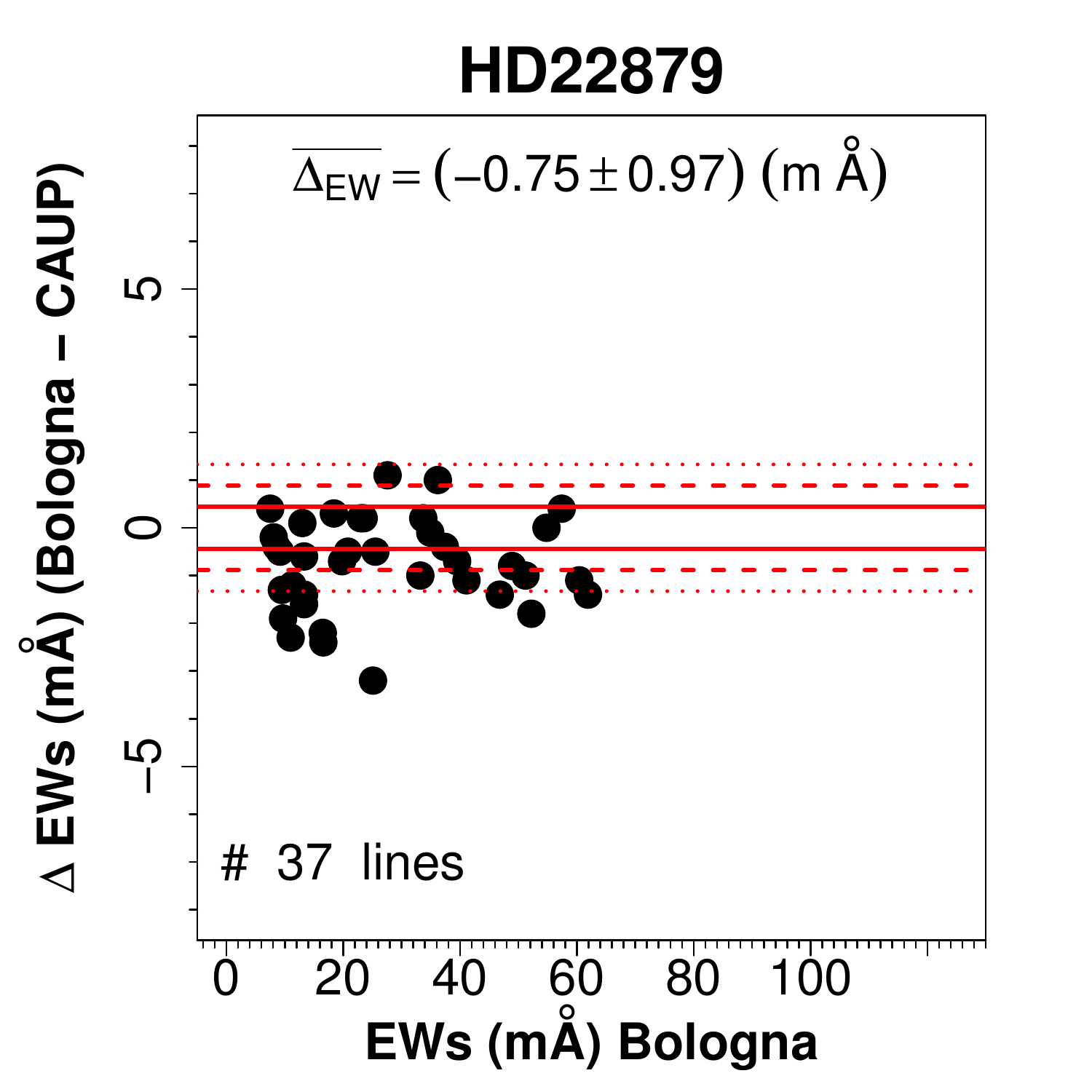}
\includegraphics[height = 4.5cm]{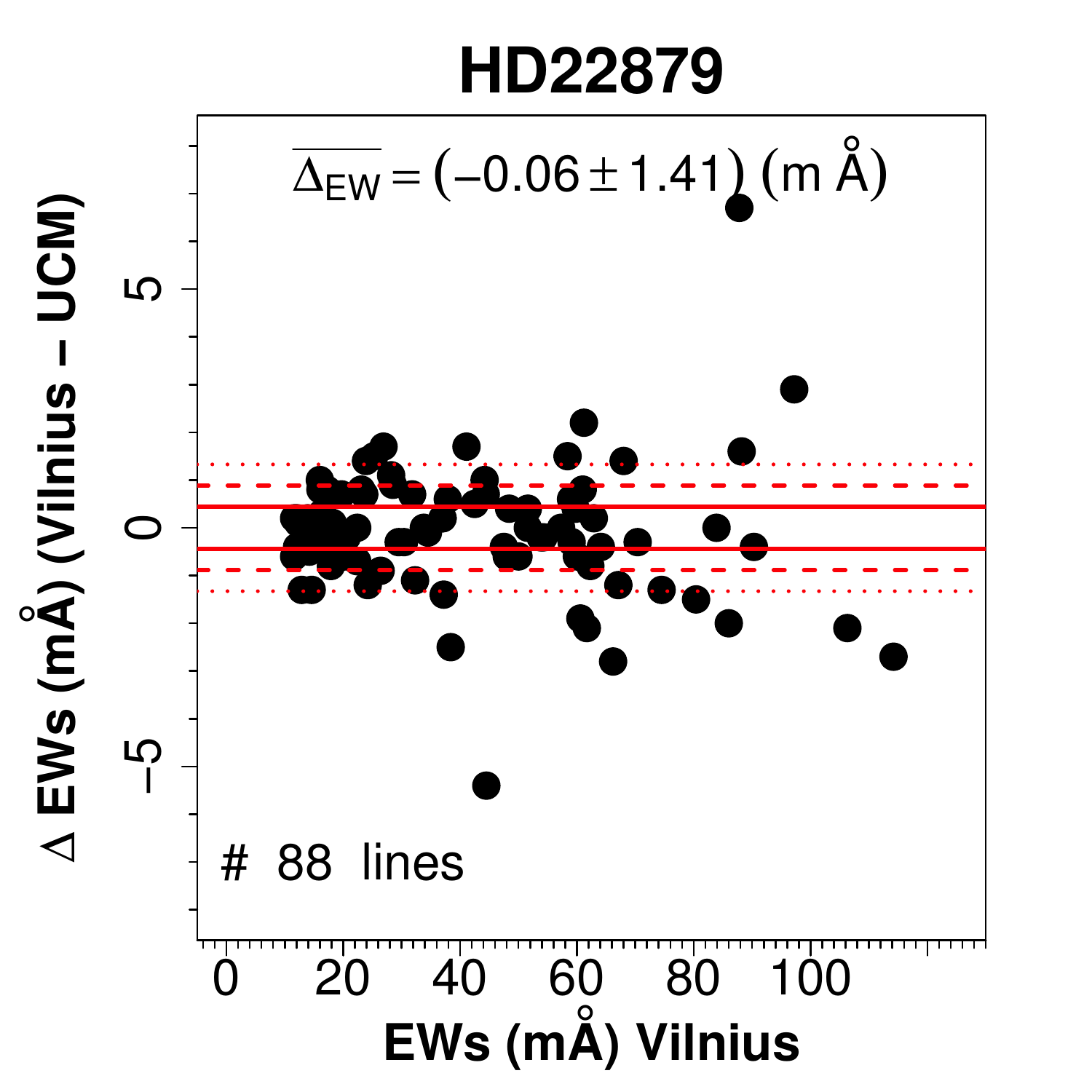}
\includegraphics[height = 4.5cm]{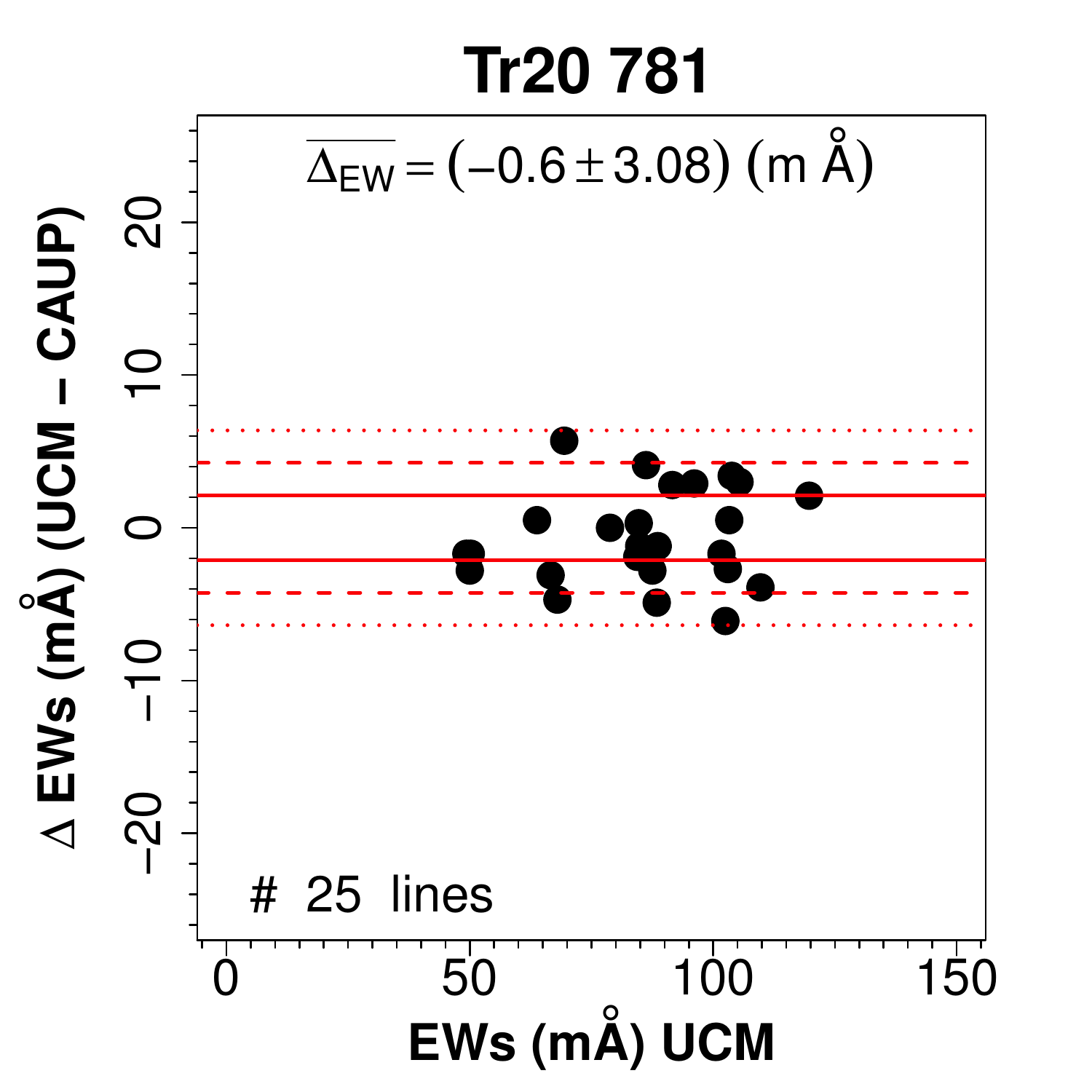}
\includegraphics[height = 4.5cm]{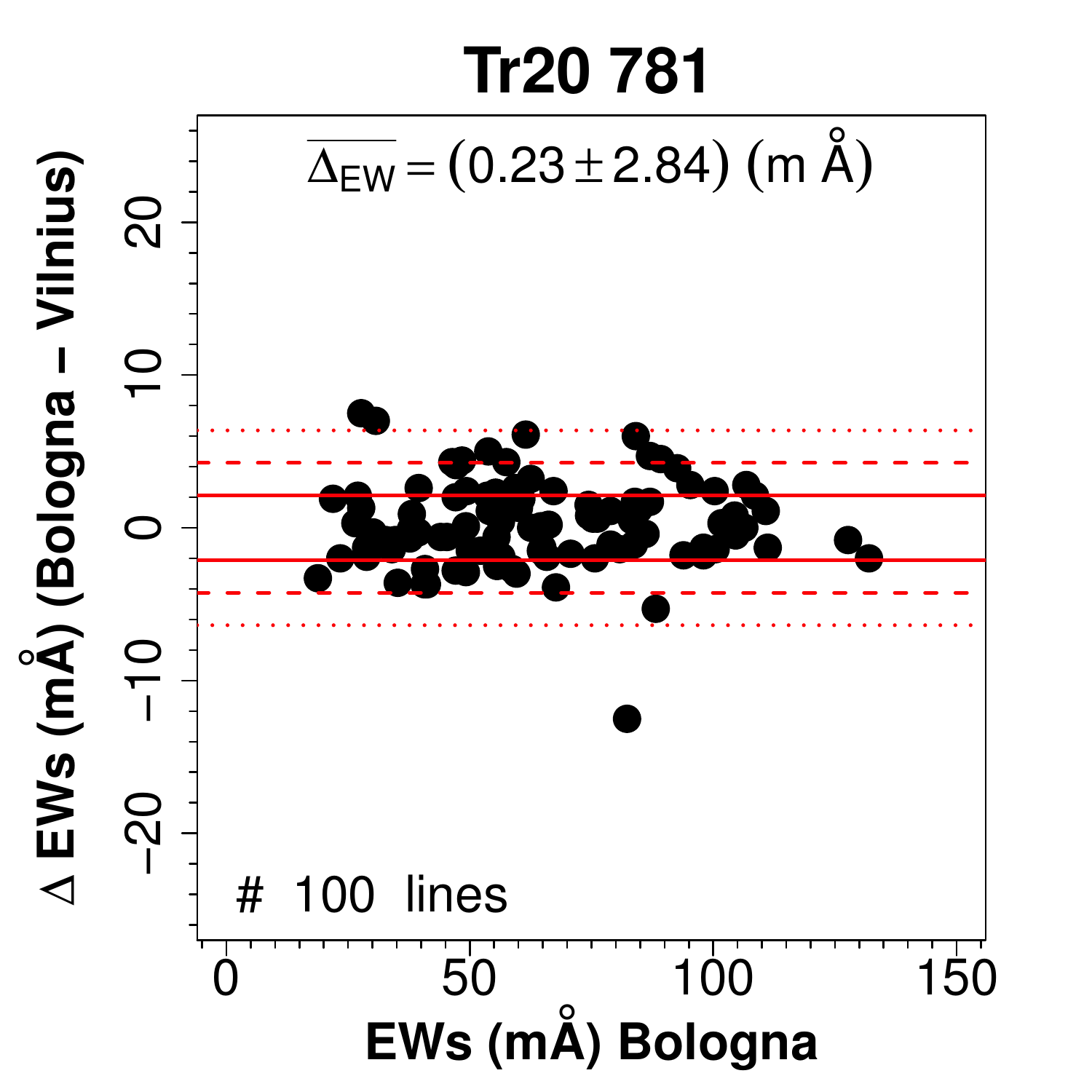}
\includegraphics[height = 4.5cm]{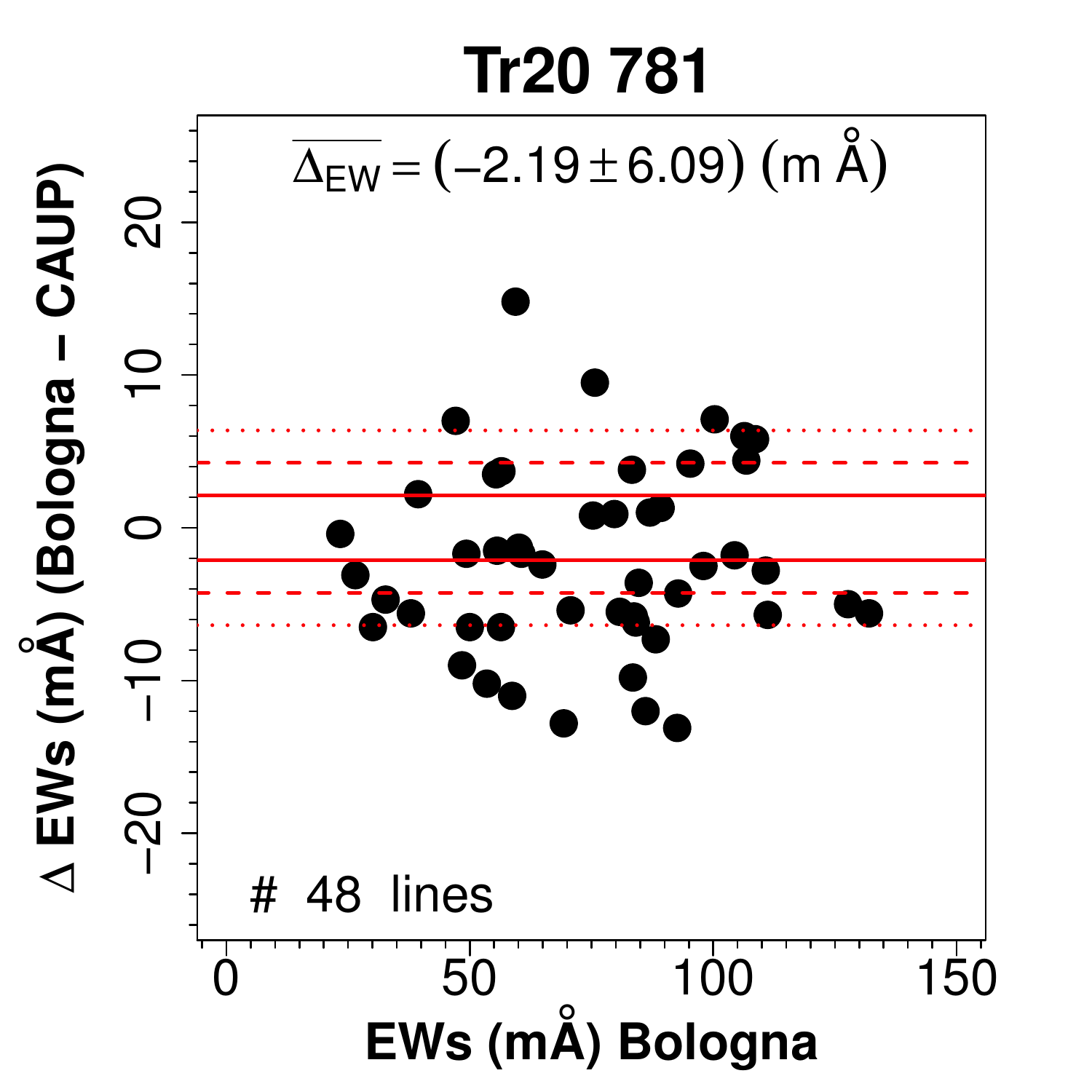}
\includegraphics[height = 4.5cm]{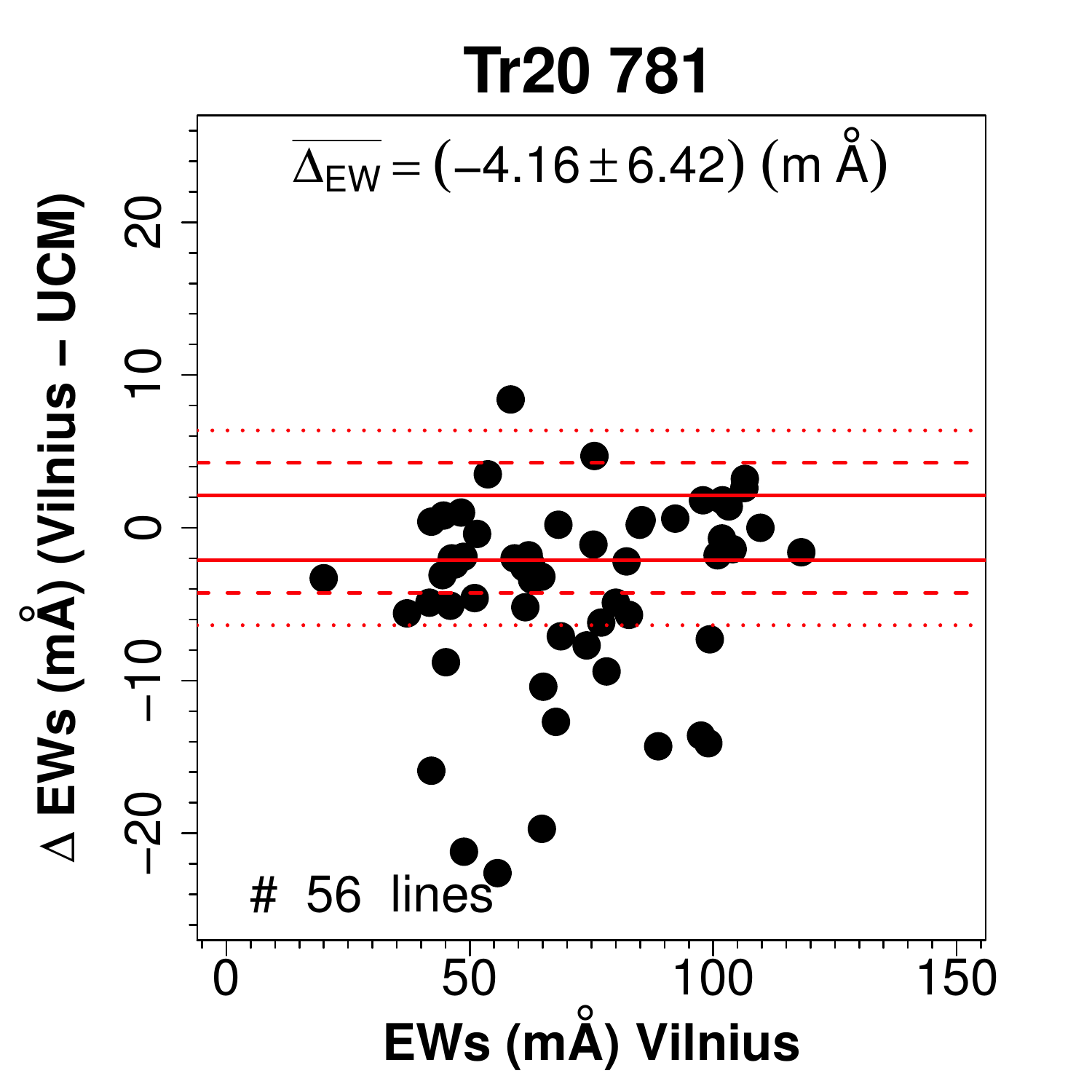}
 \caption{Comparison between equivalent widths measured by different Nodes for two stars. \emph{Top row:} Star \object{HD 22879}, a benchmark star used for calibration with $T_{\rm eff}$ = 5786 K, $\log g$ =  4.23, and [Fe/H] = $-$0.90. The median values of the S/N per pixel are 239 and 283 for the blue and red part of the spectra, respectively. The red lines indicate the typical 1$\sigma$ (solid line), 2$\sigma$ (dashed line), and 3$\sigma$ (dotted line) uncertainty of the EW computed with the \citetads[][]{1988IAUS..132..345C} formula, adopting FWHM = 0.190 \AA, pixel size = 0.0232 \AA, and S/N = 260.  \emph{Bottom row:} A clump giant in the open cluster \object{Trumpler 20} \citepads[\object{Trumpler 20 MG 781} in the numbering system of][]{2005ApJS..161..118M}, with $T_{\rm eff}$ = 4850 K, $\log g$ = 2.75, and [Fe/H] = +0.15. The median values of the S/N per pixel are 36 and 68 for the blue and red part of the spectra, respectively. The red lines indicate the typical 1$\sigma$ (solid line), 2$\sigma$ (dashed line), and 3$\sigma$ (dotted line) uncertainty of the EW computed with the \citetads[][]{1988IAUS..132..345C} formula, adopting FWHM = 0.190 \AA, pixel size = 0.0232 \AA, and S/N = 50. In each panel, the average difference of the EWs and its dispersion are also given.}\label{fig:ewcompdr2}
\end{figure*}

\subsection{Microturbulence calibration}

A Gaia-ESO microturbulence calibration is provided and recommended for those methods that do not derive this parameter from the spectrum analysis. It is used by a few Nodes in the analysis of the UVES spectra, but is more extensively used in the analysis of Giraffe spectra (Recio-Blanco et al. 2014, in prep.), because of the reduced number of clean \ion{Fe}{i} lines available for constraining this parameter.

These relations are based on the UVES science verification results (obtained as described in Appendix \ref{sec:idr1}), on the parameters of the Gaia benchmark stars described in \citetads{2014A&A...564A.133J}, and on globular cluster data from literature sources. Three relations were derived, for different types of stars, and are valid for 4000 $<$ $T_{\rm eff}$ (K) $<$ 7000, 0.0 $<$ $\log~g$ (dex) $<$ 5.0, and $-$4.5 $<$ [Fe/H] (dex) $<$ +1.0. A full discussion of these relations will be presented in Bergemann et al. (2014b, in prep.).

\section{Equivalent widths}\label{sec:eqw}

\begin{figure*}
\centering
\includegraphics[height = 6.5cm]{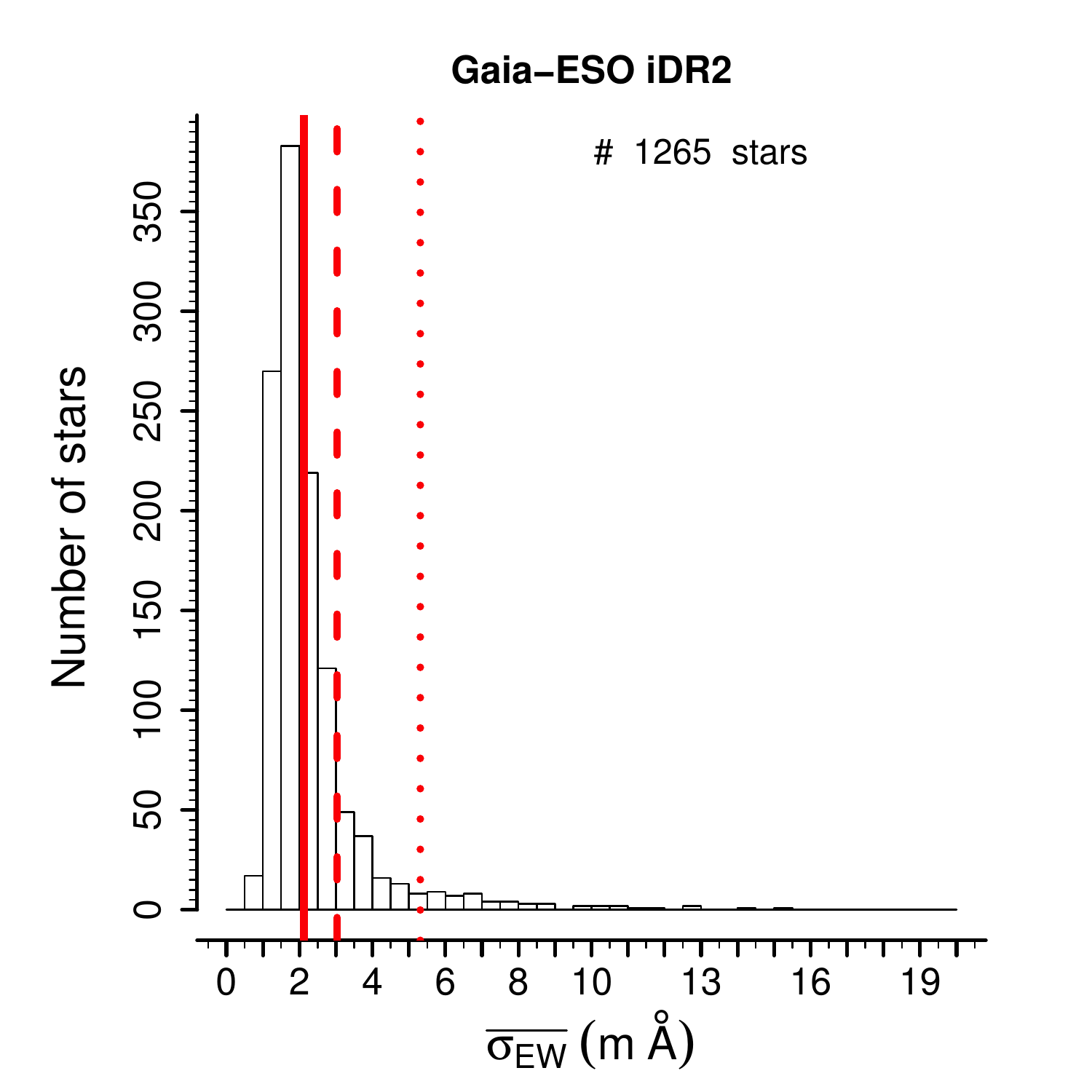}
\includegraphics[height = 6.5cm]{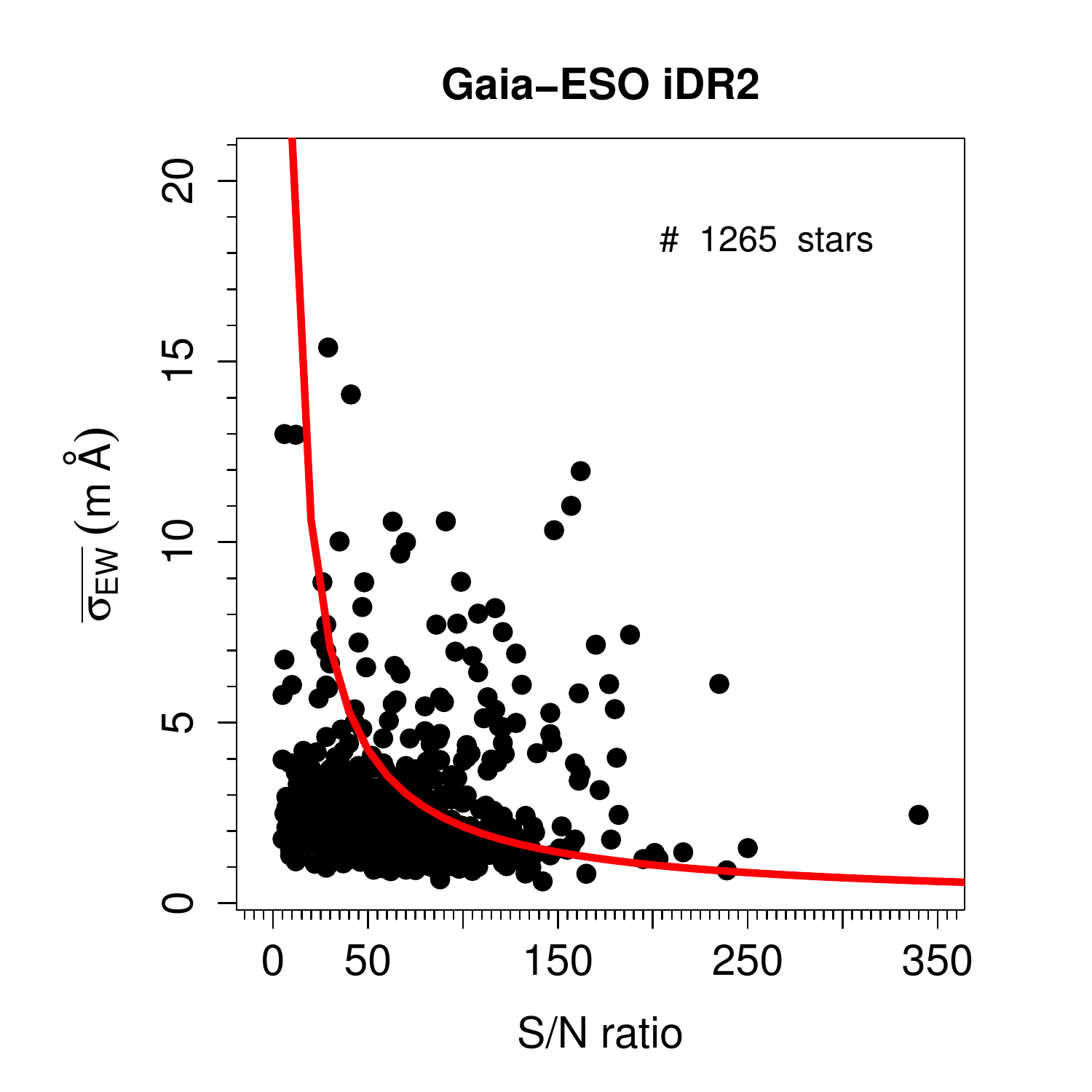}
 \caption{\emph{Left panel:} Histogram of $\overline{\sigma_{\rm EW}}$ per star, taking into account the measurements of all Nodes. Also shown are lines indicating the 2$\sigma$ uncertainty calculated with \citetads[][]{1988IAUS..132..345C} formula for S/N = 40 (dotted line at 5.31 m\AA), S/N = 70 (dashed line at 3.04 m\AA), and S/N = 100 (solid line at 2.12 m\AA). \emph{Right panel:} Dependence of $\overline{\sigma_{\rm EW}}$ with respect to the median of the S/N per pixel. Also shown is the expected 2$\sigma$ value given by the \citetads[][]{1988IAUS..132..345C} formula (as a red line).}\label{fig:ewfullcompdr2}%
\end{figure*}

Some of the analysis methodologies described in Appendix \ref{sec:nodes} rely on the measurement of EWs to determine both the stellar atmospheric parameters and elemental abundances of the stars. The measurements of these EWs are going to be released as part of the Gaia-ESO data products. \emph{Equivalent widths will be given only for lines effectively used by at least one Node in their analysis.} The tables that will be released to the community will include for each line: the average EW, the multiple-measurement dispersion, number of measurements, and flags (where applicable).

\begin{figure*}
\centering
\includegraphics[height = 7cm]{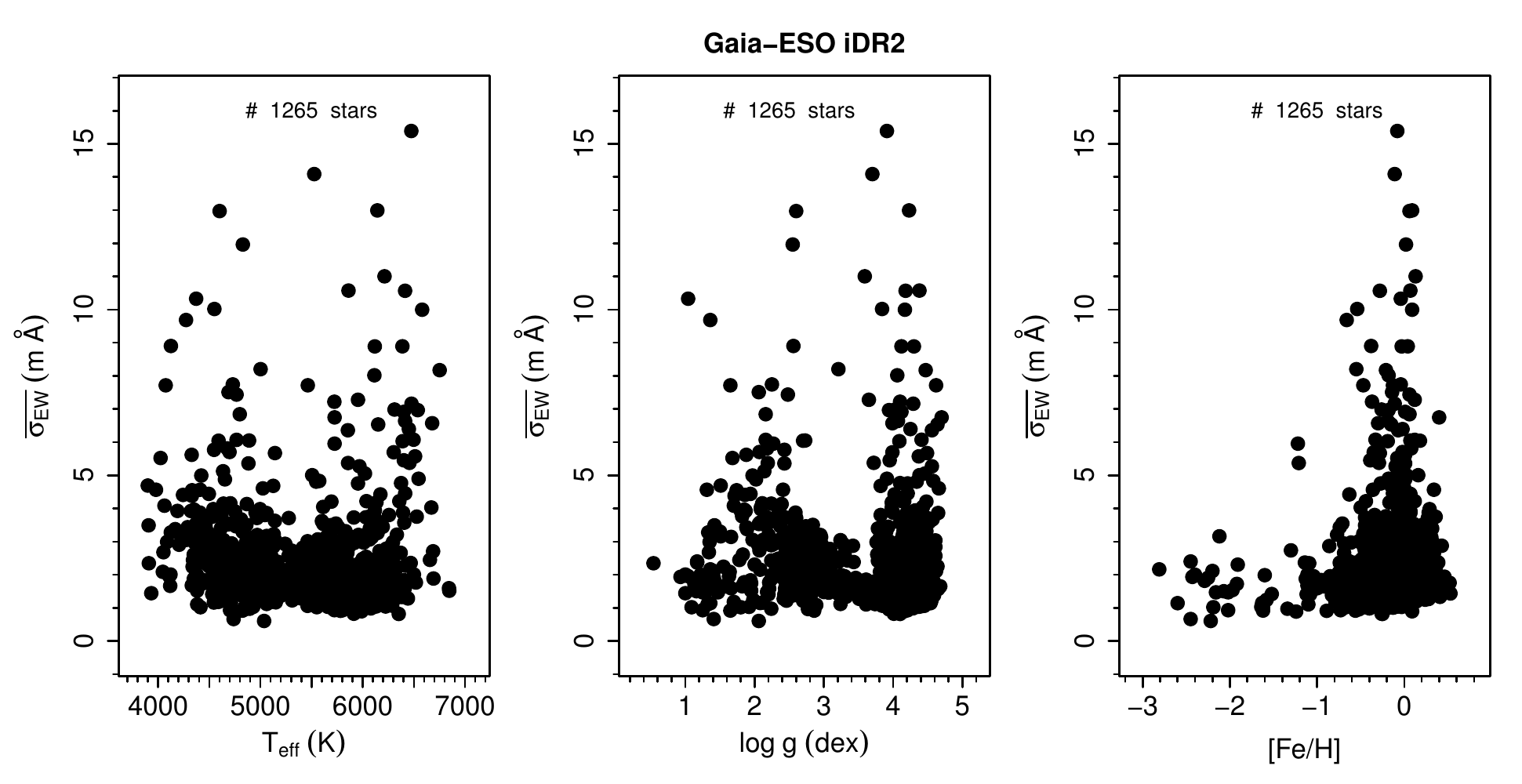}
 \caption{Mean of all the standard deviations of the EW measurements in a star, $\overline{\sigma_{\rm EW}}$, as a function of the atmospheric parameters.}\label{fig:ewparamdr2}%
\end{figure*}

In this Section, we discuss a comparison of multiple measurements of the EWs in the spectra that are part of the iDR2 data set. The EWs are available for 1265 stars, out of the 1268 observed by the Gaia-ESO Survey for which atmospheric parameters were determined.

The Nodes making use of the traditional EW-based analysis method are: Bologna, CAUP, Concepcion, EPInArBo, Li\`ege, UCM, and Vilnius. The Li\`ege Node measures EWs with the {\sf GAUFRE} code. However, their measurements were lost because of a computer problem. Therefore, the discussion in this section concentrates on the results from only two codes that measure EWs automatically: {\sf ARES} \citepads{2007A&A...469..783S} and {\sf DAOSPEC} \citepads{2008PASP..120.1332S,2010ascl.soft11002S}. Currently, only {\sf DAOSPEC} returns a value for the EW measurement error.

Figure \ref{fig:ewcompdr2} shows the comparison between the EWs of lines of different elements measured by different groups in two stars. They represent an easy and a hard case for measuring EWs. One star is a metal-poor dwarf observed with high S/N per pixel ($\sim$ 260), the other a metal-rich giant observed with relatively low S/N per pixel ($\sim$ 50). In this and other plots in this section, we compare the multiple measurement scatter with the statistical uncertainty in the EW measurement given by the \citetads[][]{1988IAUS..132..345C} ``formula'' (equation 7 of that article). This formula gives the EW uncertainty due to random noise when fitting the line profile with a Gaussian. This value is used here as a reference for the ``expected uncertainty'' but does not take into account all possible sources of error such as, for example, continuum placement.

In Fig.\ \ref{fig:ewcompdr2}, the EWs measured with the same code by different Nodes (left plots using {\sf ARES} and center-left plots using {\sf DAOSPEC}) tend to agree to within 2 or 3$\sigma$, although systematic differences might be present in some cases. When comparing the EWs measured with {\sf ARES} (CAUP and UCM Nodes) with those measured by {\sf DAOSPEC} (Bologna and Vilnius Nodes) -- center-right and right plots of Fig. \ref{fig:ewcompdr2} -- it is noticeable that the scatter increases. There seems to be no trend between the $\Delta$ EWs and the EWs themselves. Such trends could produce biases in the determination of the microturbulence.

Figure \ref{fig:ewfullcompdr2} depicts the behavior of $\overline{\sigma_{\rm EW}}$. For each spectral line used for abundance determination in a given star, the average value of the multiple determinations of its EW is computed, together with its standard deviation. For each star, we define $\overline{\sigma_{\rm EW}}$ as the mean of all the standard deviations of the lines measured in that star. Figure \ref{fig:ewfullcompdr2} shows that for the majority of the stars, the measurements tend to agree to a level that is better than the expected statistical uncertainty given by the S/N of the spectra. About 70\% of the stars have the blue part of the spectrum with median S/N per pixel below 70. For this S/N the expected 2$\sigma$ uncertainty of the EWs is of the order of 3 m\AA. As shown in the left panel of Fig. \ref{fig:ewfullcompdr2}, about $\sim$ 13.7\% of the stars have $\overline{\sigma_{\rm EW}}$ above that. In a few cases, however, it can reach up to $\sim$ 15 m\AA. A more detailed comparison with the S/N expectation -- right panel of the same figure -- shows that for about 11.7\% of the stars the quantity $\overline{\sigma_{\rm EW}}$ is above the 2$\sigma$ expectation.

The cases with higher dispersion might be related to different issues that make the measurement of EWs difficult (e.g. low temperature, high-metallicity, and/or broad lines). Other problems contributing to increase the scatter in the measurements include the different ways that the continuum is defined in each code (global vs. local continuum for {\sf DAOSPEC} and {\sf ARES} respectively), the presence of reduction artefacts, unrecognized binarity in the spectra, the residual wavelike pattern in the continuum, caused by problems with the blaze-function correction, as sometimes seen in high S/N echelle spectra\footnote{We note in particular that HD 22879, which is used as an example in Fig. \ref{fig:ewcompdr2} suffers with this issue. This will perhaps affect more seriously DAOSPEC than ARES, as DAOSPEC performs a global fit of the continuum for the whole spectrum. Therefore, the expected uncertainty computed with the \citetads[][]{1988IAUS..132..345C} formula for the EW measurements should be taken as a lower limit.}, and the free parameters in each code that need to be adjusted for the measurements. Therefore, the scatter in the measurement of EWs is not just statistical in nature.

\begin{figure}
\centering
\includegraphics[height = 7cm]{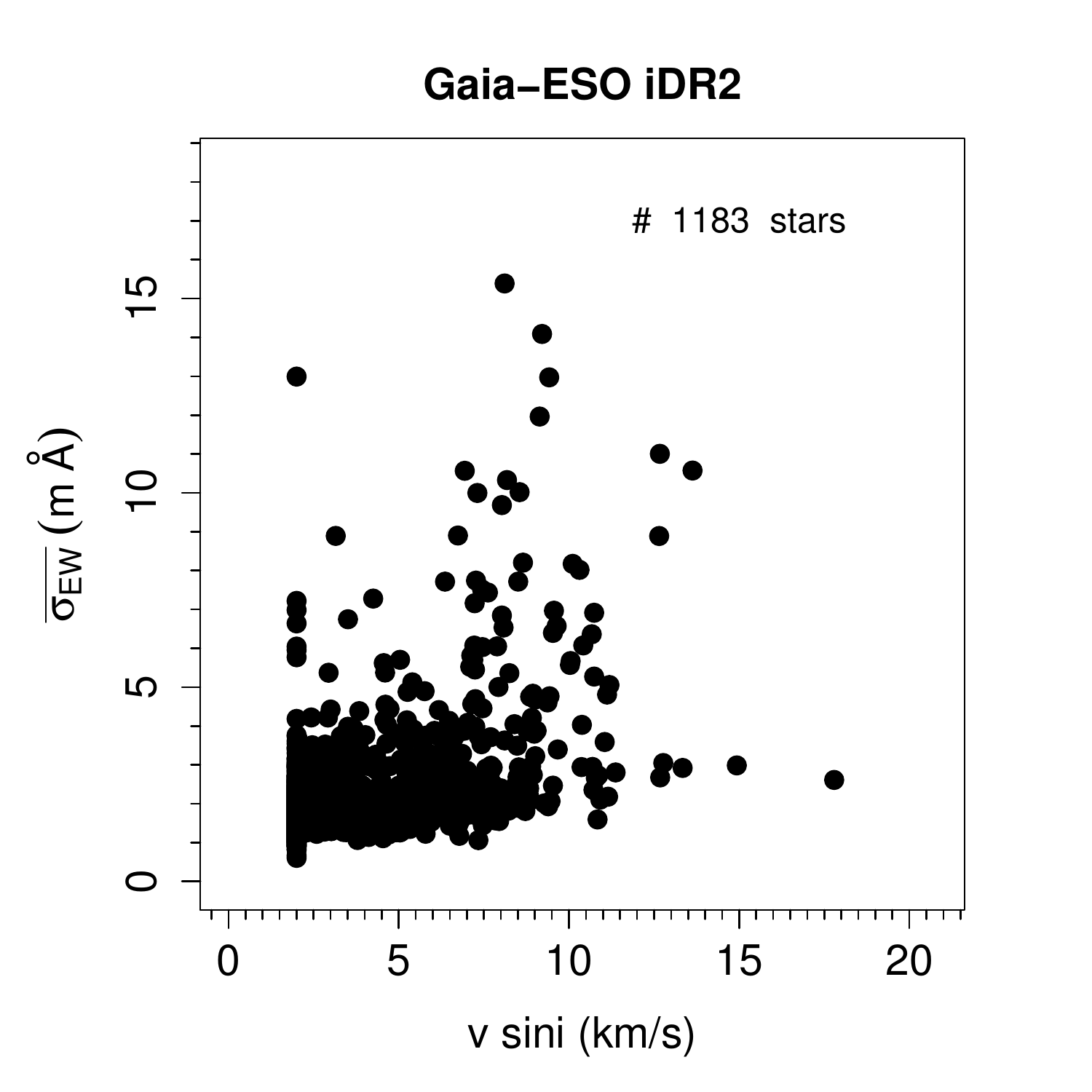}
 \caption{Mean of all the standard deviations of the EW measurements in a star, $\overline{\sigma_{\rm EW}}$, as a function of the rotational velocity of the star.}\label{fig:ewsvsinidr2}%
\end{figure}

In Fig. \ref{fig:ewparamdr2}, $\overline{\sigma_{\rm EW}}$ is plotted against the atmospheric parameters of the stars. In Fig. \ref{fig:ewsvsinidr2}, we plot $\overline{\sigma_{\rm EW}}$ against the rotational velocity ($v \sin i$) of the stars. Not all stars have an estimate of $v \sin i$, because this measurement fails in some cases (see \citeauthor{2014A&A...565A.113S} \citeyear{2014A&A...565A.113S}). The figures show that most of the stars where $\overline{\sigma_{\rm EW}}$ $>$ 5 m\AA\ tend to be metal-rich objects, some are cool, and many display high rotation. All these factors increase the uncertainty with which EWs can be measured with automatic methods.

It is not the scope of this section to delve into the details of why a perfect agreement between multiple measurements of the same line is not obtained. Both {\sf ARES} and {\sf DAOSPEC} are fully described in dedicated publications, which include comparisons between each other, and between them and other codes. We thus refer the reader to \citetads{2007A&A...469..783S}, \citetads{2008PASP..120.1332S}, and \citetads{2014A&A...562A..10C} for these detailed discussions.

It is the goal of this section to document how the measurements have been done and to discuss the quality of the results and their limitations. For the majority of the stars, the scatter in the multiple measurements compares well with the statistical uncertainty estimated with the \citetads[][]{1988IAUS..132..345C} formula. Thus, the EW  measurements for these stars do not seem to be affected by additional sources of error. For the remaining stars, multiple factors play a role, some of which were identified above. Any detected abnormality in the spectra is flagged and the information will be part of the final catalog. 

We have, however, identified the general regions of the parameter space where problems are likely to occur. We are working to improve the analysis of these stars, and expect to provide improved results for future releases. A satisfactory agreement between the multiple measurements of EWs is obtained for about 88\% of the stars discussed here. For the ones with higher scatter in the EWs, we expect also a large scatter in the comparison of the atmospheric parameters and abundances. We remind however, that not all analysis methodologies make use of EWs. As we discuss in the following sections, all values of atmospheric parameters and abundances are given together with an estimate of the method-to-method dispersion. This is a measurement of the precision of these values. Therefore values with an increased dispersion are more uncertain and should not be given the same weight as more precise results.

\begin{table*}
\caption{Reference parameters of the benchmark stars.}
\label{tab:bench}  
\centering
\begin{tabular}{lcccccccl}
\hline\hline
Star & $T_{\rm eff}$ & $\sigma_{T_{\rm eff}}$ & $\log g$ & $\sigma_{{\rm log} g}$ & [Fe/H] & $\sigma_{\rm [Fe/H]}$ & Parameter & Remark \\
        &   (K)    & (K) & (dex) & (dex) & (dex) & (dex) & Space & \\
\hline
Sun         & 5777 &  1  & 4.44 & 0.01  &  0.00 & 0.01  & MRD & Used in iDR2; only archival data   \\
Arcturus    & 4247 & 28  & 1.59 & 0.04  & -0.53 & 0.01  & MRG & Only archival data    \\
Procyon     & 6545 & 82  & 3.99 &  0.02 & -0.04 & 0.01  & MRD & Problems with order merging                      \\
18 Sco      & 5747 & 29  & 4.43 & 0.01  &  0.01 & 0.01  & MRD & Used in iDR2                      \\
61 Cyg A    & 4339 & 22  & 4.43 & 0.16  & -0.33 & 0.03  & MRD & Only archival data    \\
61 Cyg B    & 4045 & 20  & 4.53 & 0.04  & -0.38 & 0.02  & MRD & Only archival data    \\
\object{Alf Cen A}    & 5847 & 68  & 4.31 & 0.02  &  0.24 & 0.01  & MRD & Used in iDR2                       \\
Alf Cet     & 3796 & 64  & 0.91 & 0.08  & -0.45 & 0.05  & MRG & Cool star; only archival data    \\
Alf Tau     & 3927 & 39  & 1.22 & 0.10  & -0.37 & 0.02  & MRG & Cool star; only archival data    \\
Bet Ara     & 4172 & 48  & 1.01 & 0.13  & -0.05 & 0.04  & MRG  &Used in iDR2                      \\
Bet Gem     & 4858 & 55  & 2.88 & 0.05  &  0.12 & 0.01  &  MRG & Only archival data    \\
Bet Hyi     & 5873 & 38  & 3.98 & 0.02  & -0.07 & 0.01  & MRD & Used in iDR2                      \\
Bet Vir     & 6083 & 17  & 4.08 & 0.01  &  0.21 & 0.01  & MRD & Used in iDR2; problems with order merging                      \\
Del Eri     & 5045 & 59  & 3.77 & 0.02  &  0.06 & 0.01  & MRD & Used in iDR2                      \\
Eps Eri     & 5050 & 25  & 4.60 & 0.03  & -0.10 & 0.01  &  MRD & Only archival data    \\
Eps For     & 5069 & 59  & 3.45 & 0.05  & -0.62 & 0.01  & MRD & Used in iDR2                      \\
Eps Vir     & 4983 & 56  & 2.77 & 0.01  &  0.13 & 0.01  & MRG & Only archival data    \\
Eta Boo     & 6105 & 19  & 3.80 & 0.02  &  0.30 & 0.01  &  MRD & Used in iDR2; $v \sin i$ $\simeq$ 12.7 km s$^{-1}$                    \\
Gam Sge     & 3807 & 48  & 1.05 & 0.10  & -0.16 & 0.04  & MRG & Used in iDR2; cool star                     \\
Ksi Hya     & 5044 & 33  & 2.87 & 0.01  &  0.14 & 0.01  & MRG  & Used in iDR2                      \\
Mu Ara      & 5845 & 29  & 4.27 & 0.02  &  0.33 & 0.01  &  MRD & Used in iDR2                      \\
Mu Leo      & 4474 & 52  & 2.50 & 0.07  &  0.26 & 0.02  &  MRG & Used in iDR2                      \\
Tau Cet     & 5331 & 15  & 4.44 & 0.02  & -0.50 & 0.01  &  MRD  & Used in iDR2                     \\
HD 22879     & 5786 & 16  & 4.23 & 0.02  & -0.88 & 0.01  &  MRD & Used in iDR2                     \\
HD 49933     & 6635 & 18  & 4.21 & 0.03  & -0.46 & 0.01  &  MRD & Used in iDR2; $v \sin i$ $\simeq$ 10.0 km s$^{-1}$                    \\
HD 84937     & 6275 & 17  & 4.11 & 0.06  & -2.09 & 0.02  & MPS & Used in iDR2; metal-poor star; only archival data    \\
HD 107328    & 4496 & 53  & 2.11 & 0.07  & -0.34 & 0.01  &  MRG & Used in iDR2                      \\
HD 122563    & 4587 & 54  & 1.61 & 0.07  & -2.74 & 0.01  & MPS  & Used in iDR2 ; metal-poor star                     \\
HD 140283    & 5720 & 29  & 3.67 &  0.04 & -2.43 & 0.02  & MPS  & Used in iDR2 ; metal-poor star                     \\
HD 220009    & 4275 & 50  & 1.43 & 0.10  & -0.75 & 0.01  & MRG & Used in iDR2                       \\
\hline
\end{tabular}
\tablefoot{$T_{\rm eff}$ and $\log g$ are direct determinations (see Heiter et al. 2014a, in prep). Metallicities were derived by \citetads[][]{2014A&A...564A.133J}. The metallicity uncertainty listed here only reflects the standard deviation of the mean abundance of the \ion{Fe}{i} lines. Also given is the parameter space group to which the star belongs (MRD, MRG, or MPS -- see text.)}
\end{table*}
%

\section{Atmospheric parameters}\label{sec:atm}

As presented in the Appendix \ref{sec:nodes}, the methods used to derive atmospheric parameters differ from Node to Node. They range from the standard use of EWs of Fe lines to different algorithms that use libraries of observed and/or synthetic spectra.

Once the different Nodes have finalized the first step of the spectroscopic analysis, we face the challenge of putting all the results together, understanding the differences and systematics, and producing a single list with the best, recommended values of the four atmospheric parameters ($T_{\rm eff}$, $\log g$, $\xi$, and [Fe/H]).

In the analysis of Gaia-ESO data, we aim to understand both the precision and accuracy with which the atmospheric parameters can be determined. The dispersion among the results from different methodologies is a good indication of the precision of the values. The accuracy is judged using the comprehensive set of calibrators observed by the Survey, in particular the Gaia benchmark stars and a set of calibration clusters were used. In addition to those, for subsequent releases we expect to use giants that have asteroseismic-estimated gravities, determined using CoRoT light curves, to help in the calibration effort \citepads[see e.g.][]{2012MNRAS.419L..34M}. 

In the subsections that follow below, we describe how the recommended atmospheric parameters for the iDR2 data set were determined. These results will be part of the first Gaia-ESO public release. The results used in the first few Gaia-ESO science verification papers were determined in a slightly different way, as presented in Appendix \ref{sec:idr1}. We start the discussion presenting the use of the main calibrators employed in the Gaia-ESO analysis.

\begin{table*}
\caption{\small{Average difference between the Node result for the Gaia benchmark stars and the reference values in each region of the parameter space.}}\label{tab:nodediff}  
\centering
\small{
\begin{tabular}{lccc|ccc|ccc}
\hline\hline
 & \multicolumn{3}{c}{MRD} & \multicolumn{3}{c}{MRG} & \multicolumn{3}{c}{MPS} \\
Node &  $\Delta(T_{\rm eff})$ &  $\Delta(\log~g)$ & Num.  &  $\Delta(T_{\rm eff})$ &  $\Delta(\log~g)$  & Num. &  $\Delta(T_{\rm eff})$ &  $\Delta(\log~g)$  & Num. \\
 & (K) & (dex) & of stars  &  (K) &  (dex)  & of stars & (K)  &  (dex) & of stars \\
\hline
Bologna           &  46 & 0.13 & 11 & 163 & 0.40 & 7  & --  &  --  &  0  \\
CAUP               & 93  & 0.21 &  8 &  193 & 0.42 &  4 & --  &  --  &  0  \\
Concepcion     & 150 & 0.28 & 8 & 162 & 0.48  & 5  & 87 & 1.11 & 1 \\ 
EPINARBO        &  57 & 0.14 & 10 &  74 & 0.31 & 7 & 167 & 0.35 & 1 \\
IACAIP             & 131 & 0.16 & 9 & 114 & 0.22 & 7 &  82  & 0.23  & 1 \\
Liege                & 186 & 0.22 & 8 & 208 & 0.62 & 7 & -- & -- & 0 \\
LUMBA             &  81 & 0.14 & 11 & 139 & 0.39 & 5 & 165 & 0.07 & 3 \\
Nice                 &  78 & 0.26 & 11 &  82 & 0.30 & 5 &  59 & 0.20 & 3 \\
OACT              & 169 & 0.19 & 10 & 159 & 0.37 & 7 &  -- & -- & 0 \\
ParisHeidelberg   &  71 & 0.12 & 10  &  91 & 0.34 & 5 &  87 & 0.43 & 1 \\
UCM               & 123 & 0.11 & 11 & 465 & 0.94 & 6 & --  &  --  &  0 \\
ULB               & -- & -- & -- & -- & -- & -- & -- & -- & -- \\
Vilnius           &  59 & 0.09 & 11 & 184 & 0.51 & 6  & 2 & 1.10 & 1 \\
\hline\hline
\end{tabular}
}
\end{table*}
\begin{figure*}
\centering
\includegraphics[height = 6cm]{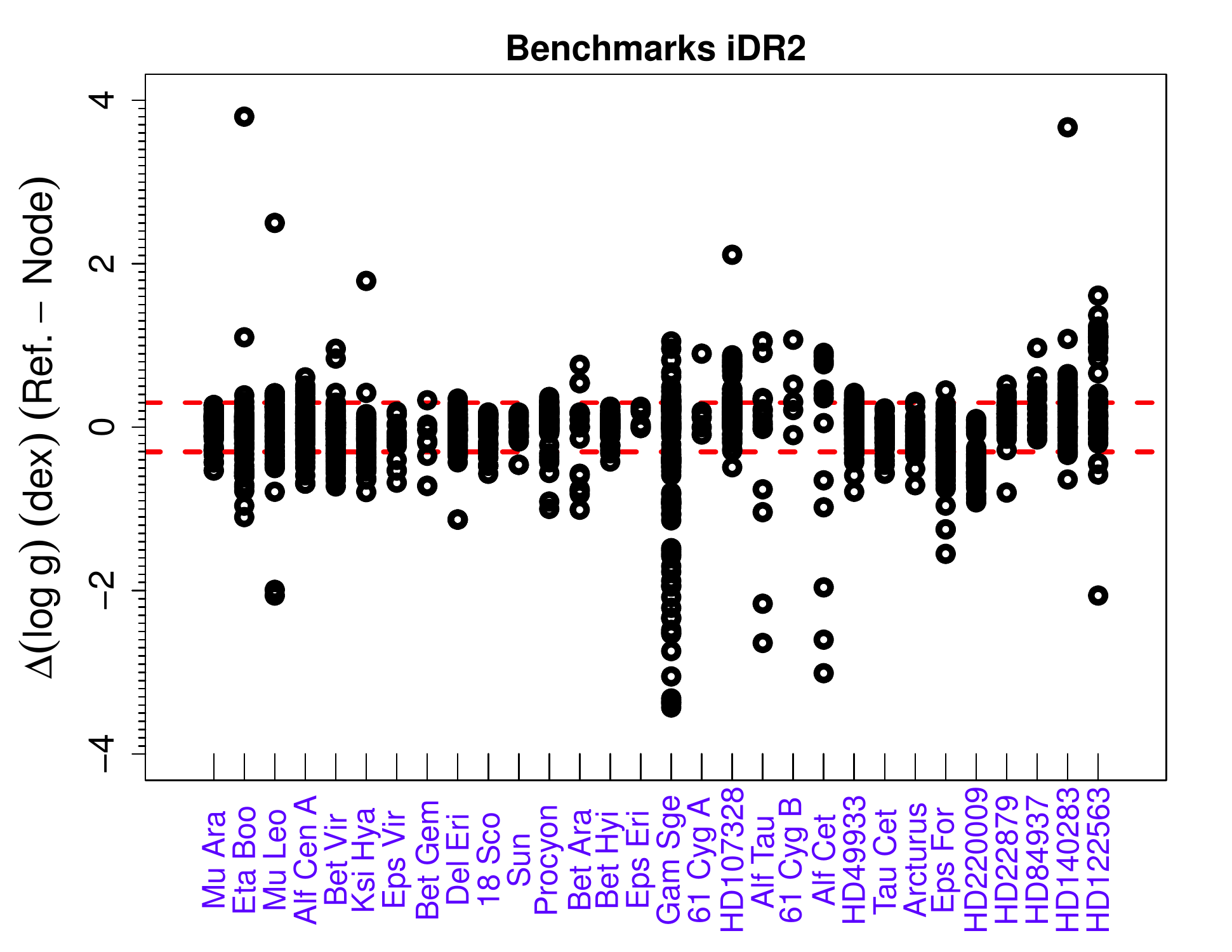}
\includegraphics[height = 6cm]{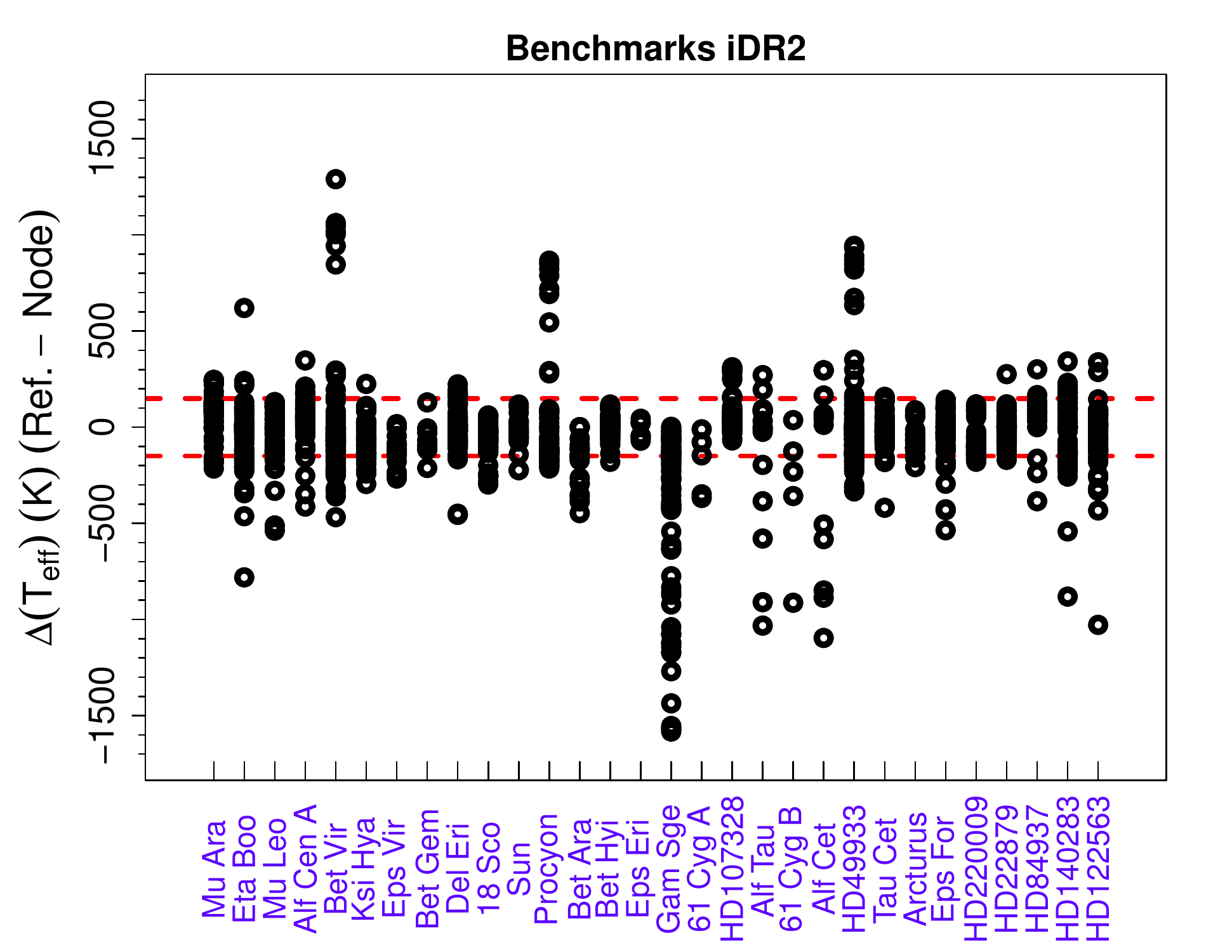}
\caption{All Node results for the 30 benchmark stars included in iDR2. The stars are ordered by decreasing metallicity from left to right. \emph{Left panel:} The difference between the reference and the Node value of $\log~g$. The red dashed lines indicate an interval of $\pm$ 0.30 dex. \emph{Right panel:} The difference between the reference and the Node value of $T_{\rm eff}$. The red dashed lines indicate an interval of $\pm$ 150 K.}\label{fig:benchidr2} 
\end{figure*}

\subsection{The Gaia benchmark stars}\label{sec:bench}

The parameters ($T_{\rm eff}$ and $\log~g$) of these well-known bright stars are available from direct methods or from calibrations that are independent of spectroscopy (see Heiter et al. 2014a, in prep.). The metallicities used here as reference were determined in \citetads{2014A&A...564A.133J} using these same parameters. Table \ref{tab:bench} compiles the reference parameters of the 30 benchmark stars available for the iDR2 analysis. The spectra analyzed include both new Gaia-ESO observations and the spectrum library of \citetads[][]{2014A&A...566A..98B}. 

The atmospheric parameter scale of the Gaia-ESO results is tied to the system defined by these benchmark stars. This is a considerable improvement with respect to the standard approach of using the Sun as the only reference. The Gaia benchmark stars are distributed across the parameter space, meaning that we can choose better references for stars that are not solar-like.

\subsubsection{The accuracy of the Node results}

We divided the benchmark stars in three groups to judge the accuracy of the results in different corners of the parameter space separately. The groups were: 1) \emph{metal-rich dwarfs}: stars with [Fe/H] $>$ $-$1.00 and $\log~g$ $>$ 3.5 (contains 11 benchmark stars); 2) \emph{metal-rich giants}: stars with [Fe/H] $>$ $-$1.00 and $\log~g$ $\leq$ 3.5 (contains 7 benchmark stars); and 3) \emph{metal-poor stars}: stars with [Fe/H] $\leq$ $-$1.00 (contains three benchmark stars). Only one group of metal-poor stars was defined because only three benchmark stars with [Fe/H] $\leq$ $-$1.00 are available. 

Some Nodes had difficulties analyzing the archival data. Because the spectra were obtained with different spectrographs, they were made available in a different format with respect to the standard Gaia-ESO one. The analysis problem was a shortcoming caused by the use of automatic pipelines designed to deal with a large amount of data in the same format. Thus, to judge the Node results accuracy for iDR2 we decided to use: 1) the results of 19 benchmark stars observed by Gaia-ESO; 2) the analysis of a FLAMES spectrum of the Sun\footnote{Obtained on the evening twilight sky and available here \url{http://www.eso.org/observing/dfo/quality/GIRAFFE/pipeline/solar.html}}; and 3) the analysis of the archival spectrum of \object{HD 84937} (one of the few metal-poor stars in this list). 

For each Node, in each of the three areas of the parameter space, we calculate what is the average quadratic difference between the reference and the derived atmospheric parameters (only $T_{\rm eff}$ and $\log~g$) of the stars. If this average quadratic difference is within $\pm$ 100 K and $\pm$ 0.20 dex of the reference values, the Node results are considered to be very accurate (in that region of the parameter space). These average differences per Node are given in Table \ref{tab:nodediff}. It is important to notice in this table that different Nodes succeeded in analyzing a different number of stars in each region of the parameter space.

\begin{figure*}
\centering
\includegraphics[height = 4.2cm]{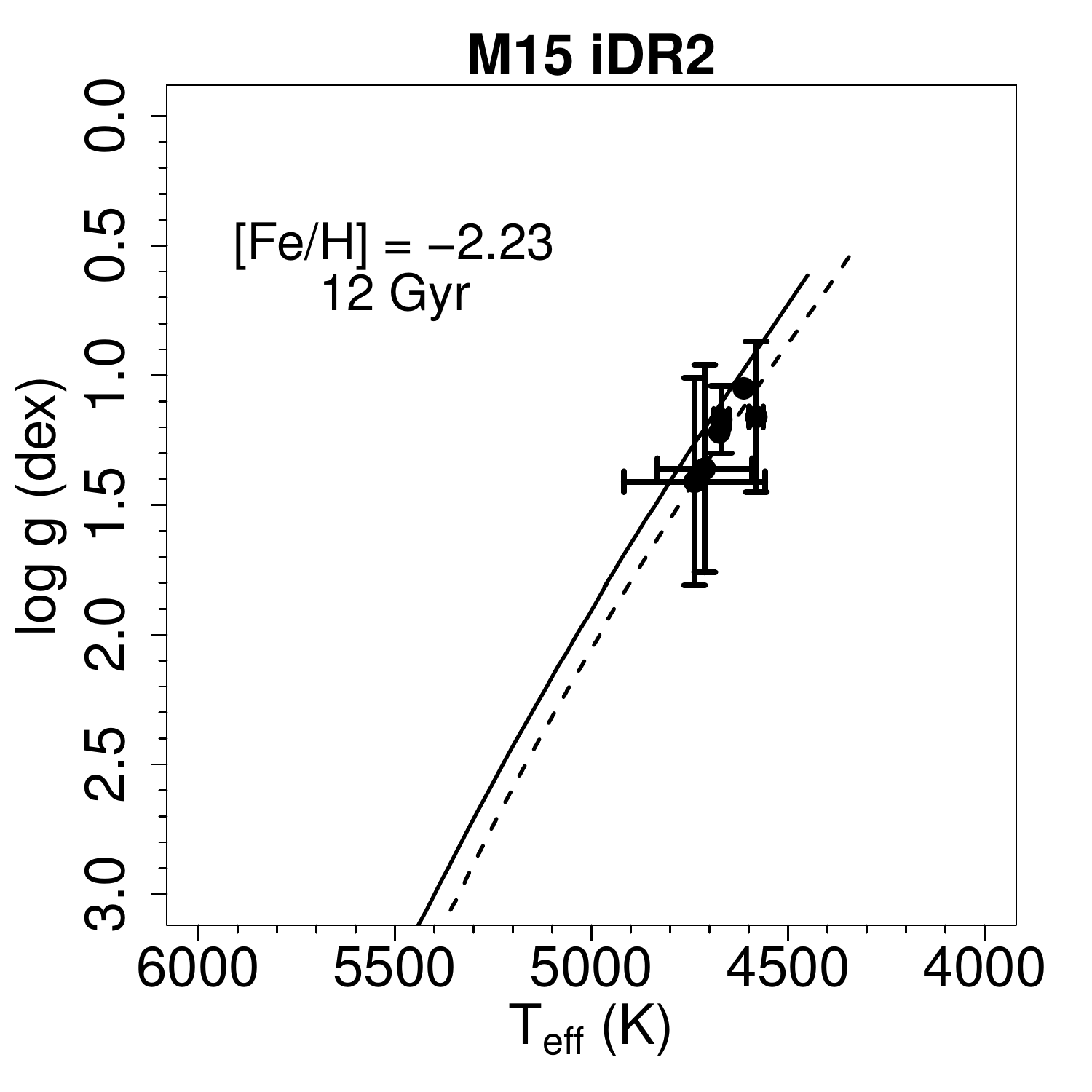}
\includegraphics[height = 4.2cm]{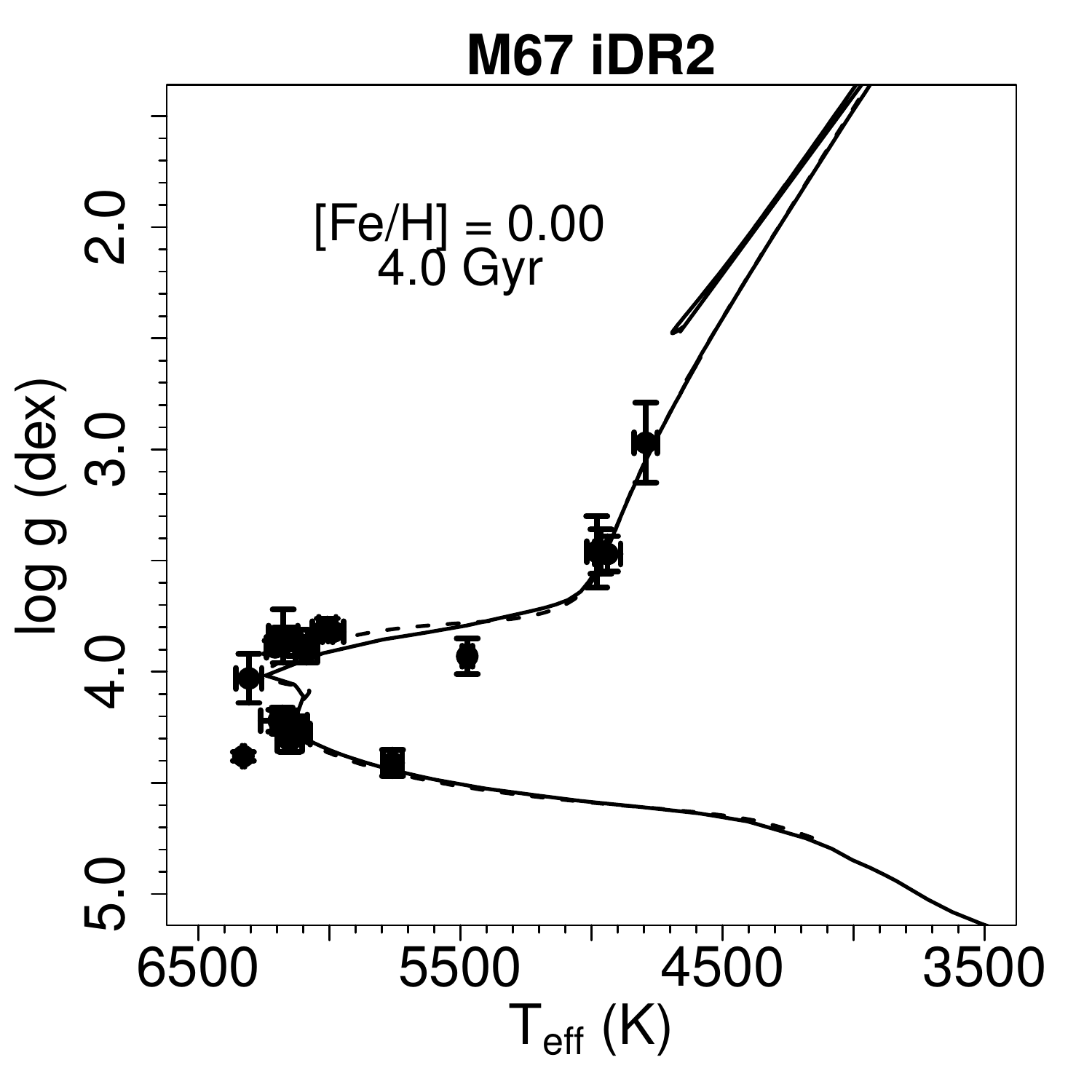}
\includegraphics[height = 4.2cm]{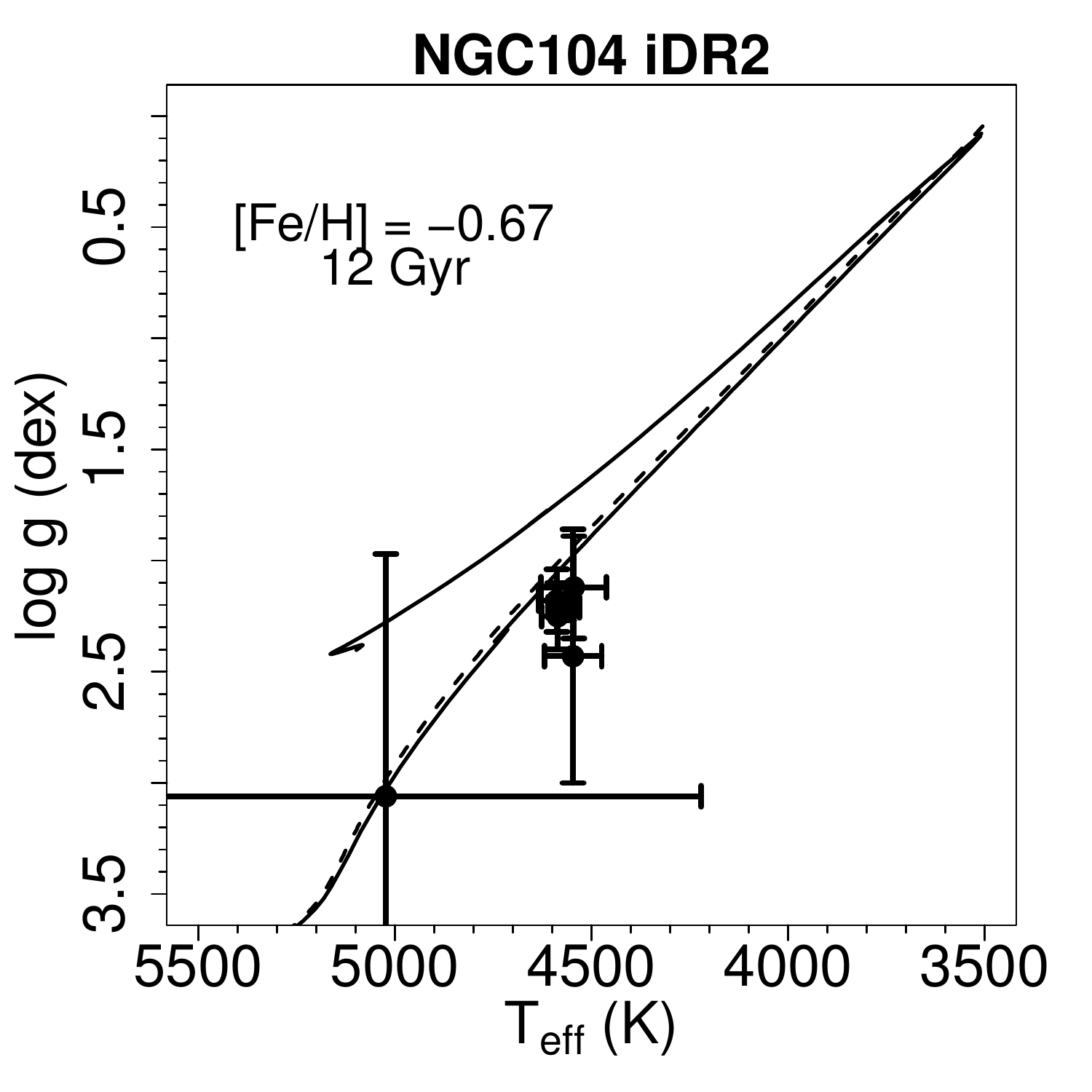}
\includegraphics[height = 4.2cm]{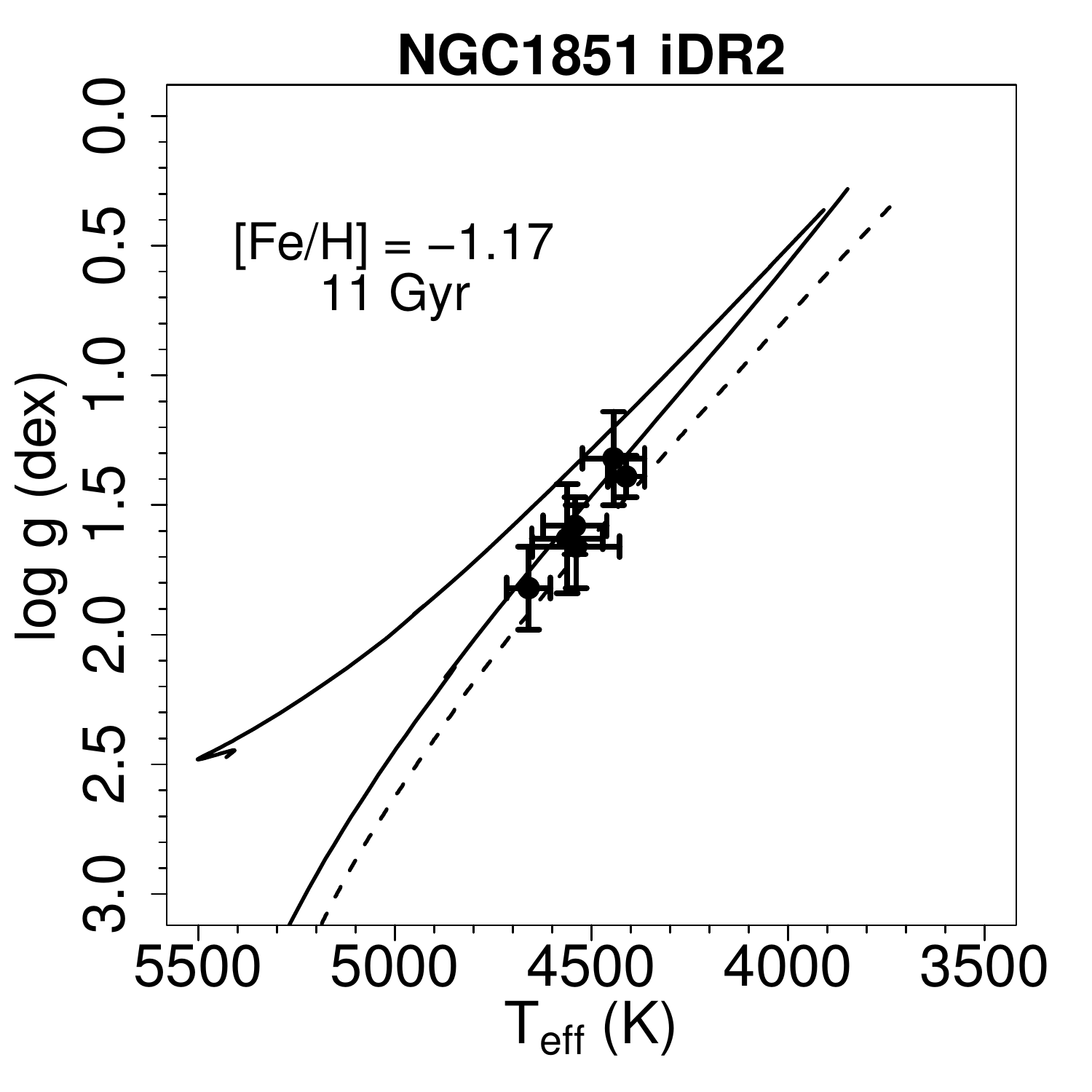}
\includegraphics[height = 4.2cm]{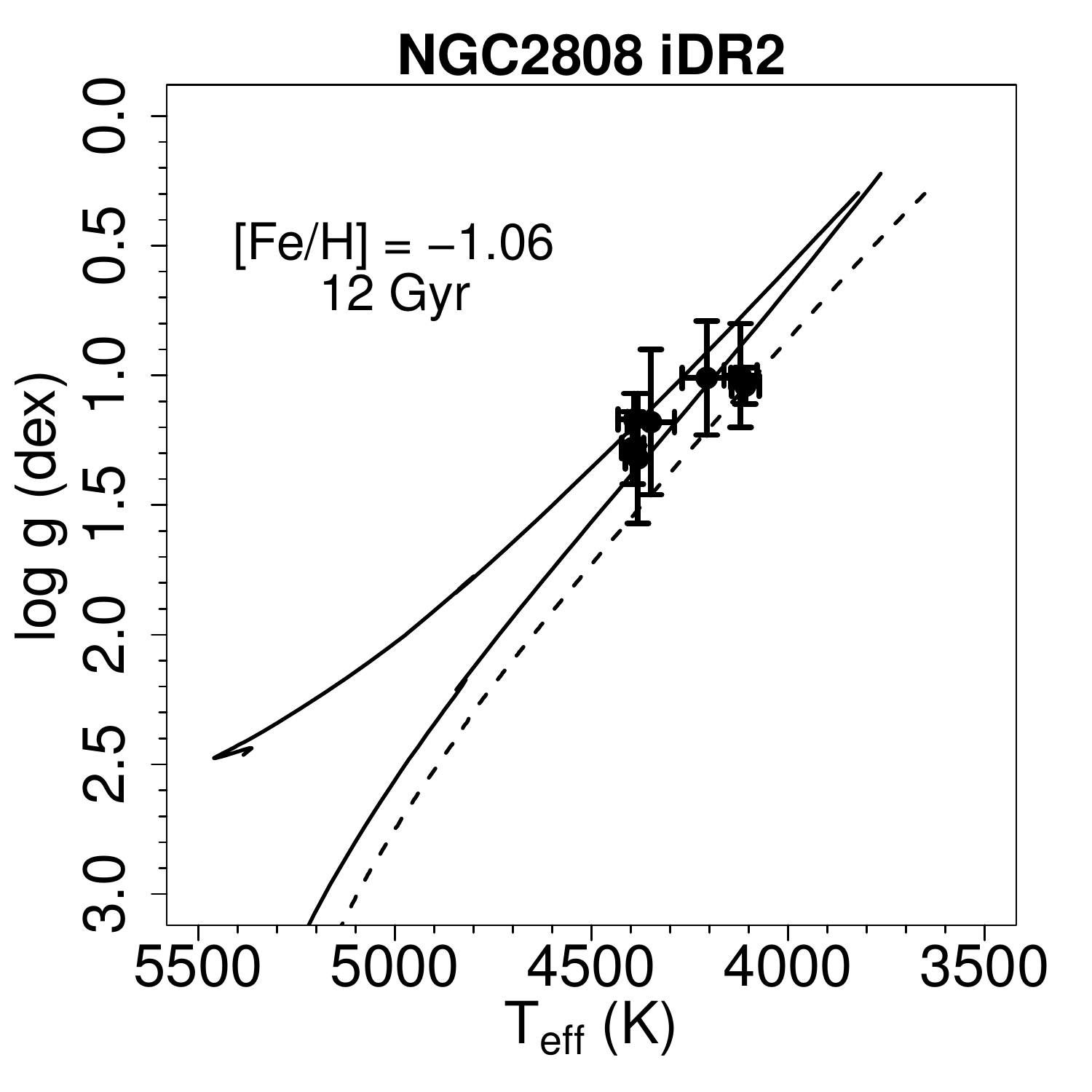}
\includegraphics[height = 4.2cm]{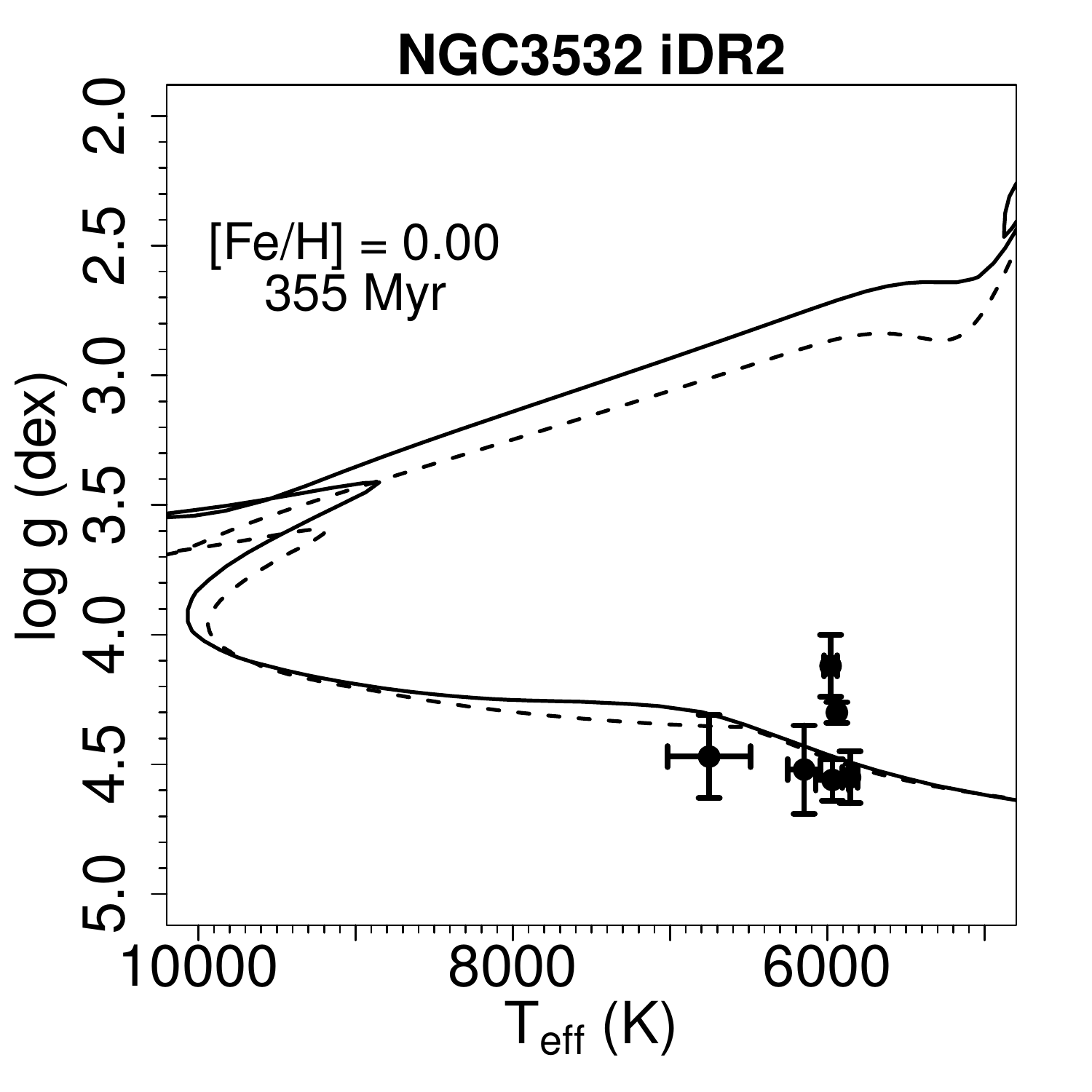}
\includegraphics[height = 4.2cm]{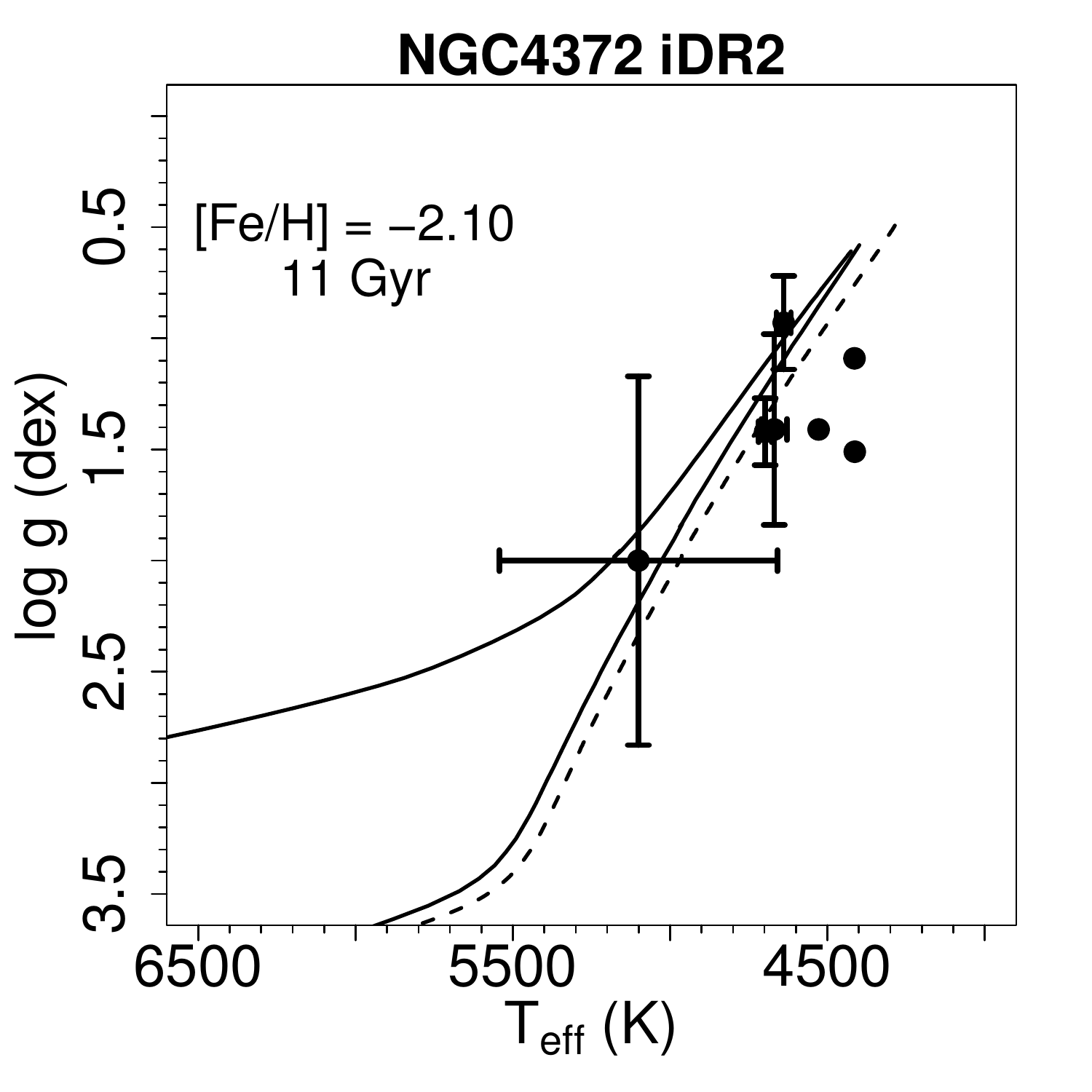}
\includegraphics[height = 4.2cm]{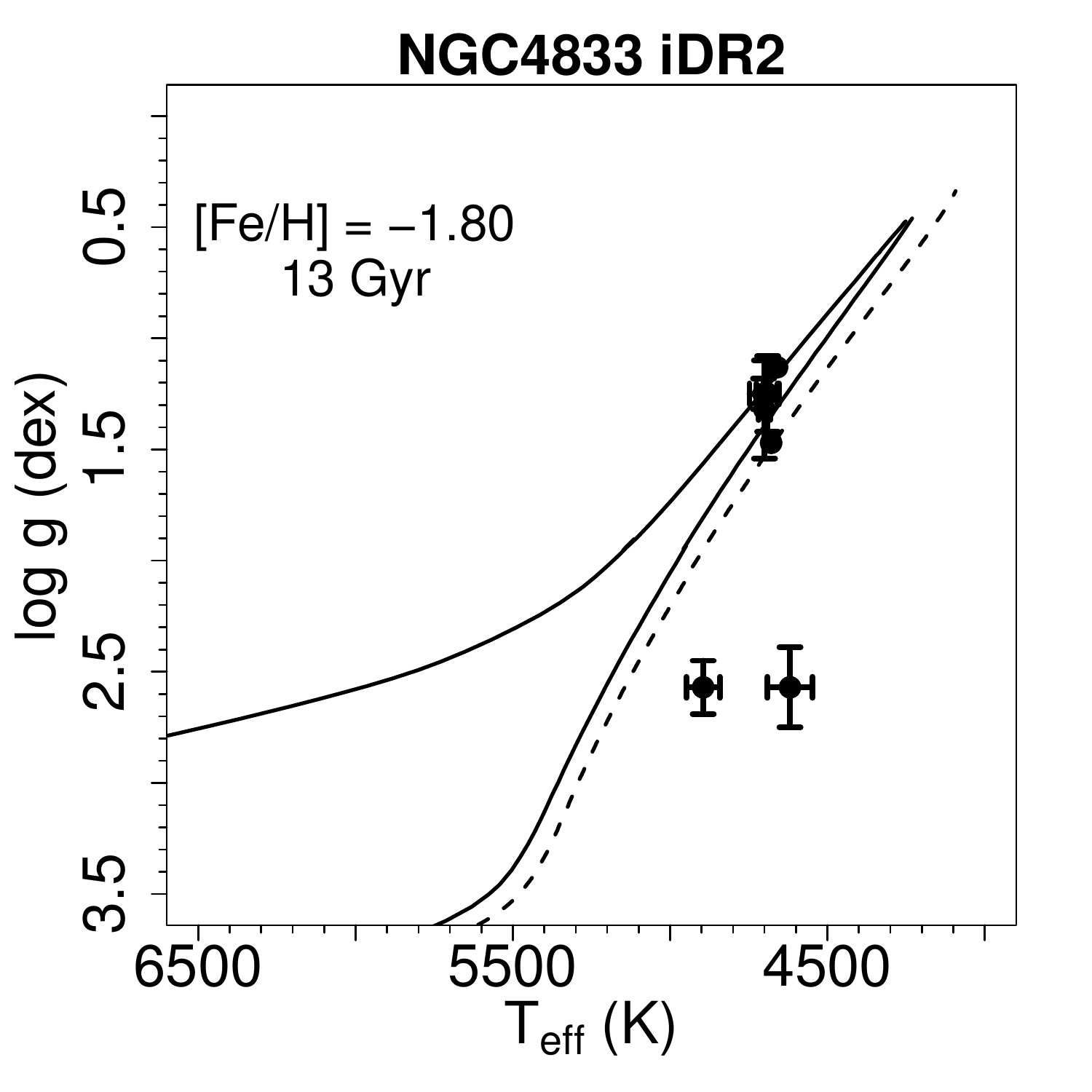}
\includegraphics[height = 4.2cm]{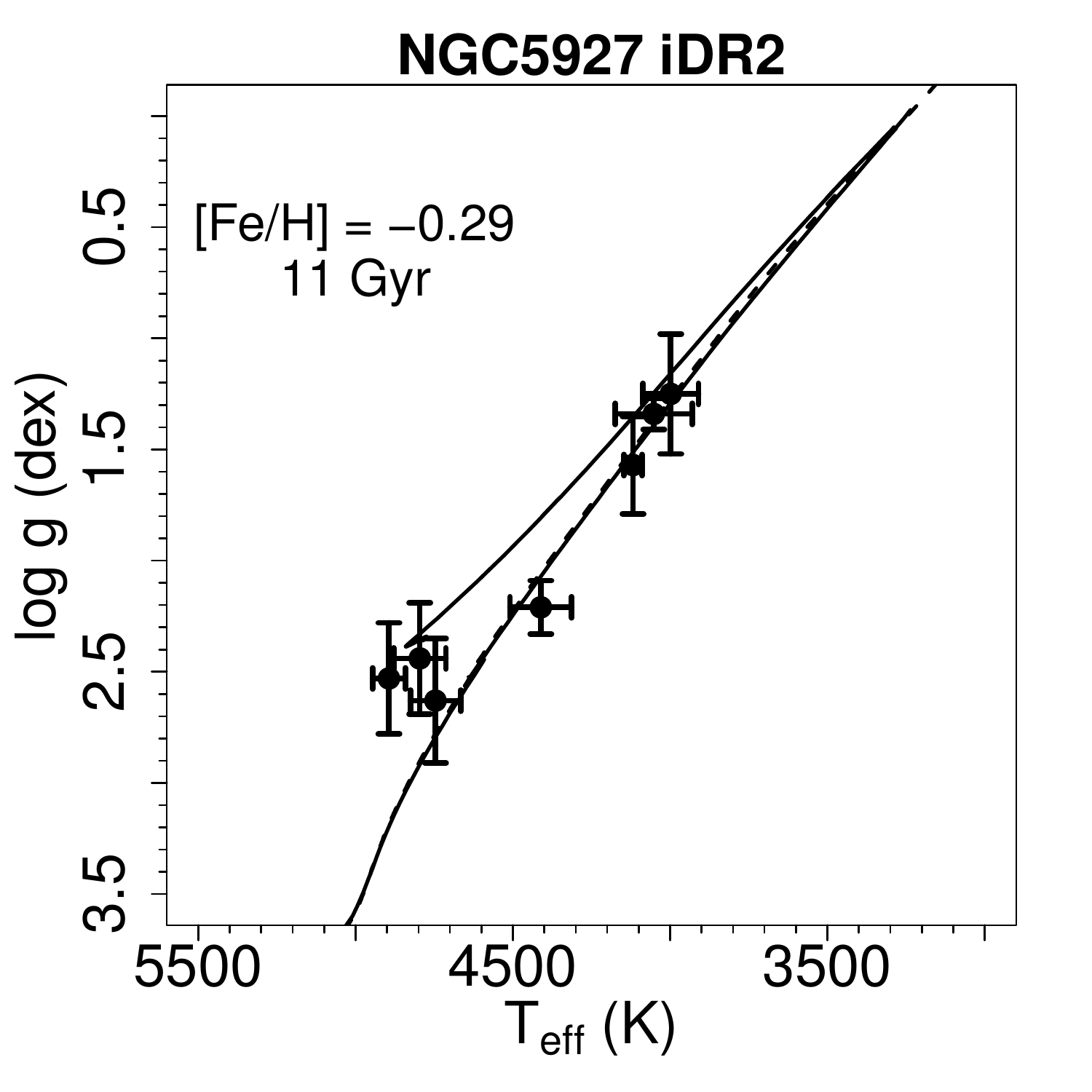}
\includegraphics[height = 4.2cm]{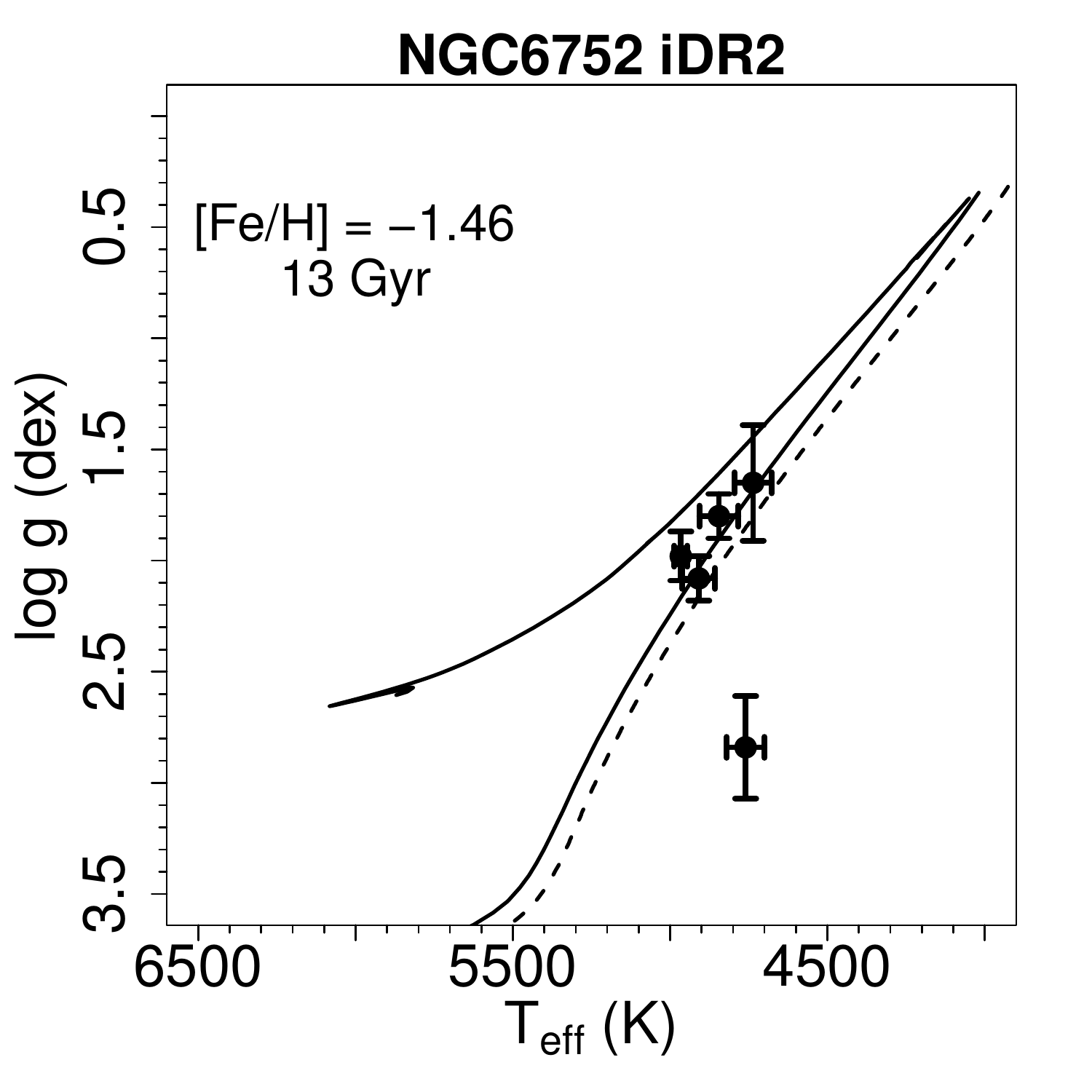}
 \caption{iDR2 recommended parameters of the stars of the calibration clusters in the $T_{\rm eff}$-$\log~g$ plane. No attempt was made to identify non-member stars, i.e. the plots include all stars observed in the field of the clusters. The ages and metallicities were taken from the catalog of \citetads[][and online updates]{1996AJ....112.1487H} for the globular clusters and from the WEBDA database for the open clusters. The isochrones were computed with the web-tool of the PARSEC group \citepads[][all with solar-scaled composition]{2012MNRAS.427..127B}, solid lines, and with BeSPP \citepads[Bellaterra Stellar Parameter Pipeline,][$\alpha$-enhanced below {[Fe/H]} = $-$0.80]{2013MNRAS.429.3645S} which uses the GARSTEC stellar evolution code \citepads{2008Ap&SS.316...99W}, dashed lines. Error bars represent the method-to-method dispersion of each atmospheric parameter (see Sect. \ref{sec:mad}).}\label{fig:clusters}%
\end{figure*}
The ULB Node uncovered problems with their analysis quite late during the process of homogenization. As there was no time to recompute the atmospheric parameters so close to the end of the analysis cycle, the Node decided to withdraw its results. A second Node (Li\`ege) was considered to produce very uncertain results for the ``metal-poor stars'' group. The results of this Node for this region of the parameter space were not used and the values are not included in Table \ref{tab:nodediff}. The OACT Node did not analyze the metal-poor benchmarks. Their method needs observed spectra of metal-poor stars among the library used as reference, but these are currently lacking. The Nodes Bologna, CAUP, and UCM encountered other problems when analyzing these benchmark stars. As weights for the MPS region of the parameter space are not available for these Nodes, their results for metal-poor stars were not used.

Systematic biases are one component that can affect the accuracy of the results, making the results seem to be less accurate. They can in principle be corrected for, so that the unbiased results would agree better with the reference atmospheric parameters. For iDR2, however, bias correction was not implemented. This improvement will be implemented for future releases. 

Figure \ref{fig:benchidr2} shows a comparison between all the Node results for the benchmark stars with respect to their reference $T_{\rm eff}$ and $\log g$ values. All results are part of these figure, this includes the analysis of single exposure spectra of the stars, many of which have low S/N per pixel ($<$ 20). So the full range in the values displayed does not translate directly to the real uncertainty of the analysis. The final accuracy was judged only on the results for the final co-added spectra. Most of the results tend to be in reasonable agreement with the reference values, but outliers are present. Clear problems appear in some special cases: \emph{i)} Gam Sge, Alf Cet, and Alf Tau are cool stars, with $T_{\rm eff}$  $\lesssim$ 4000 K, that almost all Nodes have difficulties in analyzing; \emph{ii)} Procyon and Bet Vir have spectra with reduction problems; \emph{iii)} Eta Boo and HD 49933 are relatively fast rotators ($v \sin i$ $\geqslant$ 10\,km s$^{-1}$); and \emph{iv)} the very metal-poor stars HD 84937, HD 122563, and HD 140283. This comparison already indicates the regions of the parameter space where the results derived here have increased uncertainty, i.e., very cool stars ($T_{\rm eff}$ $<$ 4000 K), metal-poor stars ([Fe/H] $\leqslant$ $-$2.0), and fast rotators.

 \subsection{Calibration clusters}\label{sec:calclus}

\begin{figure*}
\centering
\includegraphics[height = 4cm]{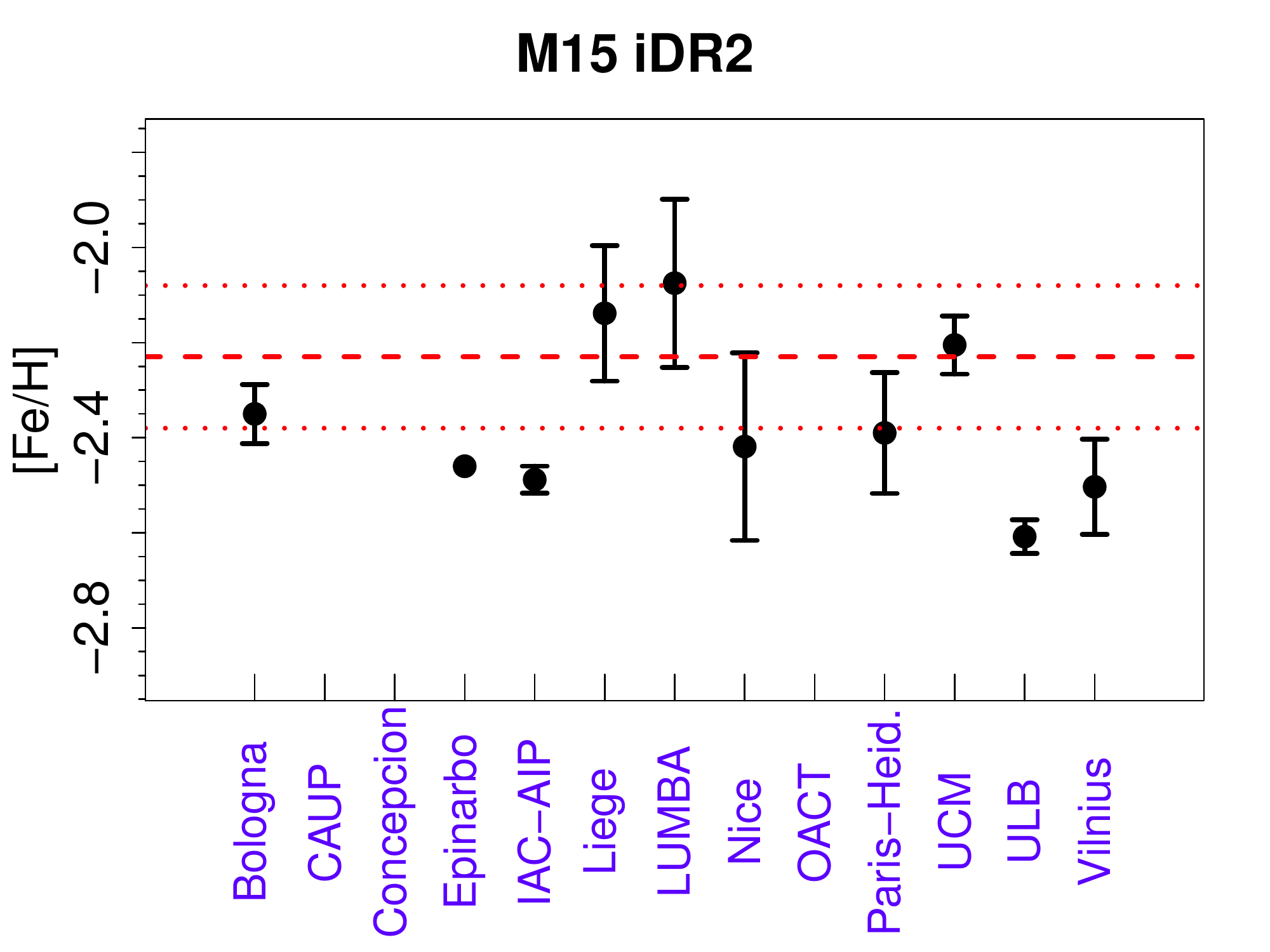}
\includegraphics[height = 4cm]{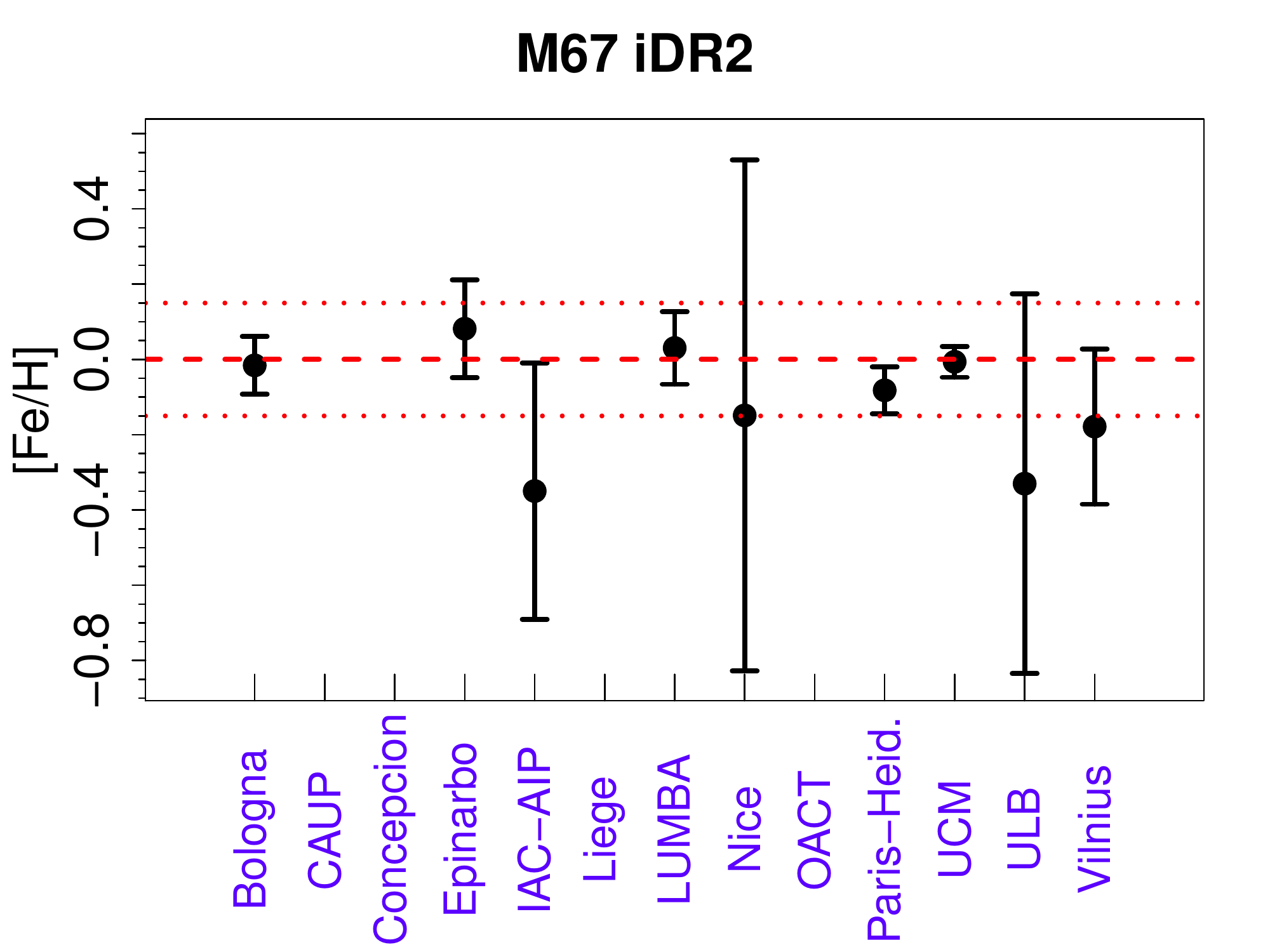}
\includegraphics[height = 4cm]{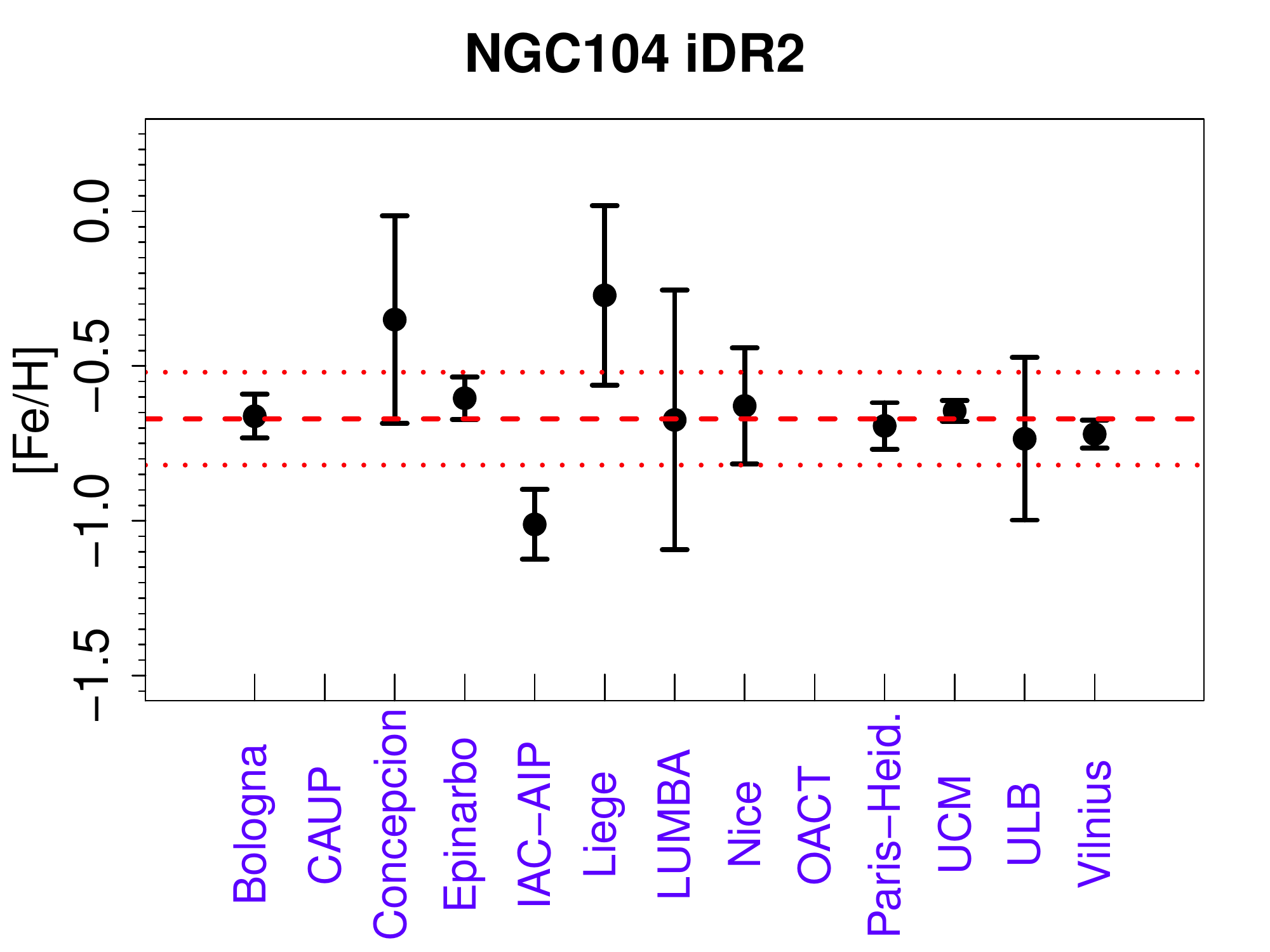}
\includegraphics[height = 4cm]{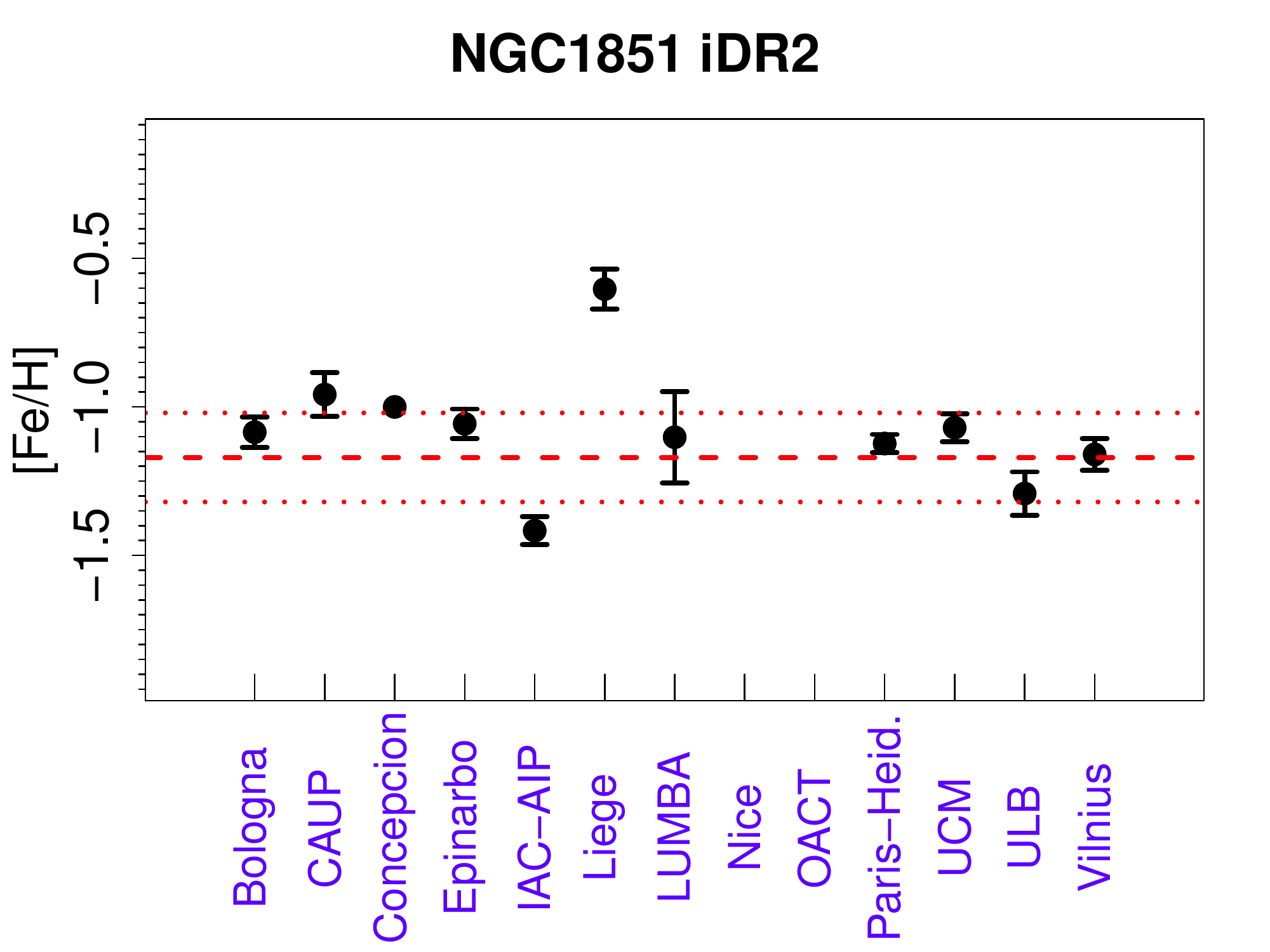}
\includegraphics[height = 4cm]{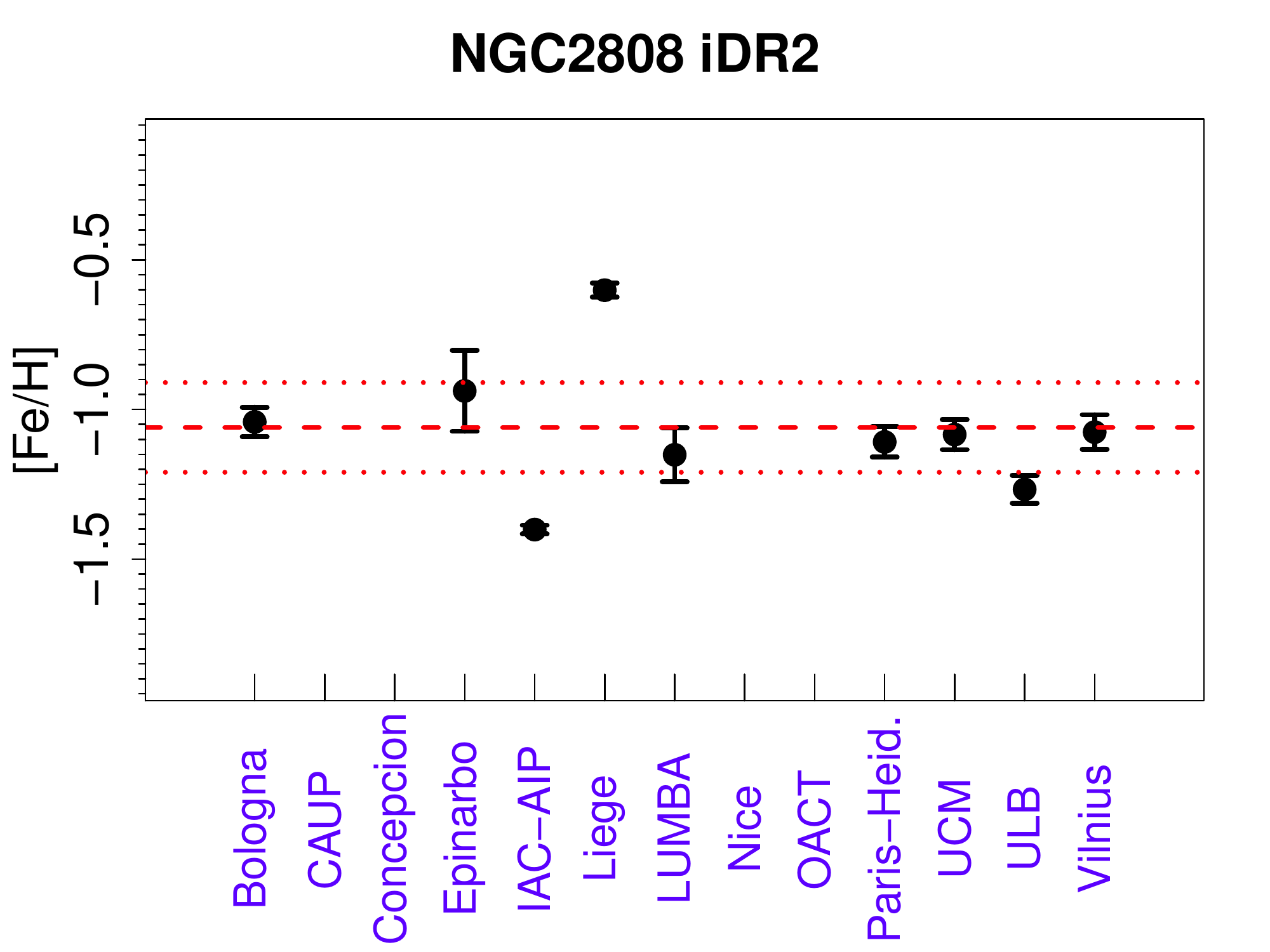}
\includegraphics[height = 4cm]{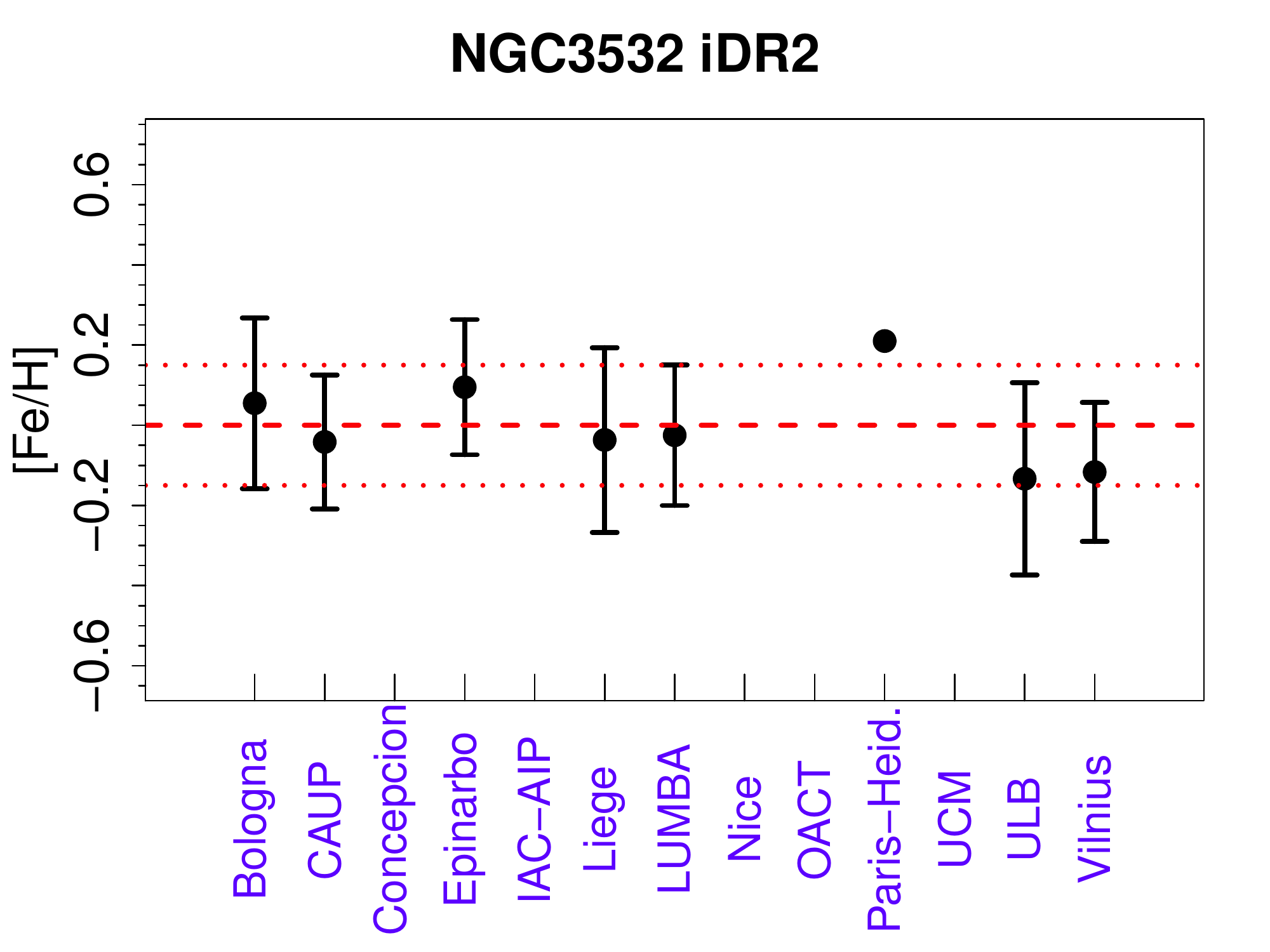}
\includegraphics[height = 4cm]{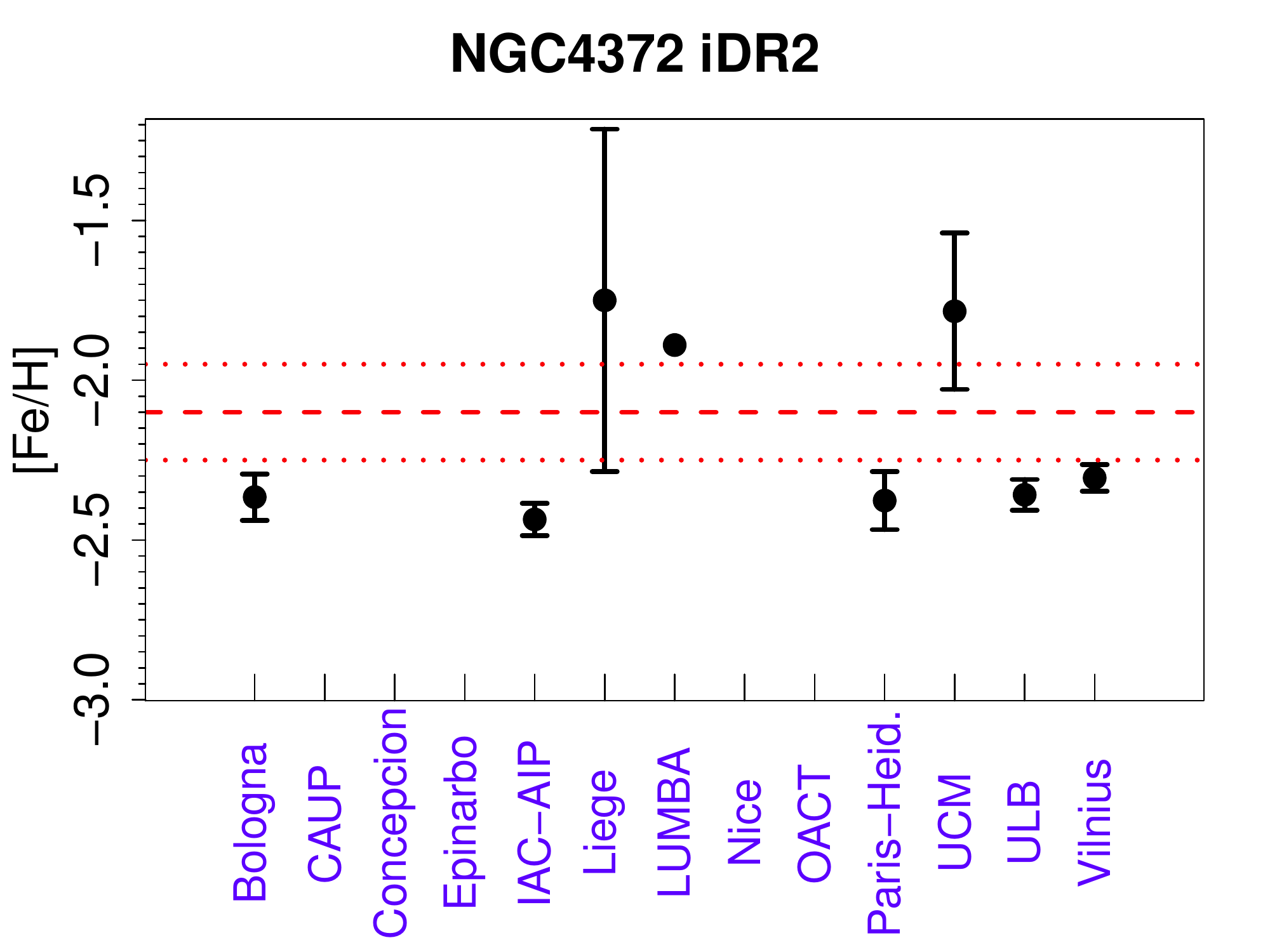}
\includegraphics[height = 4cm]{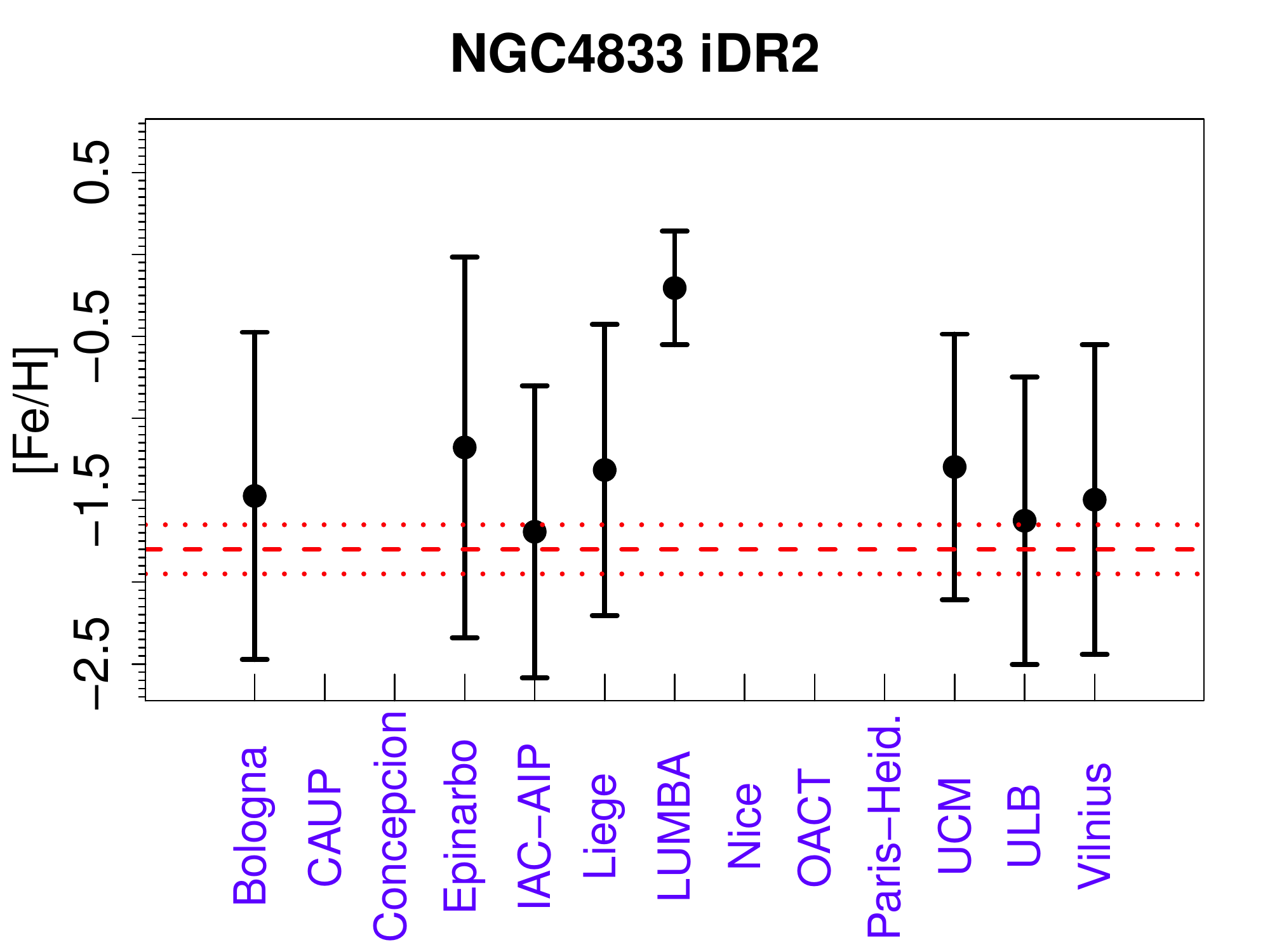}
\includegraphics[height = 4cm]{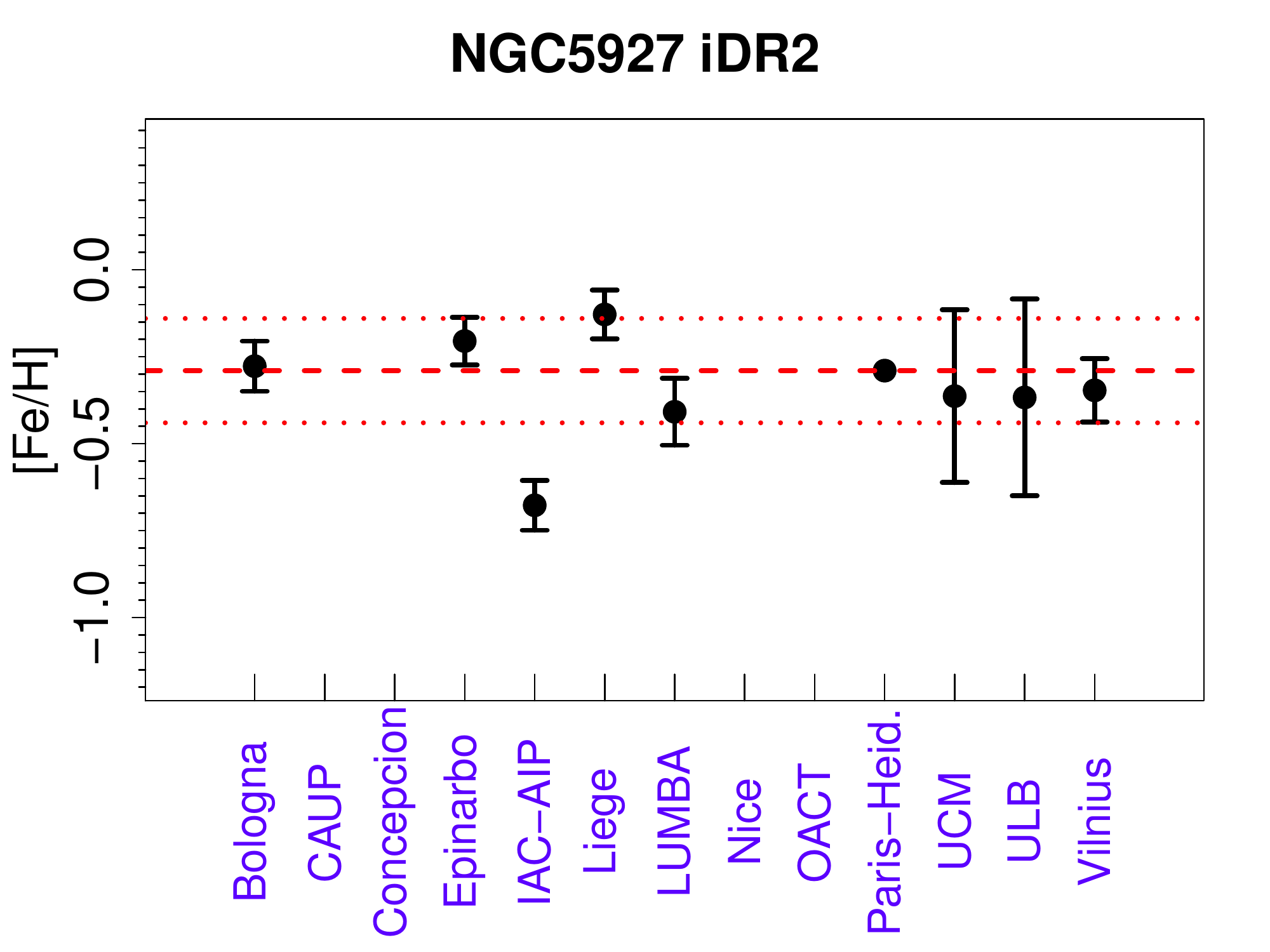}
\includegraphics[height = 4cm]{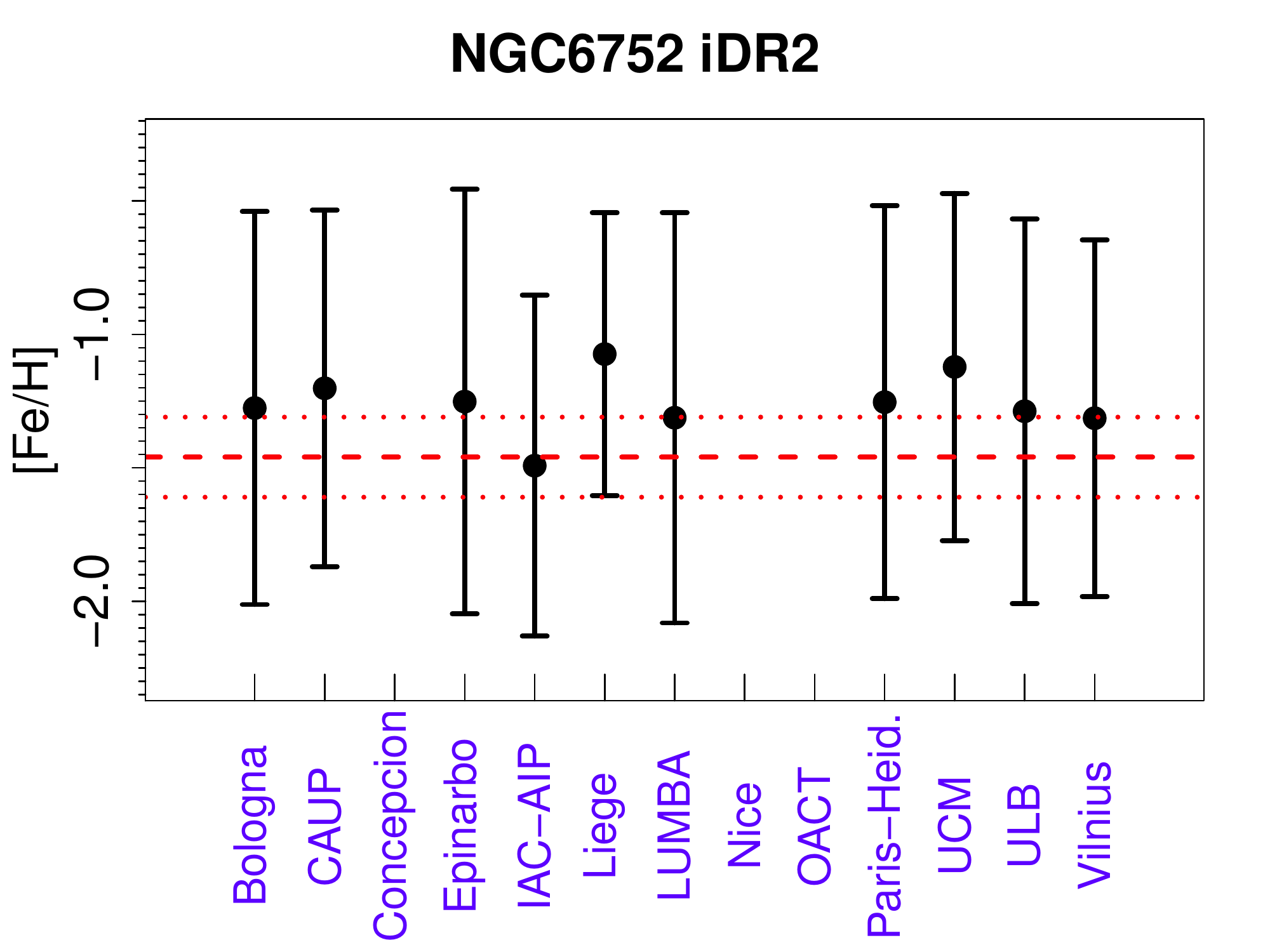}
 \caption{Average metallicity obtained by the Nodes in comparison with a literature estimate of each calibrating cluster. The error bars indicate the standard error of the mean metallicity value of all stars in the cluster field analyzed by that given Node. Large error bars are thus caused by the presence of non-member stars with very different metallicities. Different Nodes might have analyzed different number of stars in each cluster. The red dashed line is the literature metallicity of the cluster, taken from the catalog of \citetads[][and online updates]{1996AJ....112.1487H} for the globular clusters and from the WEBDA database for the open clusters. The dotted lines indicate a variation of $\pm$0.15 dex in the metallicity.}\label{fig:fehclusters}%
\end{figure*}

\begin{figure*}
\centering
\includegraphics[height = 6cm]{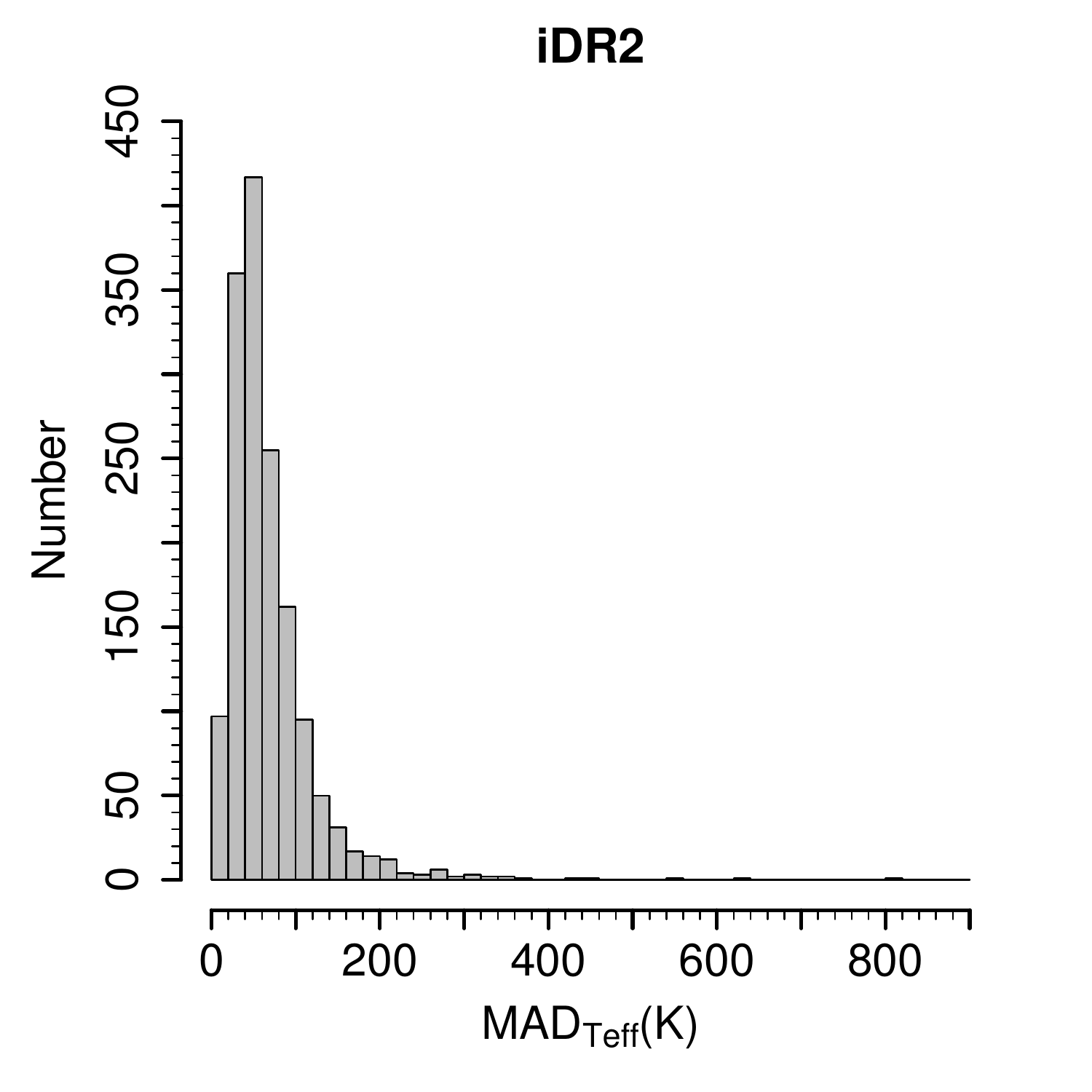}
\includegraphics[height = 6cm]{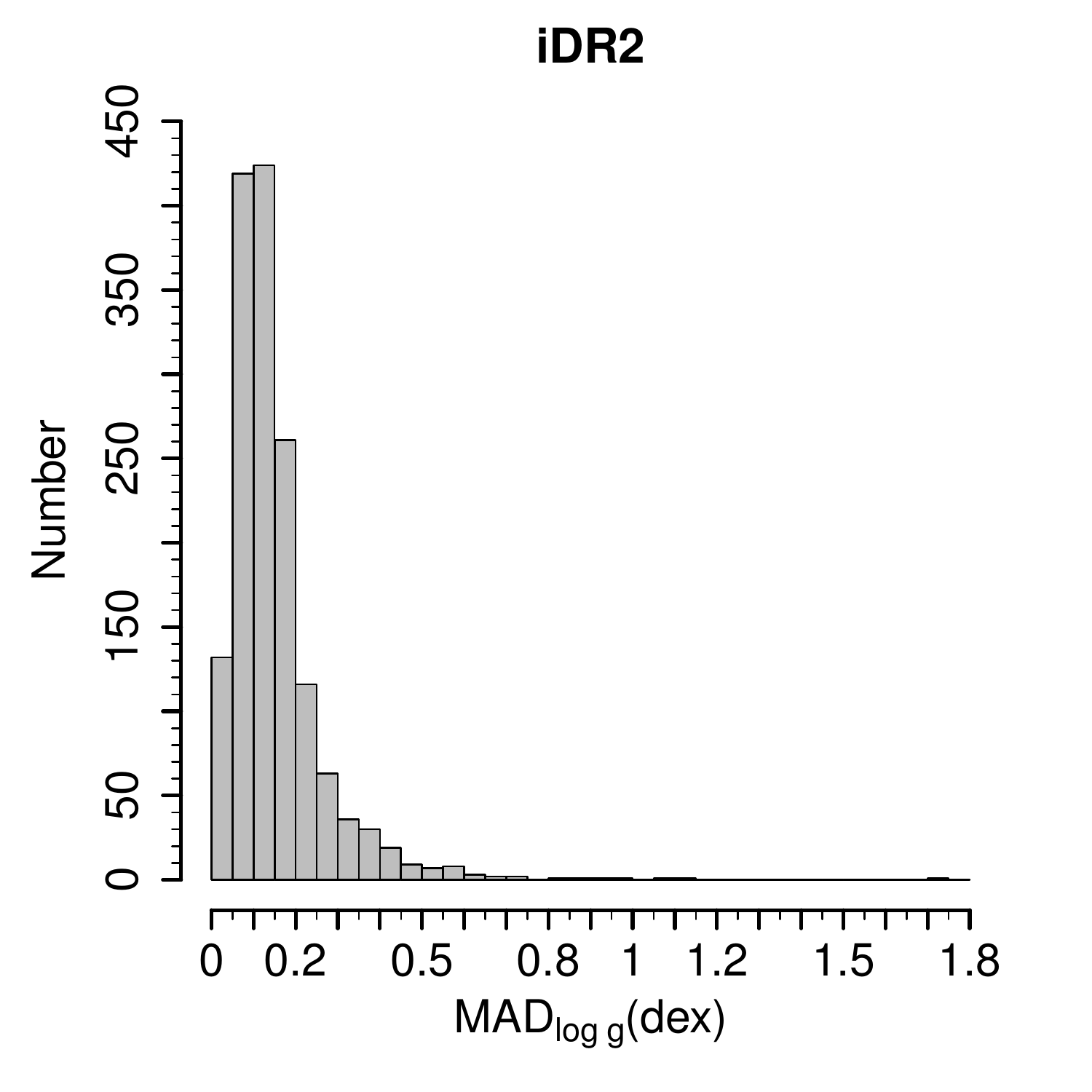}
\includegraphics[height = 6cm]{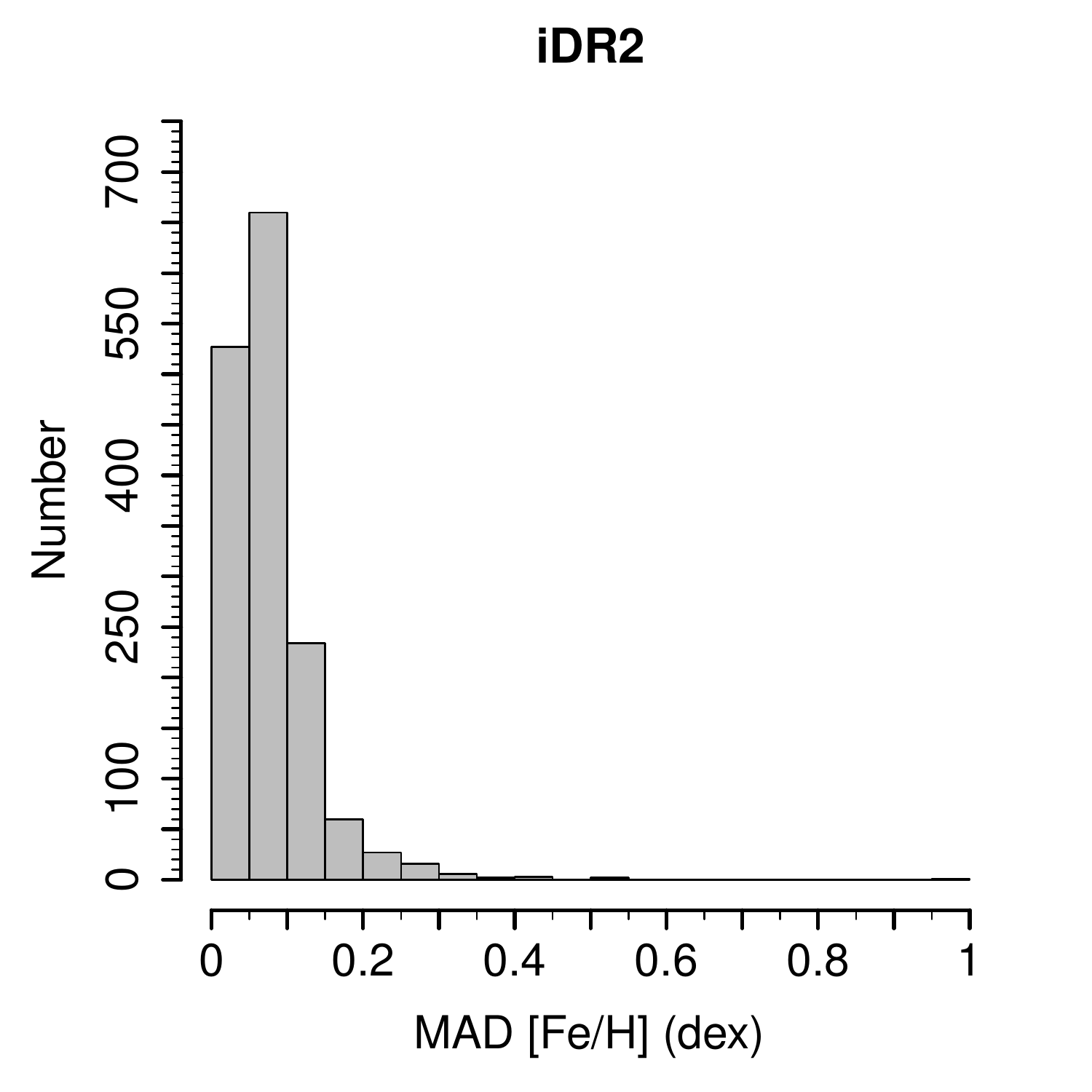}
 \caption{Histograms showing the distribution of the method-to-method dispersion of the atmospheric parameters of the 1517 results obtained in iDR2 (some stars have multiple results, as single exposure spectra were analyzed sometimes). The dispersion is only computed if at least three Nodes provided results for that given star. \emph{Left:} The dispersion of $T_{\rm eff}$. \emph{Center:} The dispersion of $\log g$. \emph{Right:} The dispersion of [Fe/H].}\label{fig:histidr2}%
\end{figure*}

A list of open and globular clusters are being observed by Gaia-ESO for calibration purposes (see Pancino et al. 2014, in prep.). Among other uses, they can serve as a second level of calibration to assess the physical consistency of the results. The calibration clusters used for iDR2 included the globular clusters \object{M 15}, \object{NGC 104}, \object{NGC 1851}, \object{NGC 2808}, \object{NGC 4372}, \object{NGC 4833}, \object{NGC 5927}, and \object{NGC 6752} and the open clusters \object{M 67} (with archival data), \object{NGC 3532}, and \object{NGC 6705} (both with Gaia-ESO data). More of these calibration clusters have been and will be observed as the Survey progresses. They will be added to the calibration effort for future releases.

The observed stars were red giants in the globular clusters, cool main-sequence stars in NGC 3532, and AB-type stars in the open cluster \object{NGC 6705}\footnote{NGC 6705 was also observed for science goals, in this case the targets were FGK-type stars.}. The AB-type stars were selected to be used as a control sample for comparison between the analysis of FGK-type stars and the analysis of OBA-type stars (see details in Blomme et al. 2014, in prep.). Unfortunately, most of these stars turned out to be fast rotators and results for them were deemed uncertain and were excluded during quality control.

The physical consistency of the atmospheric parameters of cluster stars can be judged by comparing the derived values with those expected for an isochrone calculated with the chemical composition and age of that cluster. If the results follow an unphysical relation in the diagram, they will be excluded and the Node results in that part of the parameter space disregarded. Although a few stars for which the parameters do not follow exactly the isochrones were identified, in most cases the agreement was deemed acceptable within the uncertainties. To illustrate that no grossly wrong parameters were found, Fig. \ref{fig:clusters} compares the final recommended atmospheric parameters of the stars observed in the calibration clusters with isochrones computed with literature values for age and metallicity. In these plots, we did not remove stars that might be non-members of the clusters. That is part of scientific analyses that will be presented elsewhere. The results reproduce quite well the predicted slope of the red giant branches of the clusters. The M 67 open cluster is particularly interesting, as data of main sequence, turn-off, and giant stars was available. All these evolutionary regions are very well reproduced by the results. The differences between the two isochrone sets on the red-giant branch are explained by the fact that PARSEC isochrones have solar-scaled composition, while GARSTEC isochrones are $\alpha$-enhanced below [Fe/H] = $-0.80$, to be consistent with the $\alpha$-enhancement observed in the metal-poor clusters.
\begin{figure*}[]
\centering
\includegraphics[height = 6cm]{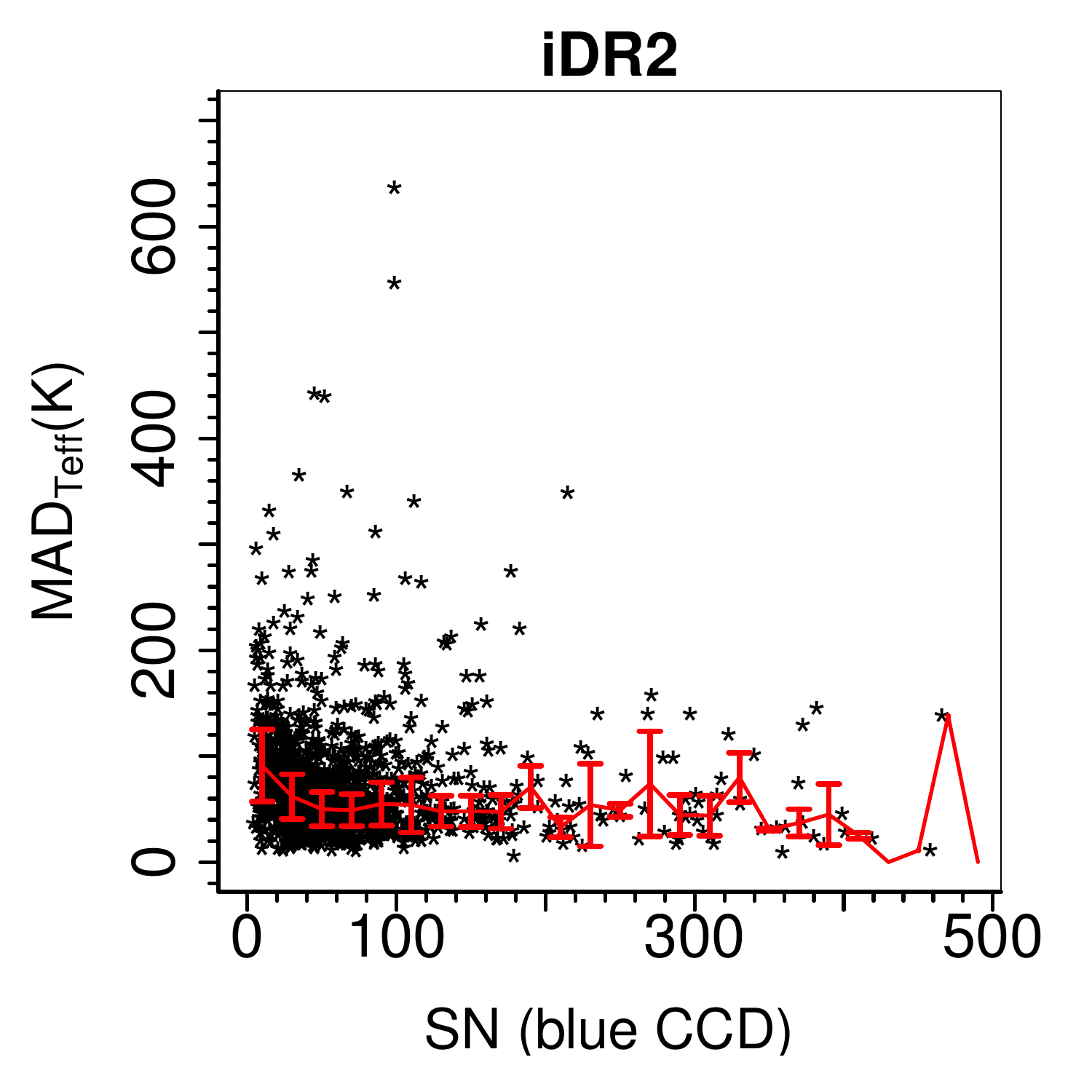}
\includegraphics[height = 6cm]{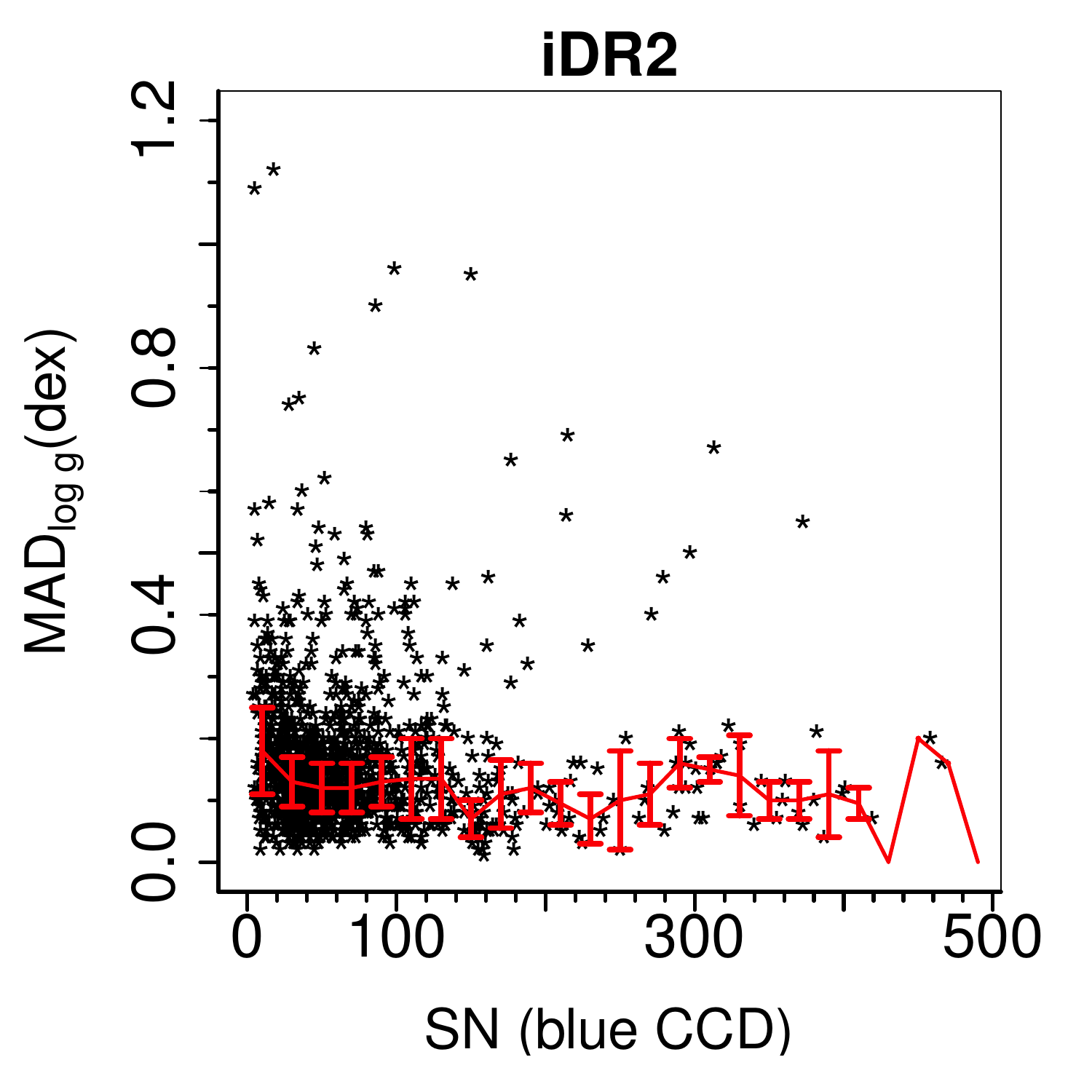}
\includegraphics[height = 6cm]{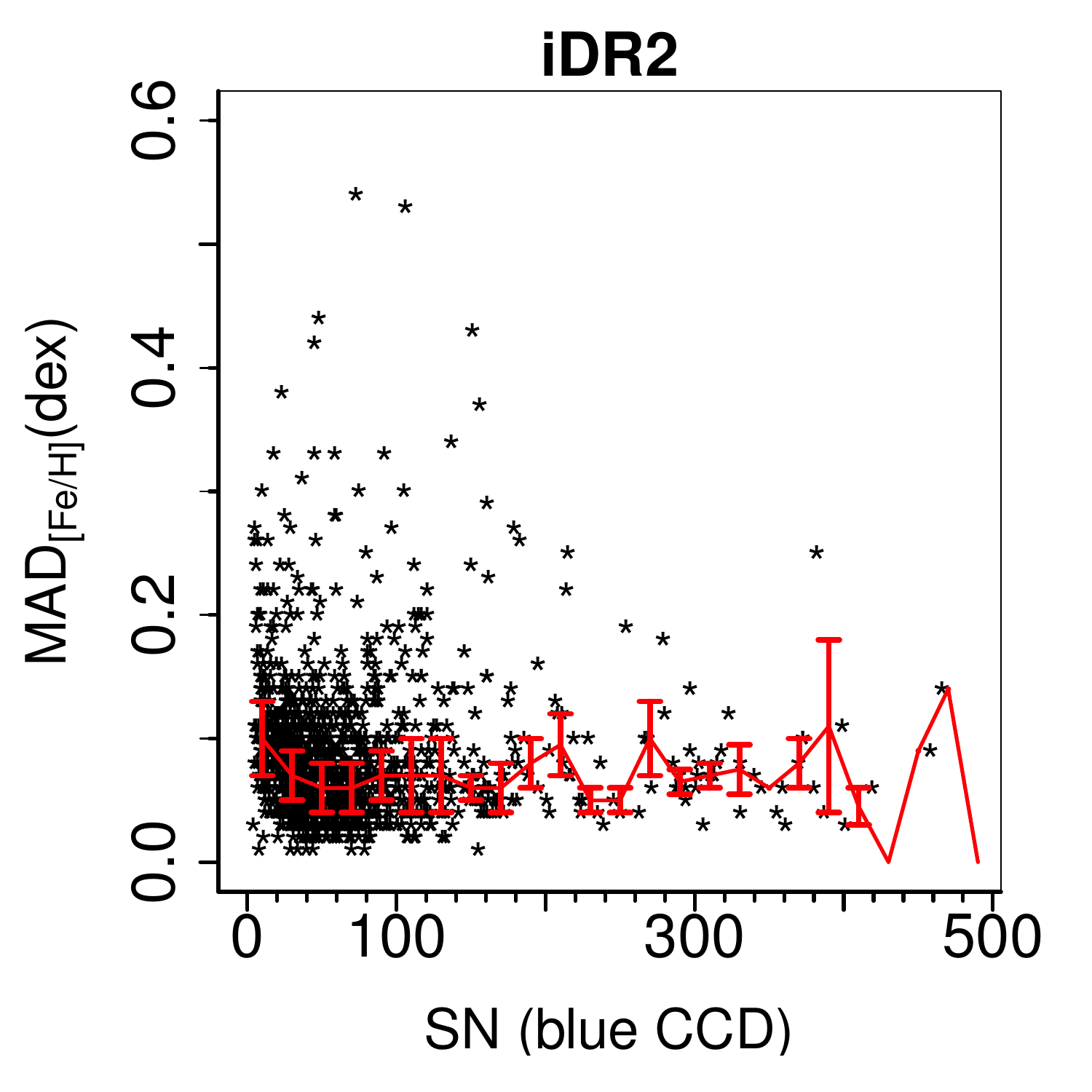}
 \caption{Dependency of the method-to-method dispersion on the median S/N per pixel of the spectra. The S/N of the bluer part of the spectrum is used as reference. The red line connects the median value in each bin of S/N (in steps of 20). The error bars in the line represent the median absolute deviation.}\label{fig:madsn}%
\end{figure*}
\begin{figure*}
\centering
\includegraphics[height = 5.5cm]{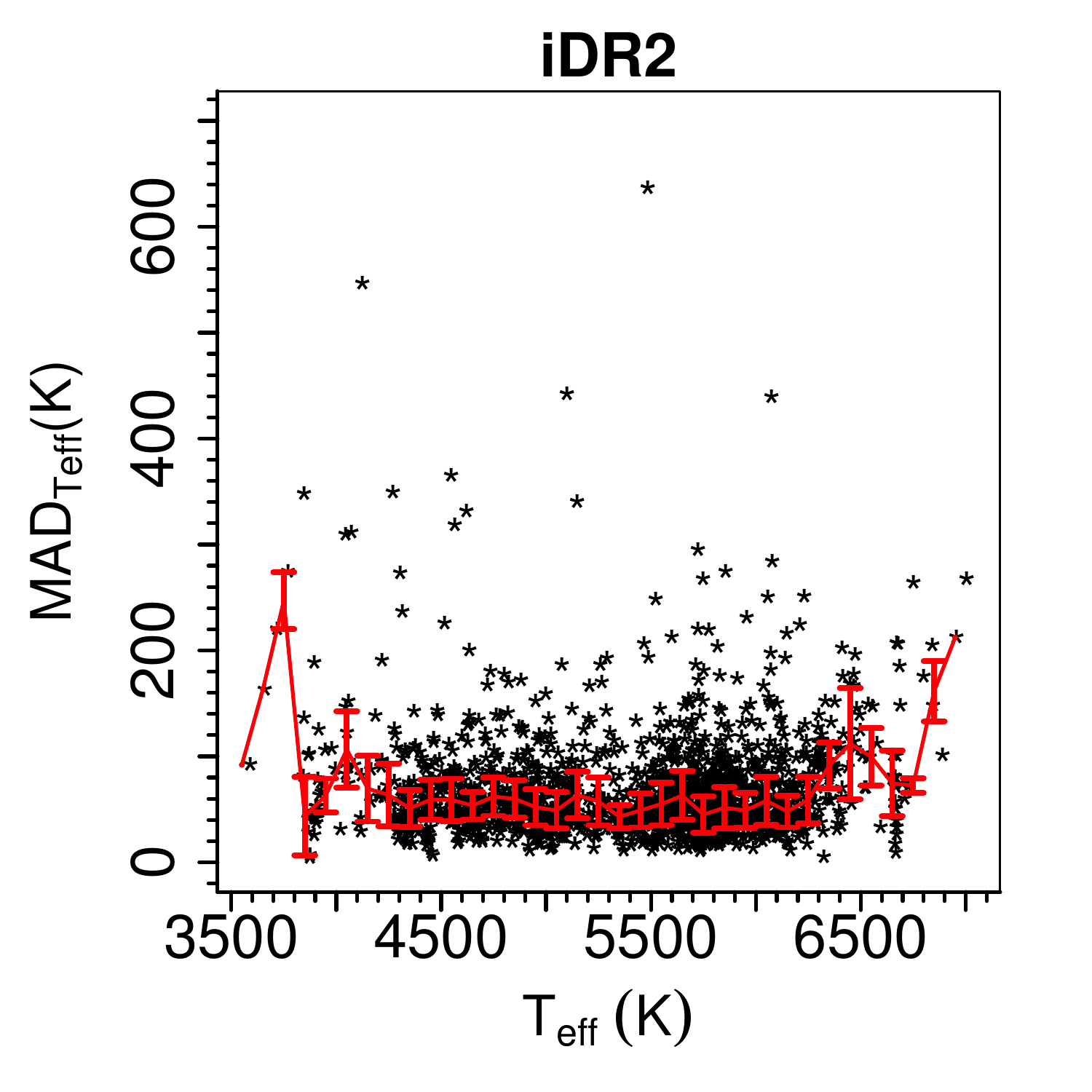}
\includegraphics[height = 5.5cm]{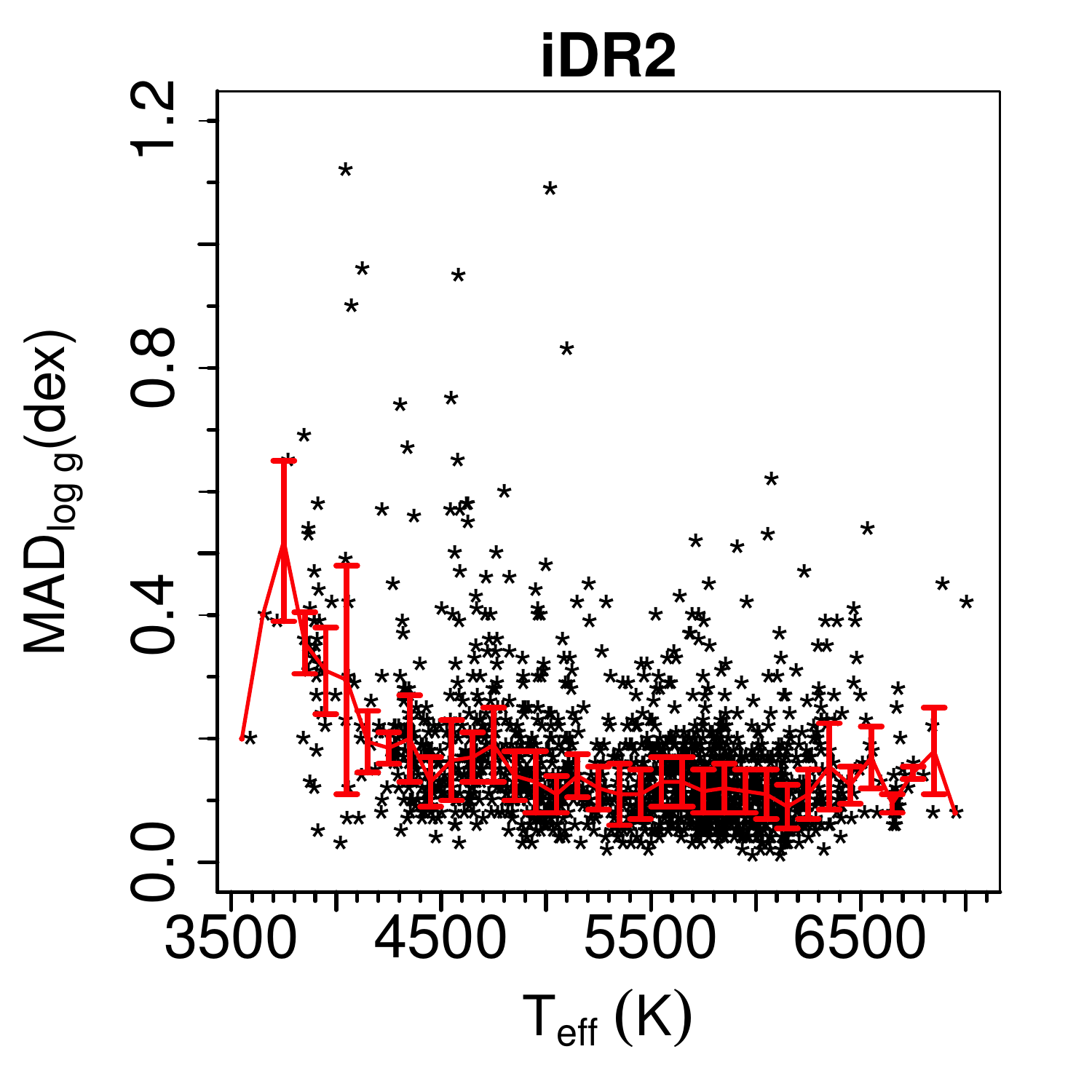}
\includegraphics[height = 5.5cm]{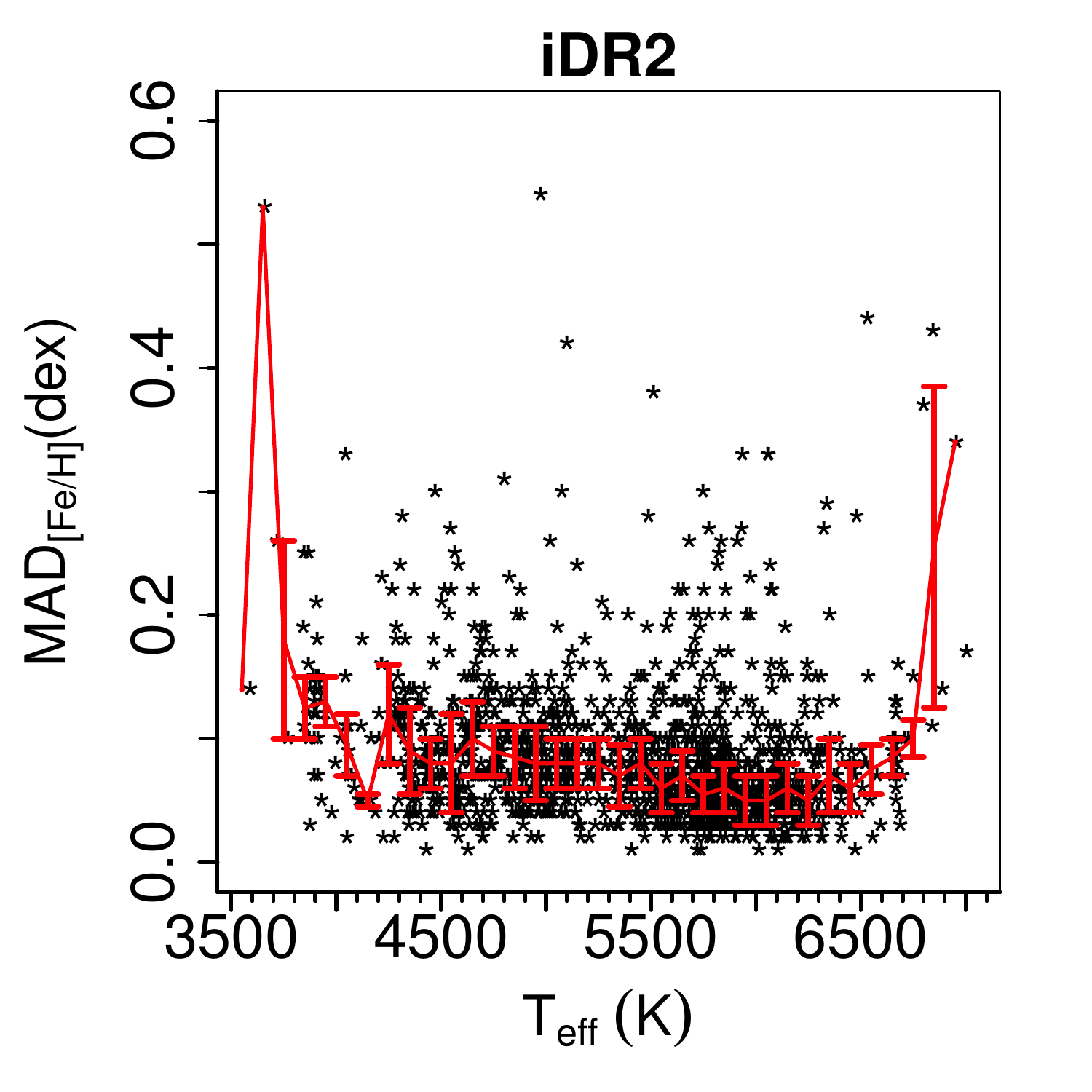}
\includegraphics[height = 5.5cm]{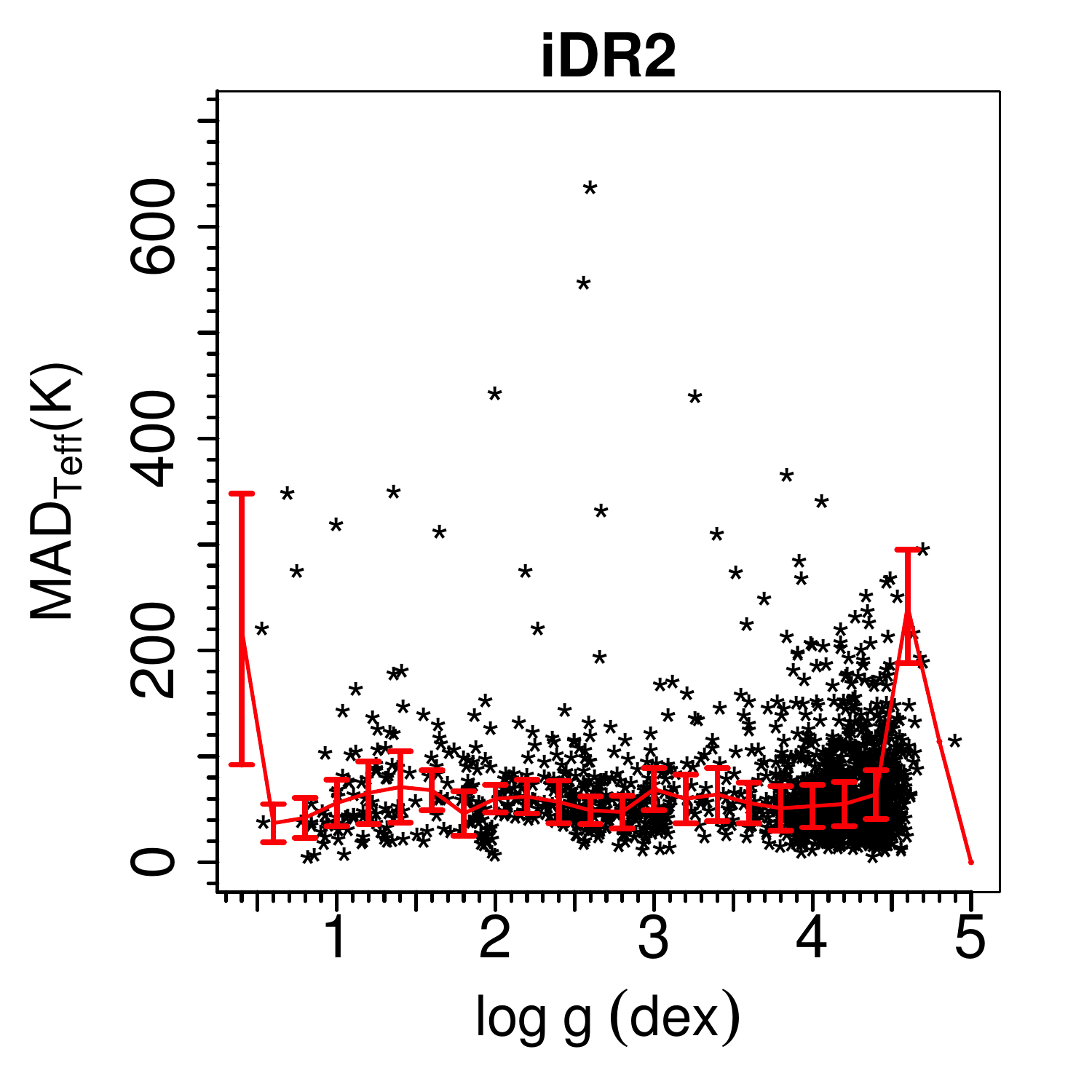}
\includegraphics[height = 5.5cm]{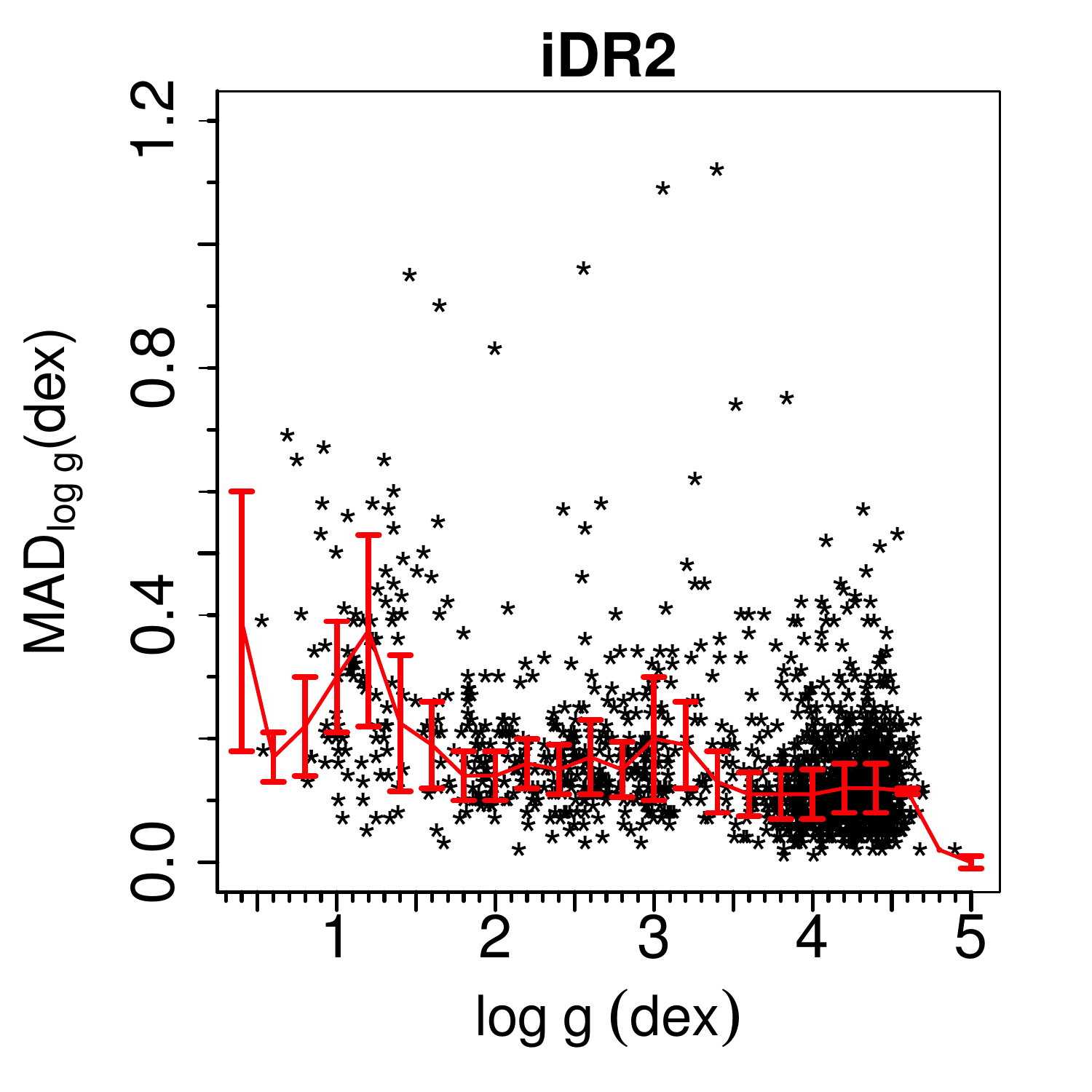}
\includegraphics[height = 5.5cm]{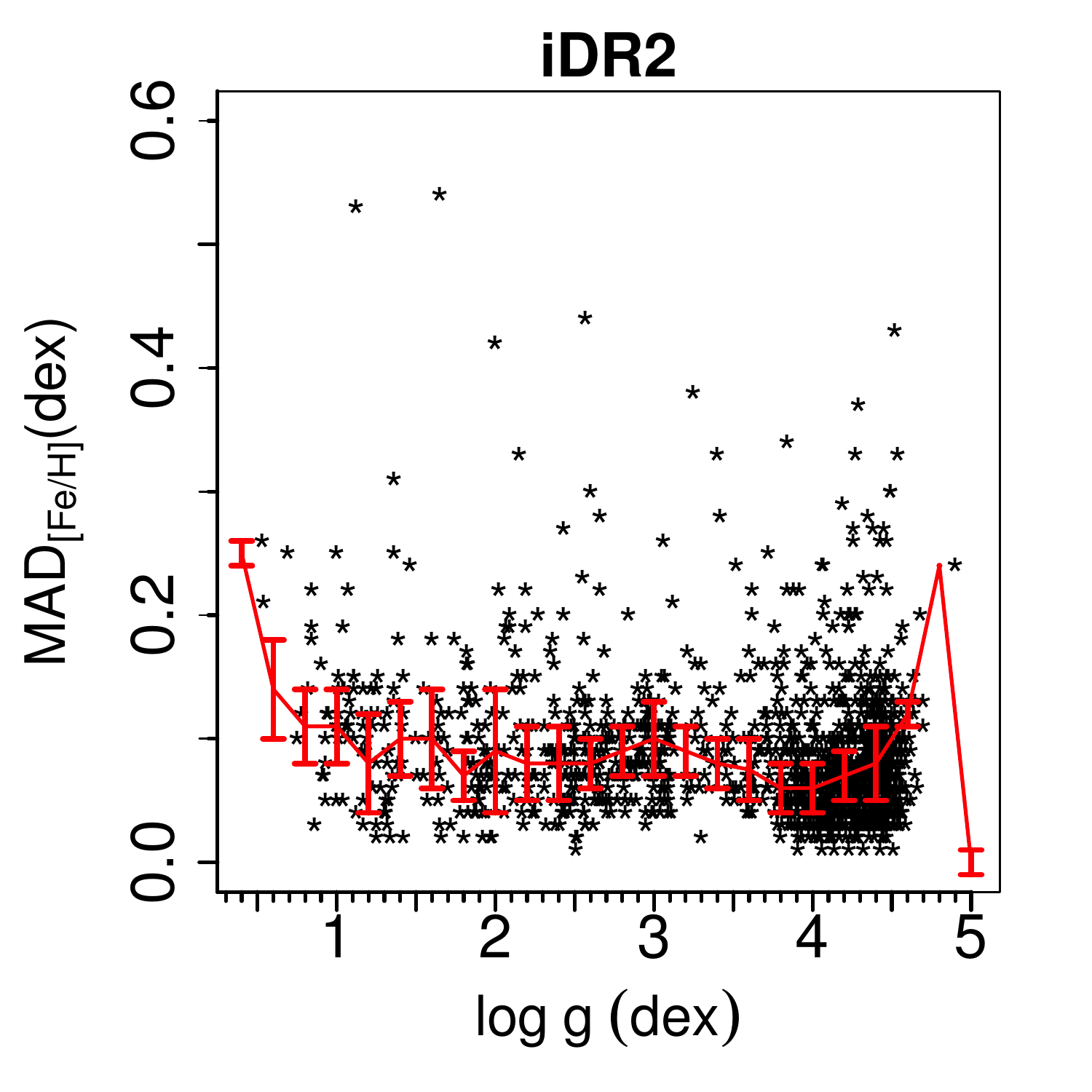}
\includegraphics[height = 5.5cm]{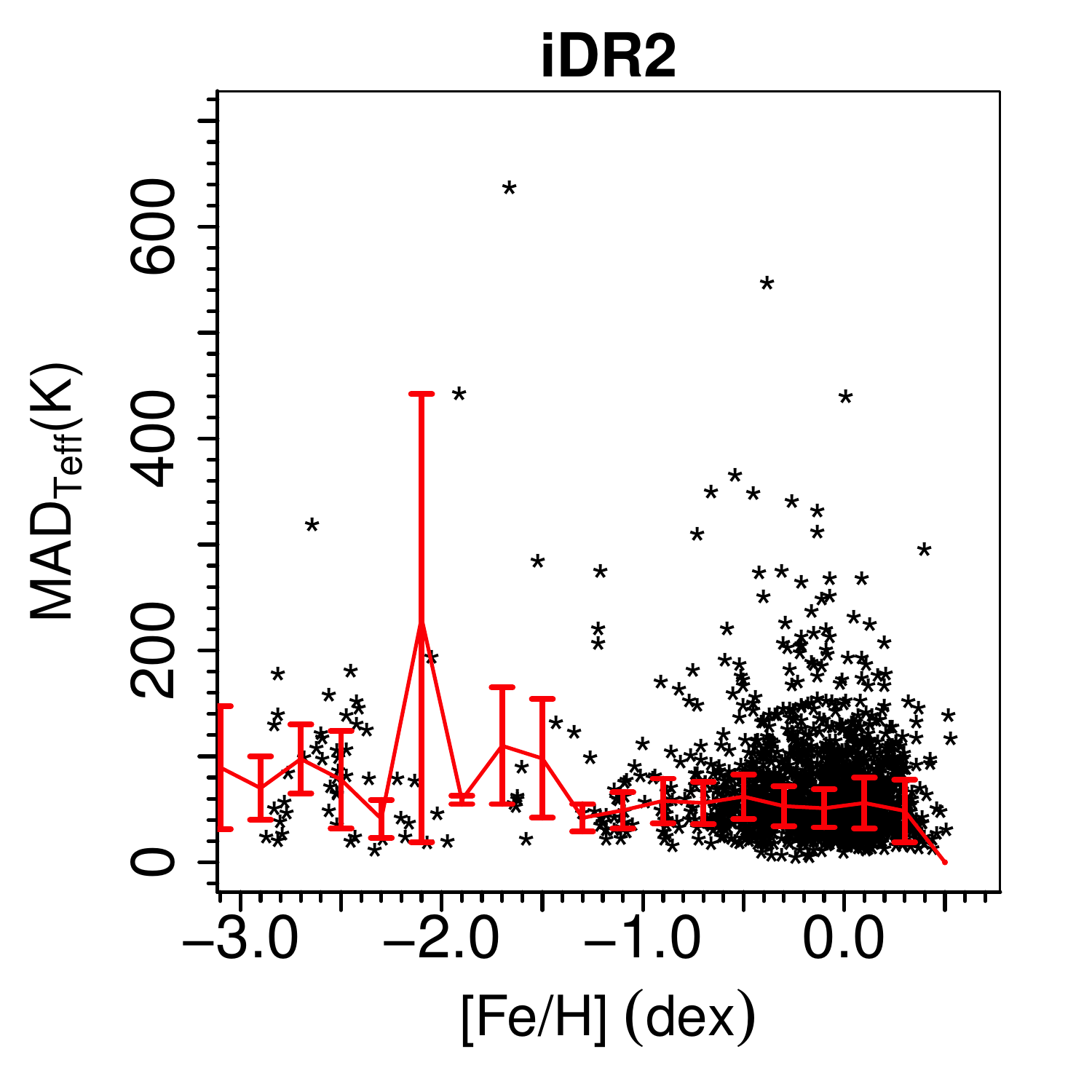}
\includegraphics[height = 5.5cm]{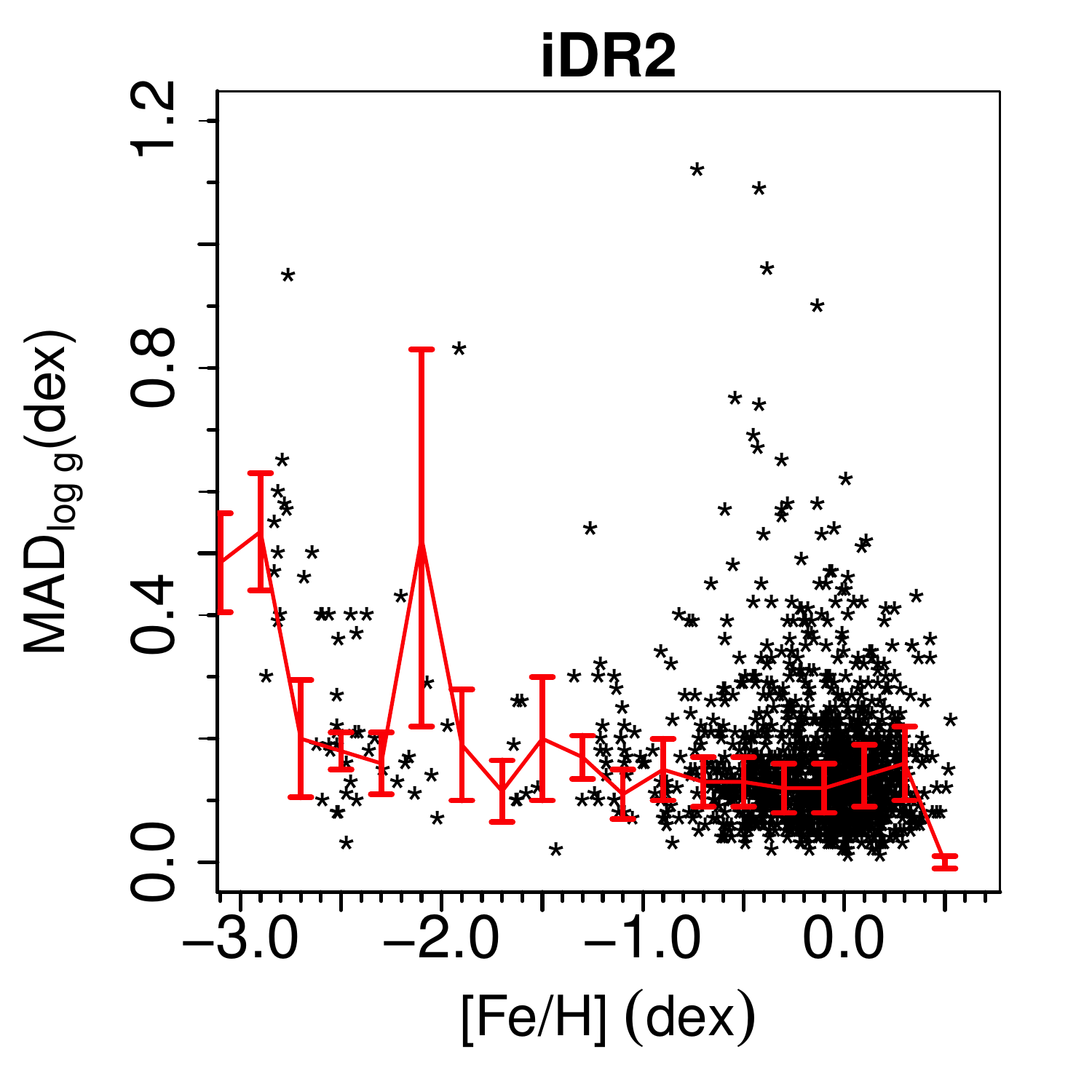}
\includegraphics[height = 5.5cm]{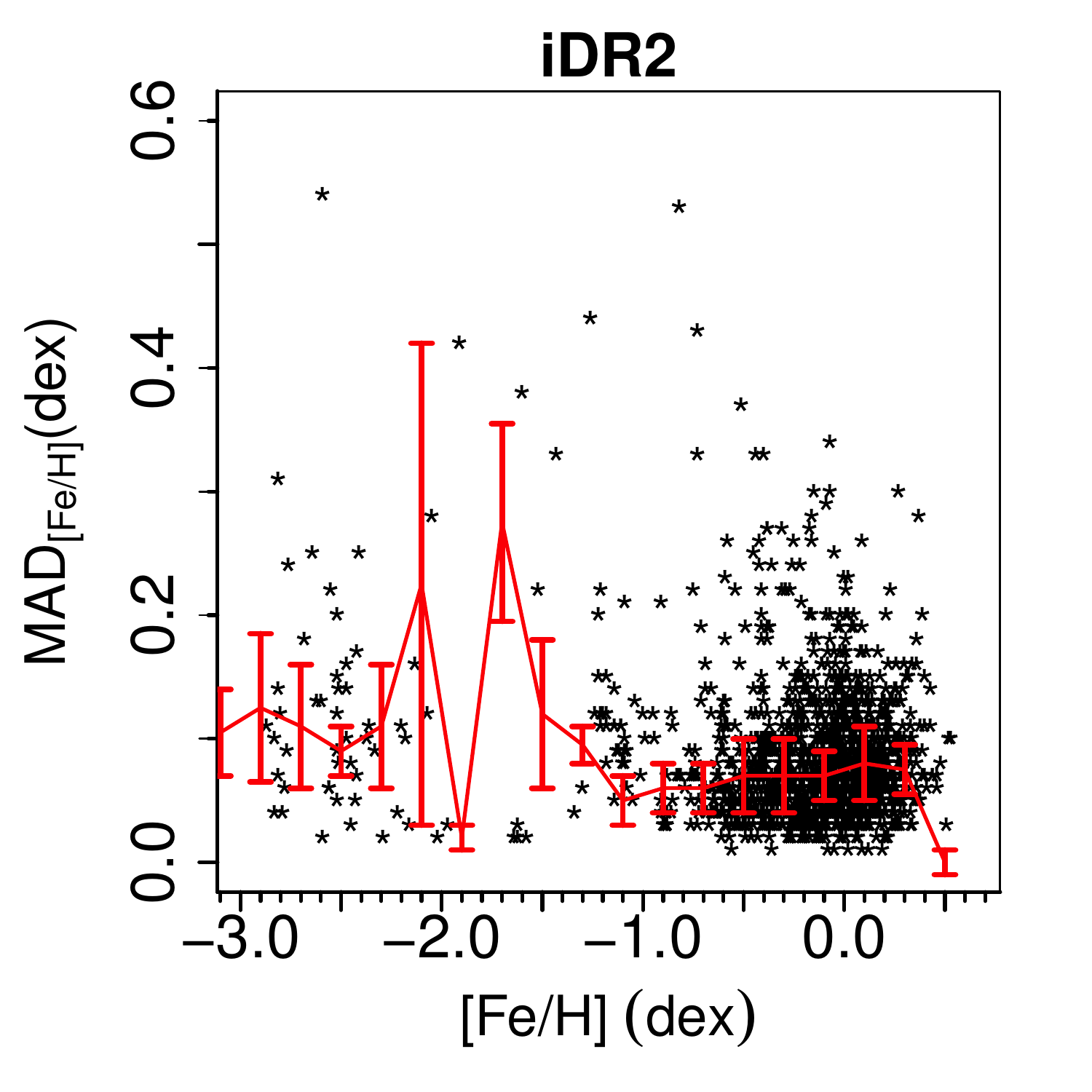}
 \caption{Dependency of the method-to-method dispersion with respect to the atmospheric parameters of the stars. The red line connects the median value in each bin of 100 K (top row), 0.20 dex (middle row), and 0.20 dex (bottom row). The error bars in the line represent the median absolute deviation.}\label{fig:madatm}%
\end{figure*}

Another consistency test is the metallicity determination for the stars in a given cluster. Assuming that all observed stars are cluster members and that there is no metallicity dispersion, it is expected that a given Node should recover very similar metallicities for all stars. Of course, these conditions are sometimes not fulfilled\footnote{For example, NGC 1851 seems to present a small dispersion in metallicity and star-to-star variations of s-process elements \citepads{2008ApJ...672L..29Y,2011A&A...533A..69C}}. In Fig. \ref{fig:fehclusters}, the mean metallicity obtained by each Node for each of the calibrating clusters is shown in comparison with the literature value. The error bars in the plot are the standard deviation of the mean. Non-members were not removed, and different Nodes were able to analyze a different number of stars in each cluster. Therefore the understanding of each Node result individually in this plot is complex, but the general behavior is very informative. In most cases, the dispersion in the metallicity values of a given Node is small and the average of the multiple Nodes agree within the dispersion bars. Cases like NGC 4833 and NGC 6752, where the dispersion within a given Node is large, are probably caused by non-members with very different metallicities from that of the cluster.

\subsection{Method-to-method dispersion}\label{sec:mad}

To compare the results of different Nodes and quantify the method-to-method dispersion of each parameter we decided to use the median and the associated MAD (median absolute deviation). The MAD is defined as the median of the absolute deviations from the median of the data and is given by:

\begin{equation}
{\rm }MAD = {\rm median}_{i} (| X_{i} - {\rm median}_{j}(X_{j})|)  \,.
\end{equation}

For the iDR2 results, the histograms of the method-to-method dispersions are shown in Fig. \ref{fig:histidr2}. The median values of the method-to-method dispersion are 55 K, 0.13 dex, and 0.07 dex for $T_{\rm eff}$, $\log g$, and [Fe/H], respectively. The third quartile of the distribution has values of 82 K, 0.19 dex, and 0.10 dex for $T_{\rm eff}$, $\log g$, and [Fe/H], respectively. These values indicate an overall excellent agreement between the multiple methods for 75\% of the results available in iDR2.

This agreement is obtained on absolute values of the parameters, not on relative ones, as we do not implement differential analyses. That all the different methods do not yield exactly the same results should not be surprising, given all the different factors involved in the analysis. Examples are the different ways to constrain the atmospheric parameters and the physics included in each different analysis code.

We recall here that the method-to-method dispersion is a measure of the precision of the results, i.e. the degree to which multiple methodologies can agree on the atmospheric parameters of a star. They are not the physical uncertainty of the values.

Figure \ref{fig:madsn} shows how the method-to-method dispersion of each atmospheric parameter ($T_{\rm eff}$, $\log~g$, and [Fe/H]) depends on the S/N of the spectrum. The plots show that there is a general trend of larger disagreements being found for smaller values of S/N, although outliers are found at any S/N value. But only for the lowest values of SN ($<$ 40) the dispersion tends to increase. Otherwise, for S/N $>$ 40 it tends to stabilize around a constant value ($\sim$ 50 K, 0.13 dex, and 0.07 dex for $T_{\rm eff}$, $\log g$, and [Fe/H], respectively). Perhaps more surprisingly, the plots also suggest that good agreement between different methods can be found even if the S/N is low, as the corners of low S/N and small dispersion in the panels are well populated.

\begin{figure*}
\centering
\includegraphics[height = 6cm]{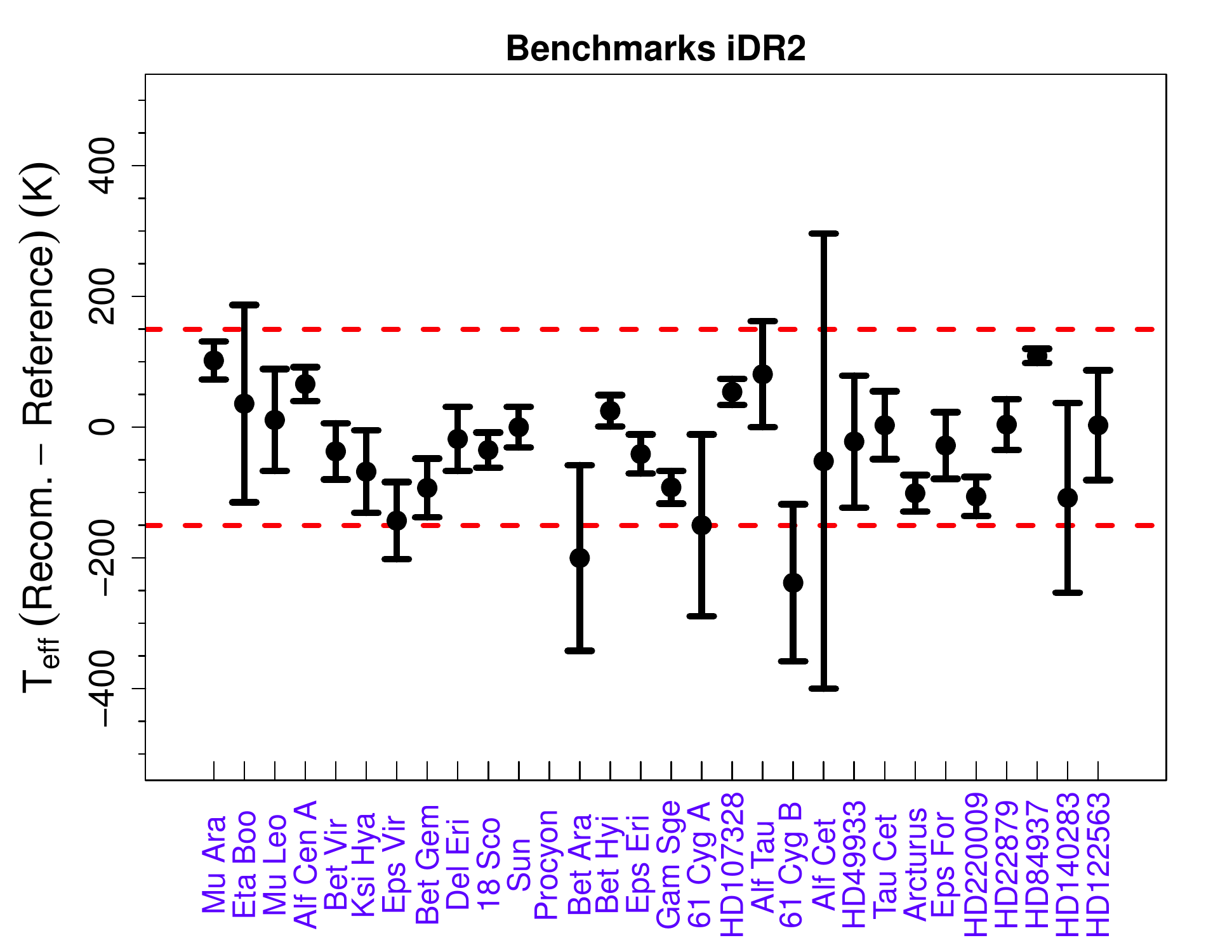}
\includegraphics[height = 6cm]{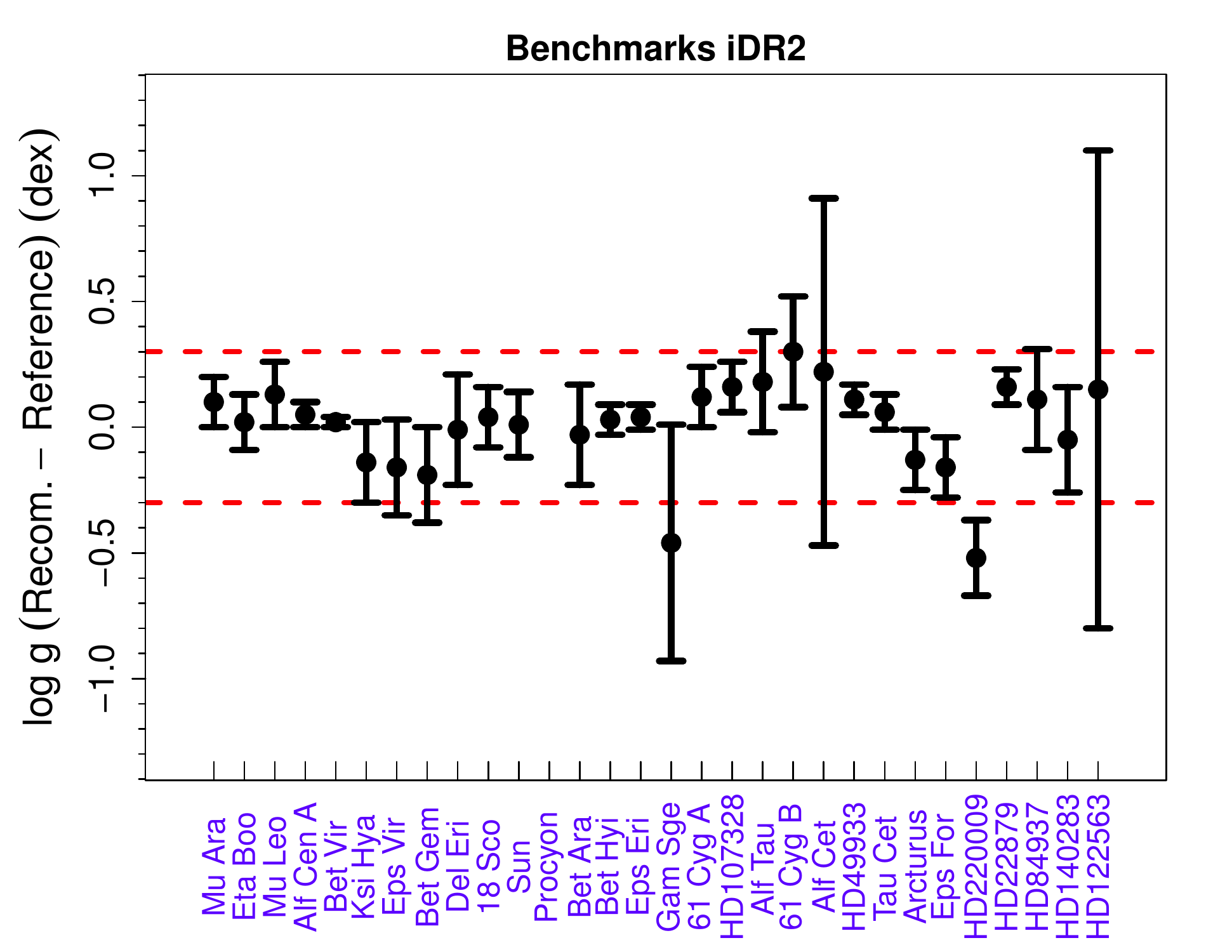}
\includegraphics[height = 6cm]{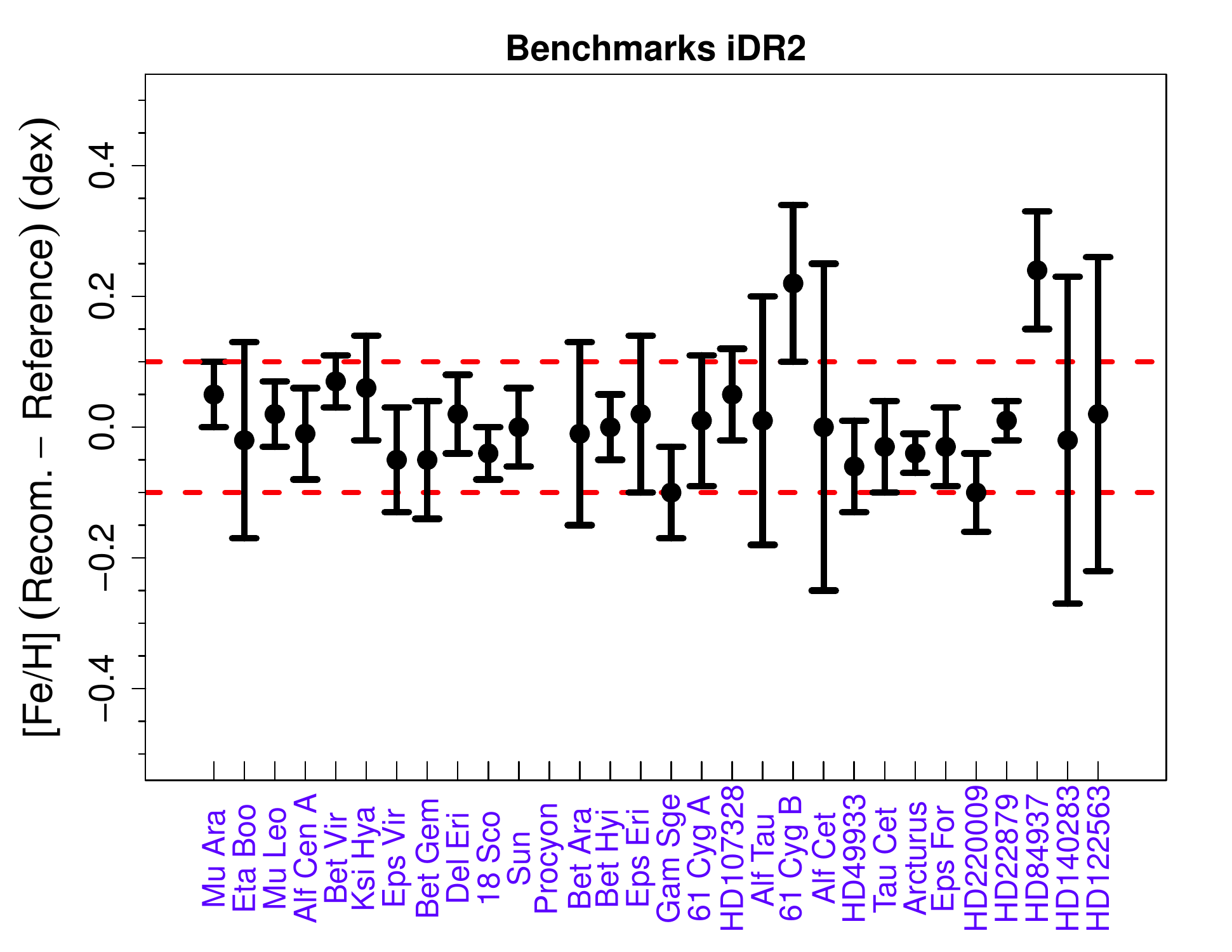}
 \caption{Difference between the recommended values of $T_{\rm eff}$, $\log g$, and [Fe/H] for the benchmark stars of iDR2 and the reference values. The error bars are the method-to-method dispersions. The stars are sorted in order of decreasing [Fe/H] (left to right). The dashed red lines indicate limits of $\pm$ 150 K for $T_{\rm eff}$, of $\pm$ 0.30 dex for $\log g$, and of $\pm$ 0.10 dex for [Fe/H].}\label{fig:benchcompidr2}%
\end{figure*}
\begin{table}[h]
\caption{\small{Node weights per region of the parameter space.}}\label{tab:weights}  
\centering
\small{
\begin{tabular}{cccc}
\hline\hline
Node & MRD & MRG & MPS \\
\hline
Bologna           & 1.000 &  0.546 &  -- \\ 
CAUP              & 0.971 &  0.495 &  -- \\
Concepcion        & 0.694 &  0.495 &  0.306 \\
EPINARBO          & 1.000 &  0.781 &   0.585 \\
IACAIP            & 0.862 & 0.901 & 0.935 \\
Liege             & 0.676 & 0.386 &  -- \\
LUMBA             & 1.000 &  0.602 &  0.758 \\
Nice              & 0.870 &  0.794 & 1.000 \\
OACT              & 0.741 &  0.585 &    -- \\
ParisHeidelberg   & 1.000 & 0.746 & 0.637 \\
UCM               & 0.893 &  0.214 &  -- \\
ULB   & -- & -- & -- \\
Vilnius           & 1.000 & 0.457 &  0.308 \\
\hline\hline
\end{tabular}
}
\end{table}

Figure \ref{fig:madatm} shows how the method-to-method dispersion of each atmospheric parameter depends on the atmospheric parameters themselves. Most of the panels do not show any significant trend. There seems to be an increase in the dispersion of the $\log~g$ values for cool stars ($<$ 4000 K) and for metal-poor stars ([Fe/H] $<$ $-$2.00 or $-$2.50), although part of it might be caused by low-number statistics. This suggests that precise results are found across almost the full parameter range of the stars analyzed here. It also indicates that to select good results, cuts in the atmospheric parameters themselves are not needed, cuts in the dispersion values are sufficient. Overall, these comparisons show that the bulk of the results are of very good quality.

\subsection{The recommended values}\label{sec:recom}

In this Section we describe the procedure used to define the recommended values of the atmospheric parameters of each star. The first step was a zeroth-order quality control of the results of each Node. Results that were excluded are those i) with very large error bars (above 900 K for $T_{\rm eff}$ and/or 1.50 dex for log $g$); ii) with microturbulence value equal to or below 0.00 km s$^{-1}$; iii) with surface gravity value above 5.00\,dex; iv) where the final Node result was the same as the input values of the method, indicating that the automatic analysis failed to converge; v) flagged as having other convergence problems.

Next, we used the results of the benchmark stars to weight the performance of each Node in the three different regions of the parameter space defined before: 1) metal-rich dwarfs, 2) metal-rich giants, and 3) metal-poor stars. For the benchmark stars in each one of these regions, we computed for each Node the average difference between the parameters it derived ($T_{\rm eff}$ and $\log~ g$) and the reference ones {(Table \ref{tab:nodediff})}.

These numbers are a measurement of the accuracy with which each Node can reproduce the reference atmospheric parameters, in each region of the parameter space. They were then used to assign weights to the Node results. If the average difference of the Node results was within 100 K for $T_{\rm eff}$ and within 0.20 dex for $\log~ g$, the Node was assigned a weight of \emph{1.00}. Thus, we are assuming that all Nodes that reproduce the values within these margins are equally accurate and their results should be fully taken into account. Nodes that are less accurate than that are assigned worse weights, in a linear scale, by dividing the average difference of its parameters by 100 K or 0.20 dex, for $T_{\rm eff}$ and $\log~ g$ respectively, and averaging these values.

\begin{table*}
\caption{Outcome of the analysis of the iDR2 data. Number of FGK-type stars observed with UVES with atmospheric parameters determined.}
\label{tab:analysisidr2} 
\centering
\begin{tabular}{lcl}
\hline\hline
Gaia-ESO type &  Number of stars & Comment \\ 
\hline 
Analyzed stars & 1447 & Gaia-ESO and archival data \\
Stars with results & 1301 & Gaia-ESO and archival data \\
Stars with results & 1268 & Only Gaia-ESO data \\
GES\_MW & 906 & Milky Way fields \\
GES\_CL & 233 & Open clusters fields \\
GES\_SD & 129 & Calibration targets \\
AR & 33 & Archival data \\
\hline
\end{tabular}
\end{table*}

The weights are computed per Node and per region of the parameter space (Table \ref{tab:weights}). The results of each star are then combined in a weighted median, taking into account the Node weight of the parameter-space region to which they belong. For that, the multiple estimates of the parameter are ranked, and the adopted value is the one interpolated to a weighted percentile of 50\%. The weighted median is given by:

\begin{equation}
wei\_median = Param_{k} + \frac{50 - P_{k}}{P_{k+1} - P_{k}} (Param_{k+1}-Param_{k}) ,
\end{equation}

\noindent where $P_{k}$ is the percentile rank of parameter \emph{k}, and is given by:

\begin{equation}
P_{k} = \frac{100}{Sum_{n}} \left(Sum_{k} - \frac{normal_{k}}{2} \right) ,
\end{equation}

\noindent where the weights are normalized on a star by star basis:

\begin{equation}
Normal_{i} = \frac{weight_{i}}{\sum\limits_{i=1}^n weight_{i}} ,
\end{equation}

\noindent the total sum of weights is then one:

\begin{equation}
Sum_{n} = \sum\limits_{k=1}^{n} weight_{k} = 1.0 , 
\end{equation}

\noindent and the partial sum of the weights is:

\begin{equation}
Sum_{i} = \sum\limits_{k=1}^{i} weight_{k}.
\end{equation}

\begin{table}
\caption{Systematic errors of the atmospheric parameters for the iDR2 data set.}
\label{tab:paramuncer} 
\centering
\begin{tabular}{crrr}
\hline\hline
Type of star &  $\sigma_{T_{\rm eff}}$ & $\sigma_{\log~g}$ & $\sigma_{[Fe/H]}$ \\ 
\hline 
Metal-rich dwarfs & 50 K & 0.10 dex & 0.05 dex \\ 
Metal-rich giants & 100 K & 0.25 dex & 0.05 dex \\
Metal-poor stars & 50 K & 0.15 dex & 0.10 dex \\
\hline
\end{tabular}
\end{table}

Thus, for iDR2 the steps to obtain the recommended parameters can be summarized as:

\begin{enumerate}
\item A zeroth order quality control is performed, removing very uncertain results.
\item The accuracy of the Node results is judged using the available benchmark stars as reference. Weights are assigned, according to how well the Nodes can reproduce the reference values in a given region of the parameter space.
\item Further consistency tests are conducted using the calibration clusters.
\item The weighted-median value of the validated results is adopted as the recommended value of that parameter.
\item The MAD is adopted as an indicator of the method-to-method dispersion (analysis precision).
\item The number of results on which the recommended value is based is also reported.
\end{enumerate}

Table \ref{tab:analysisidr2} summarizes the number of stars for which atmospheric parameters were determined during iDR2. The analysis of about 10\% of the stars was not completed for different reasons (e.g. high-rotation, double-lined signatures, too low S/N, emission lines). 

A comparison of the recommended values of the atmospheric parameters of the benchmark stars (computed as described above) with the reference values is shown in Fig. \ref{fig:benchcompidr2}. It can be seen that the recommended atmospheric parameters of the benchmark stars agree well with the reference values for the majority of the stars, i.e. within $\pm$ 150 K for $T_{\rm eff}$, $\pm$ 0.30 dex for $\log ~ g$, and $\pm$ 0.10 dex for [Fe/H]. The results become more uncertain than that for cooler stars ($T_{\rm eff}$ $\lesssim$ 4200 K), as seen for HD 220009, Bet Ara, 61 Cyg B, Alf Cet, and Gam Sge.

The comparison with the benchmark stars together with the results for the clusters (see Fig. \ref{fig:clusters}) illustrates the general good quality of the Gaia-ESO recommended results. These final recommended results are the ones whose use we advise for scientific publications. In Fig. \ref{fig:hridr2} we show the final $T_{\rm eff}$-$\log g$ plane of the stars included in the iDR2 results.

\subsection{Systematic errors}\label{sec:paramuncer}

We estimate the systematic errors of the atmospheric parameters in iDR2 using the Gaia benchmark stars. These errors are a measurement of the systematic difference between reference and recommended values of the atmospheric parameters. In other words, they are the biases and measure the average accuracy of the Gaia-ESO atmospheric parameters. These values are provided in addition to the method-to-method dispersion, as they quantify a different kind of uncertainty of the results.

The systematic errors were computed in the three regions of the parameter space defined before (i.e. metal-rich dwarfs, metal-rich giants, and metal-poor stars). They are the average of the absolute value of the difference between the reference and recommended parameters for the benchmark stars in each of these regions. To avoid reporting unrealistic small values, we adopt as lower limit values of 50 K, 0.10 dex, and 0.05 dex for $T_{\rm eff}$, $\log~g$, and [Fe/H], respectively. We do that because: 1) we are reporting average values; 2) the benchmark stars were observed with much higher S/N than the typical Gaia-ESO target, and these values could be S/N dependent, and 3) the reference parameters themselves have errors, which were not taken into account in this calculation. The final values are listed in Table \ref{tab:paramuncer}.

\subsection{The effect of the number of Nodes}\label{sec:number}

\begin{figure*}
\centering
\includegraphics[height = 6cm]{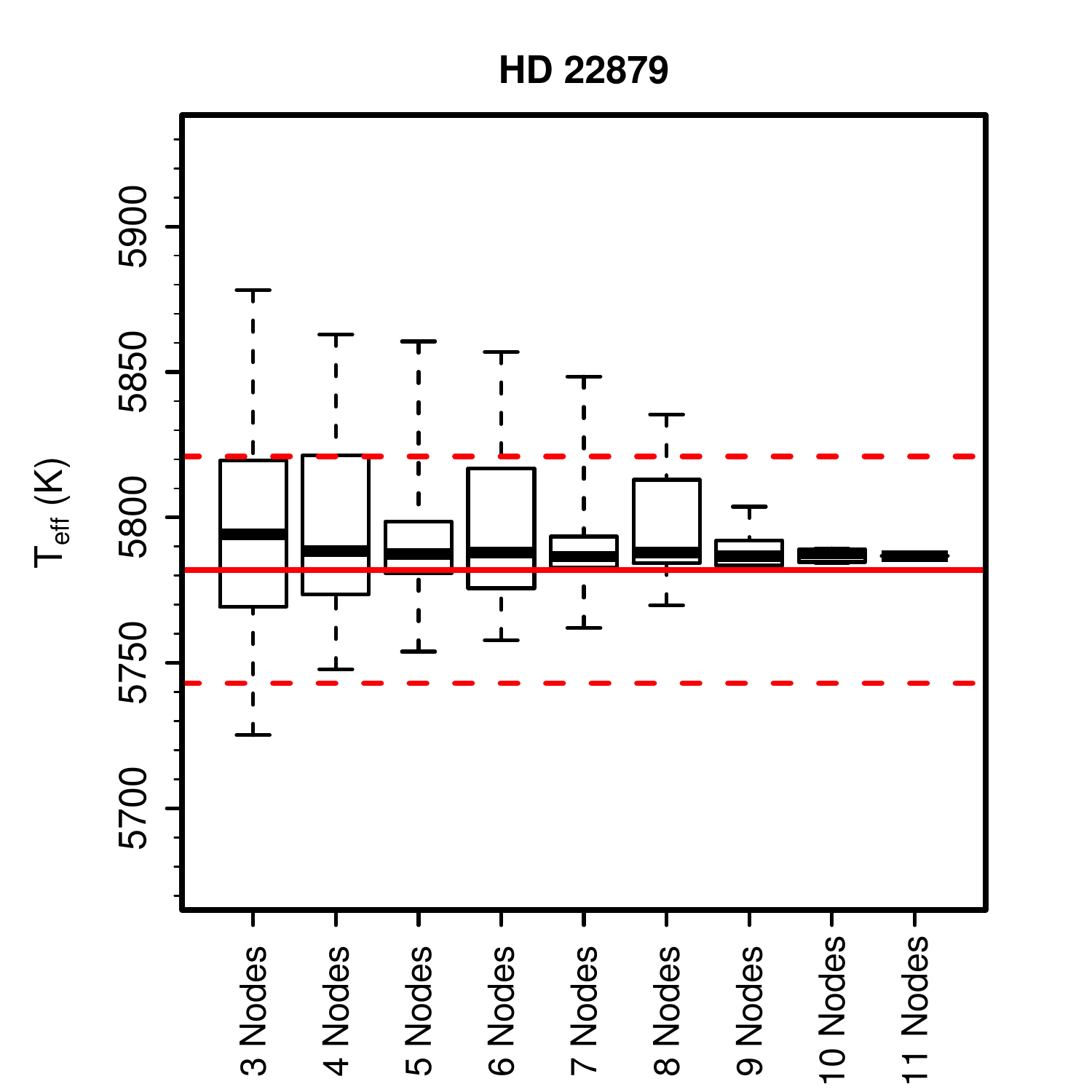}
\includegraphics[height = 6cm]{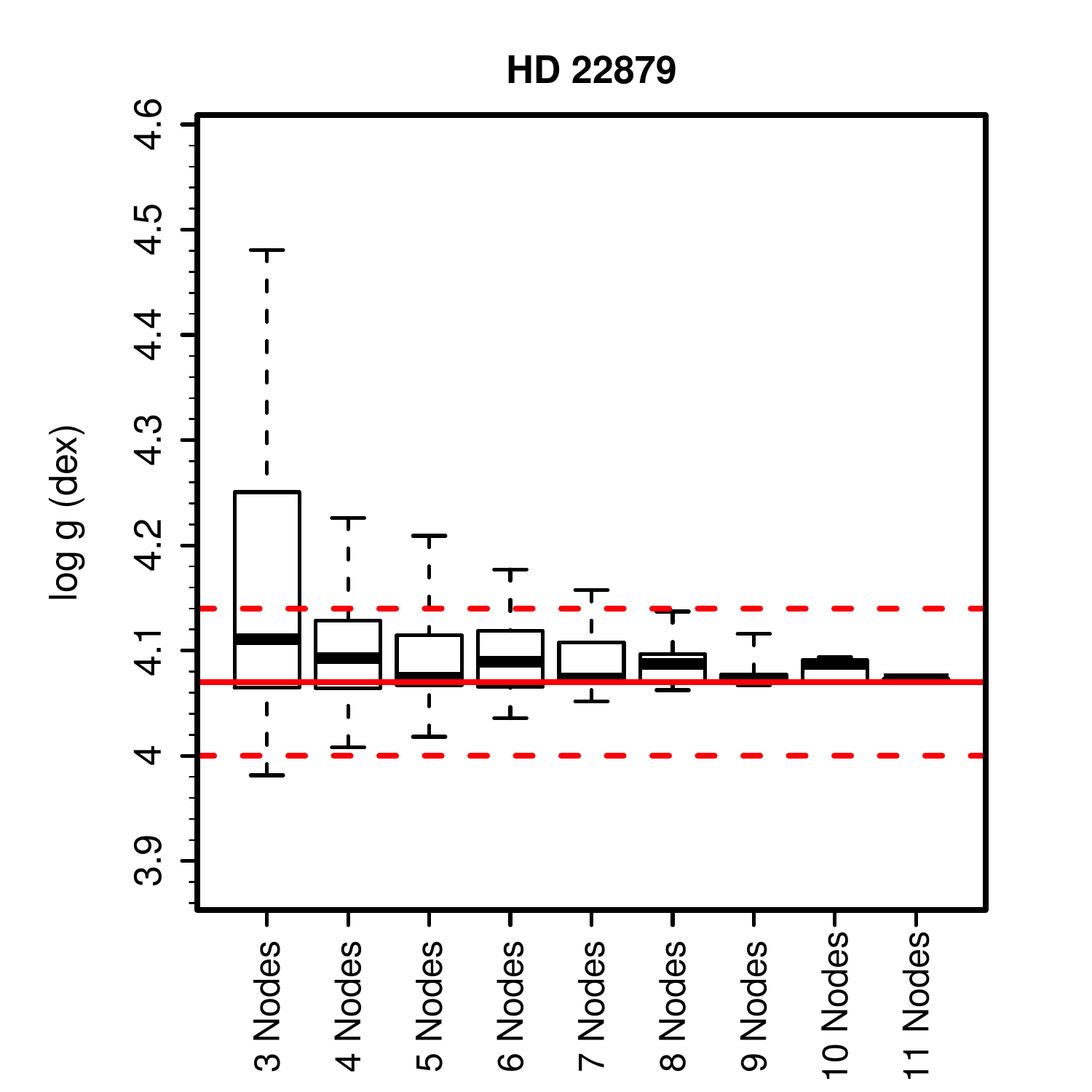}
\includegraphics[height = 6cm]{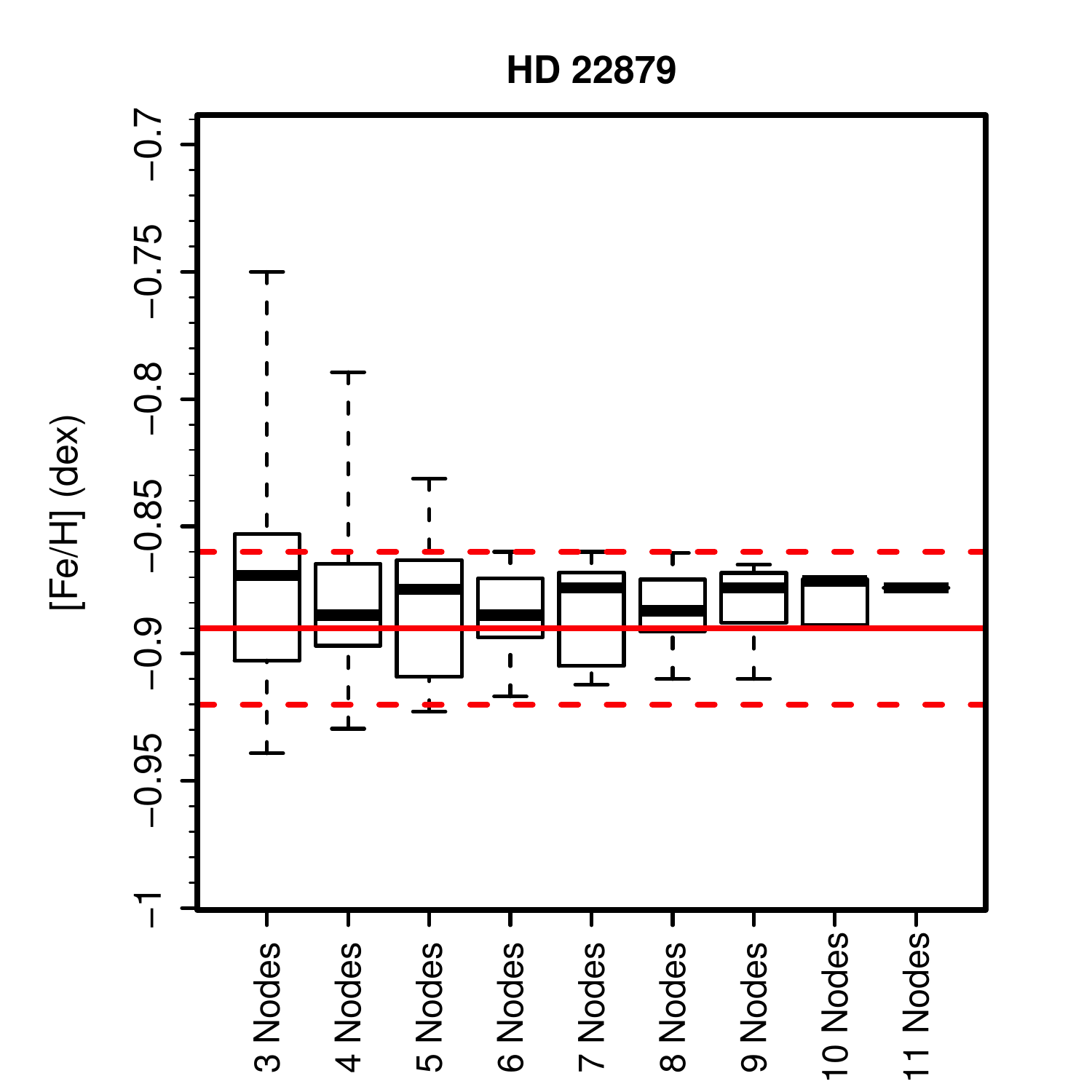}
\includegraphics[height = 6cm]{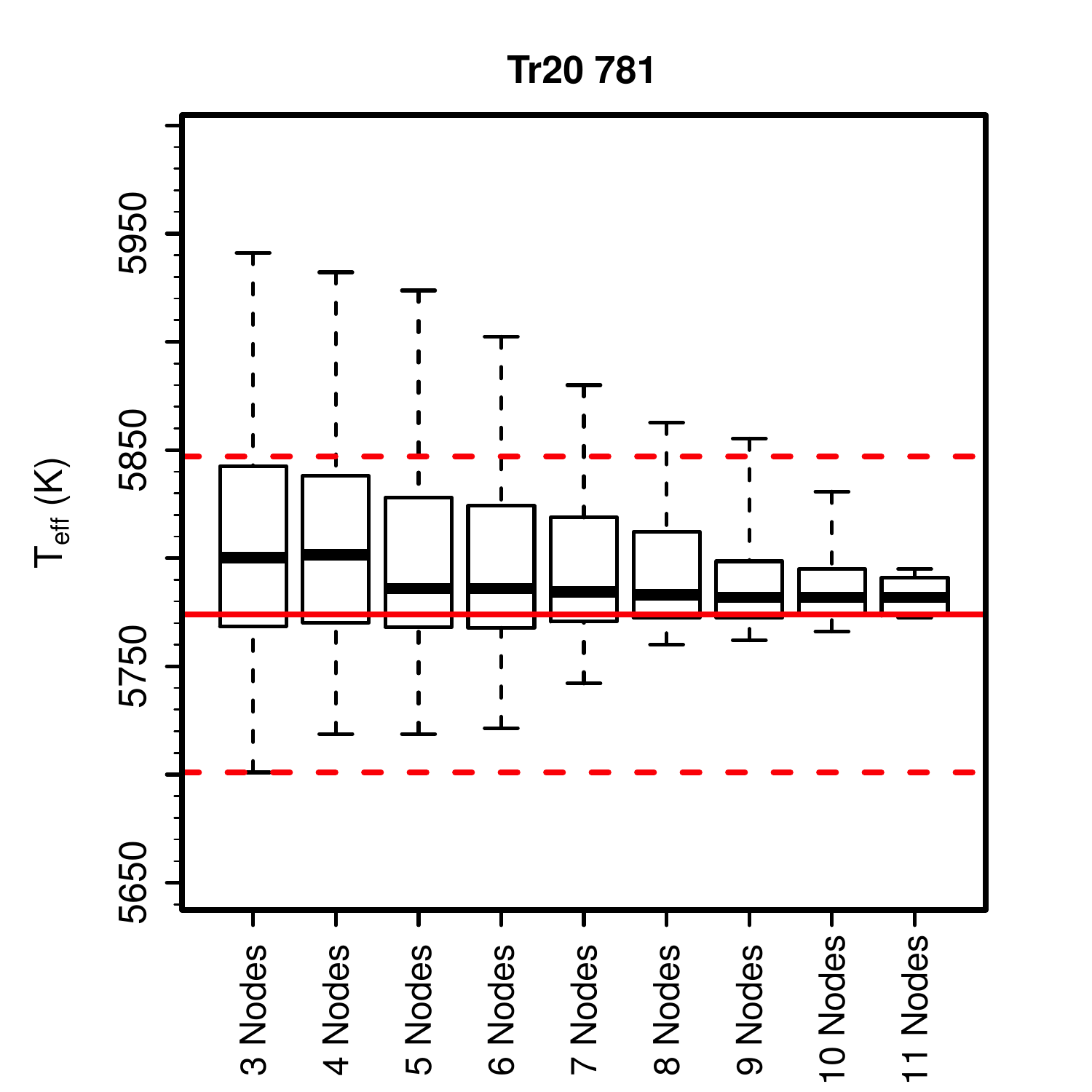}
\includegraphics[height = 6cm]{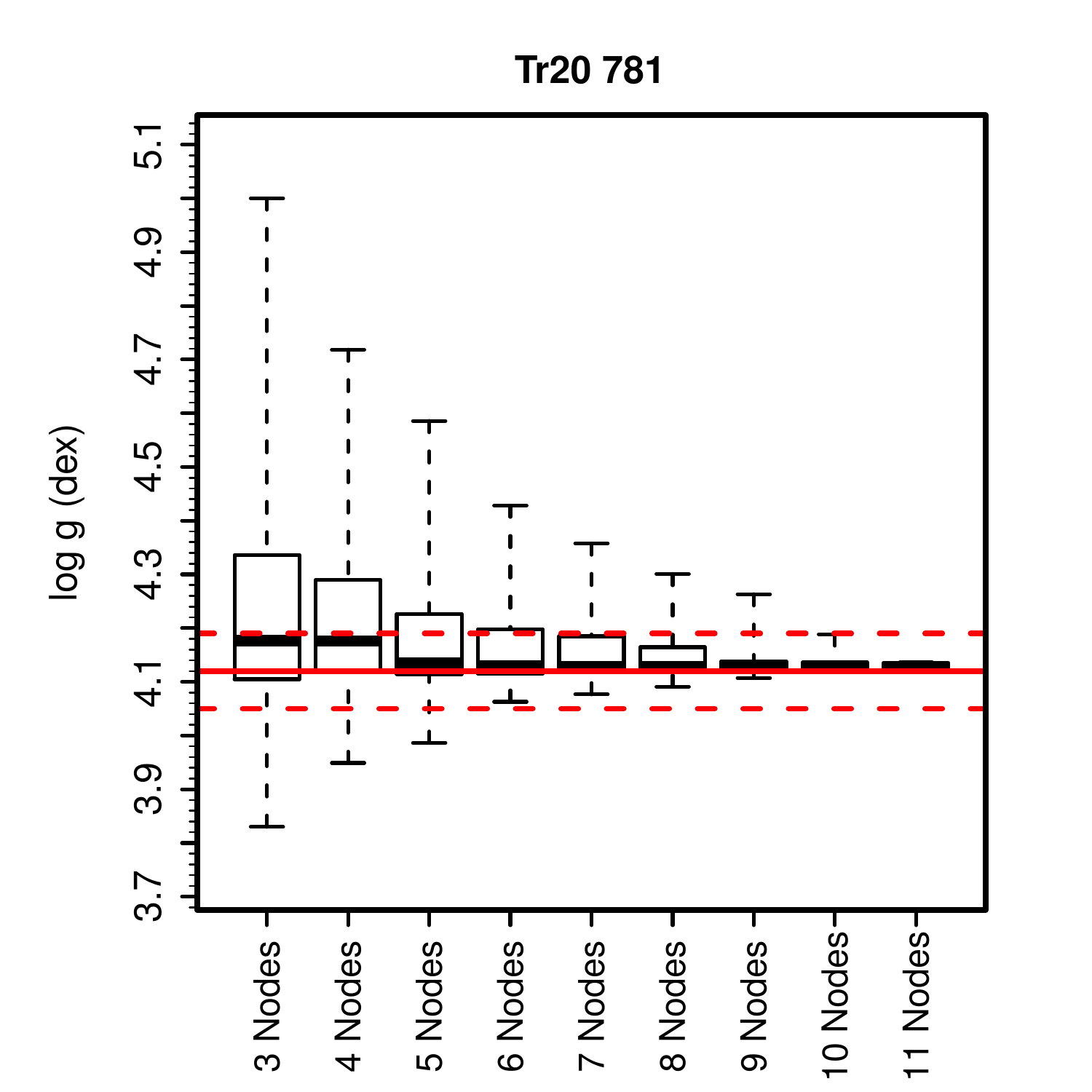}
\includegraphics[height = 6cm]{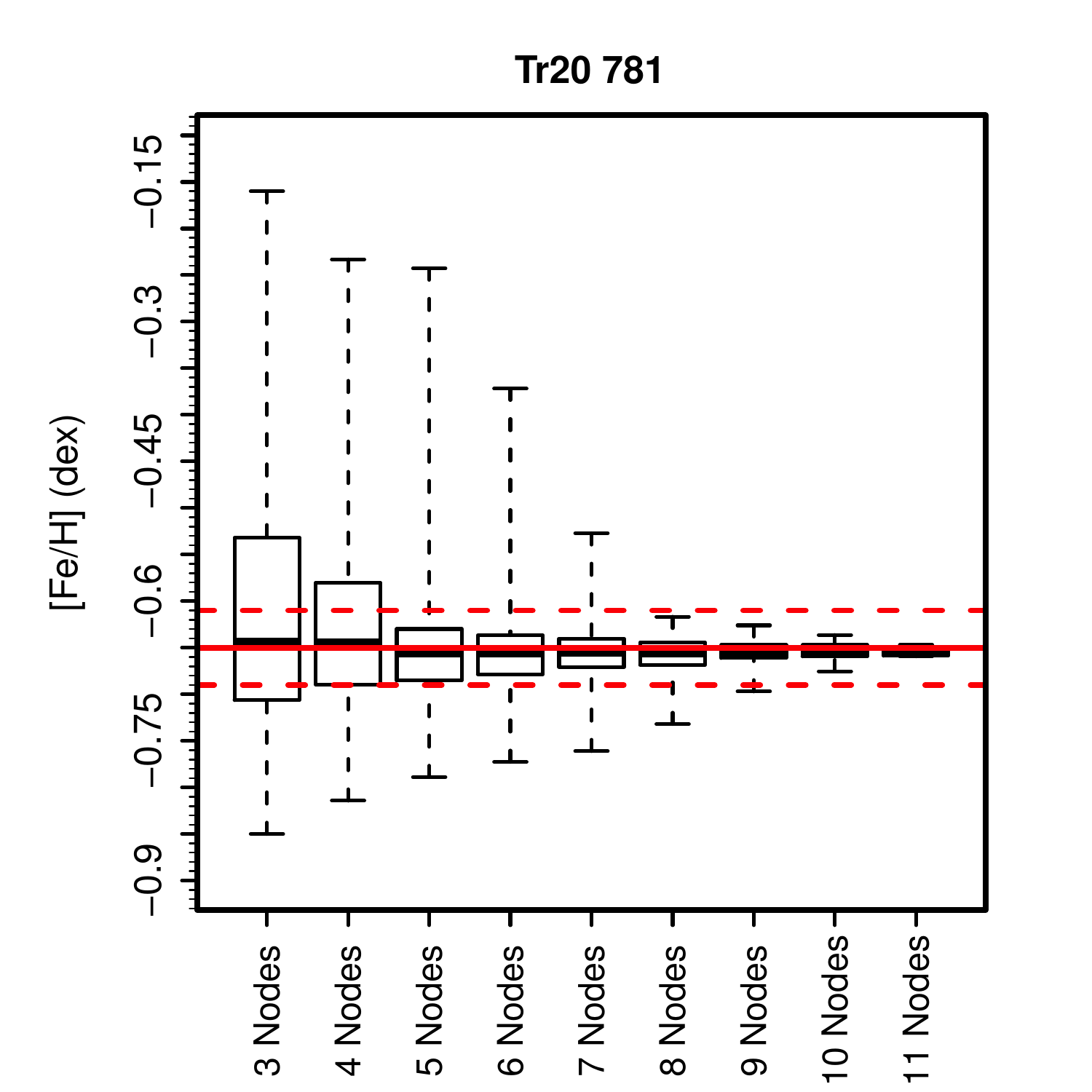}
 \caption{Histograms of the median values of randomly selected Node results. These histograms display the effect on the recommended parameters caused by changing the number of Nodes that contribute to the final value (see text for full explanation). The red solid line indicate the final recommended parameter, the red dashed lines indicate the final 1$\sigma$ method-to-method dispersion. The top panels show the case of HD 22879, the bottom panels the case of Tr20 781. In each boxplot, the thicker solid line indicates the median of the distribution, the box extends from the first to the third quartile, and the dashed lines extend to the extreme values.}\label{fig:numnodes}
 \end{figure*}

The number of Nodes contributing to the final recommended parameters varies from star to star. There are different reasons for that, including difficulties for a given method to deal with a certain kind of star. That raises the question of how homogeneous the results are as a whole.

To answer that question, we ran the following test for the stars which have results from 10 or more Nodes. First, we randomly select a number of results for that star. Second, we compute what would be the final recommended parameters based on only those selected results, using the same weighted-median approach. We repeat the random selection 1000 times, to build a distribution of the final results (and to understand which results are more likely and what is the full range of possible values). The exercise was repeated varying the number of Nodes contributing to the final results from 3 to 11 (12 is the maximum number of Nodes).

The results are plotted in Fig. \ref{fig:numnodes} for two stars, HD 22879 and Tr20 781 (also discussed in Section \ref{sec:eqw}). For each case of different number of Nodes, a boxplot with the distribution of the final weighted medians is shown. The red lines indicate the recommended parameter (using all available results) and its method-to-method dispersion. 

The comparison shows that: 
\begin{enumerate}
\item Irrespective of the number of Nodes used, most of the time, the weighted median of the random selection will agree with the final recommended value within the uncertainties.
\item Nevertheless, the fewer the number of Nodes used, the broader the distributions get. Meaning that the chance of a spurious final recommended parameter increases.
\item When the number of Nodes increases, the distribution tends to get narrower.
\end{enumerate}

These comparisons indicate that, if outliers are not present, the majority of the recommended results based on few Nodes will agree well with those based on many Nodes. Fluctuations on the final value are mostly within the uncertainties. This is a very important result lending confidence to our final recommended values. It stems from the effort to tie the final parameter scale to the Gaia benchmark stars. The results as a whole are homogeneous, within their quoted precision.

Nevertheless, some outlier Node results might be present. In the presence of outliers, the chance of the recommended value getting less accurate increases the fewer the number of Nodes used. For a large number of results, the median is a very robust measurement not affected by the presence of outliers, but not when only a few results are available. 

The strength of using multiple analyses is highlighted here, as they help to uncover which are the outliers and minimize their effect on the final recommended value. A higher number of Nodes is also needed to better constrain the confidence on the precision of the final result.

\begin{figure*}
\centering
\includegraphics[height = 6cm]{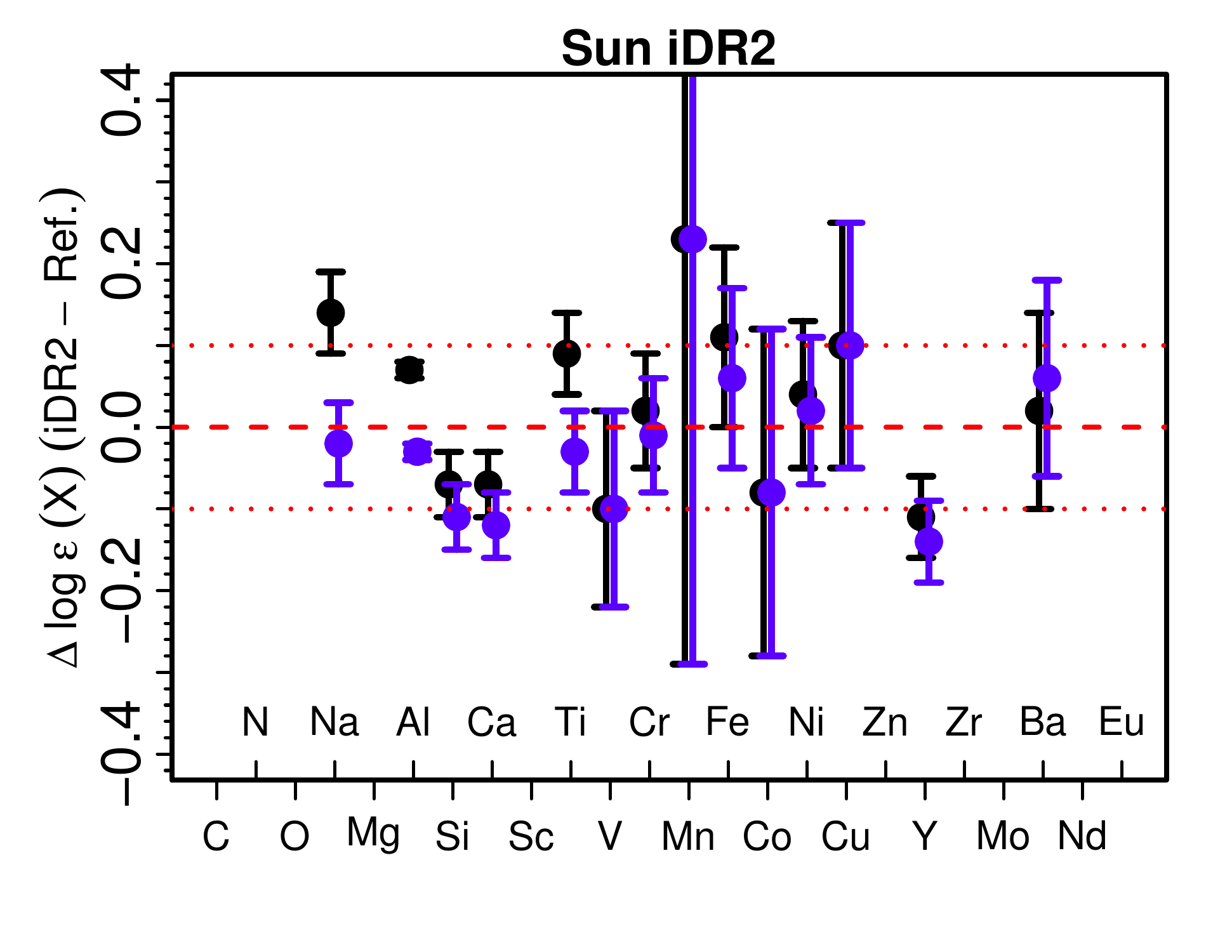}
\includegraphics[height = 6cm]{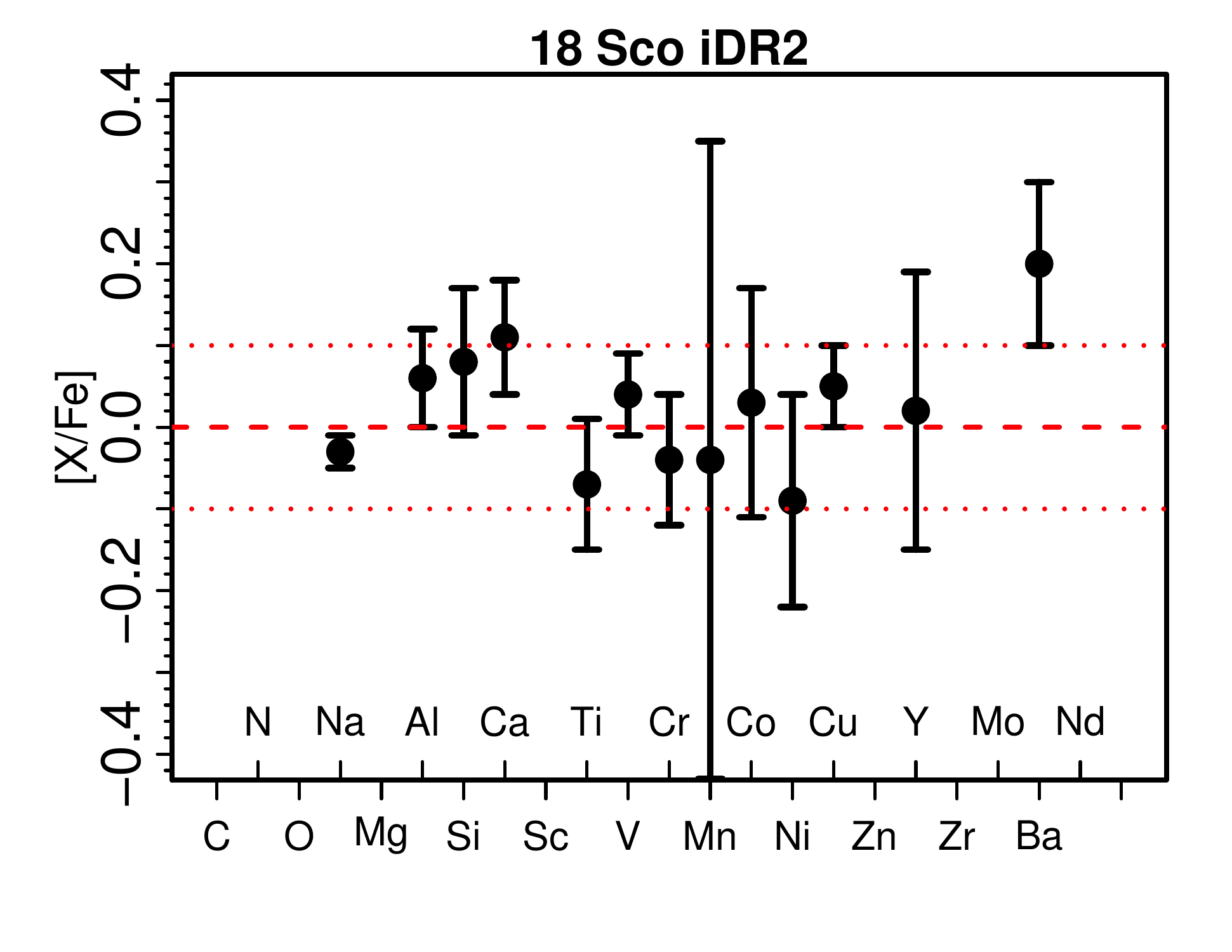}
\includegraphics[height = 6cm]{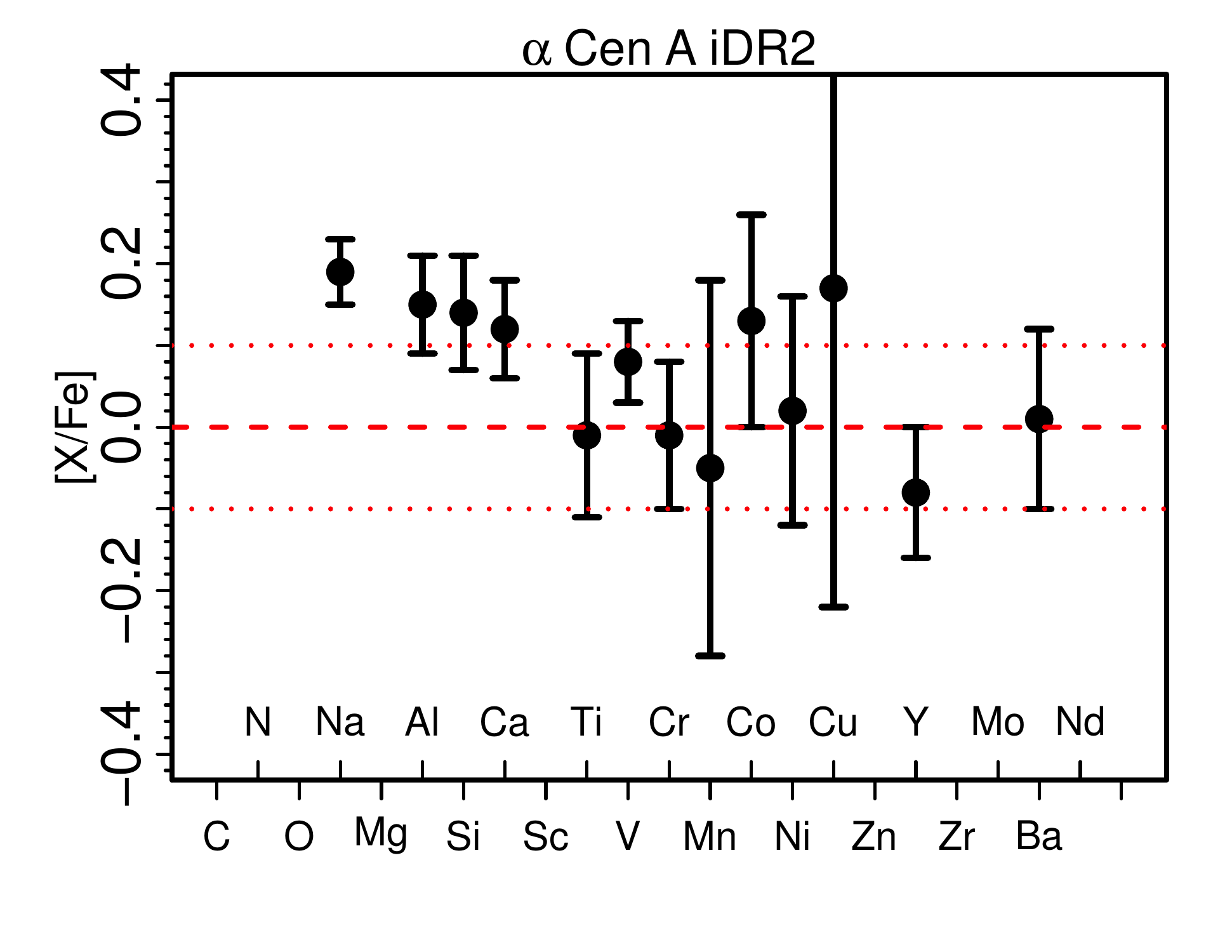}
\includegraphics[height = 6cm]{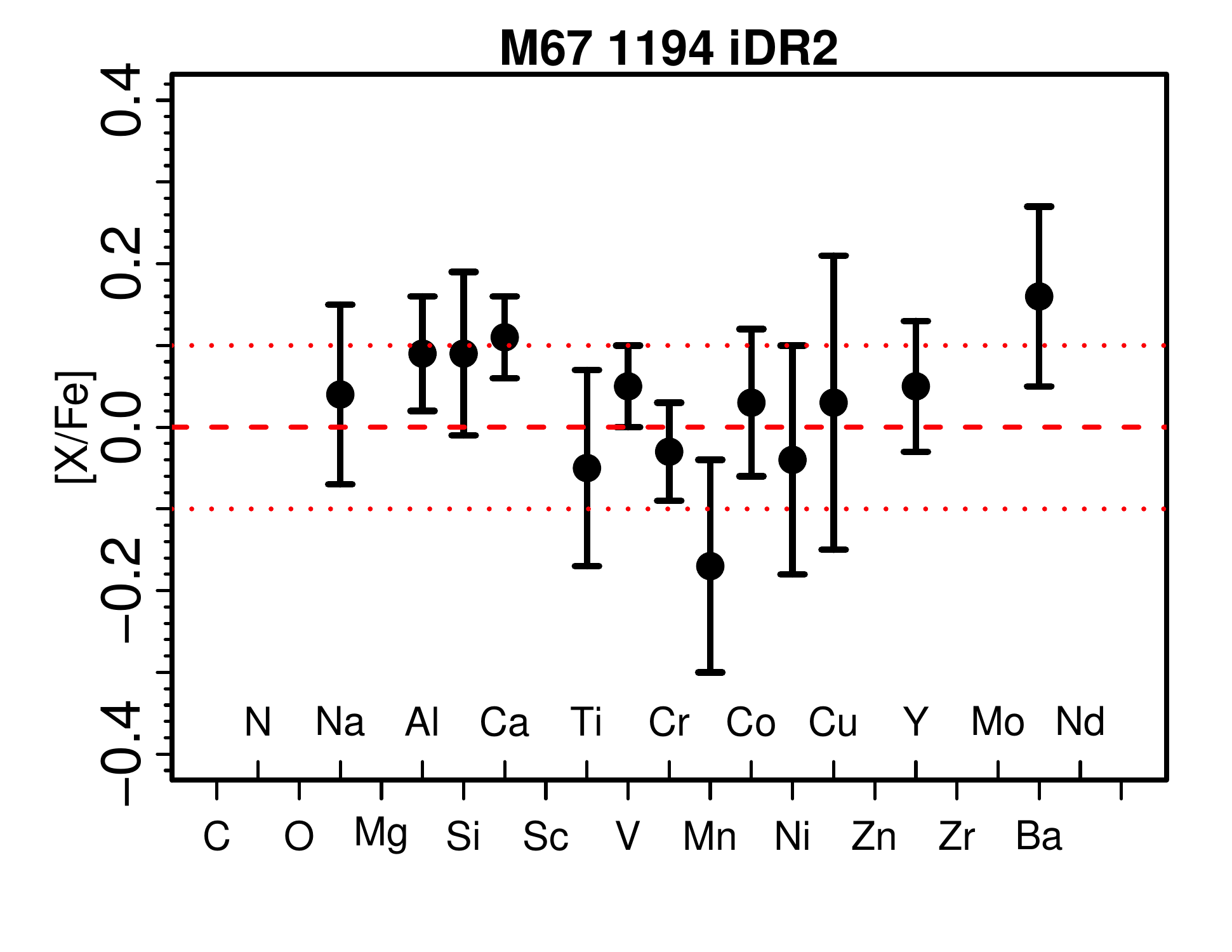}
 \caption{Abundance pattern of the Sun and selected solar analogues. \emph{Top left:} The results for the FLAMES solar spectrum for which $T_{\rm eff}$ = 5826 K, $\log g$ = 4.50, [Fe/H] = $-$0.03, and $\xi$ = 1.05 km s$^{-1}$ were derived. In black the iDR2 results are compared to the solar abundances of \citetads{2007SSRv..130..105G} and in blue to \citetads{1998SSRv...85..161G}. \emph{Top right:} The abundance pattern of the solar twin 18 Sco as derived here in comparison to the reference solar abundances computed in this work. For this star the following atmospheric parameters were derived: $T_{\rm eff}$ = 5782 K, $\log g$ = 4.39, [Fe/H] = 0.05, and $\xi$ = 1.04 km s$^{-1}$. \emph{Bottom left:} The abundance pattern of the solar analogue $\alpha$ Cen A. The reference solar abundances are the ones derived here. For this star the following atmospheric parameters were derived: $T_{\rm eff}$ = 5781 K, $\log g$ = 4.26, [Fe/H] = 0.25, and $\xi$ = 1.21 km s$^{-1}$. \emph{Bottom right:} The abundance pattern of the solar twin in the M 67 open cluster, star YBP 1194, as derived here. The reference solar abundances are the ones derived here. For this star the following atmospheric parameters were derived: $T_{\rm eff}$ = 5759 K, $\log g$ = 4.41, [Fe/H] = $-$0.01, and $\xi$ = 1.06 km s$^{-1}$. In all plots, the error bars are the method-to-method dispersion. The dashed line represents $\Delta$ $\log \epsilon$(X) = 0.00 dex or [X/Fe] = 0.00 dex, the dotted lines represent $\Delta$ $\log \epsilon$(X) = $\pm$0.10 dex or [X/Fe] = $\pm$0.10 dex. Abundances from this work are from the neutral species, except for Sc, Y, Zr, Ba, Nd, and Eu, for which they are from the ionized species, and from C and N for which they are from molecules.}\label{fig:abunsun}%
\end{figure*}

To select the best quality results, cuts on the precision should be enough for most applications. When further accuracy is needed, we recommend a cut based on the number of Nodes providing results. This cut will select out most of the results that have a higher chance of being far from the correct parameter value. The plots and the tests seem to indicate that by using five Nodes we can ensure that the majority of the results ($>$ 50\%) will be close to the real value.

\section{Abundances}\label{sec:abun}

\begin{table*}
\caption{Solar abundances derived in iDR2 in comparison to the solar abundances of \citetads{2007SSRv..130..105G}. Abundances from this work are from the neutral species, except for Sc, Y, and Ba, for which they are from the ionized species.}
\label{tab:sunabun} 
\centering
\begin{tabular}{ccc|ccc}
\hline\hline
Element & Abundance & Abundance & Element & Abundance & Abundance \\ 
     & This work & Grevesse et al. &  & This work & Grevesse et al. \\
\hline 
    C & -- & 8.39 & Mn  & 5.62 $\pm$ 0.52 & 5.39 \\
N & -- & 7.78 & Fe & 7.56 $\pm$ 0.11 & 7.45 \\
O & -- & 8.66 & Co & 4.84 $\pm$ 0.20 & 4.92 \\
 Na & 6.31 $\pm$ 0.05 & 6.17 & Ni & 6.27 $\pm$ 0.09 & 6.23 \\ 
 Mg & -- & 7.53 & Cu &  4.31 $\pm$ 0.15 & 4.21 \\ 
 Al & 6.44 $\pm$ 0.01 & 6.37 & Zn & -- & 4.60 \\ 
 Si & 7.44 $\pm$ 0.04 & 7.51 & Y & 2.10 $\pm$ 0.05 & 2.21 \\ 
 Ca & 6.24 $\pm$ 0.04 & 6.31 & Zr & -- & 2.58 \\ 
 Sc & 3.29 $\pm$ 0.11 & 3.17 & Mo & -- & 1.92 \\ 
 Ti & 4.99 $\pm$ 0.05 & 4.90 & Ba & 2.19 $\pm$ 0.12 & 2.17 \\
  V & 3.90 $\pm$ 0.12 & 4.00 & Nd  & -- & 1.45 \\ 
    Cr & 5.66 $\pm$ 0.07 & 5.64 & Eu & -- & 0.52 \\ 
\hline
\end{tabular}
\end{table*}

For iDR2, abundances were computed by eight different Nodes\footnote{The Nodes are: Bologna, CAUP, Concepcion, EPInArBo, LUMBA, Paris-Heidelberg, UCM, and Vilnius} based solely on their own set of atmospheric parameters. This was the case because, as shown during the analysis of iDR1, there is no significant difference between these and abundances computed based on the recommended atmospheric parameters (as discussed in Appendix\ \ref{sec:idr1}).

As for the atmospheric parameters, the final recommended abundances are weighted medians from the values obtained by the Nodes. We combined the abundances on a line-by-line basis, adopting the same Node weights defined before for the atmospheric parameters. The following conditions were applied to select the results before the abundances were combined, to guarantee that information was available to robustly estimate the precision of the results:

\begin{enumerate}
\item Only elemental species analyzed by at least three Nodes were considered.
\item The Node abundances of a given species, at a given star, were combined only if that star was analyzed by at least three Nodes.
\item Each spectral line was only considered if at least 3 Nodes provided abundances based on that line.
\item When information of the EWs was available, only lines with 5 $\leq$ EW (m\AA) $\leq$ 120 were used. Exceptions were sodium (5 $\leq$ EW (m\AA) $\leq$ 140) and barium (5 $\leq$ EW (m\AA) $\leq$ 250) \footnote{As pointed out by the referee, selecting the lines for deriving abundances and for atmospheric parameters based on the measurements themselves may bias the results. At the lower EW limit, lines for which the inference overestimates the EW have a higher chance to be included than those with underestimated EW (and the opposite for the upper edge). This choice will lead to insignificant biases in high S/N data, but might become important for low S/N data and/or when the spectral lines are very weak.}.
\item If, for a given species at a given star, abundances from 20 or more different spectral lines were available, we removed the ones that are flagged as blended in the Gaia-ESO line list. 
\item If, before applying the weighted median, the total number of spectral lines with abundances (for a given species at a given star) is more than 20, a 2$\sigma$ clipping from the mean value was applied. (The total number of lines is counted across all Nodes, therefore if eight Nodes provide abundances for 5 lines each, it counts as 40 lines for the clipping.)
\item The weighted median abundance of each spectral line is computed.
\item The median value of multiple lines is adopted as the recommended abundance.
\end{enumerate}

The exceptions are C, N, and O. These abundances were computed by one single Node (Vilnius), using the recommended atmospheric parameters. This choice was made to properly take into account the chemical equilibrium of the molecules. Carbon abundances were computed from C$_{2}$ molecules, nitrogen from CN molecules, and oxygen from the forbidden line at 6300 \AA\ (see Sect. \ref{sec:linelist} for the line list description). 

The iDR2 results include abundances computed in this way for the following 24 elements: C, N, O, Na, Mg, Al, Si, Ca, Sc, Ti, V, Cr, Mn, Fe, Co, Ni, Cu, Zn, Y, Zr, Mo, Ba, Nd, and Eu, for a least a few stars\footnote{Abundances of Mo are available for 66 stars, of Nd for 111 stars, of Zr for 159 stars, and of Eu for 228 stars. All other abundances are available for more than 920 stars.}. Of the 1268 stars with atmospheric parameters observed by the Survey, we derived abundances of at least 15 different elements for 1079 and for at least ten elements for 1203 stars. This sample of FGK-type stars is already one of the largest of its kind where abundances for so many elements have been determined. We stress that the list of abundances includes elements formed in different nucleosynthetic channels, i.e. s-process, r-process, Fe-peak, light, and $\alpha$-elements, providing an unprecedented data set of great scientific value.

Abundances of a number of additional elements were provided by some Nodes, but were finally excluded when the conditions listed above were applied. These abundances are not part of the iDR2 recommended results, because without multiple determinations it is not possible to estimate how precise they are. The list includes Li, S, Sr, La, Ce, Pr, Sm, Gd, and Dy. These abundances might still be used by the Gaia-ESO consortium for scientific applications, but they are not in the final Gaia-ESO iDR2 abundance scale, but in the scale defined by the individual Node results which they are part of. Whenever such abundances are used in a publication, this difference will be stressed.

\subsection{The Sun and solar analogues}

In Fig. \ref{fig:abunsun} we show the abundance pattern of the Sun, and the solar twins/analogues \object{18 Sco} \citepads{1997ApJ...482L..89P}, $\alpha$ Cen A, and M 67 1194\footnote{Star \object{NGC 2682 YBP 1194} with identification number from \citetads{2008A&A...484..609Y}.} \citepads{2011A&A...528A..85O}, as computed here. The solar abundance pattern is compared to both the solar abundances of \citetads{1998SSRv...85..161G} and \citetads{2007SSRv..130..105G}. The other stars are compared to the solar abundances derived in this work (the solar abundances are given in Table \ref{tab:sunabun}). With a few exceptions, the solar values derived here agree with reference solar abundances to within $\pm$ 0.10 dex. The abundances of the three other stars mostly agree with the solar ones also to within $\pm$ 0.10 dex. For some elements solar abundances are not listed, either because of weak lines (e.g. CNO) or because they were computed by a reduced number of Nodes, and therefore did not fulfil the criteria discussed above for combining the abundances. Where solar abundances were not derived, we recommend the use of those from \citetads{2007SSRv..130..105G} for compatibility with the adopted model atmospheres.

\subsection{Trends with metallicity}

\begin{figure*}
\centering
\includegraphics[height = 4.55cm]{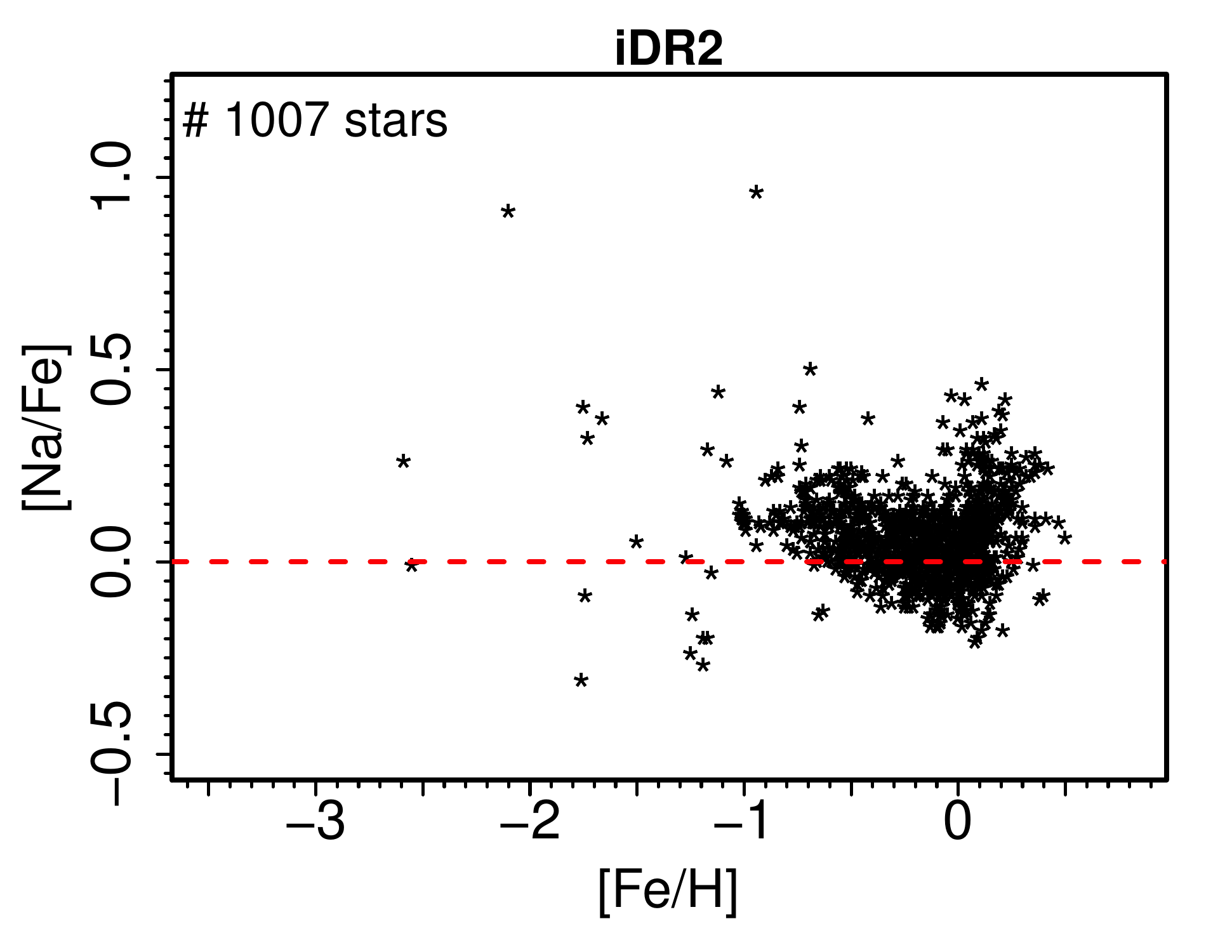}
\includegraphics[height = 4.55cm]{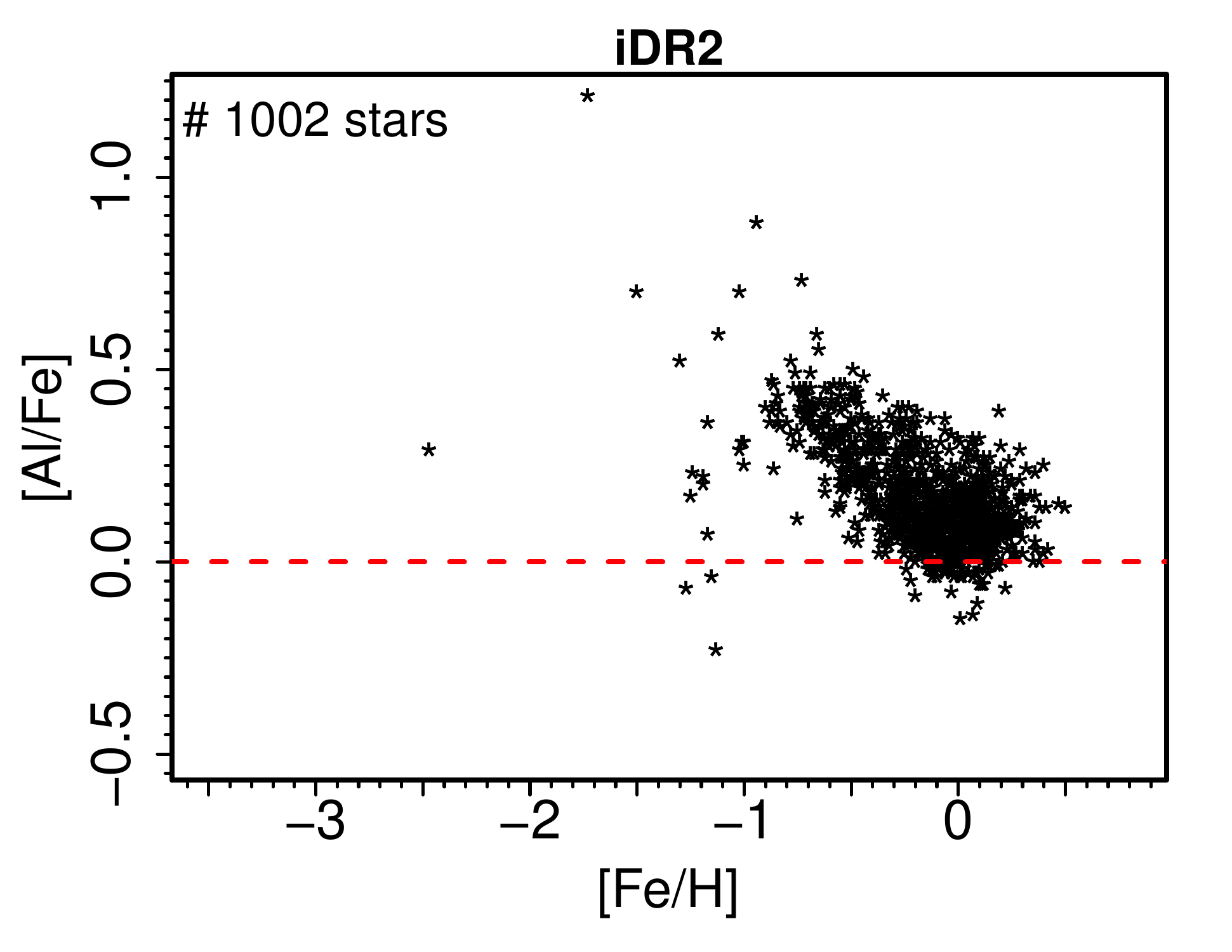}
\includegraphics[height = 4.55cm]{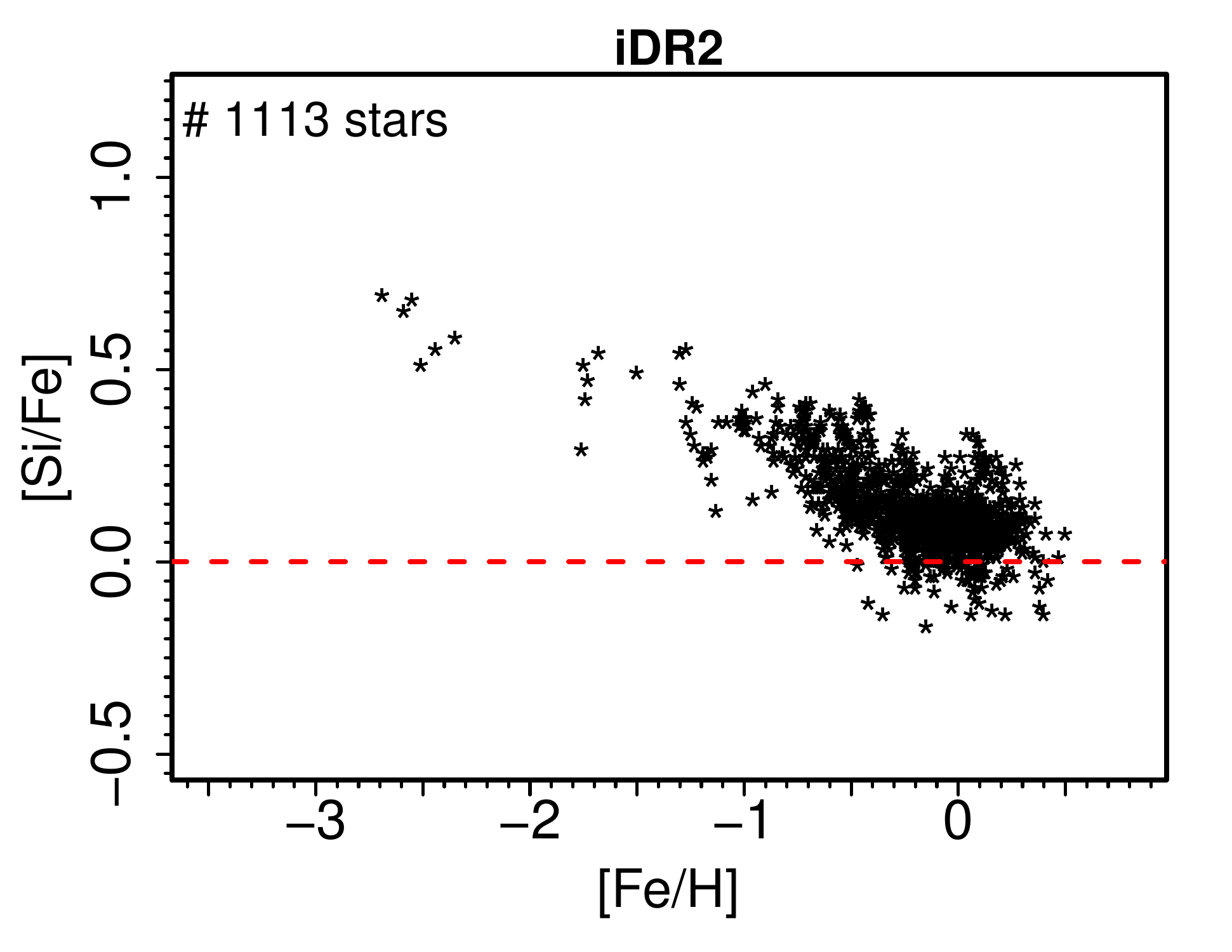}
\includegraphics[height = 4.55cm]{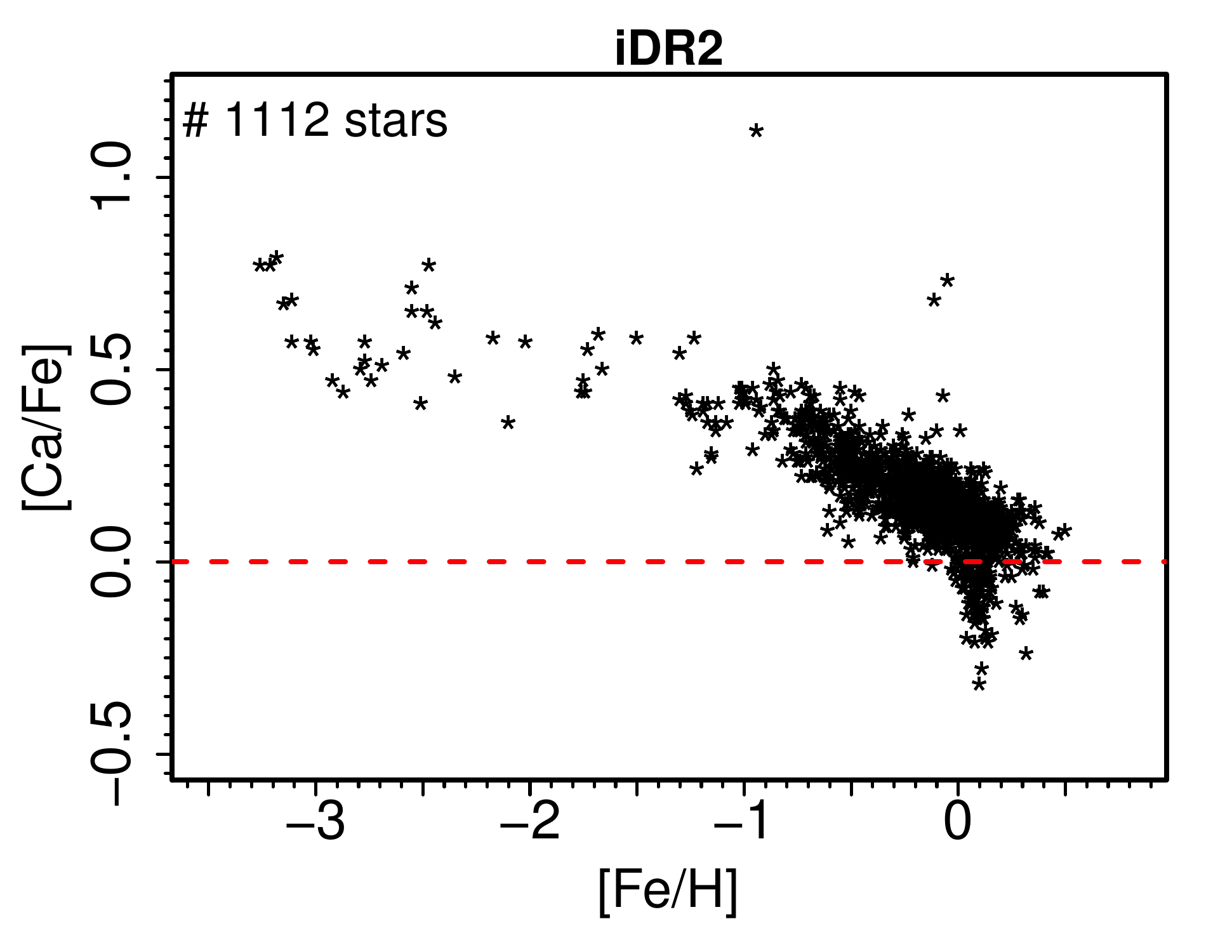}
\includegraphics[height = 4.55cm]{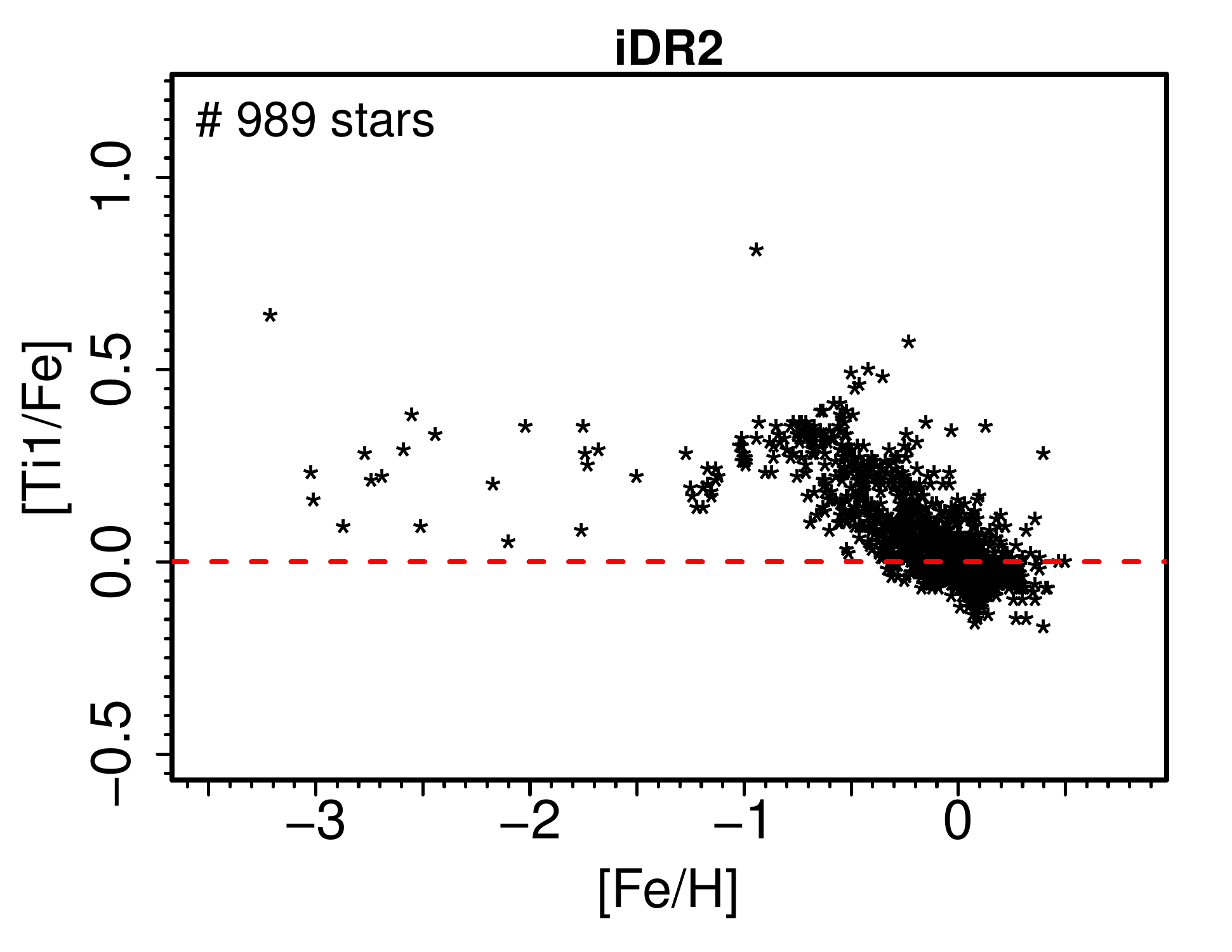}
\includegraphics[height = 4.55cm]{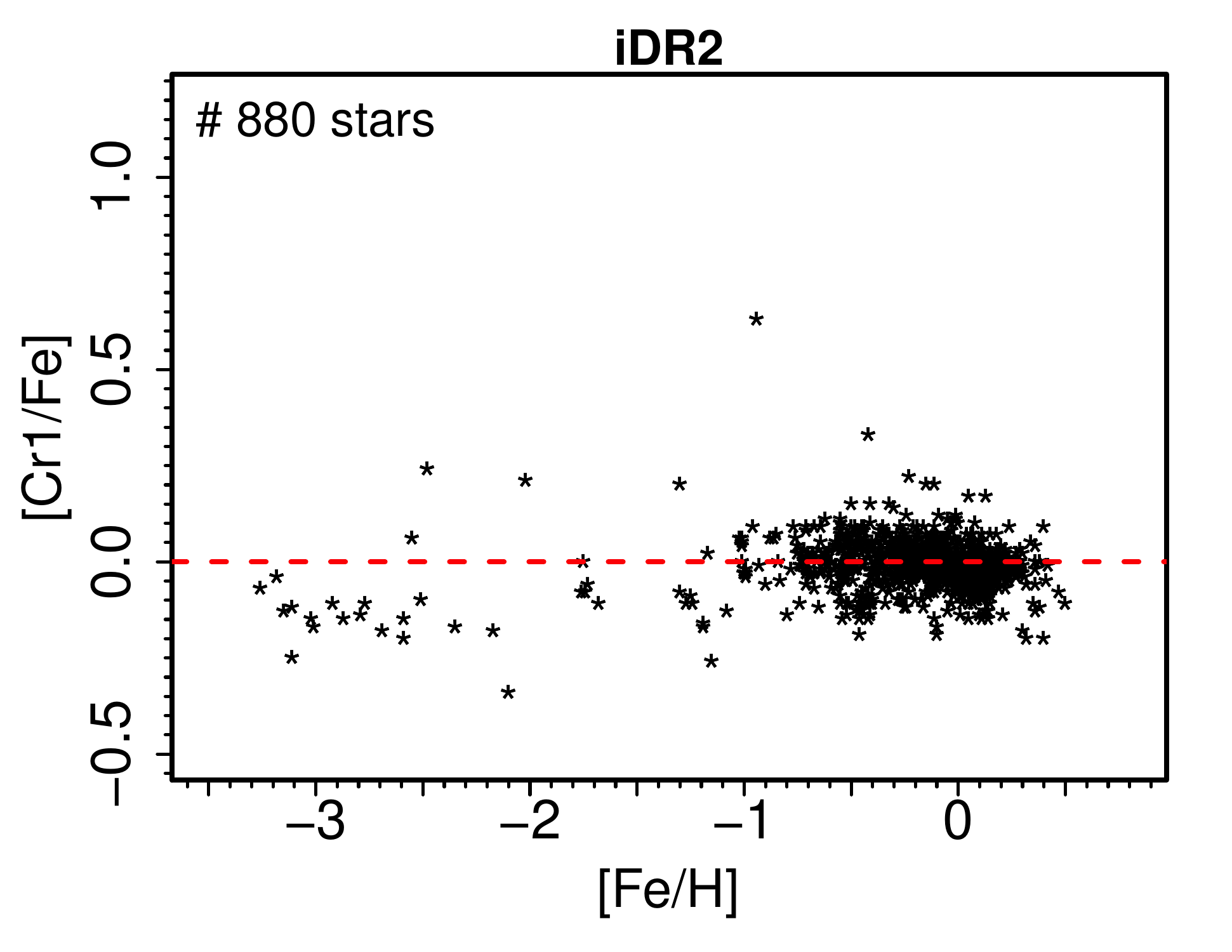}
\includegraphics[height = 4.55cm]{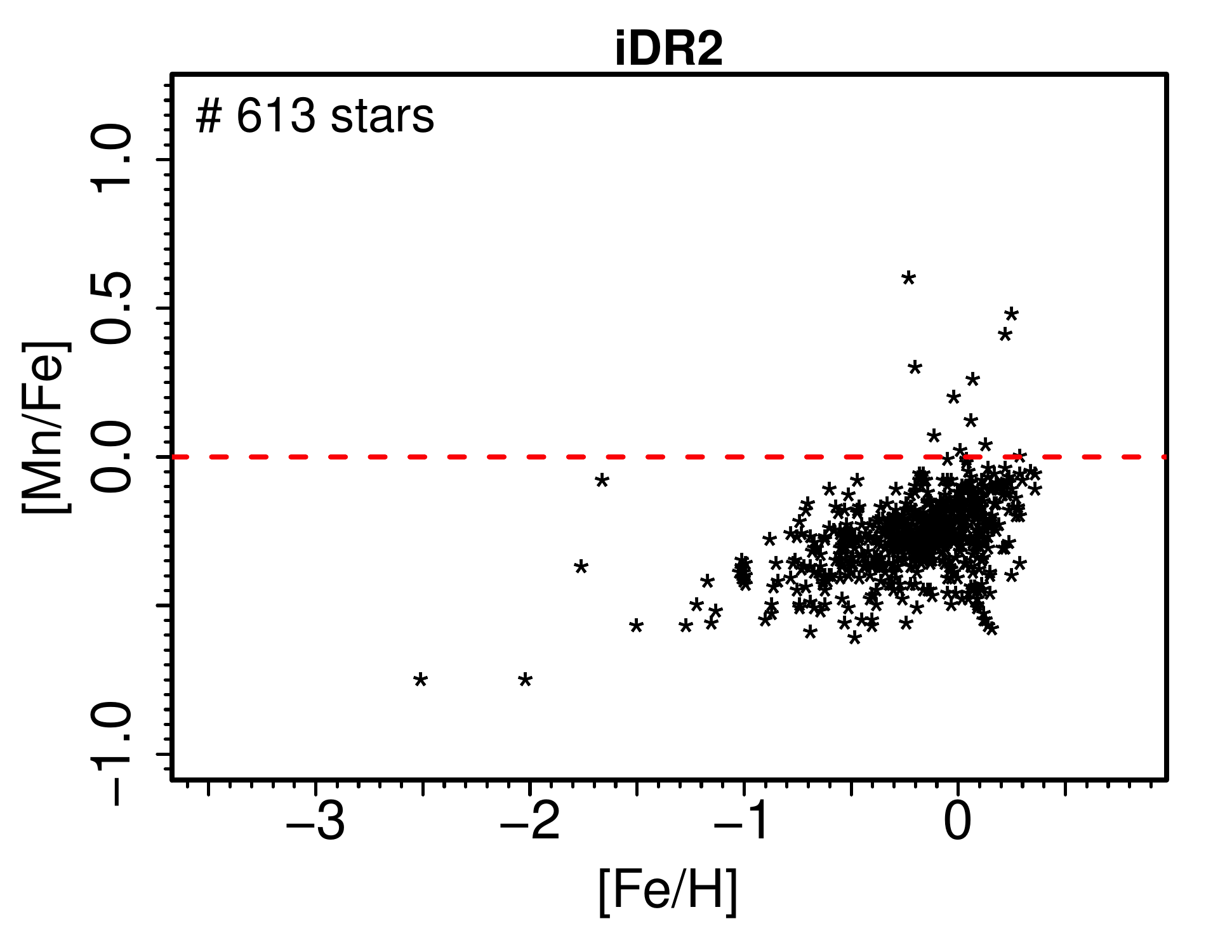}
\includegraphics[height = 4.55cm]{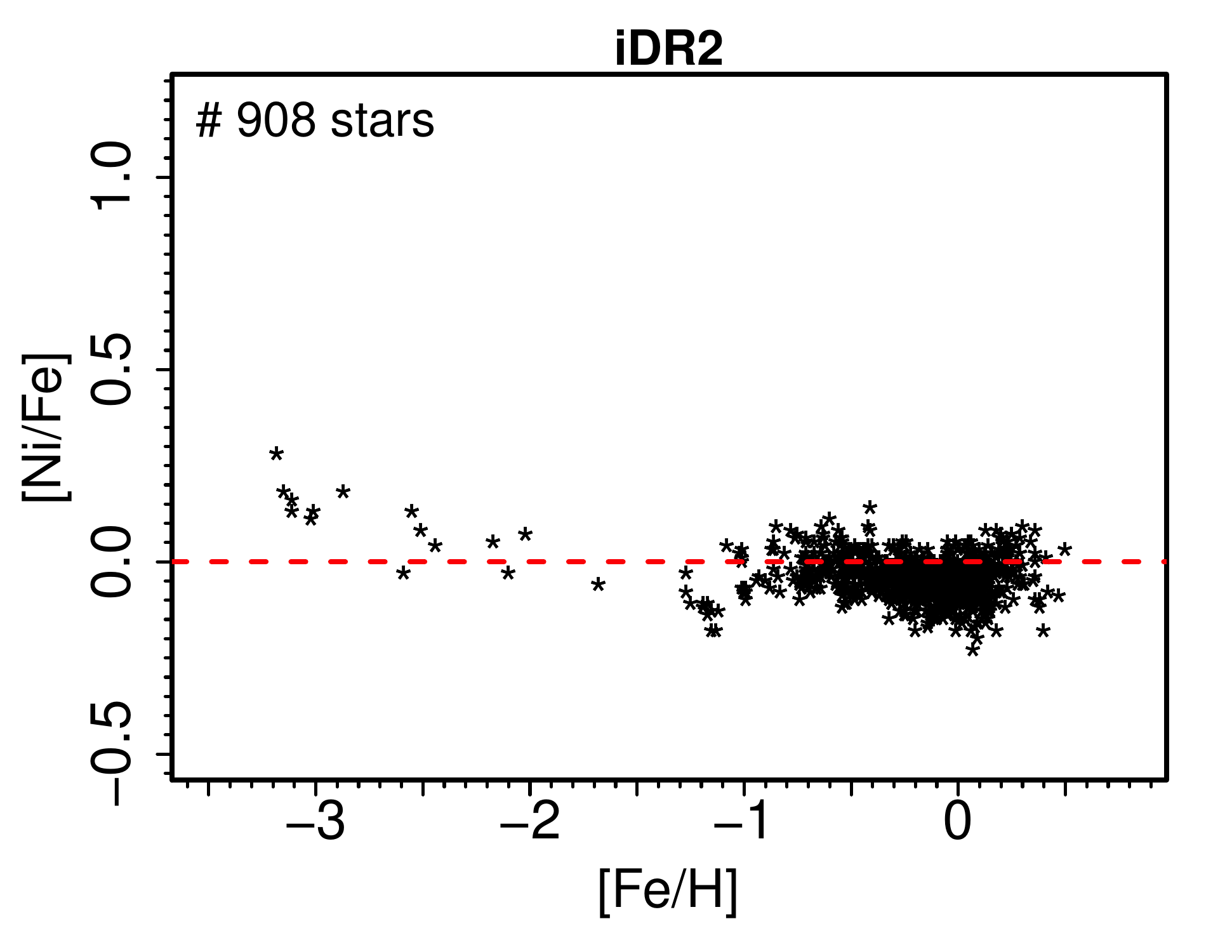}
\includegraphics[height = 4.55cm]{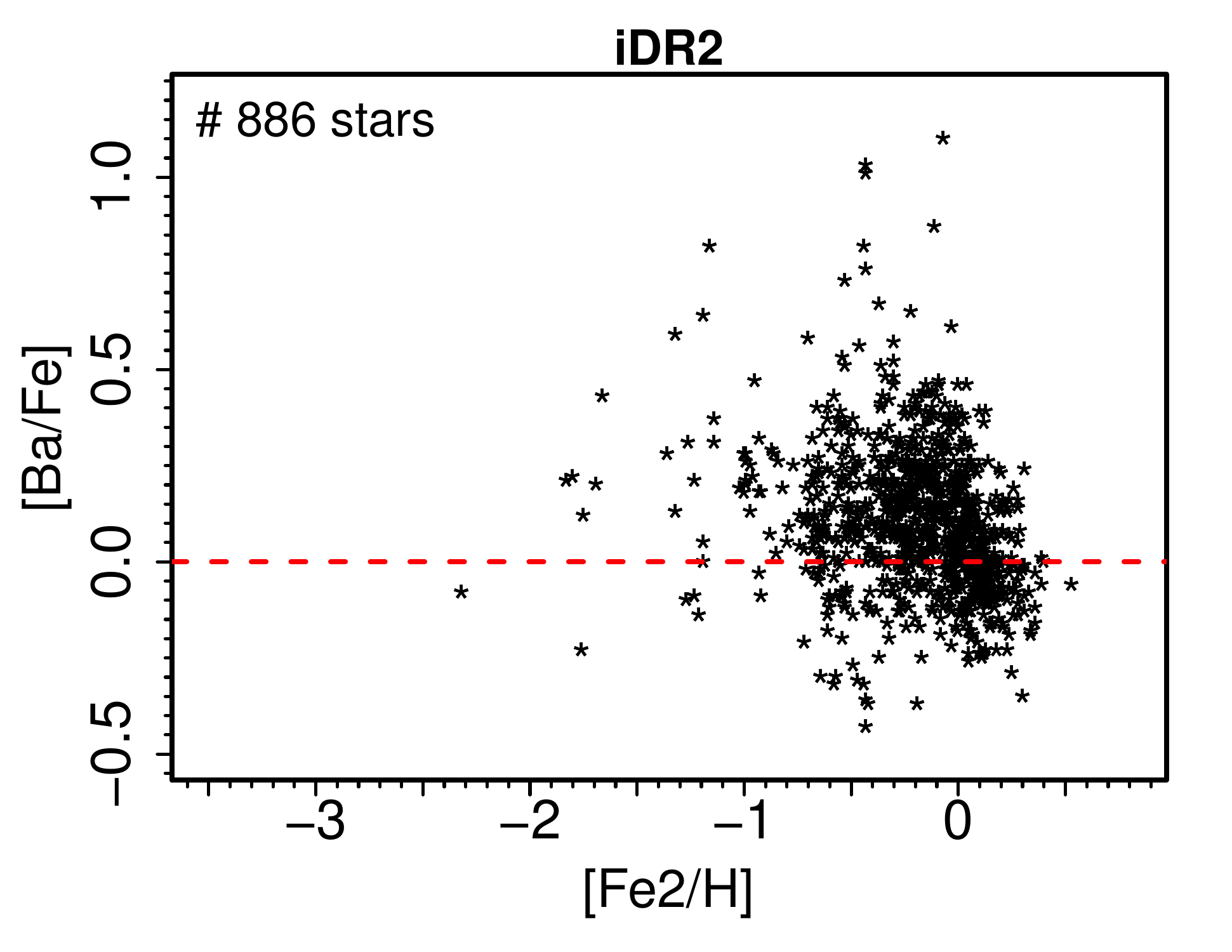}
 \caption{Abundances trends with metallicity for a few selected elements. Only results where the method-to-method dispersion is below 0.20 dex are plotted. Note that the \ion{Mn}{i} plot has a different scale. All abundances shown are from the neutral species, except for the Ba plot, where \ion{Ba}{ii} and \ion{Fe}{ii} are used.}\label{fig:trends}%
\end{figure*}

In Fig. \ref{fig:trends} we show the trend with metallicity for the [X/Fe] ratio of a few selected elements. All elements display a behavior with metallicity in agreement with what has been established by earlier works \citepads[see e.g.][and references therein]{1993A&A...275..101E,2004AJ....128.1177V,2005A&A...438..139S,2006MNRAS.367.1329R,2012A&A...545A..32A,2014A&A...562A..71B}. In these plots, we selected only the best quality results, excluding abundances where the method-to-method dispersion is above 0.20 dex. The inclusion of these extra abundances tends to increase the scatter in each plot. The figures are included only to illustrate which abundances have been derived, and that the general behavior seems correct. The proper scientific discussion requires a full investigation which is not the goal of this release paper.

\subsection{Iron abundances and metallicities}

\begin{figure*}
\centering
\includegraphics[height = 6cm]{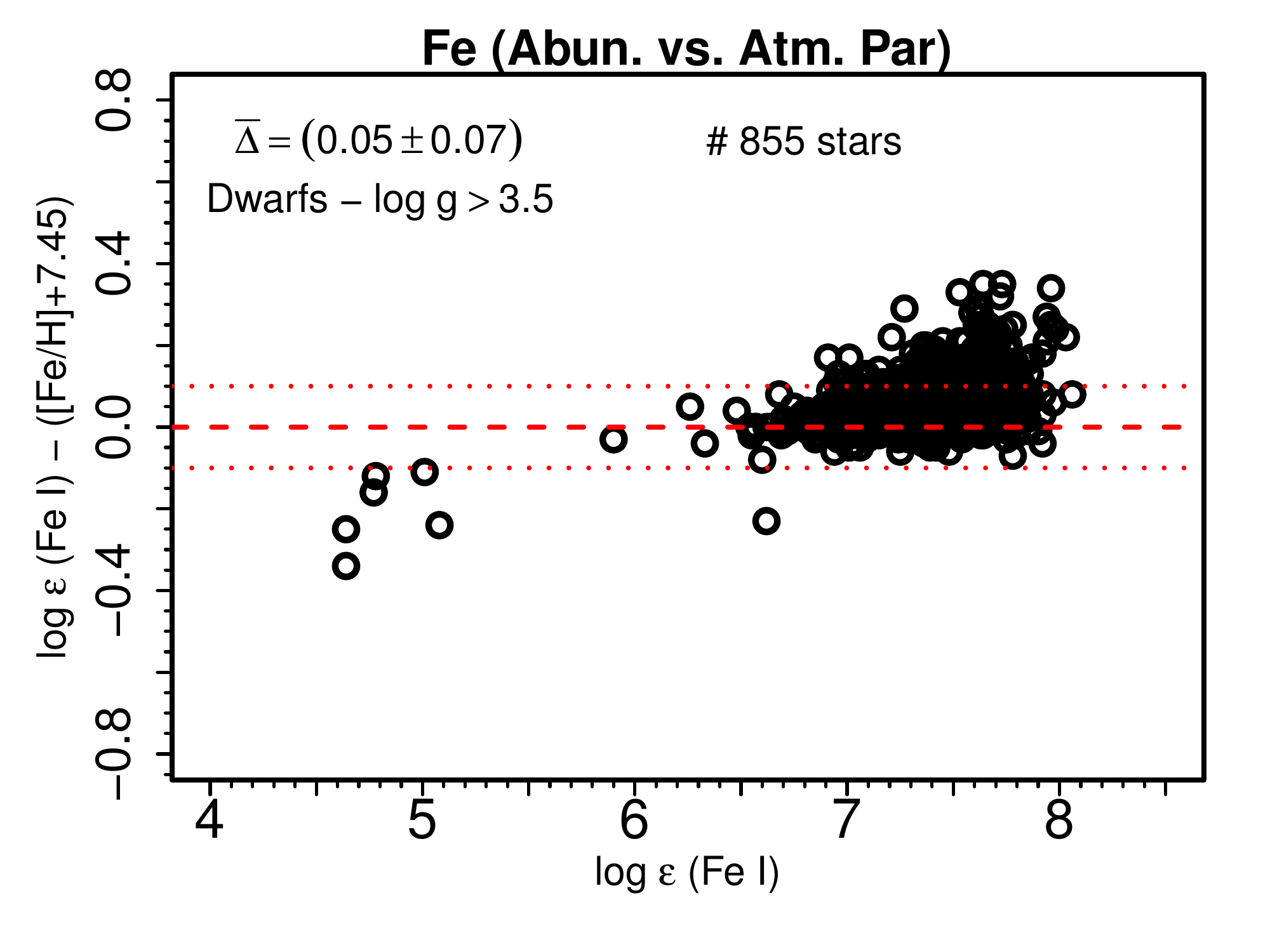}
\includegraphics[height = 6cm]{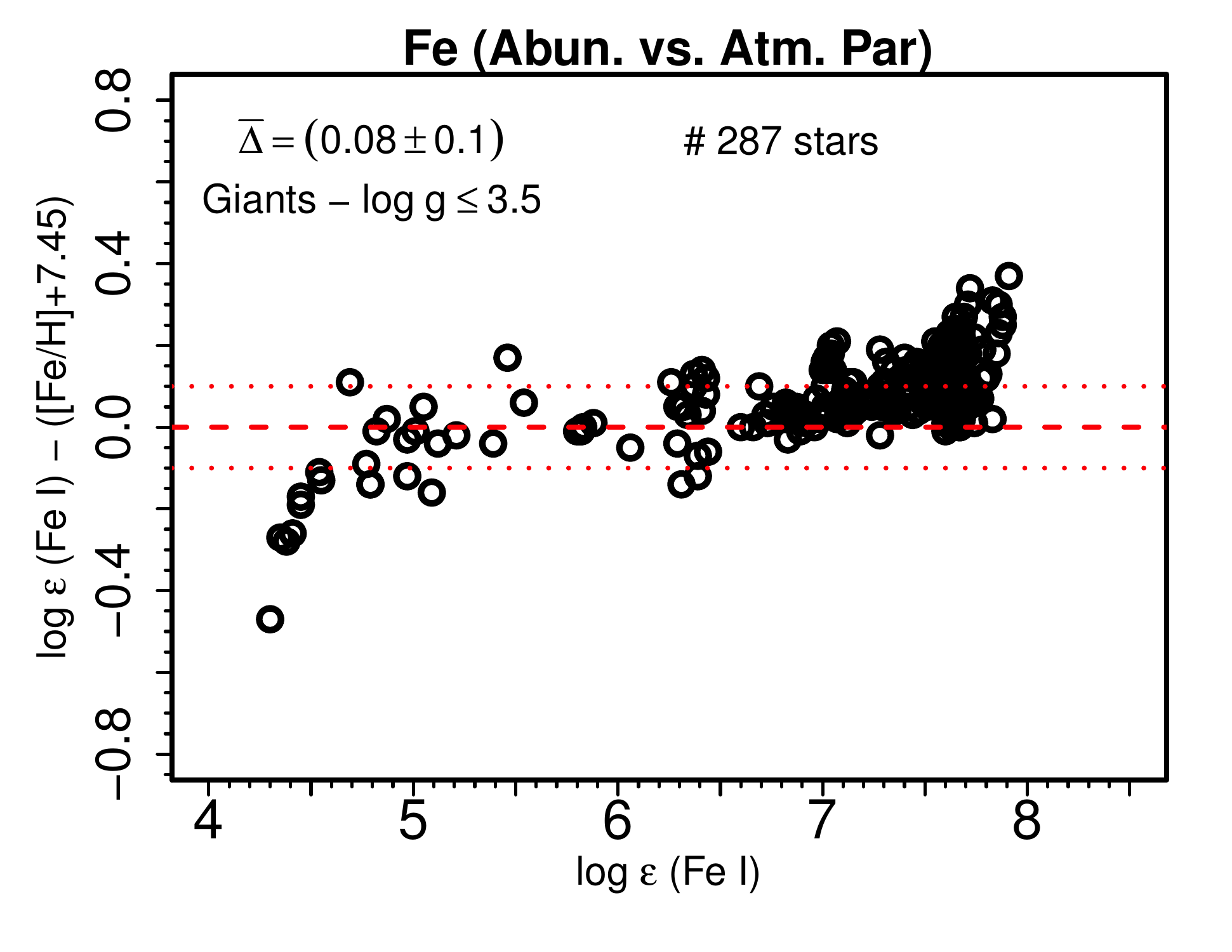}
 \caption{Comparison between the abundances of \ion{Fe}{i} and the metallicity as an atmospheric parameter [Fe/H], in dwarfs (left panel) and giants (right panel). The metallicity [Fe/H] is put in the $\log \epsilon$ scale by adding the solar Fe abundance from \citetads{2007SSRv..130..105G}, $\log \epsilon$(Fe)$_{\odot}$ = 7.45. Only results where the method-to-method dispersion of \ion{Fe}{i} is 0.20 dex or less are shown. The dotted lines indicate a difference of $\pm$ 0.10 dex, the dashed line represents \ion{Fe}{i} equal to [Fe/H].}\label{fig:fe1feh}%
\end{figure*}
\begin{figure*}
\centering
\includegraphics[height = 6cm]{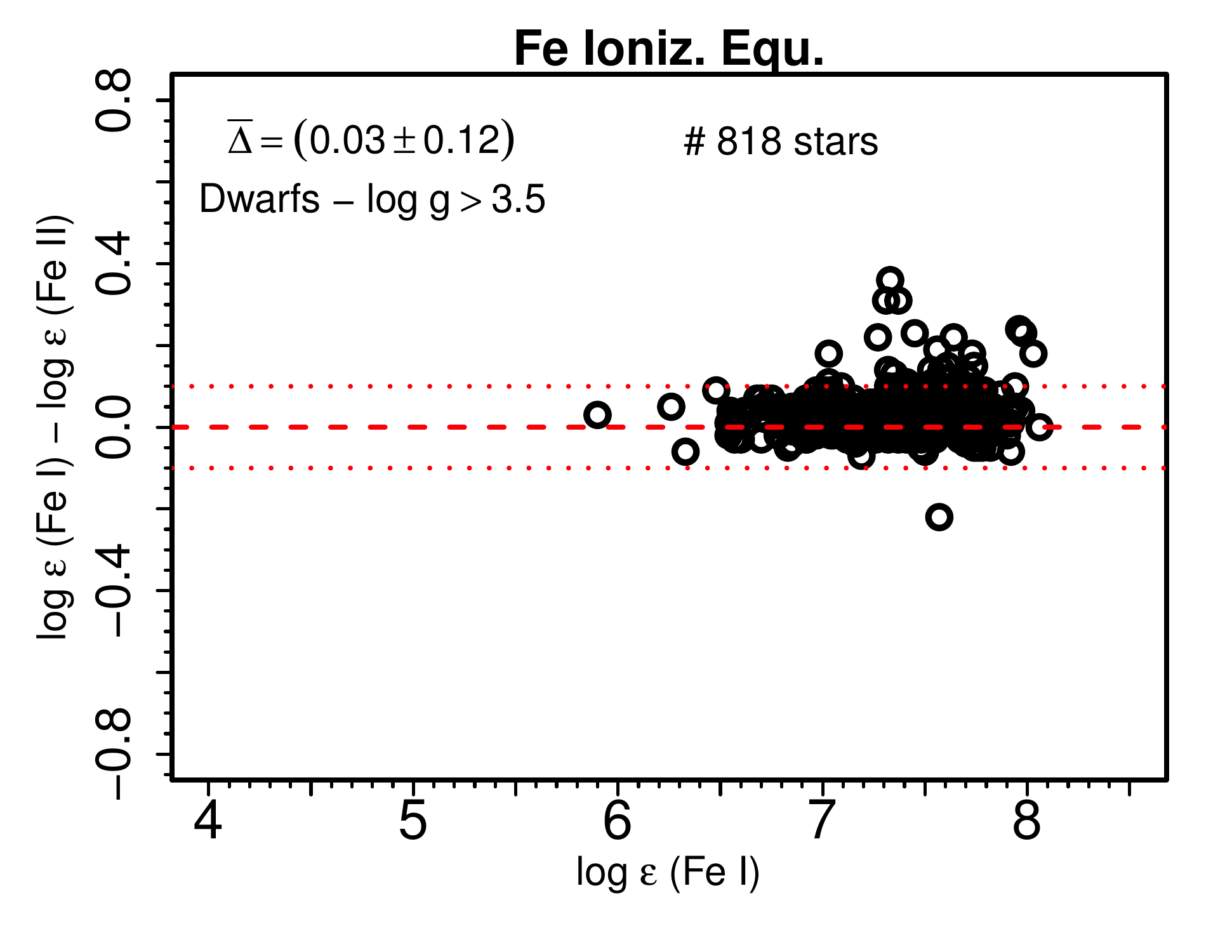}
\includegraphics[height = 6cm]{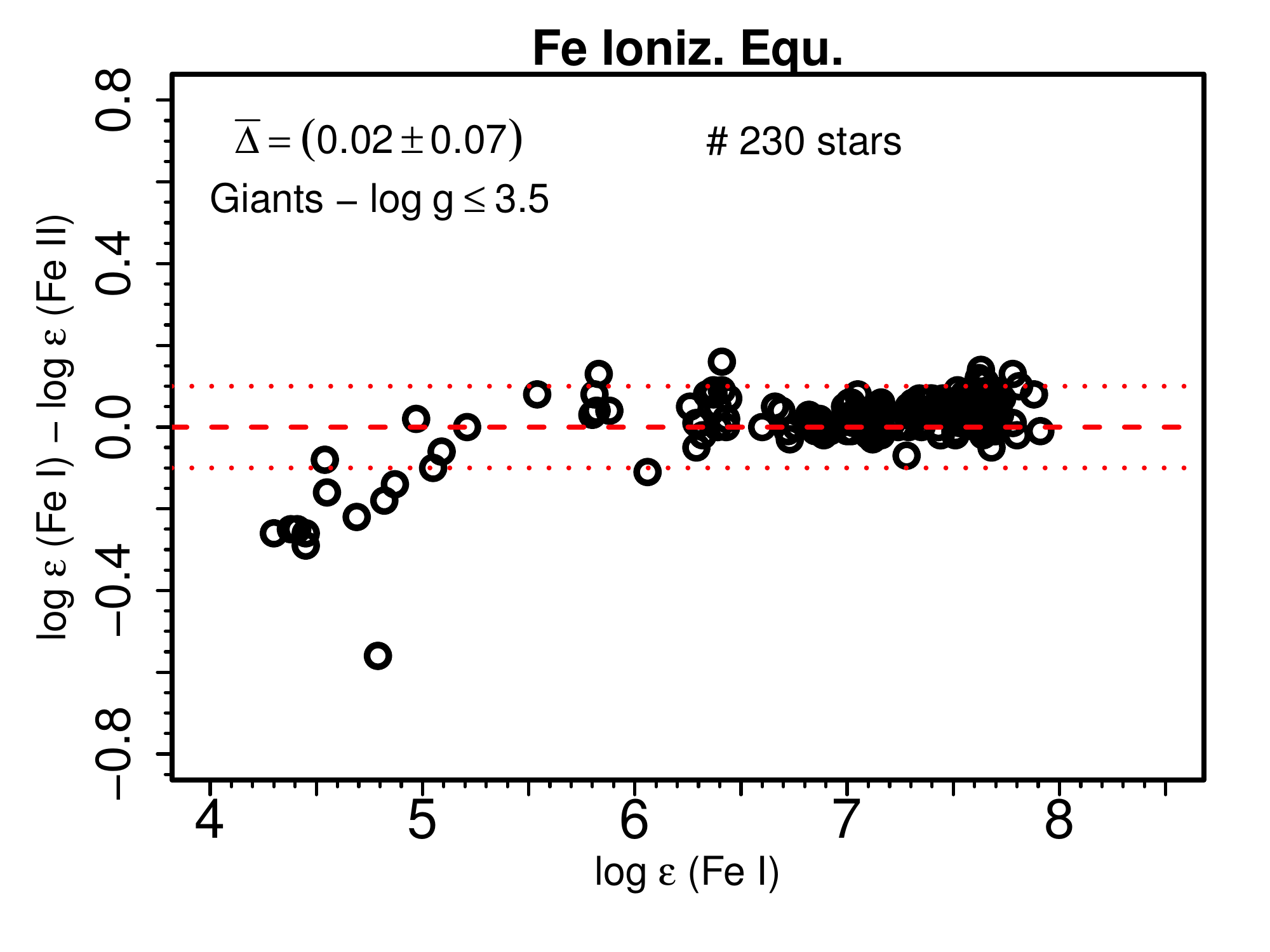}
 \caption{Comparison between the abundances of \ion{Fe}{i} and \ion{Fe}{ii}, to check for the ionization equilibrium, in dwarfs (left panel) and giants (right panel). Only results where the method-to-method dispersion of both \ion{Fe}{i} and \ion{Fe}{ii} is 0.20 dex or less are shown. The dotted lines indicate a difference of $\pm$ 0.10 dex, the dashed line represents \ion{Fe}{i} equal to \ion{Fe}{ii}.}\label{fig:fe1fe2}%
\end{figure*}

The data products determined here include both an [Fe/H] value determined during the derivation of the atmospheric parameters, and abundances derived from \ion{Fe}{i} and \ion{Fe}{ii} lines. The [Fe/H] value is a combination of the values used by the Nodes to constrain the metallicity of the model atmosphere adopted for a given star. For some methodologies, this metallicity is the Fe abundance, while for others it is a global value of the metal content, referred to as [M/H]. For deriving the recommended value of the metallicity as an atmospheric parameter, no distinction was made between [Fe/H] and [M/H]. For about 75\% of the stars, results of eight or more Nodes were used to compute [Fe/H].

The abundances derived from \ion{Fe}{i} and \ion{Fe}{ii} lines are calculated using the line-by-line abundances of the Nodes. Only five of the Nodes provided abundances of the iron lines. Since the final values of the Fe abundances and of [Fe/H] are computed in different ways, it is important to check if they are consistent. Such a comparison is shown in Fig. \ref{fig:fe1feh}. Only stars where the method-to-method dispersion of the \ion{Fe}{i} abundances is below 0.20 dex are shown. Dwarfs and giants are displayed separately, to check for possible systematic effects in stars of different evolutionary stages.

\begin{figure*}
\centering
\includegraphics[height = 4.5cm]{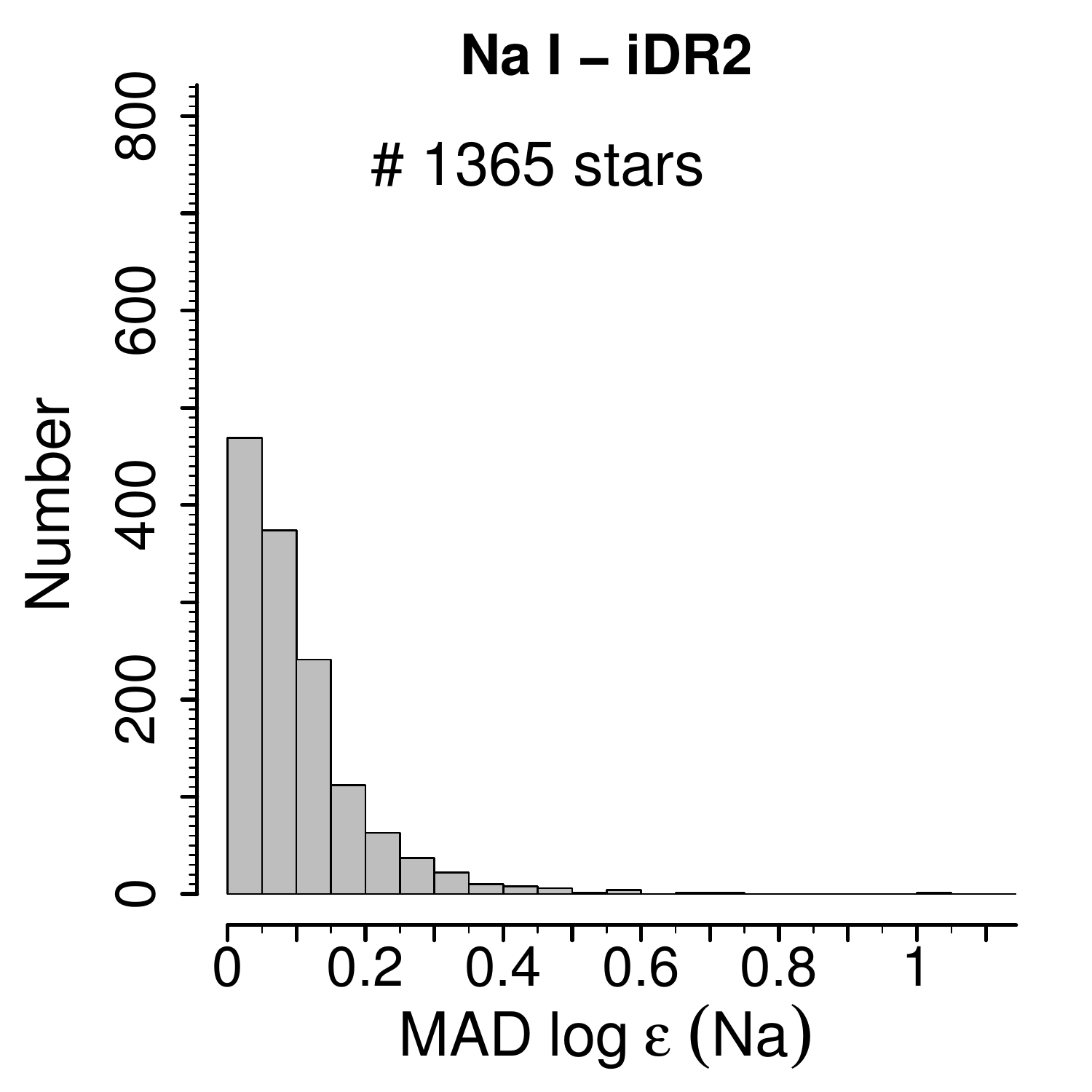}
\includegraphics[height = 4.5cm]{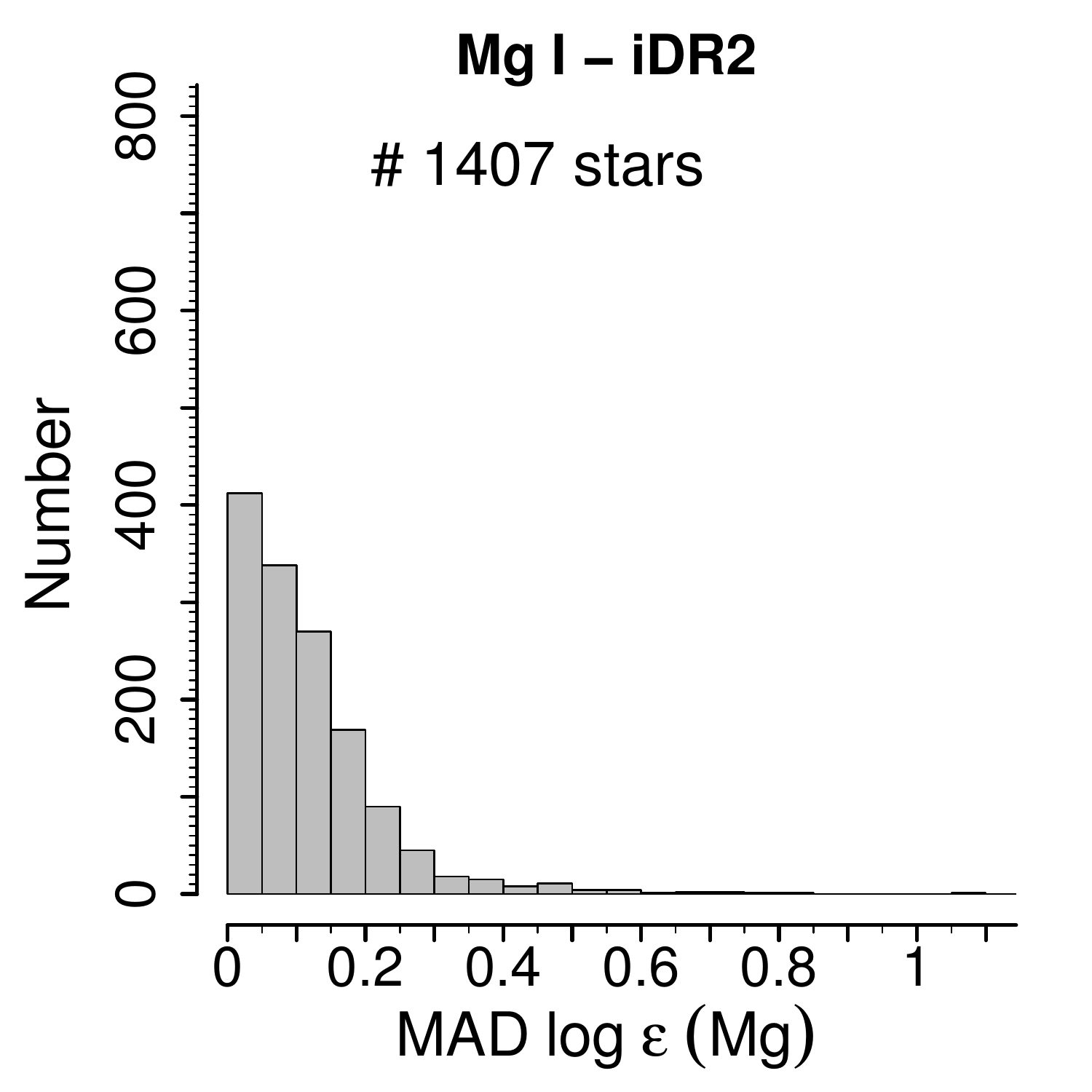}
\includegraphics[height = 4.5cm]{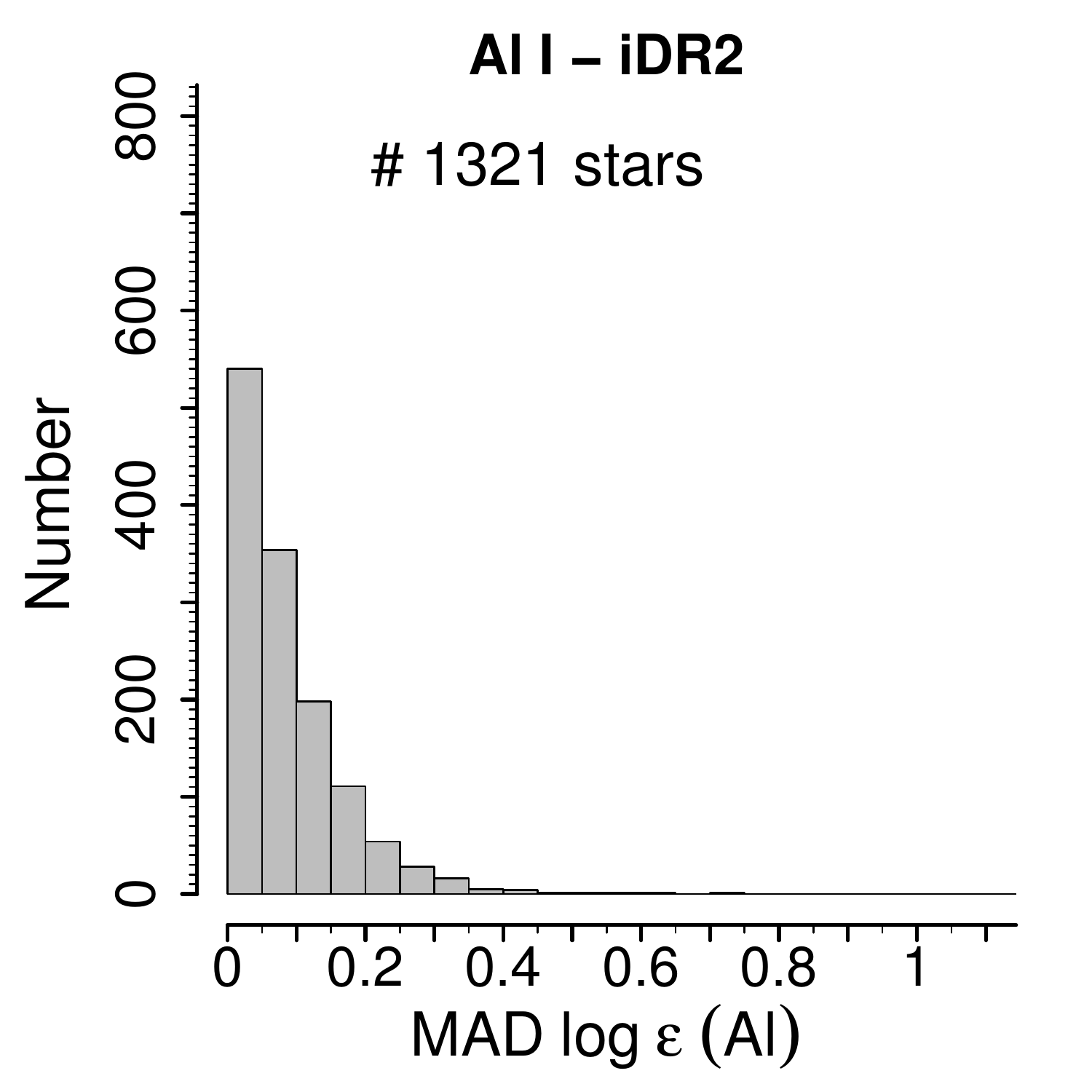}
\includegraphics[height = 4.5cm]{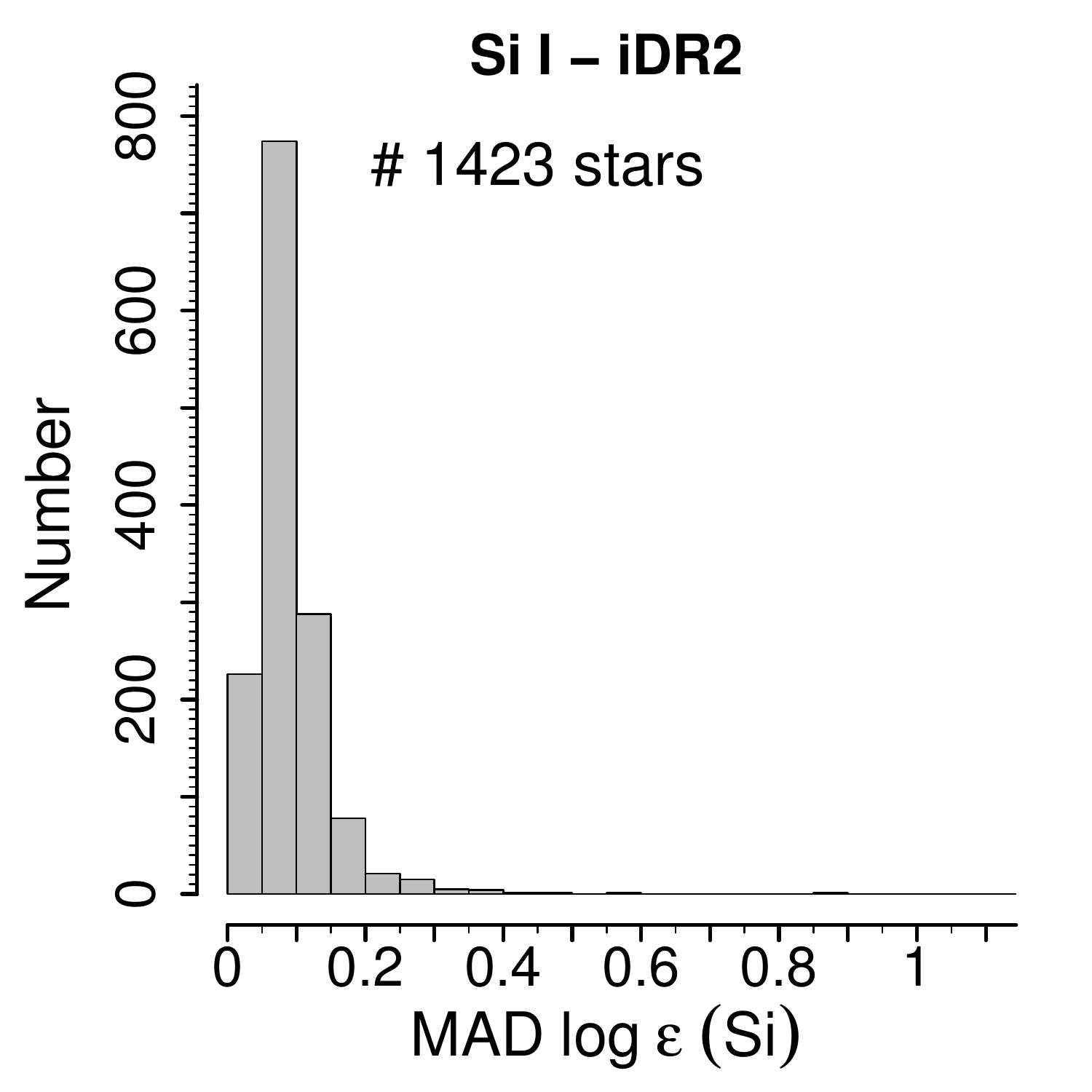}
\includegraphics[height = 4.5cm]{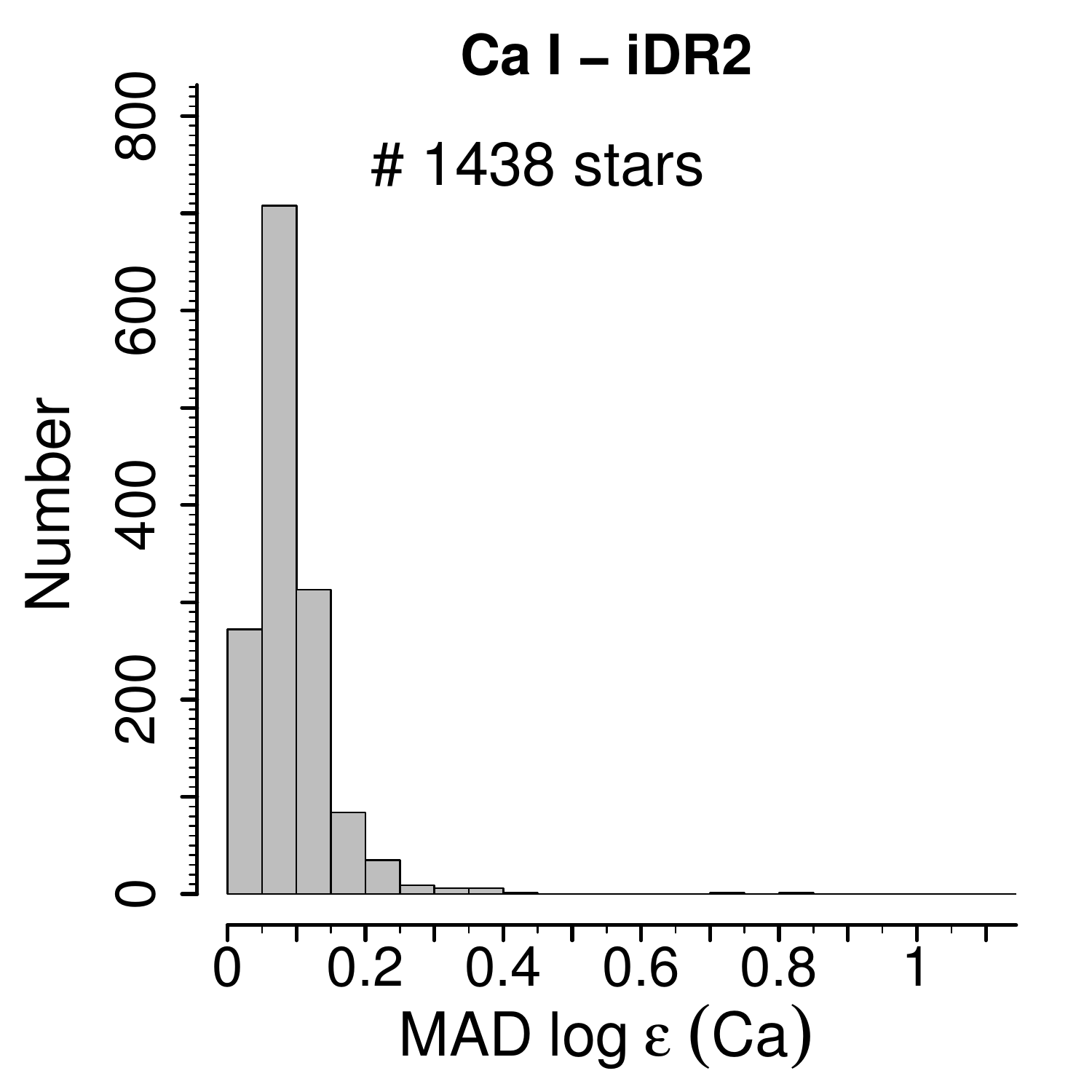}
\includegraphics[height = 4.5cm]{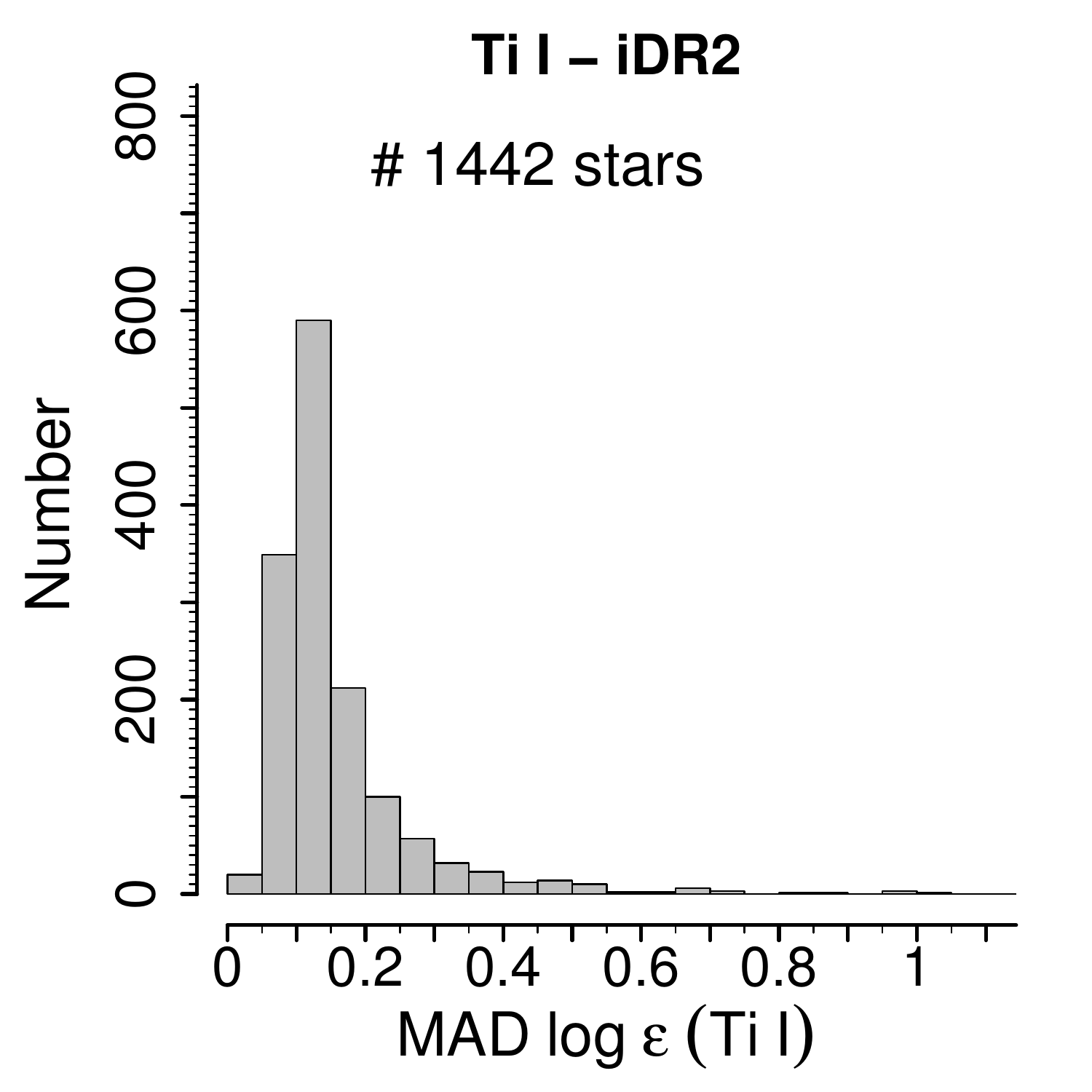}
\includegraphics[height = 4.5cm]{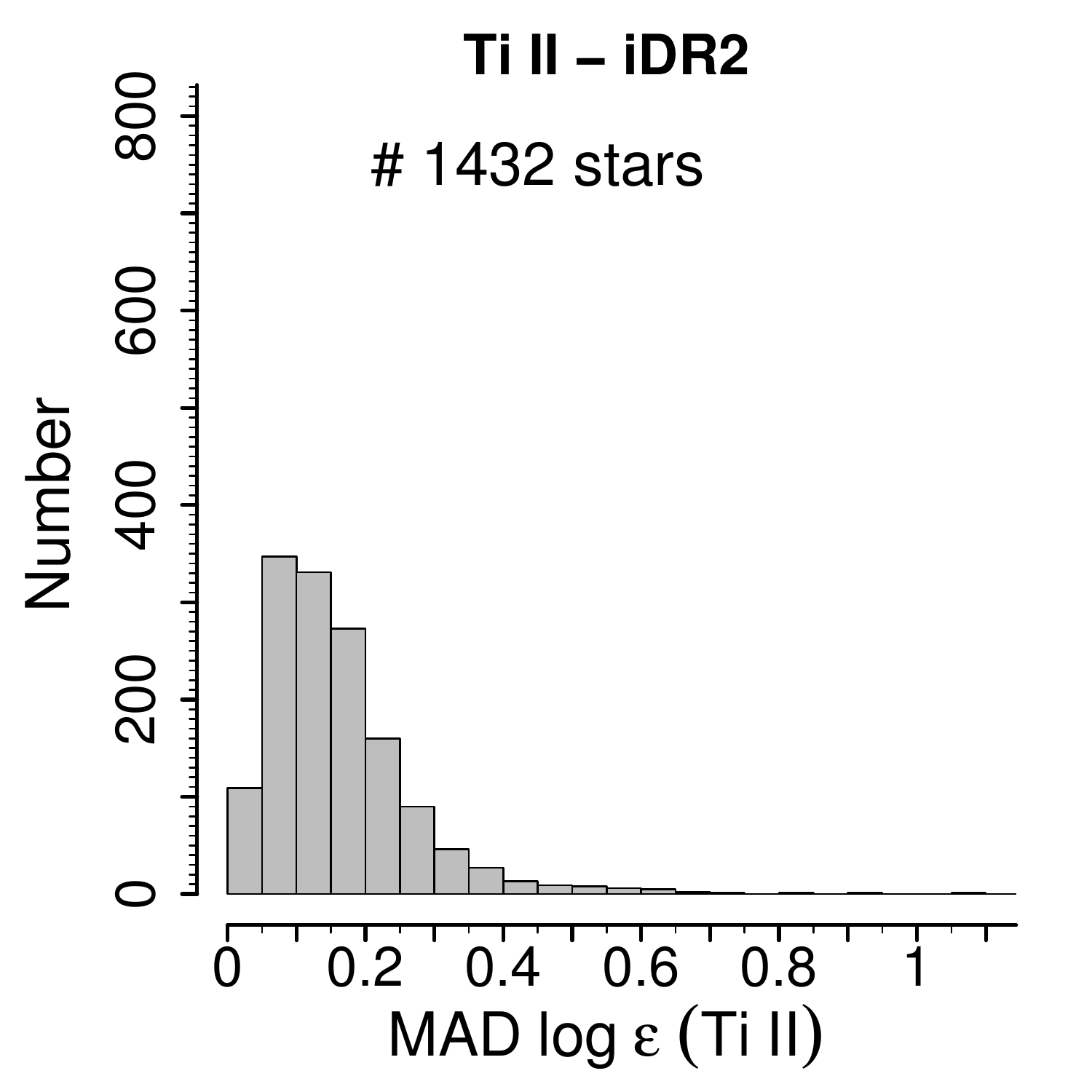}
\includegraphics[height = 4.5cm]{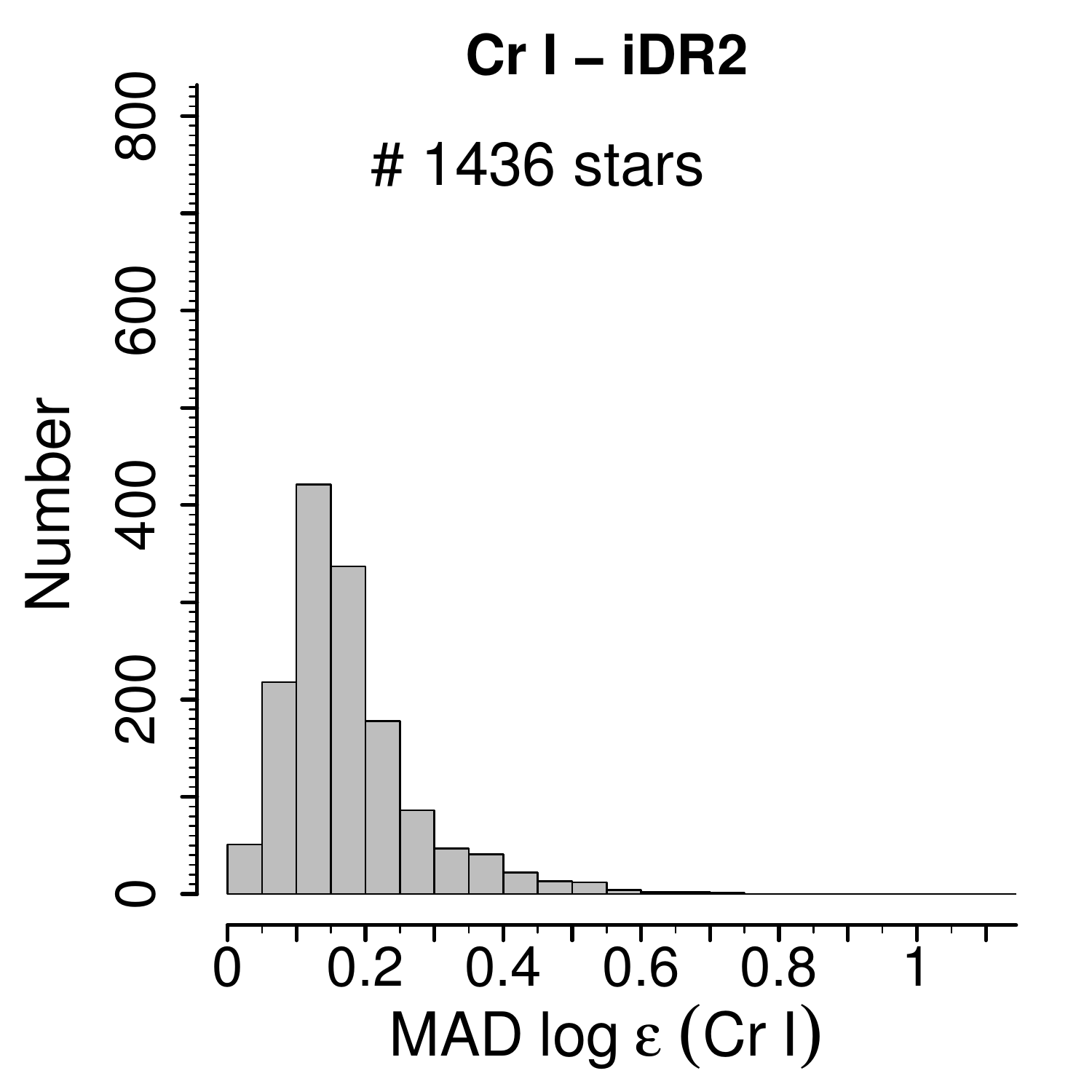}
\includegraphics[height = 4.5cm]{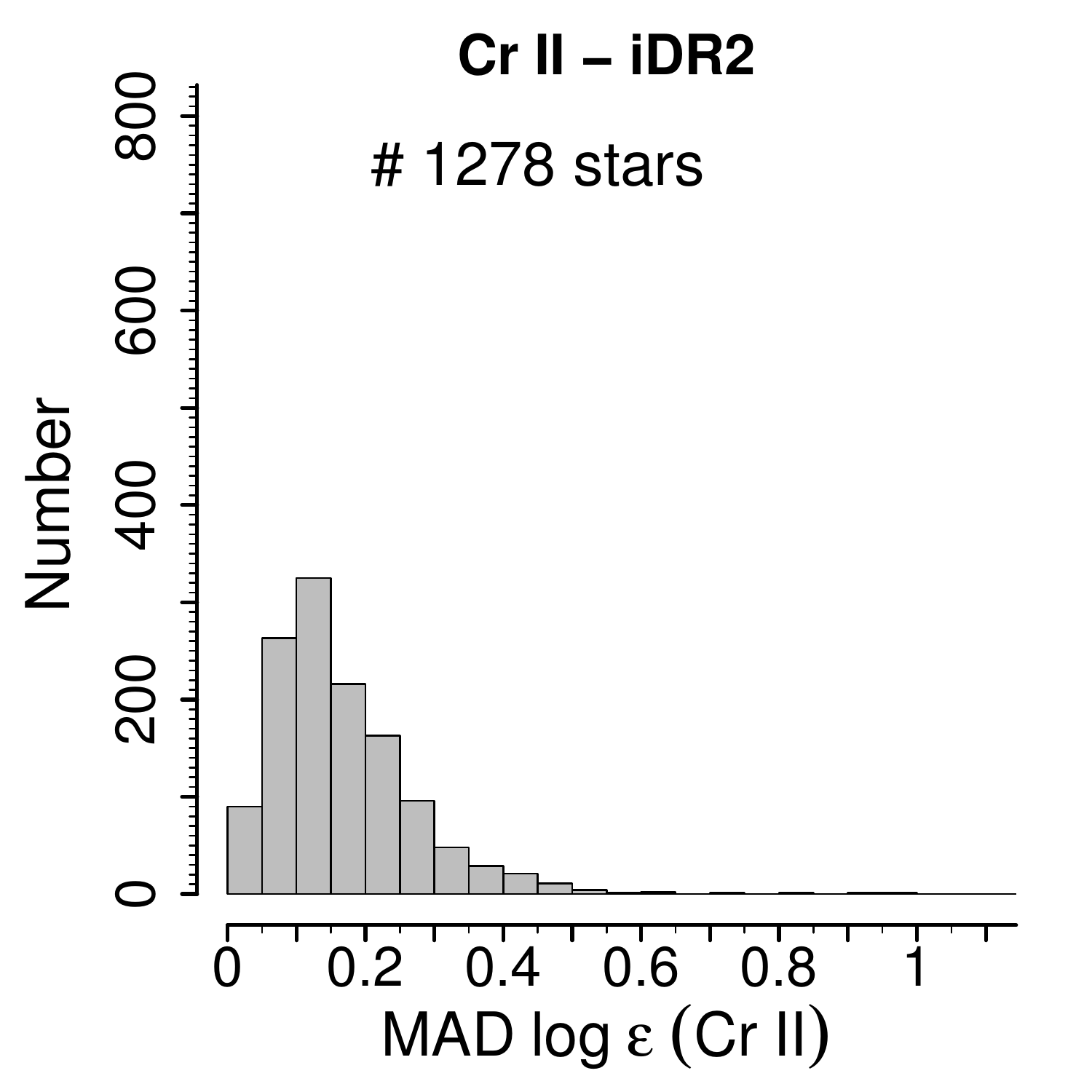}
\includegraphics[height = 4.5cm]{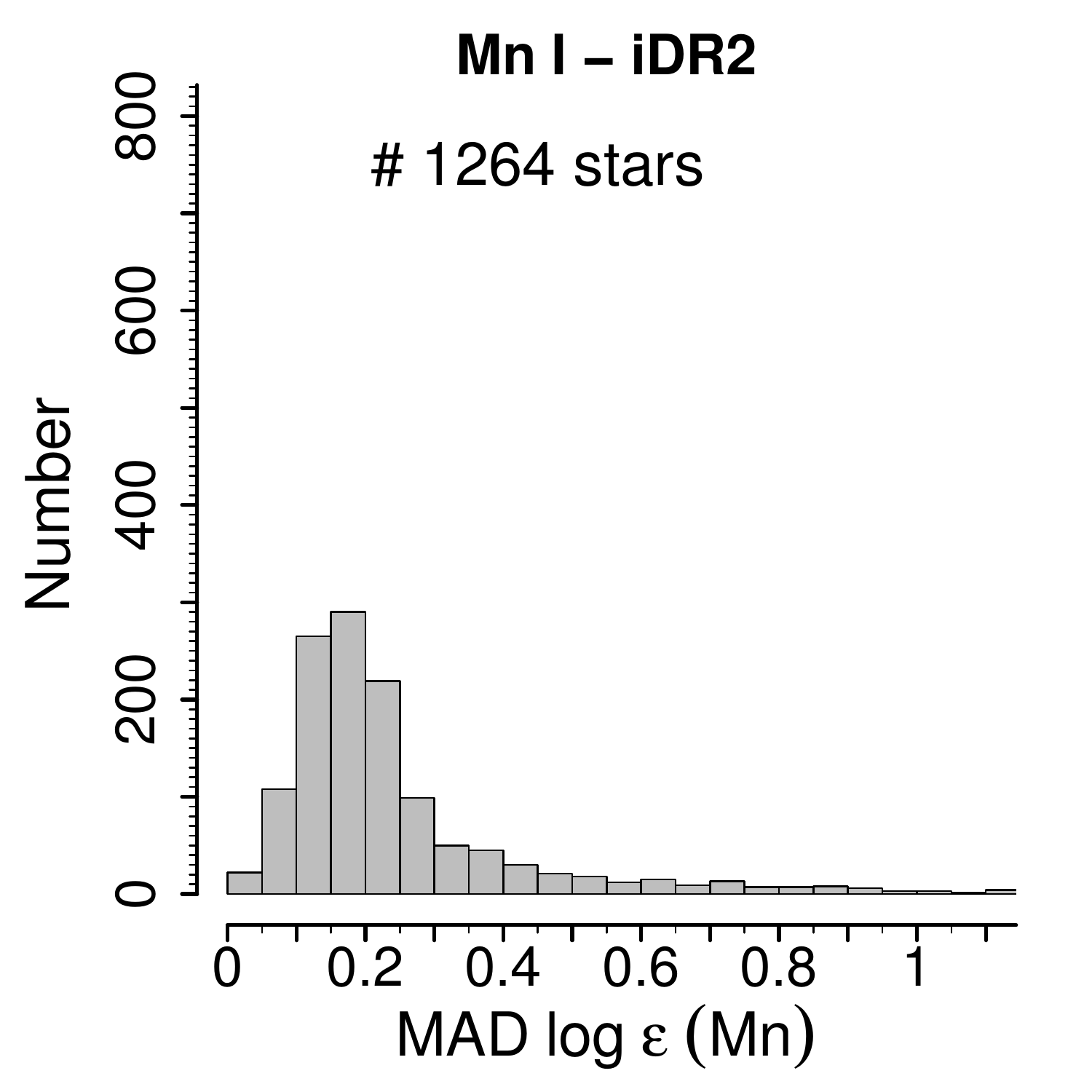}
\includegraphics[height = 4.5cm]{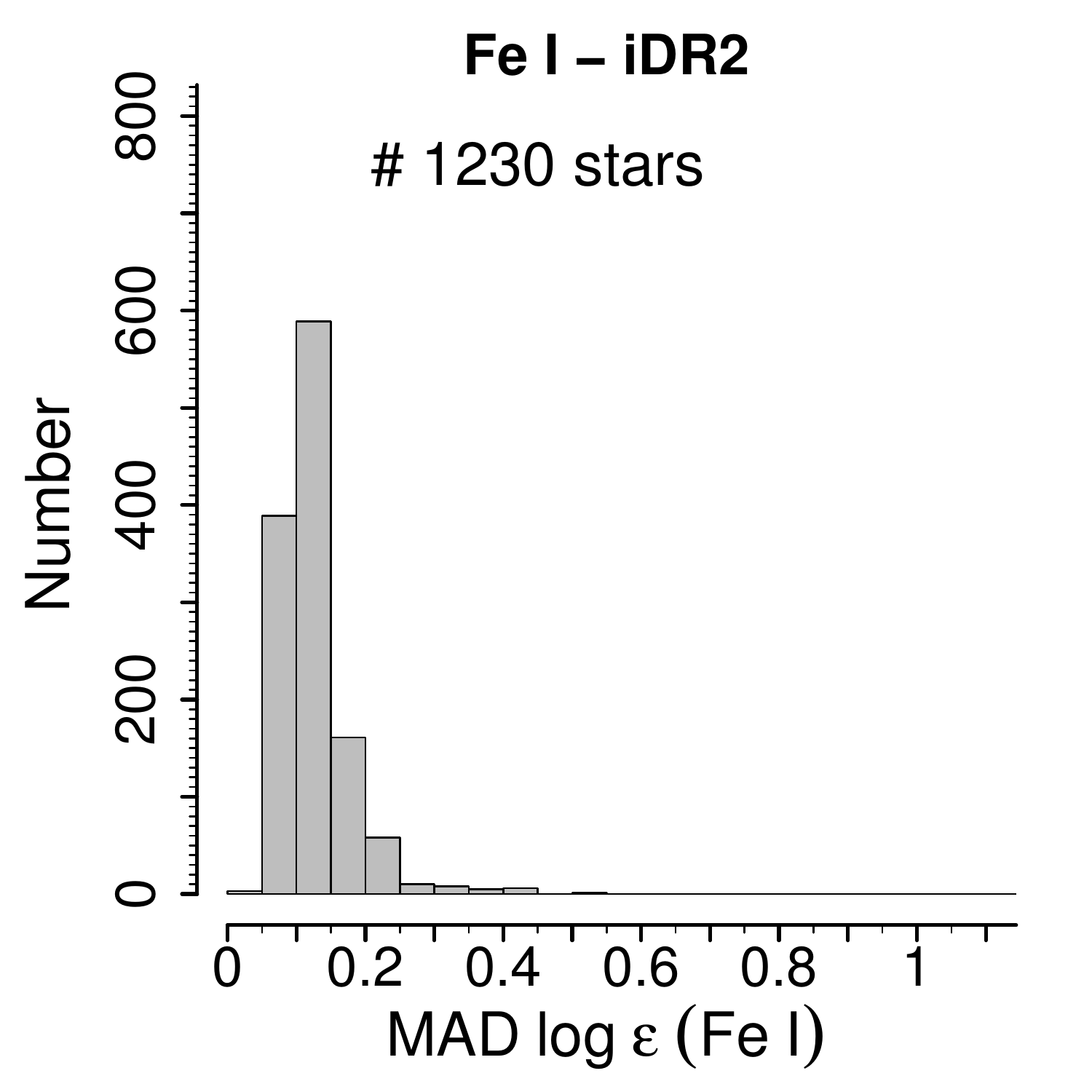}
\includegraphics[height = 4.5cm]{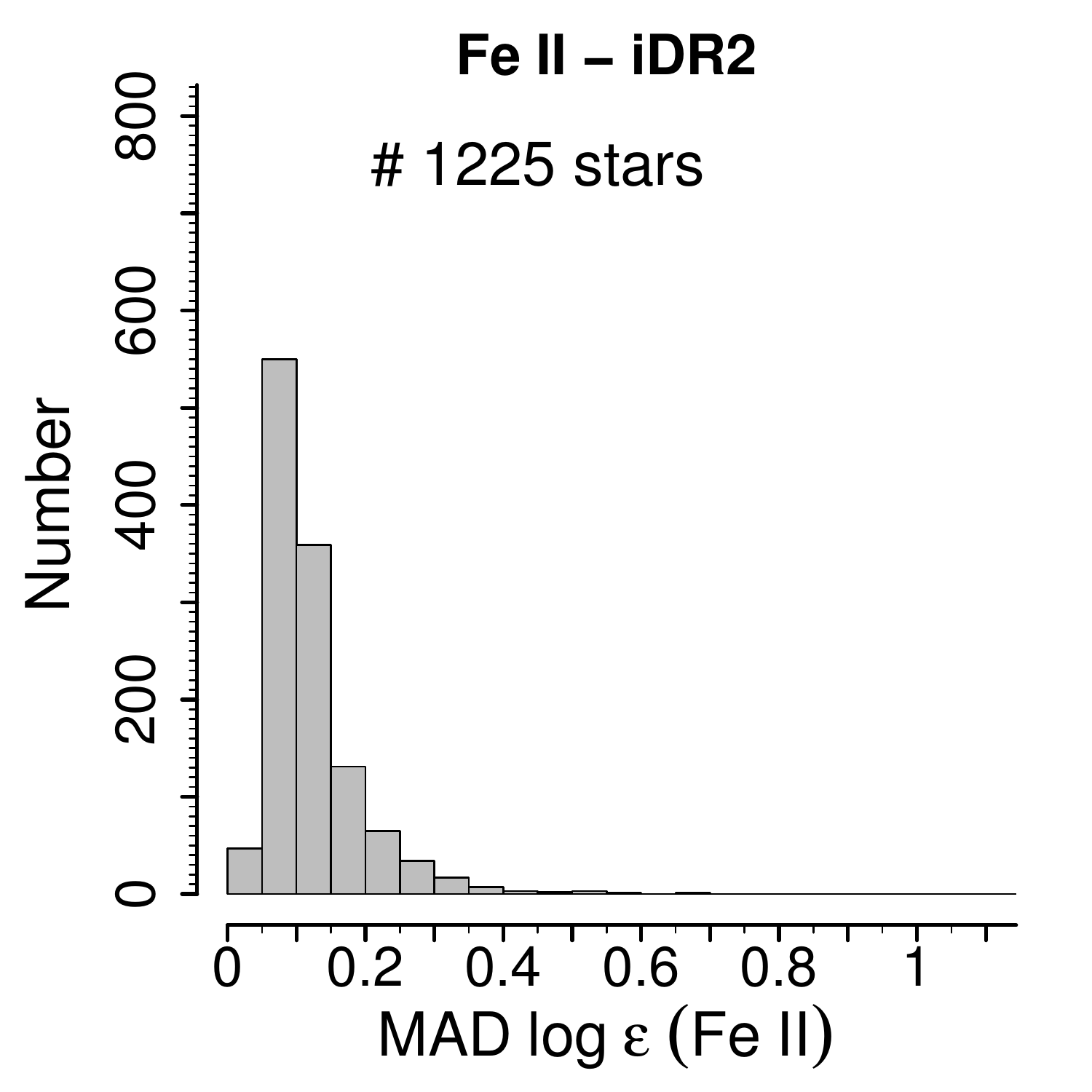}
\includegraphics[height = 4.5cm]{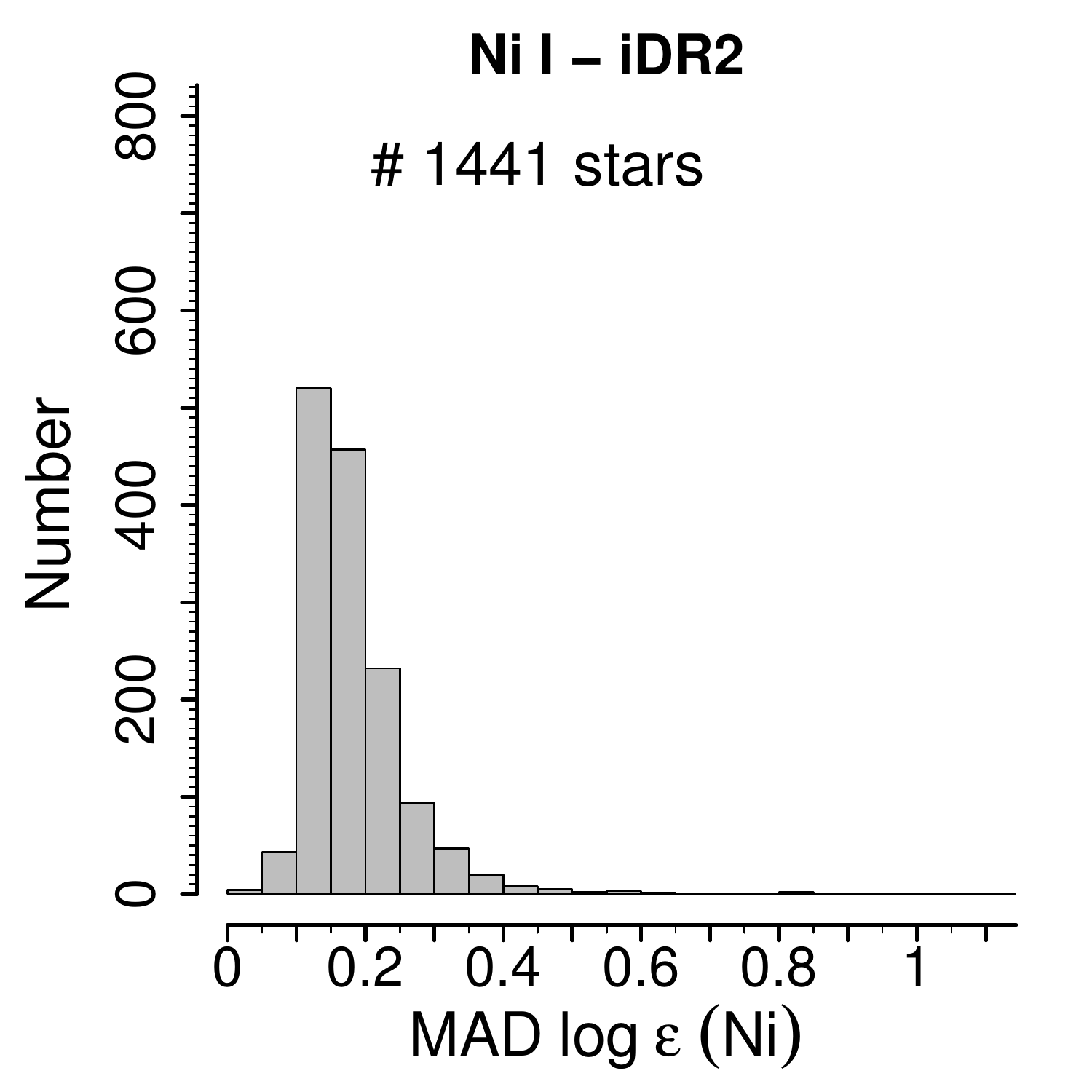}
\includegraphics[height = 4.5cm]{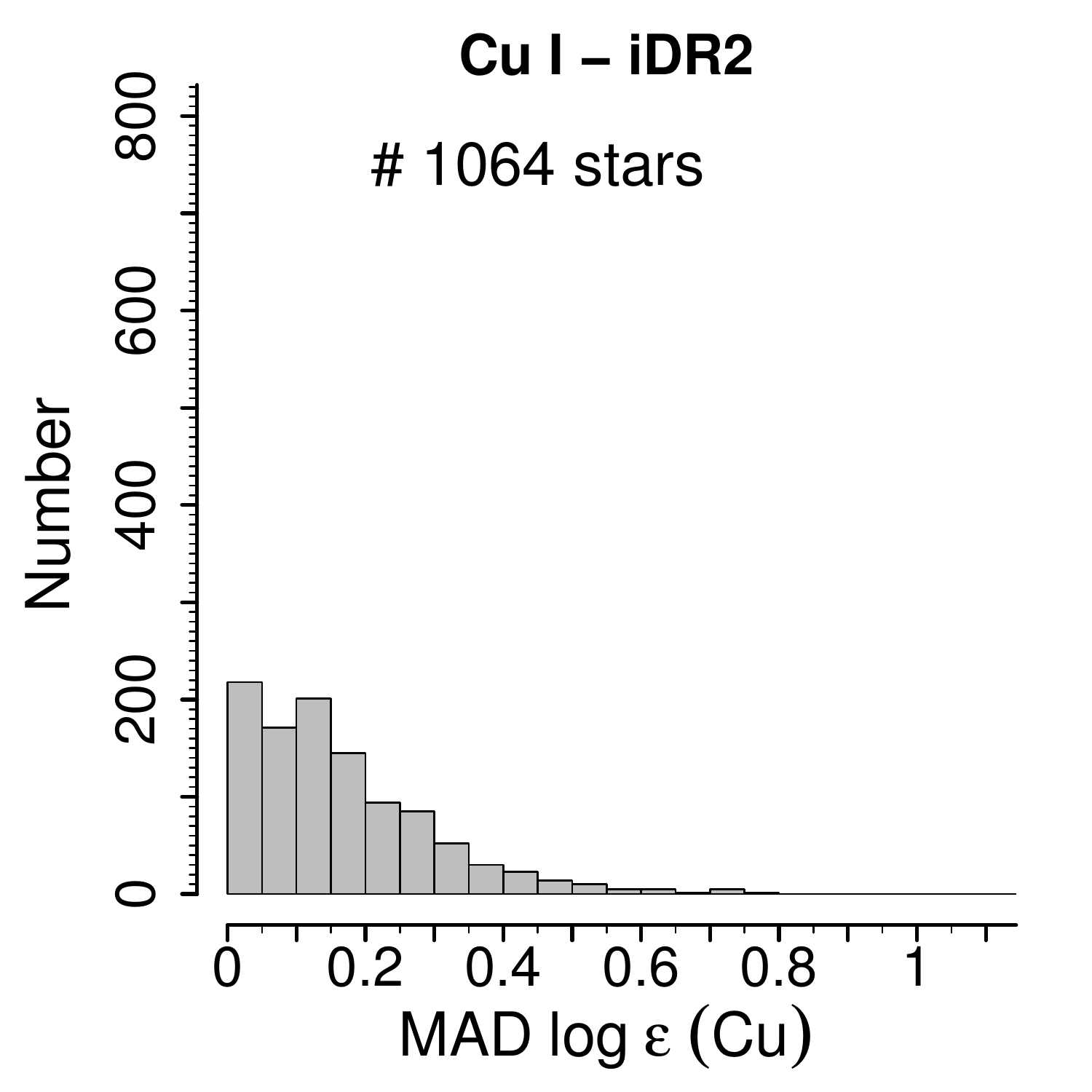}
\includegraphics[height = 4.5cm]{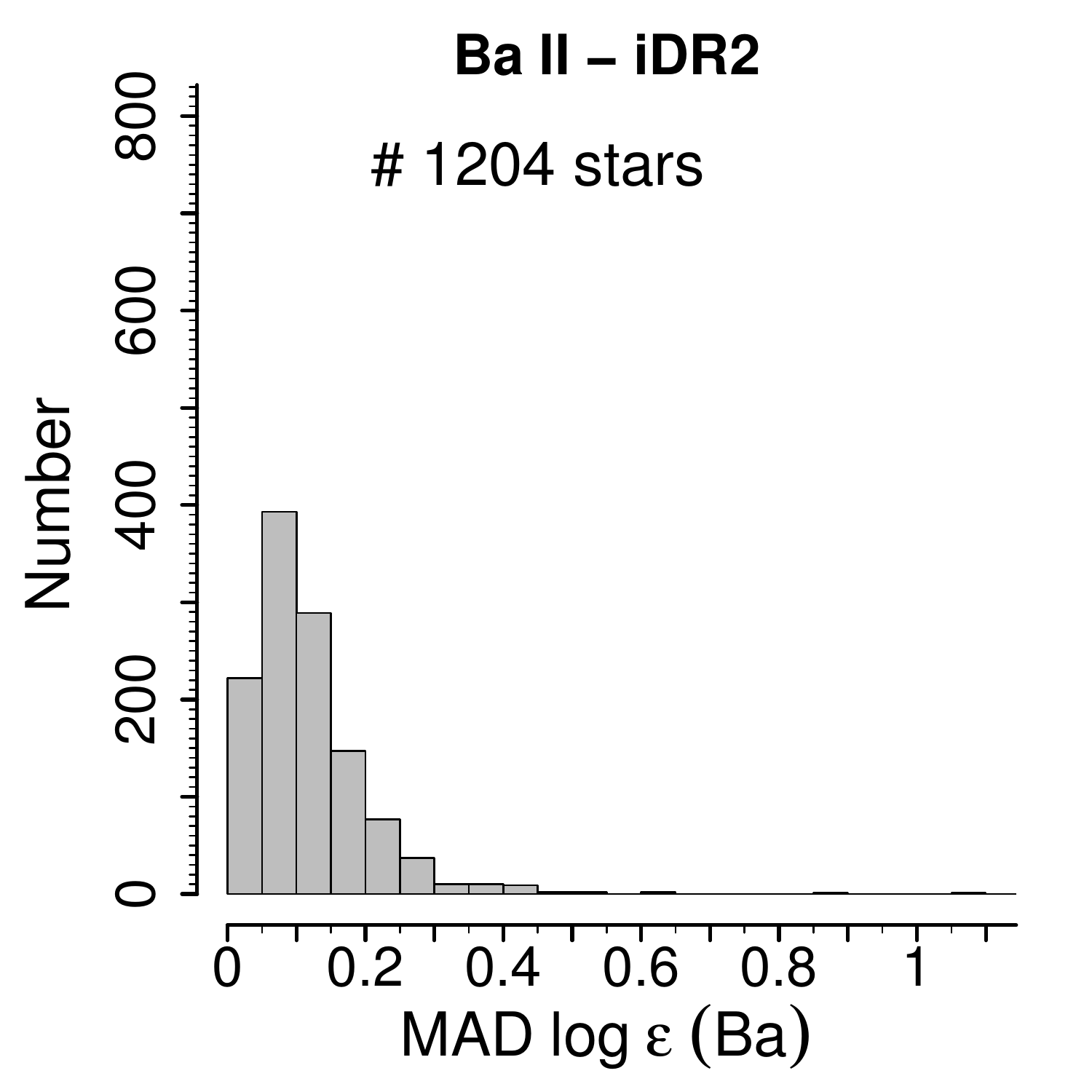}
\includegraphics[height = 4.5cm]{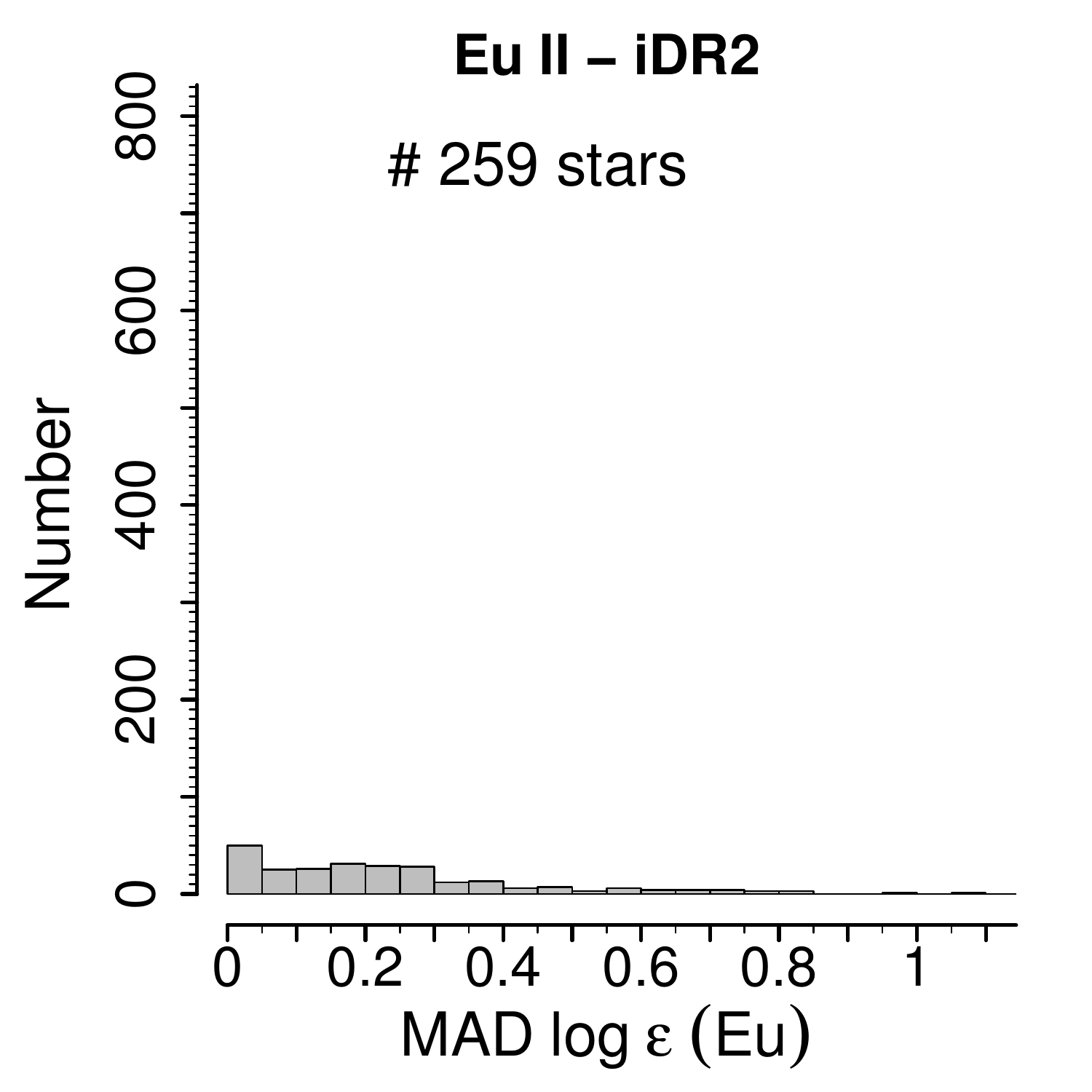}
 \caption{Histograms with the method-to-method dispersion of selected species included in the iDR2 results.}\label{fig:madelem}%
\end{figure*}

If the [Fe/H] values are used to compute Fe abundances, adopting $\log \epsilon$(Fe)$_{\odot}$ = 7.45 from \citetads{2007SSRv..130..105G}, an average offset between 0.05 and 0.08 dex is found with respect to the listed \ion{Fe}{i} abundance, with a scatter of the order of 0.10 dex. We notice that this offset is of the similar magnitude as the difference between the Solar \ion{Fe}{i} abundance derived here ($\log \epsilon$(Fe)$_{\odot}$ = 7.56, Table \ref{tab:sunabun}) and the value from \citetads{2007SSRv..130..105G}. In other words, if the stellar \ion{Fe}{i} abundance is used together with our Solar Fe abundance, the [Fe/H] would agree with the [Fe/H] value of the atmospheric parameters. Thus, although an average offset between our Fe abundances and the one of Grevesse et al. is present, it seems to be consistent throughout the whole sample. For the most metal-poor stars the values seem to be in disagreement. Therefore, we again advise that care is needed when using the results for the metal-poor stars. Some of the most metal-rich stars (dwarfs and giants) also show a worse agreement. These are difficult cases to analyze because of the increased importance of line blends, and should also be treated with care. We are working on improving the analysis for future releases, and expect improvements for these stars.

Figure \ref{fig:fe1fe2} compares the average Fe abundances obtained from \ion{Fe}{i} and \ion{Fe}{ii} lines in stars where the method-to-method dispersion of both \ion{Fe}{i} and \ion{Fe}{ii} is below 0.20 dex. Dwarfs and giants are displayed separately, but the general behavior is similar. A good agreement exists between \ion{Fe}{i} and \ion{Fe}{ii} values for almost all stars. Average offsets are small ($\sim$ 0.02-0.03 dex) and the scatter also seems to be within the uncertainties except, once again, for the most metal-poor stars of the sample. Ionization equilibrium is, however, an invalid assumption for metal-poor giants (because of non-LTE effects and possible departures of real atmospheres from model ones). As was seen previously in Table \ref{tab:nodediff}, most of the EW methods (the ones that enforce ionization equilibrium) failed in the analysis of the metal-poor benchmark stars. Therefore, their results in this region of the parameter space were not used. The EW methods that did manage to perform the analysis show a huge difference between the $\log~g$ derived enforcing ionization equilibrium and the fundamental $\log~g$ of the benchmark stars. The methods that do not enforce ionization equilibrium (the ones that look for best fitting synthetic spectra) reproduce better the real $\log~g$ of the benchmark stars. Therefore, the lack of agreement between \ion{Fe}{i} and \ion{Fe}{ii} is likely to be the correct behavior, and not a problem.

\subsection{Method-to-method dispersion}

The method-to-method dispersion of the abundances can be used as an indicator of the precision with which the results were derived. In Fig. \ref{fig:madelem} we show the histogram of the MADs of a few selected elements. In most cases, the majority of the results show very good agreement among the multiple determinations. Usually, the majority of the results have MAD below 0.20 dex. The agreement becomes worse for ionized species (like \ion{Ti}{ii}, \ion{Cr}{ii}, \ion{Ba}{ii}, and \ion{Eu}{ii}) and/or those that have important hyperfine structure (like Mn and Cu). 

\begin{figure*}
\centering
\includegraphics[height = 5cm]{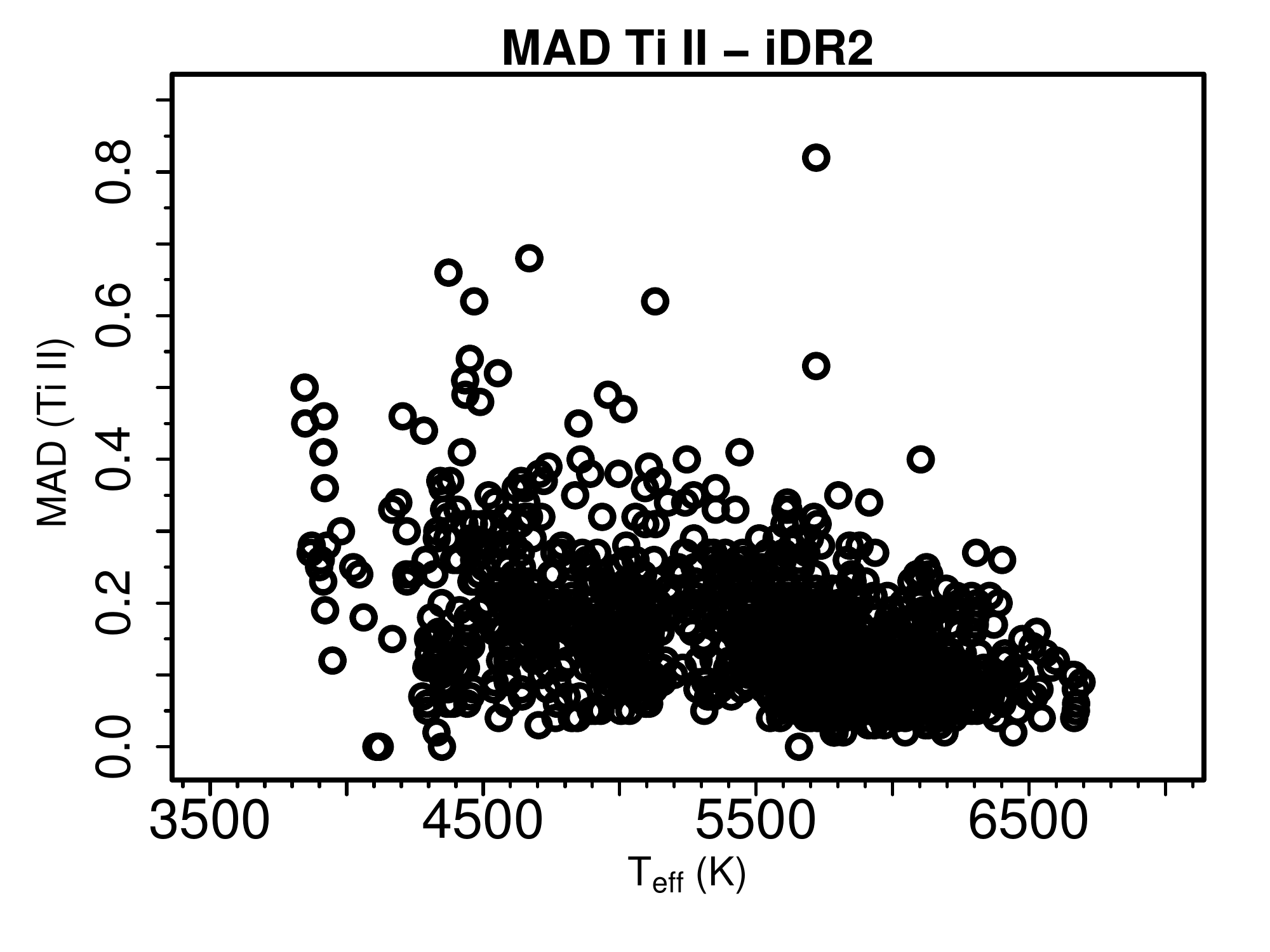}
\includegraphics[height = 5cm]{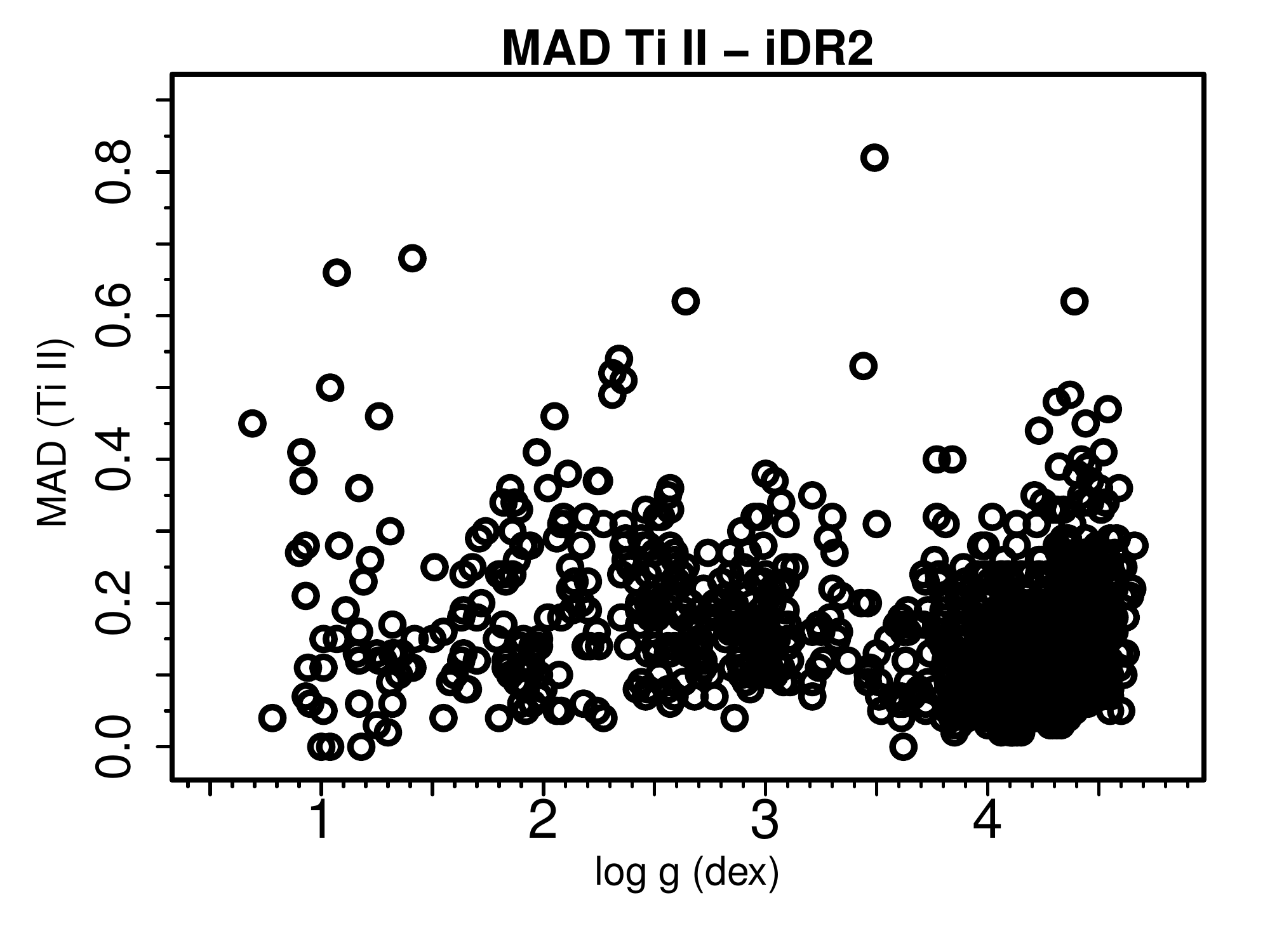}
\includegraphics[height = 5cm]{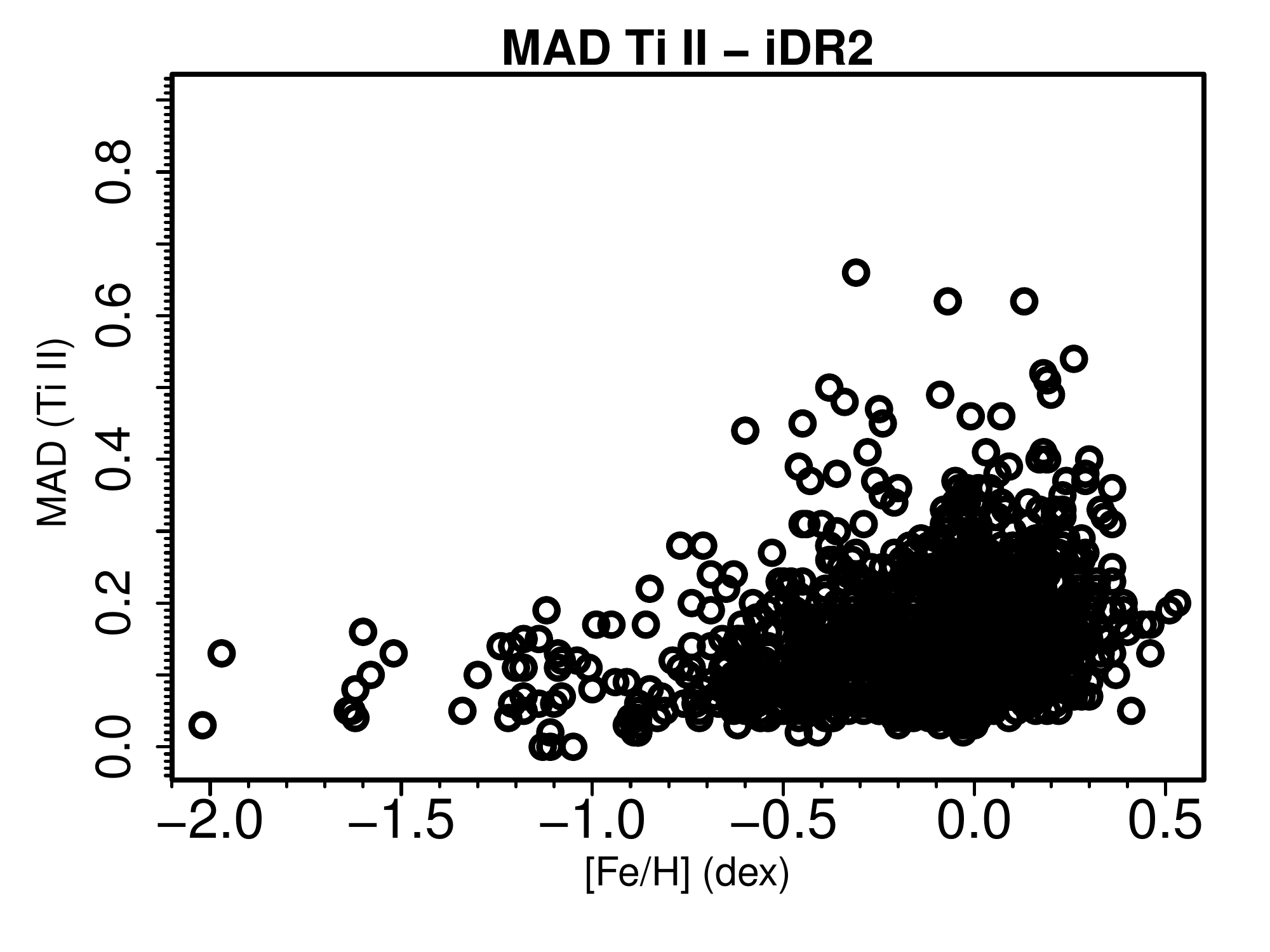}
 \caption{Method-to-method dispersion of the \ion{Ti}{ii} abundances as a function of the atmospheric parameters ($T_{\rm eff}$, $\log~g$, and [Fe/H]). The dispersion of other ionized species behave in a similar way.}\label{fig:madion}%
\end{figure*}

In Fig. \ref{fig:madion} we show the behavior of the method-to-method dispersion of \ion{Ti}{ii} as a function of the atmospheric parameters. No correlation is apparent in these plots and the behavior is the same for the other ionized species. The surface gravity is the parameter that is harder to constrain and is the one that shows larger difference between the methods (see Section 7.3). Deviations from the ionization equilibrium, problems with the lines of FeII (which are usually weaker and/or blended), and issues with atomic data are probably behind the increased method-to-method dispersion of these elements. We are working on improving the $\log~g$ determination for future releases. In particular, the Survey observed stars in the CoRoT fields, for which asteroseismic $\log~g$ values are being determined, and those will be used as reference for the next releases. With better surface gravity values we expect more precise abundances of the ionized species.

For the iDR2 abundances, we adopt the MADs as the typical uncertainties. This is akin to using the standard deviation of multiple lines of the same element, as commonly done in the literature. The third quartile of the method-to-method dispersion distribution is below 0.15 dex for \ion{Na}{i}, \ion{Al}{i}, \ion{Si}{i}, \ion{Ca}{i}, \ion{Sc}{ii}, and \ion{Ba}{ii}. These are the elements for which the quality of the abundances is higher. For the other elements, the third quartile of the method-to-method dispersion distribution is between 0.15 and 0.20 dex for \ion{Mg}{i}, \ion{Ti}{i}, \ion{V}{i}, \ion{Zr}{i}, \ion{Mo}{i}; between 0.20 and 0.25 dex for \ion{Sc}{i}, \ion{Ti}{ii}, \ion{Cr}{i}, \ion{Cr}{ii}, \ion{Ni}{i}, \ion{Cu}{i}, \ion{Y}{ii}, \ion{Zr}{ii}. For the remaining species, \ion{Mn}{i}, \ion{Co}{i}, \ion{Zn}{i}, \ion{Nd}{ii}, and \ion{Eu}{ii}, it is between 0.25 and 0.35 dex. These last elements have more uncertain abundances and should be used with care.

The most robust abundances are those where the method-to-method dispersion is smaller, as that means that the abundances computed by different groups agreed well. Therefore, we recommend that abundances for scientific purposes are chosen carefully, taking the method-to-method dispersion into account. An upper value of 0.20 dex seems to be a reasonable compromise between number of stars and precision. More stringent cuts should be considered if needed.

\section{Summary}\label{sec:end}

This paper describes the analysis of high-resolution UVES spectra of FGK-type stars obtained by the Gaia-ESO Survey. The analyses of other type of stars and/or spectra are described elsewhere (Blomme et al. 2014, in prep., Lanzafame et al. 2014, submitted, Recio-Blanco et al. 2014, in prep.). These data are used to derive precise and accurate values of atmospheric parameters and detailed elemental abundances. 

Multiple methods are used to determine these quantities. A single pipeline would be internally more homogeneous but, in a broad survey like Gaia-ESO, might introduce different systematics in different regions of the parameter space. The parameter scale is tied to the one defined by the Gaia benchmark stars, a set of well-studied stars with fundamental atmospheric parameters determined independently from spectroscopy. In addition, a set of open and globular clusters is used to evaluate the physical soundness of the results. Each of the implemented methodologies is judged against the benchmarks to define weights in three different regions of the parameter space: i) metal-rich dwarfs; ii) metal-rich giants; and iii) metal-poor stars. The final recommended results are the weighted medians of those from the individual methods. We quantify the precision of the results by means of the method-to-method dispersion, a unique Gaia-ESO product. These results are only possible thanks to the massive combined efforts of all the scientists involved in the spectrum analysis and would be hard to quantify outside such a large collaboration.

The work described here is part of the analysis effort conducted to prepare the upcoming public catalog of Gaia-ESO advanced data products. The analysis of two iDRs has been completed. These internal data releases happen roughly every 6 months, when a new analysis cycle is launched. The full analysis cycle takes between 3 and 4 months to be completed. The data products from the iDR3 analysis (at the time of writing still ongoing) will be included in the public release alongside the results of the iDR2 analysis.

Only the best recommended parameters and abundances, processed as described here, and later subjected to the final Survey-wide homogenization (see Fran\c cois et al. 2014, in prep), will be present in the public catalog. The identification, presentation, and discussion of individual scientific topics based on these results is left to the many scientific teams involved in the Gaia-ESO Survey. Because of that, we refrained from presenting in-depth scientific discussions using the results described here. The tables with the public release results will be available through the ESO data archive\footnote{\url{http://archive.eso.org/wdb/wdb/adp/phase3_spectral/form?phase3_collection=GaiaESO}}, as is already the case for the first batch of reduced Gaia-ESO spectra, but also through a dedicated Gaia-ESO Survey science archive\footnote{\url{http://ges.roe.ac.uk/index.html}}  hosted by the Wide Field Astronomy Unit (WFAU) of the Institute for Astronomy, Royal Observatory, Edinburgh, UK. This science archive is designed to provide functionalities beyond the ones available at the ESO archive.

In the latest internal release (iDR2), atmospheric parameters of 1301 FGK-type stars were derived. For 75\% of these stars, the multiple determinations of $T_{\rm eff}$, $\log g$, and [Fe/H] for the same star agree to better than 82 K, 0.19 dex, and 0.10 dex, respectively. The tests and comparisons presented here indicate that care is needed with the results of both cool ($T_{\rm eff}$ $<$ 4000 K) or metal-poor stars ([Fe/H] $<$ $-$2.00). Abundances for 24 elements were derived in at least a few stars: C, N, O, Na, Mg, Al, Si, Ca, Sc, Ti, V, Cr, Mn, Fe, Co, Ni, Cu, Zn, Y, Zr, Mo, Ba, Nd, and Eu. We derived abundances of at least 15 different elements for 1079 stars and for at least ten elements for 1203 stars. For the abundances, the majority of the multiple determinations agree to better than 0.20 dex.

The list includes abundances of elements formed in all nucleosynthetic channels. This highlights the unique value of the sample being analyzed here. This is only possible thanks to the high-quality, high-resolution, and large wavelength coverage of the UVES data. The exciting potential of these results is exemplified by the variety of early science papers being produced by the Gaia-ESO collaboration \citepads[e.g.][]{2014A&A...565A..89B,2014arXiv1407.1510C,2014A&A...561A..94D,2014A&A...563A.117F,2014A&A...563A..44M,2014A&A...567A..55S}.

The value of the Gaia-ESO science products will be further enhanced when the results of the Gaia mission \citepads{2001A&A...369..339P} become available. Gaia will provide parallaxes, proper motions, and spectrophotometric metallicities for $\sim$ 10$^{9}$ stars and radial velocities for $\sim$150 million stars. The chemical information coming from the Gaia spectra are, however, limited: metallicities ([Fe/H]) and abundances for a few elements, mostly $\alpha$-elements (Ca, Si, Ti), will be obtained for $\sim$ two million stars brighter than $V$ $\leqslant$ 12-13 mag \citepads{2005MNRAS.359.1306W}. Abundances of elements formed by other nucleosynthetic channels (s-process, r-process, Fe peak elements, light elements) in fainter stars, covering a larger volume in the Galaxy, require additional observations from ground-based observatories, such as the ones being carried out within the on-going Gaia-ESO Survey.

The sample of high-resolution spectra of FGK-type stars discussed here is already among the largest ones of its kind analyzed in a homogeneous way. The results will enable significant advances in the areas of stellar evolution and Milky-Way formation and evolution.

\begin{acknowledgements}
R.Sm., through an ESO fellowship, has received funding from the European Community's Seventh Framework Programme 
(FP7/2007-2013) under grant agreement No. 229517. A.J.K. and U.H. acknowledges support by the Swedish National Space Board (SNSB) through several grants. D.G. gratefully acknowledges support from the Chilean BASAL Centro de Excelencia en Astrofisica y Tecnologias Afines (CATA) grant PFB-06/2007. E.C., H.G.L., L.Sb. and S.D. acknowledge financial support by the Sonderforschungsbereich SFB\,881 ``The Milky Way System'' (subprojects A2, A4, A5) of the German Research Foundation (DFG). I.S.R. gratefully acknowledges the support provided by the Gemini-CONICYT project 32110029. L.Sb. and S.D. acknowledge the support of Project IC120009 "Millennium Institute of Astrophysics (MAS)" of Iniciativa Cient\'{\i}fica Milenio del Ministerio de Econom\'{\i}a, Fomento y Turismo de Chile. M.V. acknowledges financial support from Belspo for contract PRODEX COROT. P.B. acknowledges support from the PNCG of INSU CNRS. Part of the computations have been performed on the 'Mesocentre SIGAMM' machine, hosted by Observatoire de la Cote d'Azur. P.d.L., V.H. and A.R. acknowledge the the support of the French Agence Nationale de la Recherche under contract ANR-2010-BLAN- 0508-01OTP and from the ``Programme National de Cosmologie et Galaxies'' (PNCG) of CNRS/INSU, France. S.G.S, E.D.M., and V.Zh.A. acknowledge support from the Funda\c c\~ao para a Ci\^encia e Tecnologia (Portugal) in the form of grants SFRH/BPD/47611/2008, SFRH/BPD/76606/2011, SFRH/BPD/70574/2010, respectively. S.Vi. gratefully acknowledges the support provided by FONDECYT reg. n. 1130721. T.B. was funded by grant No. 621-2009-3911 from The Swedish Research Council. T.Mo. acknowledges financial support from Belspo for contract PRODEX GAIA-DPAC. We acknowledge the support from INAF and Ministero dell' Istruzione, dell' Universit\`a' e della Ricerca (MIUR) in the form of the grant "Premiale VLT 2012". This work was partly supported by the European Union FP7 program through ERC grant number 320360 and by the Leverhulme Trust through grant RPG-2012-541. The results presented here benefit from discussions held during the Gaia-ESO workshops and conferences supported by the ESF (European Science Foundation) through the GREAT Research Network Program. This research has made use of the SIMBAD database, operated at CDS, Strasbourg, France, of NASA's Astrophysics Data System, of the compilation of atomic lines from the Vienna Atomic Line Database (VALD), and of the WEBDA database, operated at the Department of Theoretical Physics and Astrophysics of the Masaryk University.

\end{acknowledgements}

\bibliographystyle{aa} 
\bibliography{../smiljanic} 

\appendix

\section{Nodes and methods}\label{sec:nodes}

The UVES data of late-type stars are analyzed in parallel by 13 different Nodes. The details of each analysis methodology and the codes employed are described in the subsections below. Table \ref{tab:nodsum} summarizes some characteristics of the methodology employed by each Node.

\subsection{Bologna}\label{sec:bologna}

The Bologna Node employs the classical EW method for determining atmospheric parameters and abundances. The atmospheric parameters are determined by erasing any trend of the abundances of the iron lines with excitation potential and with EW, and by minimizing the difference between the abundances given by ionized and neutral iron lines. A final health check of the method is provided by verifying that no significant trend of iron abundances with wavelength is present. Abundances are derived for each absorption line of the species of interest.

To measure EWs, the automated {\sf FORTRAN} code {\sf DAOSPEC} \citepads{2008PASP..120.1332S,2010ascl.soft11002S} is used. {\sf DAOSPEC} is designed to measure EWs in high-resolution (R $\geq$ 15\,000) high-S/N stellar spectra ($\geq$ 30). Upon request, the code normalizes the spectrum by adjusting, iteratively, polynomials to the residuals spectrum (i.e., a spectrum obtained by removing all measured lines from the original spectrum). {\sf DAOSPEC} provides: a global uncertainty of the fit in the form of an average root mean square (r.m.s.) of the residuals spectrum; a radial velocity measurement (with its 1$\sigma$ spread and the number of lines on which it is based); and the EWs with their uncertainty and quality parameters.

{\sf DAOSPEC} can be somewhat difficult to configure, especially when many spectra with different properties (i.e., S/N, line crowding, full width half maximum -- FWHM -- and exact spectral coverage) need to be measured in a short time, as is the case for Gaia-ESO. Therefore, the code is executed through a wrapper that configures automatically many of its parameters, providing all the statistics and graphical tools to explore the results and correct the deviant cases. This wrapper program is called {\sf DOOp} \citepads[{\sc D}AOSPEC {\sc O}ption {\sc O}ptimizer {\sc p}ipeline,][]{2014A&A...562A..10C}. 

Finally,  to derive automatically the atmospheric parameters and elemental abundances the code {\sf GALA}\footnote{{\sf GALA} is freely distributed at the Cosmic-Lab project Web site, \url{http://www.cosmic-lab.eu/Cosmic-Lab/Products.html}} is used \citepads{2013ApJ...766...78M}. {\sf GALA} is based on the Kurucz suite of abundance calculation codes \citepads{2005MSAIS...8...14K,2004MSAIS...5...93S}. {\sf GALA} can run starting from a random first guess of the atmospheric parameters and converges rapidly to meaningful solutions for spectra with the resolution, S/N, and wavelength coverage of the UVES spectra analyzed here. {\sf GALA} performs a rejection of too weak or too strong absorption lines (the limits are set around the $\log$ (EW/$\lambda$) $\simeq$ $-$4.7 and $-$5.9, depending on the star), selects only lines having a certain measurement error (cutting above 5-20\%, depending on the spectrum), and performs a sigma-clipping rejection in abundance (set to 2.5$\sigma$). {\sf GALA} provides uncertainties on the atmospheric parameters and on the derived abundances, both in the form of a 1$\sigma$ spread of the abundances of each line (together with the number of used lines for each species) and in the form of errors on the abundances induced by the uncertainties on the atmospheric parameters \citepads[using the prescription of][in the case of the present analysis]{2004A&A...416.1117C}.

\subsection{Catania}\label{sec:catania}

The Catania Node uses the code {\sf ROTFIT}, developed by \citetads{2003A&A...405..149F,2006A&A...454..301F} in {\sf IDL}\footnote{{\sf IDL} (Interactive Data Language) is a registered trademark of ITT Visual Information Solutions.} software environment. The code originally performed only an automatic MK spectral classification and $v \sin i$ measurement minimizing the $\chi^2$ of the residual ($observed - reference$) spectra. The reference spectra come from an adopted spectrum library and are artificially broadened by convolution with rotational profiles of increasing $v \sin i$. The code was later updated for evaluating the atmospheric parameters $T_{\rm eff}$, $\log g$, and $\rm [Fe/H]$ with the adoption of a list of reference stars with well known parameters \citepads[e.g.,][]{2009A&A...504..829G}. 

Unlike codes based on the measurements of EWs and curves of growth, {\sf ROTFIT} can be applied to the spectra of FGK-type stars with relatively high rotational velocity ($v \sin i$ $\geq$ 20 km\,s$^{-1}$), where the severe blending of individual lines either hampers or absolutely prevents the use of the above methods. Nevertheless, the analysis was limited to stars with $v \sin i$ $\leq$ 300 km\,s$^{-1}$.

A reference library composed of 270 high-resolution (R = 42\,000) spectra of slowly-rotating FGKM-type stars available in the ELODIE archive \citepads{2001A&A...369.1048P} was used. For most of these reference stars, basically those with spectral type in the range from mid-F to late-K, the atmospheric parameters have been redetermined by L. Spina using the EPInArBo methodology (see Section \ref{sec:epinarbo}). For the remaining few stars, either the recent values tabulated in the PASTEL catalog \citepads{2010A&A...515A.111S} or derived in the works of \citetads{2012ApJ...748...93R} and \citetads{2012ApJ...757..112B}, for the M-type dwarfs, were used. Although the parameter space is not regularly sampled, the reference stars cover all the regions relevant for the analysis of FGK-type stars with [Fe/H] $\ge -2.0$.

Segments of the spectra with 100~{\AA} each are analyzed independently. Spectral regions heavily affected by telluric lines and the cores of Balmer lines, that can be contaminated by chromospheric emission, are excluded. The final stellar parameters $T_{\rm eff}$, $\log g$, $\rm [Fe/H]$, and $v \sin i$, are the averages of the results of each $i$-th spectral segment weighted according to the $\chi_i^{2}$ of the fit and to the amount of information contained in the segment, which is expressed by the total line absorption $f_i=\int(F_{\lambda}/F_{\rm C}-1)d\lambda$. The uncertainties  of $T_{\rm eff}$, $\log g$, $\rm[Fe/H]$, and $v \sin i$ are the standard errors of the weighted means added in quadrature to the average uncertainties of the stellar parameters of the reference stars evaluated as $\pm$\,50\,K, $\pm$\,0.1\,dex, $\pm$\,0.1\,dex, and $\pm$\,0.5\,km\,s$^{-1}$ for $T_{\rm eff}$, $\log g$, $\rm [Fe/H]$, and $v \sin i$, respectively. Moreover, the MK spectral type and luminosity class of the star is also provided. They are defined as those of the reference star which more frequently matches the target spectrum in the different spectral regions.

\subsection{CAUP}\label{sec:caup}

The CAUP Node determines the stellar atmospheric parameters ($T_{\rm eff}$, $\log g$, $\xi$) and the metallicity automatically, with a method used in previous works now adapted to the Gaia-ESO Survey \citepads[e.g.][]{2008A&A...487..373S,2011A&A...526A..99S}. The method is  based on the excitation and ionization balance of iron lines using [Fe/H] as a proxy for the metallicity. The list for the iron lines used to constrain the parameters was selected from the Gaia-ESO line list using a new procedure described in detail in \citetads{2014A&A...561A..21S}.

The EW of the lines are automatically determined using the {\sf ARES}\footnote{{\sf ARES} can be downloaded 
at \url{http://www.astro.up.pt/\~sousasag/ares/}} code \citepads{2007A&A...469..783S} following the approach of \citetads{2008A&A...487..373S,2011A&A...526A..99S} to adjust {\sf ARES} according to the S/N of each spectrum.

The stellar parameters are computed assuming LTE using the 2002 version of {\sf MOOG} \citepads{1973PhDT.......180S} and the MARCS grid of models. For that purpose, the interpolation code provided with the MARCS grid was modified to produce an output model readable by {\sf MOOG}. Moreover, a wrapper program was implemented to the interpolation code to automatize the method.

As damping prescription, the Uns$\ddot{{\rm o}}$ld approximation multiplied by a factor recommended by the Blackwell group (option 2 within {\sf MOOG}) was used. The atmospheric parameters are inferred from the previously selected 
$\ion{Fe}{i}$-$\ion{Fe}{ii}$ line list. A minimization algorithm, the Downhill Simplex Method \citepads{1992nrca.book.....P}, is used to  find the best parameters. In order to identify outliers caused by incorrect EW values, a 3$\sigma$ clipping of the 
$\ion{Fe}{i}$ and $\ion{Fe}{ii}$ lines is performed after a first determination of the stellar parameters. After this clipping, the procedure is repeated without the rejected lines. The uncertainties in the stellar parameters are determined as in previous works \citepads[][]{2008A&A...487..373S,2011A&A...526A..99S}.

Individual abundances are derived using the same 
tools and methodology as described above, but using the 2010 version of {\sf MOOG} \citepads[see][for details]{2009A&A...497..563N, 2012A&A...545A..32A}. The line list for elements other than Fe was selected through the cross-matching between the line list used by \citetads[][]{2012A&A...545A..32A} and the line list provided by Gaia-ESO. The atomic data from the Gaia-ESO Survey was adopted. The errors for the abundances represent the line-to-line scatter. 

\subsection{Concepcion}\label{sec:concepcion}

The Concepcion Node uses the abundances from \ion{Fe}{i} and \ion{Fe}{ii} lines to obtain atmospheric parameters using the classical EW method. The atmospheric parameters are determined by satisfying the excitation and ionization equilibrium, and by minimizing trends of abundance with EW. The spectroscopic optimization of all the atmospheric parameters is achieved simultaneously.

The EWs are determined with the automatic code {\sf DAOSPEC} (see description in Section \ref{sec:bologna}). The code adopts a saturated Gaussian function to fit the line profile and a unique value for the FWHM for all the lines. The input values of FWHM are derived manually using the IRAF\footnote{IRAF is distributed by the National Optical Astronomy Observatory, which is operated by the Association of Universities for Research in Astronomy (AURA) under cooperative agreement with the National Science Foundation.} task splot, leaving {\sf DAOSPEC} free to re-adjust the values according to the global residual of the fitting procedure. The measurement of EWs is repeated by using the optimized FWHM value as a new input value until convergence is reached at a level of 5\%. The EWs are measured after a re-normalization of the continuum, done to remove any residual trends introduced during the data reduction. 

{\sf GALA} is used to determine the atmospheric parameters and elemental abundances (see description of {\sf GALA} in Section \ref{sec:bologna}). Starting from an initial guess of atmospheric parameters, {\sf GALA} converges rapidly to a meaningful solution. Finally, it computes accurate internal errors for each atmospheric parameter and abundance. When the initial set of parameters are poorly known or in cases of large uncertainties, the Guess Working-Block of {\sf GALA} is used. This Working-Block verifies the initial parameters quickly by exploring the parameters space in a coarse grid, saving a large amount of time. In addition, the errors in the EW measurement obtained from {\sf DAOSPEC} are provided as an input, so the best model atmosphere is computed taking into account the abundance uncertainties of the individual lines. 

\subsection{EPInArBo}\label{sec:epinarbo}

The EPInArBo (ESO-Padova-Indiana-Arcetri-Bologna) Node performs the spectral analysis with the codes {\sf DOOp} and {\sf FAMA} \citepads[{\sc F}ast {\sc A}utomatic {\sc M}oog {\sc A}nalysis,][]{2013A&A...558A..38M}\footnote{FAMA is available from \url{http://cdsarc.u-strasbg.fr/viz-bin/qcat?J/A+A/558/A38}}. The former (described in Sect.~\ref{sec:bologna}) makes it more convenient to measure EWs in hundreds of spectra in a single batch. The latter is an automation of the 1D-LTE code {\sf MOOG} and allows the determination of stellar parameters and individual element abundances. 

The EWs are measured after a re-normalization of the continuum. Each line is measured using a Gaussian fit. Equivalent widths in the range between 20-120 m\AA, for the \ion{Fe}{i} and \ion{Fe}{ii} lines, and in the range between 5-120 m\AA, for the other elements, were used. 

{\sf FAMA} uses the EWs of \ion{Fe}{i} and \ion{Fe}{ii} to derive stellar parameters ($T_{\rm eff}$, $\log g$, [Fe/H], and $\xi$). A set of  first-guess parameters are first produced using the available photometric data and information from the target selection, using the following steps: 
\begin{enumerate}
\item A first guess estimation of $T_{\rm eff}$ is given by the \citetads{1999A&AS..140..261A} and \citetads{2010A&A...512A..54C} relations for both cluster and field stars;
\item The cluster parameters, such as distance, age, and reddening, available in the reports prepared by 
Gaia-ESO working group 4 (cluster stars target selection, see Bragaglia et al. 2014, in prep.) are used to fix the surface gravity;
\item For the field stars, the information available from target selection is used (i.e., whether the star was a turn-off dwarf or bulge/inner-disk giant) to set a first guess gravity. 
\end{enumerate}

The stellar parameters are obtained by searching iteratively for the three equilibria (excitation, ionization, and trend between $\log$ n(Fe I) and $\log$ (EW/$\lambda$)), i.e. with a series of recursive steps starting from a set of initial atmospheric parameters and arriving at a final set of atmospheric parameters which fulfills the three equilibrium conditions. 

The convergence criterion is set using the information on the quality of the EW measurements, i.e., 
the minimum reachable slopes are linked to the quality of the spectra, as expressed by the dispersion $\sigma_{\rm FeI}$ around the average value $< \log$ n(FeI) $>$. This is correct in the approximation that the main contribution to the dispersion is due to 
the error in the EW measurement rather than to inaccuracy in atomic parameters, as e.g.,  the oscillator strengths ($\log gf$).

\subsection{IAC-AIP}\label{sec:iacaip}

The IAC-AIP Node employs the optimization code {\sf FERRE} to identify the combination of atmospheric parameters of a synthetic model that best matches each observed spectrum. {\sf FERRE} searches for the best solution in a $\chi^2$ sense using the Nelder-Mead algorithm \citep{NelderMead65}, and the model evaluation is sped up by holding a pre-computed grid in memory and interpolating within it. The algorithm is the same described by \citetads{2006ApJ...636..804A} for the analysis of SDSS/SEGUE data and by  \citetads{2014ApJS..211...17A} for the analysis of SDSS/APOGEE spectra. Model interpolations are carried out with cubic Bezier splines, whose accuracy has been studied in detail by \citetads{2013MNRAS.430.3285M}. For each spectrum, five searches initialized at randomly chosen points on the parameter space are performed, and the best solution is retained.

The adopted grid of model spectra was not the one described in Sect. \ref{sec:grid}. It was calculated using the code {\sf Turbospectrum} \citepads{1998A&A...330.1109A,2012ascl.soft05004P} based on MARCS model atmospheres with the VALD3 line list \citepads[][]{2011BaltA..20..503K}, with updates on  $\log gf$ values according to the Gaia-ESO line list version 3.0. The parameter range covered by the grid is: $T_{\rm eff}$ = 3000 -- 7000~K, $\log g$ = 0.0 -- 5.0, [Fe/H] = $-$2.5 -- +1.0, $v \sin i$ = 1 -- 128 km s$^{-1}$, $\xi$ = 0.5 -- 4 km s$^{-1}$, and [$\alpha$/Fe] = $-$0.4 -- +0.4. The model spectra were smoothed by Gaussian convolution to the resolving power of the observations (R = 47\,000). To speed up the analysis, the [$\alpha$/Fe] is tied to the overall metallicity of each star, i.e. with enhanced [$\alpha$/Fe] ratios at low metallicity, while $\xi$ is tied to both $T_{\rm eff}$ and $\log g$ according to the Gaia-ESO microturbulence relation for the iDR1 analysis. 

All of the available UVES orders, before merging by the data reduction software, from both CCDs are used, excluding only regions with many telluric lines and the core of the H$\alpha$ line. The continuum for both the observations and the models is set by cutting the spectra into 2\AA\, wide chunks, dividing each chunk by its mean value, and all spectra are weighted according to their variance. All observations are shifted to rest wavelength. For observations with one radial velocity in the header, this value was used. If two values were present (one for each CCD), the average value was used. In case no velocity was available, a cross correlation using a hot template star ($T_{\rm eff}$ = 7000~K, $\log g$ = 2) spun up to 50km s$^{-1}$ was used to derive the radial velocity. If this failed, a value of 0.0 km s$^{-1}$ was used.

\subsection{Li\`ege}\label{sec:liege}

The Li\`ege Node performs the analysis using the {\sf GAUFRE} tool \citepads{2013EPJWC..4303006V}. {\sf GAUFRE} is a {\sf C++} code that performs the determination of atmospheric parameters and abundances in an automatic way. The tool is made up of several sub-programs with specific tasks \citepads[see][for details]{2013EPJWC..4303006V}. For the Gaia-ESO Survey UVES spectra, {\sf GAUFRE-EW} is used. This sub-program determines $T_\mathrm{eff}$, $\log g$, [M/H], and $\xi$, in an iterative way using the EWs of Fe lines. 

The starting point is the normalization of the spectrum and the measurement of the EWs of every line present in the input line list (when detectable). The program selects a spectral range of 3-4 \AA~ around the line center and the spectrum is then fitted with a polynomial function in order to determine the continuum and the line position. At this stage, several parameters, as the degree of the function and the amplitude of the spectral range to fit, can be defined by the user.

The program then feeds {\sf MOOG} with the measured EWs and an appropriate MARCS model atmosphere. Within the errors, the {\sf MOOG} results must satisfy four conditions: fulfill the Fe ionization and excitation equilibria, show no dependence between the \ion{Fe}{i} abundances and $\log$ (EW/$\lambda$), and finally yield a mean metallicity identical to that of the adopted model atmosphere. The appropriate MARCS model atmosphere is derived by interpolating within the MARCS grid.

The program iterates until the four conditions are fulfilled. The Downhill Simplex Method \citepads{NelderMead65,2002nrc..book.....P} is adopted for estimating at each iteration step the new set of atmospheric parameters. The starting point of the process can be determined by the user. Here, photometric temperatures using  \citetads[][]{2005ApJ...626..465R} and, when available, $\log g$ from asteroseismology were adopted. When no information from photometry or asteroseismology is available, the starting point is set to $T_\mathrm{eff}$ = 5000 K, $\log g$ = 4.0 dex, [M/H] = 0.0 dex, and $\xi$ = 1.0 km s$^{-1}$.

The uncertainty in $T_\mathrm{eff}$ is derived from the standard deviation of the least-square fit of the \ion{Fe}{i} abundance vs.\ excitation potential. The uncertainty in $\log g$ is determined by propagating the uncertainty in $T_\mathrm{eff}$. The uncertainty in $\xi$ is calculated based on the standard deviation of the least-square fit of the \ion{Fe}{i} abundance vs.\ $\log$ (EW/$\lambda$). The uncertainty in [Fe/H] takes into account the uncertainties in $T_\mathrm{eff}$, $\log g$, $\xi$, and the line-to-line scatter of the \ion{Fe}{i} abundances. 

\subsection{LUMBA}\label{sec:lumba}

The LUMBA (Lund-Uppsala-MPA-Bordeaux-ANU) Node uses a stellar parameter and abundance pipeline (hereafter referred to as {\sf SGU}) that is based upon the {\sf SME} (Spectroscopy Made Easy) spectrum synthesis program \citepads{1996A&AS..118..595V}\footnote{\url{http://www.stsci.edu/\~valenti/sme.html}}. {\sf SME} is a suite of {\sf IDL} and {\sf C++} routines developed to compute theoretical spectra and perform a $\chi^2$ fit to observed spectra. The code assumes LTE and plane-parallel geometry. Chemical equilibrium for molecules is determined as described in \citetads{1998ApJ...498..851V}. 

A detailed description of the {\sf SGU} pipeline will be published elsewhere (Bergemann et al. 2014c, in prep).  Briefly, in the {\sf SGU} pipeline, synthetic spectra are computed in pre-defined wavelength segments, which are $5$ to $20\, \AA$ wide. The selected line list is a reduced version of the Gaia-ESO version 3.0 line list and includes the atomic and molecular blends relevant for the analysis of FGKM-type stars. Basic stellar parameters are determined iteratively, exploring the full parameter space in $T_{\rm eff}, \log g$, [Fe/H], micro- and macro-turbulence. The number of iterations varies, depending on the stellar parameters, value of the goodness-of-fit test ($\chi^2$), and convergence. The main purpose of {\sf SGU} is to control the sequence of steps which defines the parameters to solve for, in the current iteration, and specify the wavelength regions to include in the test statistics. Usually, $3$ to $4$ steps for dwarfs and subgiants, and $2$ steps for giants are used. The wavelength regions (referred to as ``masks'') to be included in the $\chi^2$ fit also vary depending on the step. The masks cover the lines of \ion{H}{i} (H$_\beta$ and H$_\alpha$), \ion{Mg}{i} triplet at $5170\, \AA$, and a carefully-selected set of Fe lines. In total, about $60$ diagnostic \ion{Fe}{i} and \ion{Fe}{ii} transitions are used. The merged not normalized Gaia-ESO spectra are used with a run-time continuum normalization.  For the abundance analysis, special masks were developed, which cover the lines of selected elements. For iDR1, atmospheric parameters were computed assuming LTE. For iDR2, the pipeline was modified to include NLTE corrections in Fe (Bergemann et al. 2014d, in prep). That resulted in improved stellar parameters (especially $\log~g$) for low-metallicity stars. Further, the effects were quite small for more metal-rich stars. Abundances are determined in the last step using stellar parameters from the previous runs. 

Errors in the other stellar parameters are estimated from internal {\sf SME} errors based on S/N and Fe line-to-line scatter (but in many cases, lines of different elements were used to derive stellar parameters, including H and Mg), combined with the spread in differences between our results for the benchmark stars library and those values that have been deemed acceptable. 

\subsection{Nice}\label{sec:nice}

The Nice Node analysis is based on the automated stellar parametrization pipeline developed for the AMBRE Project \citepads{2012A&A...542A..48W}. At the core of the pipeline is the stellar parametrization algorithm {\sf MATISSE} (MATrix Inversion for Spectrum SynthEsis), developed at the Observatoire de la C\^{o}te d'Azur primarily for use in the Gaia RVS (Radial Velocity Spectrometer) stellar parametrization pipeline \citepads{2006MNRAS.370..141R}, and the Gaia-ESO synthetic spectrum grid (see Sect.~\ref{sec:grid}).

{\sf MATISSE} is a local multi-linear regression method that simultaneously determines the stellar parameters ($\theta$) of an observed spectrum $O(\lambda)$ by the projection of the spectrum onto vector functions $B_{\theta}(\lambda)$. A $B_{\theta}(\lambda)$ function is constructed as an optimal linear combination of the local synthetic spectra $S(\lambda)$. The stellar parameters determined by the Nice Node are $T_{\rm eff}$, $\log g$, a global metallicity [M/H], and a global $\alpha$-element abundance over iron ([$\alpha$/Fe]: $\alpha$ = O, Ne, Mg, Si, S, Ar, Ca, and Ti).

To minimize the impact of mismatches between the observed and synthetic spectra, a solar flux spectrum \citepads{1998assp.book.....W} and an Arcturus spectrum \citepads{2000vnia.book.....H} are compared with corresponding Gaia-ESO synthetic spectra in the UVES spectral range. About 24\% of the UVES domain is discarded due to telluric/instrumental contamination. A further 4\% is discarded for differences between the observed and synthetic normalized fluxes greater than 10\% for the Sun or 20\% for Arcturus. These limits reject grossly discrepant spectral features (errant lines or blatant mismatched regions) between the observed and synthetic spectra. The resulting comparison prior to any normalization optimisation shows for the remainder that 95\% (resp. 80\%) of the pixels have less than 5\% difference between the Sun (resp. Arcturus) and the corresponding synthetic spectrum, while 94\% of the pixels have flux differences less than 10\% in the case of Arcturus. As MATISSE uses all the available pixels for the parameter determination, any few discrepant pixels remaining after the full iterative normalisation have little effect on the result.

The final wavelength domain totals 1447~\AA\ between 4790~\AA\ and 6790~\AA\ with 18\,080 pixels at a sampling of 0.08~\AA/px. The synthetic spectra are convolved with a Gaussian kernel (FWHM = 0.2254~\AA) for a resolution range from $R$ $\sim$ 21\,000 (4790~\AA) to $R$ $\sim$ 30\,000 (6790~\AA). The observed spectra are convolved to the same resolution using a transformation FWHM based on the measured spectral FWHM and grid FWHM.

The Nice pipeline consists of spectral processing ($v_{\rm rad}$ correction; cleaning/slicing/convolution; normalization to synthetic spectra), and stellar parameter determination by {\sf MATISSE} \citepads[SPC stage in Fig.~4 of][]{2012A&A...542A..48W}. At each iteration of these last two stages, improved estimates of the stellar parameters provide new synthetic spectra for use in the normalization until there is convergence on the final stellar parameters.

Calibration and validation of the pipeline was undertaken using three key samples: the Gaia-ESO Benchmarks (see Sect.~\ref{sec:bench}); the spectral library of \citetads{2014A&A...564A.133J}; and the AMBRE:UVES\#580 PASTEL data set (Worley et al. 2014, submitted), a sample of 2273 slit spectra that have high quality spectroscopic stellar parameters cited in the PASTEL catalog. These three samples were used to calibrate the convolution and normalization in the spectral processing by comparison of processed spectra with synthetic spectra and by comparing the {\sf MATISSE} parameters with the accepted parameters for each sample. 

\subsection{Paris-Heidelberg}\label{sec:paris}

The Paris-Heidelberg Node uses the automatic parameter determination and abundance analysis code {\sf MyGIsFOS} \citepads{2014A&A...564A.109S}. {\sf MyGIsFOS} strictly replicates a ``traditional'', or ``manual'', parameter determination and abundance analysis method in a fully automated fashion. To do so, {\sf MyGIsFOS} determines EWs and abundances for a number of \ion{Fe}{i} and \ion{Fe}{ii} features, and looks for the atmospheric parameters (T$_{\rm eff}$, $\log g$, $\xi$) that satisfy the excitation and ionization equilibrium, and that minimizes trends of abundance with EW. {\sf MyGIsFOS} uses a pre-computed grid of synthetic spectra instead of relying on on-the-fly synthesis or on a priori EW measurements. By fitting against synthetic spectra, {\sf MyGIsFOS} can use moderately blended features in abundance measurements, or treat directly HFS-affected lines. Its working can be summarized as follows:

\begin{enumerate}
\item A grid of synthetic spectra varying (in the most general case) in T$_{\rm eff}$, $\log g$, $\xi$, [Fe/H], and [$\alpha$/Fe] is provided to the code together with the input spectra (for which initial guess parameters have to be provided), and a set of spectral ``regions'' to be used either as pseudo-continuum ranges (for normalization) or as spectral features of various kinds (e.g. \ion{Fe}{i} lines).
\item The observed spectrum, and each spectrum in the synthetic grid, are pseudo-normalized using the continuum intervals, then the synthetic grid is collapsed (by interpolation) at the initial guess values for T$_{\rm eff}$, $\log g$, $\xi$ and [$\alpha$/Fe], leaving a grid whose sole dimension is [Fe/H].
\item The provided \ion{Fe}{i} and \ion{Fe}{ii} lines are fitted (by $\chi^2$-minimization) against the collapsed grid, deriving best-fit \ion{Fe}{i} abundances for each line. EWs are also measured in the process. In a series of nested loops, the aforementioned diagnostics (excitation and ionization equilibrium, etc.) are evaluated, and if needed, the stellar parameters are altered, and the whole process repeated, until convergence is achieved.
\item To measure abundances of other elements, the respective features are fitted against the same grid, collapsed at the final values of T$_{\rm eff}$, $\log g$, $\xi$, and thus varying in [Fe/H]. The best fitting metallicity value is used as the element [X/H] \citepads[this is in principle inconsistent but leads to generally accurate values, see][]{2014A&A...564A.109S}. A special case is the one of $\alpha$-elements, which are measured first after T$_{\rm eff}$, $\log g$ and $\xi$ have been set, and used to estimate the last grid parameter, [$\alpha$/Fe]. The derived value of [$\alpha$/Fe], if different enough from the estimated, triggers a new estimation of the other parameters. Finally, all the other elements are measured.
\end{enumerate}

After processing, the output is examined for signs of problems: non-converging objects are checked individually and eventually re-run. {\sf MyGIsFOS} does not estimate or vary the spectrum broadening: the grid is provided broadened at the nominal resolution of R = 47\,000. Stars showing extra-broadening (essentially moderately rotating objects) are detected by inspecting the quality of line fits, and reprocessed with appropriate broadening.

For GESviDR1Final, T$_{\rm eff}$ was {\em not} iterated within {\sf MyGIsFOS}, since this was not yet implemented. Instead, T$_{\rm eff}$ was determined from the available photometry by applying the \citetads{2009A&A...497..497G} relations. Full T$_{\rm eff}$ iteration is now in place and was used in the analysis of iDR2. In addition, {\sf MyGIsFOS} is using the Gaia-ESO grid of synthetic spectra that does not include a $\xi$ dimension, but relies on a single, pre-calibrated $\xi$ value depending on T$_{\rm eff}$, $\log g$, and [Fe/H]. Thus, {\sf MyGIsFOS} is not determining $\xi$ for the Gaia-ESO data. In the future, when a new grid of synthetic spectra with the $\xi$ dimension is available, also this quantity will be determined.

\subsection{UCM}\label{sec:ucm}

The UCM Node employs the automatic code {\sf StePar} \citepads[][]{2012A&A...547A..13T} to determine the stellar atmospheric parameters ($T_{\rm eff}$, $\log{g}$, $\xi$) and metallicity. {\sf StePar} computes the stellar atmospheric parameters using {\sf MOOG} (v.2002). Although designed to make use of a grid of Kurucz ATLAS9 plane-parallel model atmospheres \citepads{1993KurCD..13.....K}, {\sf StePar} has been now modified to operate with the spherical and non-spherical MARCS models. 

The atmospheric parameters are inferred from a previously selected $\ion{Fe}{i}$-$\ion{Fe}{ii}$ line list. The code iterates until it reaches the excitation and ionization equilibrium and minimizes trends of abundance with $\log{(EW / \lambda)}$. {\sf StePar} employs a Downhill Simplex Method \citepads{1992nrca.book.....P}. The function to minimize is a quadratic form composed of the excitation and ionization equilibrium conditions. The code performs a new simplex optimization until the metallicity of the model and the iron abundance are the same. 

Uncertainties for the stellar parameters are derived as described in \citetads{2012A&A...547A..13T}. In addition, a 3$\sigma$ rejection of the $\ion{Fe}{i}$ and $\ion{Fe}{ii}$ lines is performed after a first determination of the stellar parameters. {\sf StePar} is then re-ran without the rejected lines.

The EW determination of all the lines was carried out with the {\sf ARES} code. The approach of \citetads{2008A&A...487..373S} to adjust the parameters of {\sf ARES} according to the S/N of each spectrum was followed. Regarding the individual abundances, two line lists were prepared: one for dwarfs ($\log~g$ $\geq$ ~ 4.0) and one for giants ($\log~g$ $\leq$ ~ 4.0). To get the individual abundances, the EWs are fed to {\sf MOOG} and then a 3$\sigma$-clipping for each chemical element is performed. 

\subsection{ULB}\label{sec:ulb}

The ULB Node uses the {\sf BACCHUS} (Brussels Automatic Code for Characterizing High accUracy Spectra) code which consists of three different modules respectively designed to derive EWs, stellar parameters, and abundances. The current version relies on (i) a grid of MARCS model atmospheres, (ii) a specific procedure for interpolating among the model atmosphere thermodynamic structure within the grid \citep{MasseronPhD}, and (iii) the radiative transfer code {\sf Turbospectrum}. 

The stellar parameters determination relies on a list of selected Fe lines. The first step consists in determining accurate abundances for the selected lines using the abundance module for a given set of $T_{\rm eff}$ and $\log g$ values. The abundance determination module proceeds in the following way:  (i) a spectrum synthesis, using the full set of (atomic and molecular) lines, is used for local continuum level finding (correcting for a possible spectrum slope); (ii) cosmic and telluric rejections are performed; (iii) local S/N is estimated; (iv) a series of flux points contributing to a given absorption line is selected. Abundances are then derived by comparing the observed spectrum with a set of convolved synthetic spectra characterized by different abundances.  Four different diagnostics are used: $\chi^2$ fitting, core line intensity comparison, global goodness-of-fit estimate, and EW comparison. A decision tree then rejects the line, or accepts it keeping the best matching abundance.

The second step consists in deducing the EWs of Fe lines using the second module. One asset of the code is precisely this {\it computation} of EWs from best-matching synthetic spectra, because the EW of only the considered line is taken into account (excluding the contribution from nearby, blending lines). Indeed, EWs are computed not directly on the observed spectrum, but internally from the synthetic spectrum with the best-matching abundance. This way, the information about the contribution of blending lines is known, allowing a clean computation of the EW of the line of interest.

The last step of the procedure consists in injecting the derived EWs in {\sf Turbospectrum} to compute abundances for a grid of 27 neighbor model atmospheres (including three values of effective temperature, three of gravity, and three of microturbulence velocity), covering the parameter space of interest. For each model, the slopes of abundances against excitation potential, against EWs, as well as \ion{Fe}{i} and \ion{Fe}{ii} lines abundances are computed. 

The final parameters are determined by requesting that the ionization equilibrium is fulfilled, and that simultaneously null slopes for
abundances against excitation potential and against EWs are obtained. The whole procedure is iterated once per star, after a first guess of stellar parameters has been refined and a new seed model computed. 

\subsection{Vilnius}\label{sec:vilnius}

The Vilnius Node uses a traditional EW based method for the stellar parameters determination. Effective temperature is derived by minimizing the slope of abundances obtained from \ion{Fe}{i} lines with respect to the excitation potential.
Surface gravity is determined by forcing the measured \ion{Fe}{i} and \ion{Fe}{ii} lines to yield the same [Fe/H] value. Microturbulence is determined by forcing \ion{Fe}{i} abundances to be independent of the EWs of the lines. A custom wrapper software was developed to measure EWs, and compute the main atmospheric parameters and abundances automatically.  

Equivalent widths were measured using the {\sf DAOSPEC} software. The atomic and molecular data provided by the Gaia-ESO line list group were used. Only lines corresponding to the best quality criteria (flags provided together with the line list) were used. Different subsamples of lines were used for giant stars and for metal-poor stars.

The stellar atmospheric parameters were computed using {\sf MOOG} (v.2010) and the MARCS atmospheric models. The interpolation code provided with the MARCS grid was modified to make possible an automatic selection of the required sets of models, and the extraction of the final interpolated model in the WEBMARCS format for {\sf MOOG}. 

The wrapper code performs an iterative sequence of abundance calculations using a simultaneous quadratic minimization of: (i) abundance dependency on the line excitation potential, (ii) difference between neutral and ionized iron abundances, and (iii) scatter of neutral iron abundances. Iterations were performed on each step until a stable solution was reached. The minimization procedure was based on the Nelder-Mead method \citep{NelderMead65}. During this iterative procedure, the code searches for possible outliers in abundances determined using different lines. Every resulting abundance for every single line that departed from the mean by more than 2$\sigma$ was flagged as outlier and was omitted from further calculations.

A starting point was selected randomly in a vicinity of $T_{\rm eff}$ = 5500 K, $\log g$ = 4.0, [Fe/H] = $-$0.5 and $\xi$ = 1.5 km s$^{-1}$. The final values of atmospheric parameters for a specific star do not depend on the starting point of the calculations. The final abundances of all other elements were derived omitting possible outliers using a 2$\sigma$ criteria.

The uncertainties of the stellar parameter were determined using error estimations of the line profile fitting and the standard deviations of the abundances. The estimation of uncertainty for the effective temperatures was done by obtaining the boundary temperature values of the possible satisfactory parameter space, using the error of the linear regression fit. The uncertainty of the gravity was obtained using the possible boundary values of $\log g$, using the standard deviations of the abundances from \ion{Fe}{i} and \ion{Fe}{ii} lines. The uncertainty of the microturbulence velocity is obtained by employing the error of the standard deviation of the neutral iron abundances. The [Fe/H] standard deviation is adopted as the metallicity uncertainty. 

\section{The science verification analysis}\label{sec:idr1}
\begin{table*}
\caption[]{\label{tab:releaseidr1} Number of FGK-type stars observed with UVES and part of the iDR1 data set.}
\centering
\begin{tabular}{lcl}
 \hline \hline
  Gaia-ESO Type & Stars & Comments \\
 \hline
 Total & 508 &  Gaia-ESO only, no archival data. \\
 GES\_MW & 305 &  Stars from Milky Way fields. \\
 GES\_CL   & 133 & Stars from open cluster fields. \\
 GES\_SD & 70 & Calibration targets. \\
\hline
\hline
\end{tabular}
\end{table*}
\begin{figure*}[t]
\centering
\includegraphics[height = 6cm]{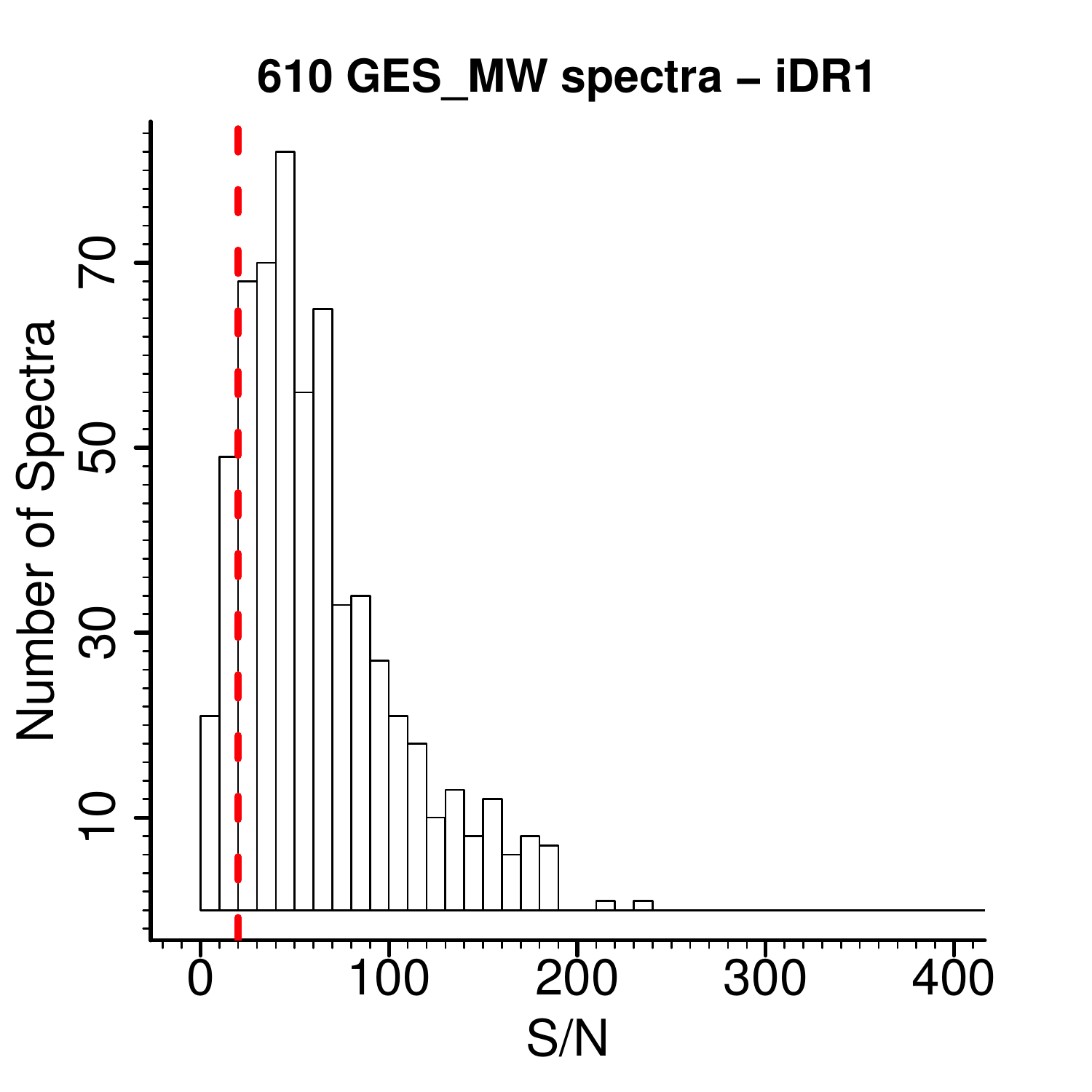}
\includegraphics[height = 6cm]{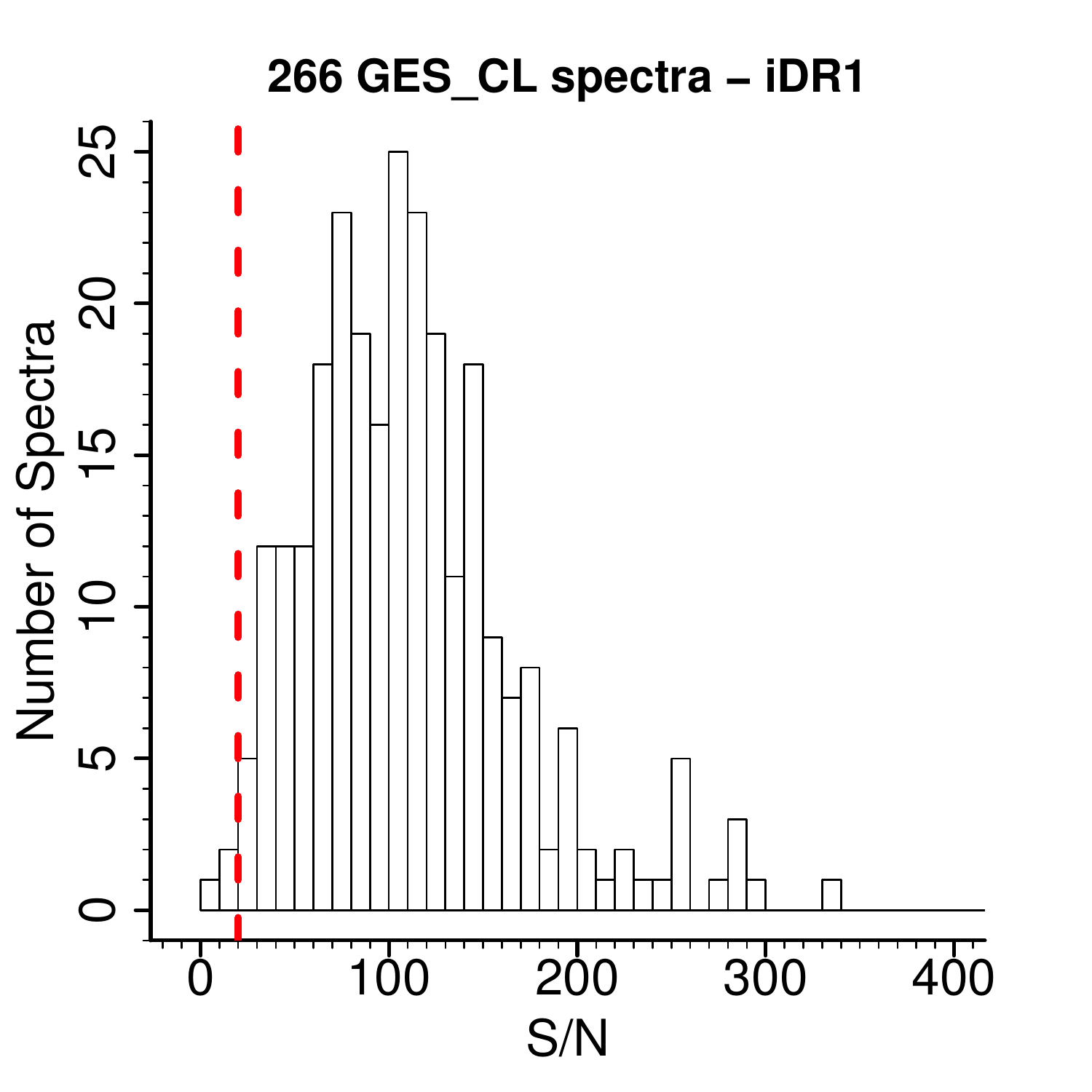}
\includegraphics[height = 6cm]{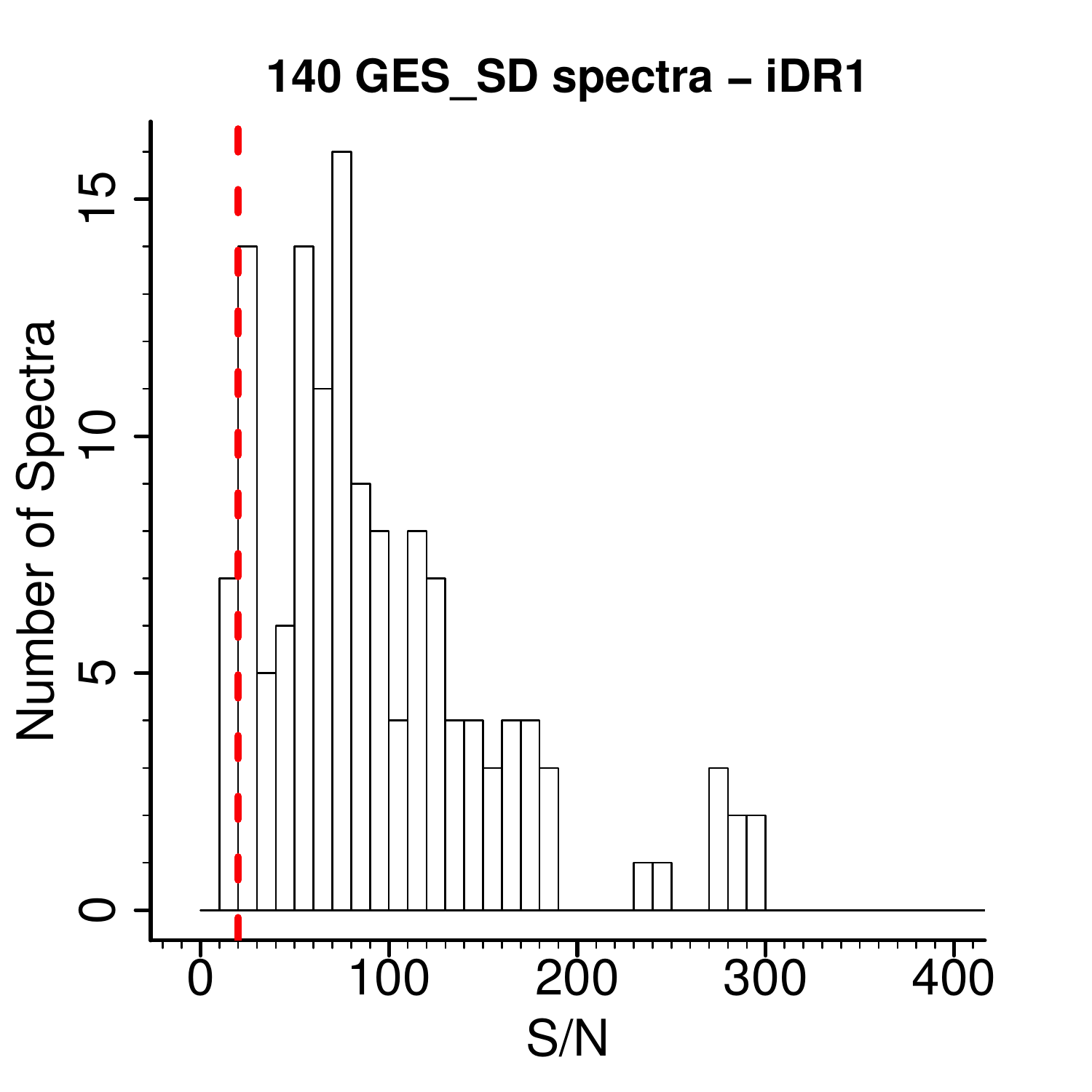}
\caption{Distribution of the median S/N of the spectra in iDR1 (508 FGK-type stars) observed with UVES. Each of the two UVES spectrum parts (from each CCD) is counted separately (thus, two spectra per star). The red dashed line indicates S/N = 20. The samples of the solar neighborhood (GES\_MW), open clusters (GES\_CL), and calibration targets (GES\_SD) are shown separately.}\label{fig:sndistidr1}
\end{figure*}

The science verification analysis was the first full analysis cycle of the Survey. The first few papers with Gaia-ESO data are based on results of this first analysis \citepads[e.g.][]{2014A&A...565A..89B,2014arXiv1407.1510C,2014A&A...561A..94D,2014A&A...563A.117F,2014A&A...563A..44M,2014A&A...567A..55S}. We therefore believe it is important to document the details, achievements, and shortcomings of this analysis. We document in particular the differences between this analysis and the analysis of iDR2, described in the main text. The data analyzed was part of the first internal data release (iDR1), described below:

\begin{itemize}
\item[$\bullet$] {\bf Internal Data Release 1 (iDR1):} This data release consists of spectra obtained up to the end of June 2012 and includes spectra of 576 FGK-type stars observed with UVES. Of these stars, 68 are part of young open clusters (age $<$ 100\,Myr). They were not analyzed by WG11 but by the working group responsible for pre-main-sequence stars (Lanzafame et al. 2014, submitted). For the moment, the results have been released only internally to the Gaia-ESO collaboration and are referred to as \emph{GESviDR1Final} (Gaia-ESO Survey verification internal data release one). We point out that the reduced spectra for part of the stars observed in the first six months are already available through the ESO data archive\footnote{\url{http://archive.eso.org/wdb/wdb/adp/phase3_spectral/form?phase3_collection=GaiaESO}}.
\end{itemize}

\begin{figure*}[t]
\centering
\includegraphics[height = 10cm]{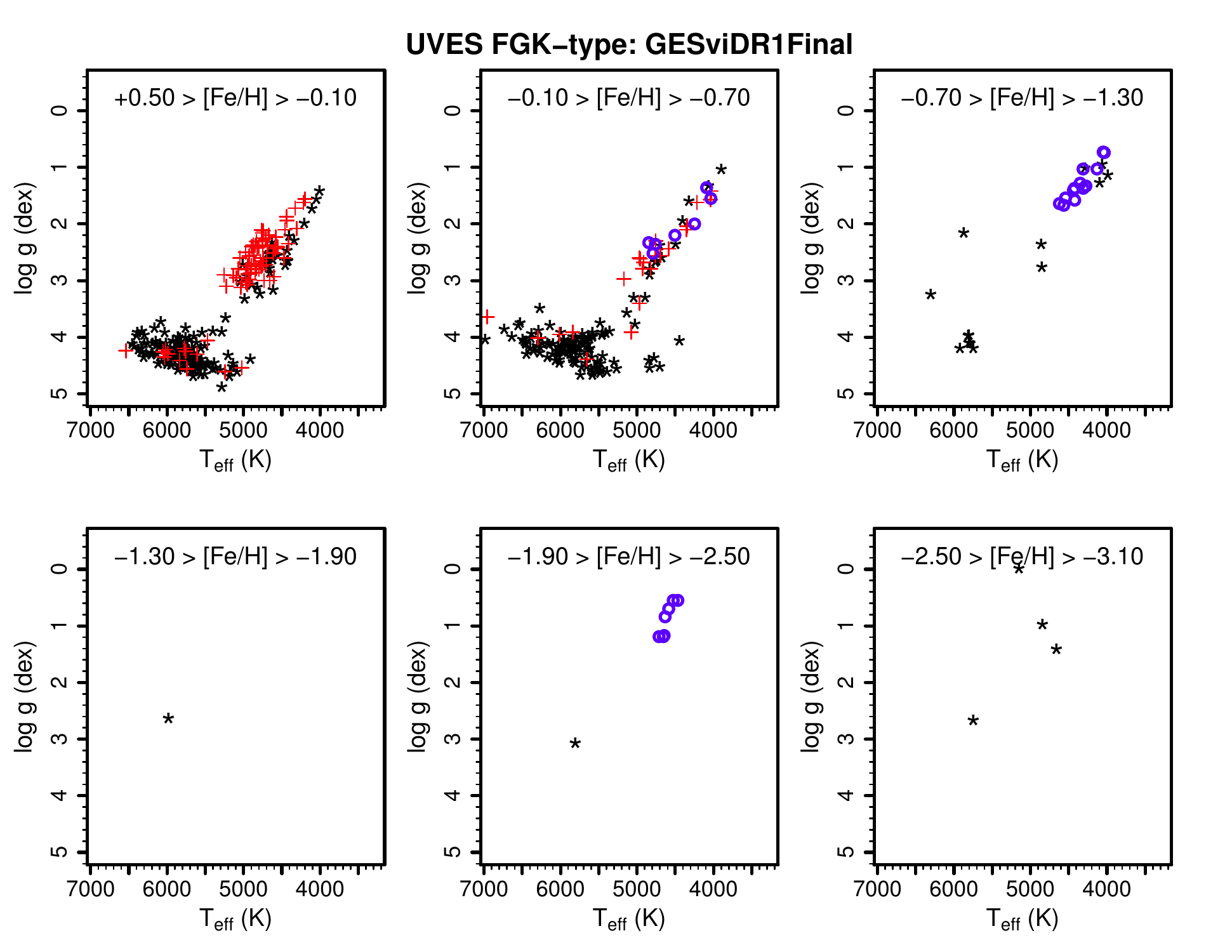}
 \caption{Distribution of 421 FGK-type stars from GESviDR1Final in the $T_{\rm eff}$-$\log g$ plane. The stars were observed with UVES during the first 6 months of the Survey and had atmospheric parameters determined as described in this paper. The panels are divided according to metallicity. Black stars represent field stars, red crosses stars observed in open-cluster fields, and blue circles stars observed in globular-cluster fields.}\label{fig:sampleidr1}%
\end{figure*}

The S/N distribution of the iDR1 data are shown in Fig. \ref{fig:sndistidr1}. Table \ref{tab:releaseidr1} summarizes the number of stars part of iDR1. Figure \ref{fig:sampleidr1} shows how the stars targeted in the first 6 months of Gaia-ESO observations are distributed in the $T_{\rm eff}$-$\log g$ plane. Atmospheric parameters were determined for 421 stars out of the 508 in the sample. For the remaining stars the analysis failed for different reasons (low S/N, fast rotation, reduction artefacts, etc). Flags will be provided indicating the reason of the failure.

In the Sections that follow, we discuss separately the data products determined in the analysis of the iDR1 data, i.e. EWs (Appendix \ref{sec:eqwidr1}), stellar atmospheric parameters (Appendix \ref{sec:atmidr1}), and elemental abundances (Appendix \ref{sec:abunidr1}). The differences between this analysis and the one of iDR2 are highlighted.

\subsection{EWs in iDR1}\label{sec:eqwidr1}

The Nodes {that determined EWs for the iDR1 data set were:} Bologna, CAUP, Concepcion, EPInArBo, UCM, ULB, and Vilnius. Three codes were used to measure EWs automatically: {\sf ARES}, {\sf BACCHUS} (T.~Masseron, unpublished, see Sect. \ref{sec:ulb}), and {\sf DAOSPEC}. Of these codes, BACCHUS was not included in the iDR2 discussion.

We include here figures similar to the ones discussed in the main text about iDR2. A comparison between these plots can show the evolution of the measurements between one iDR and the next. 

\begin{figure*}
\centering
\includegraphics[height = 4.5cm]{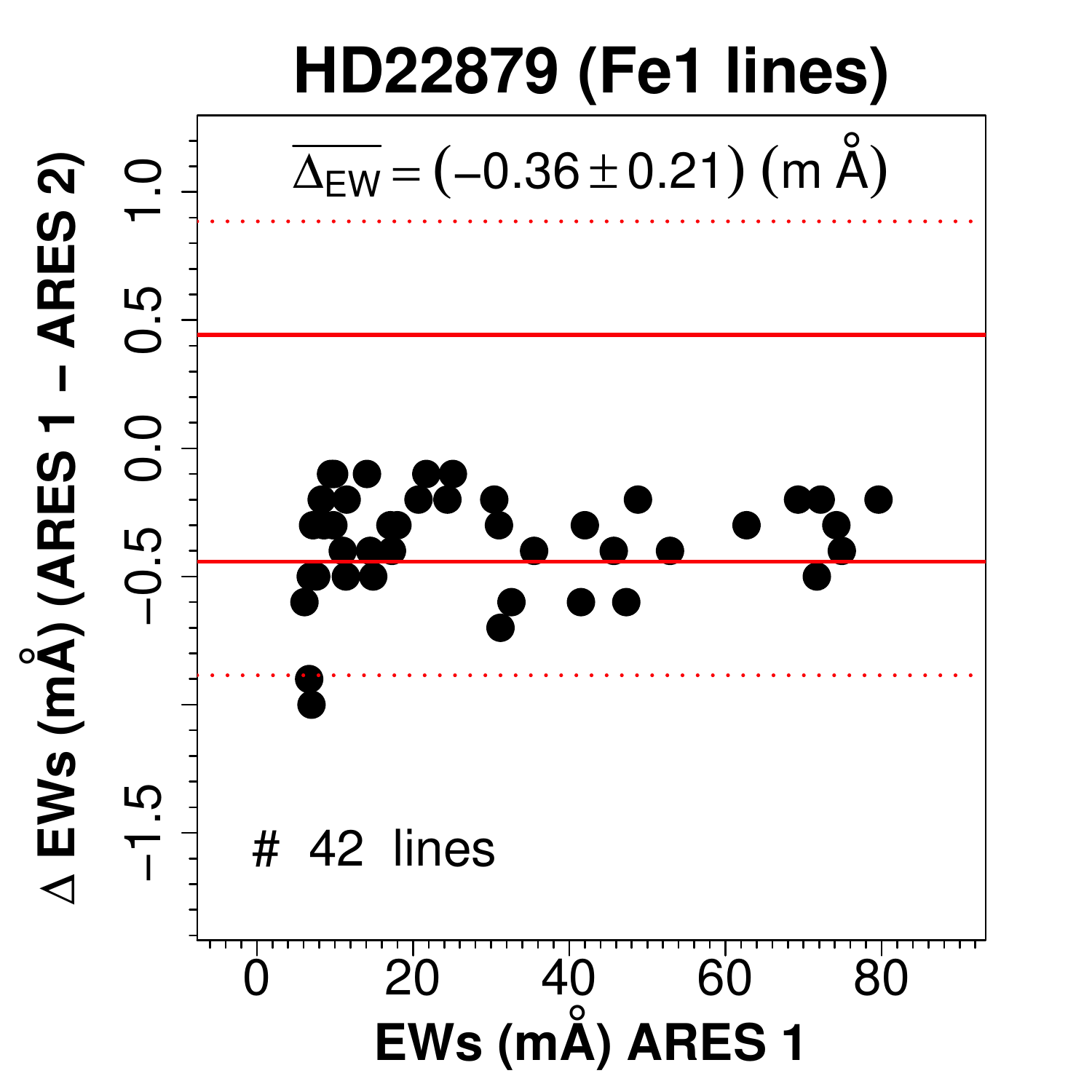}
\includegraphics[height = 4.5cm]{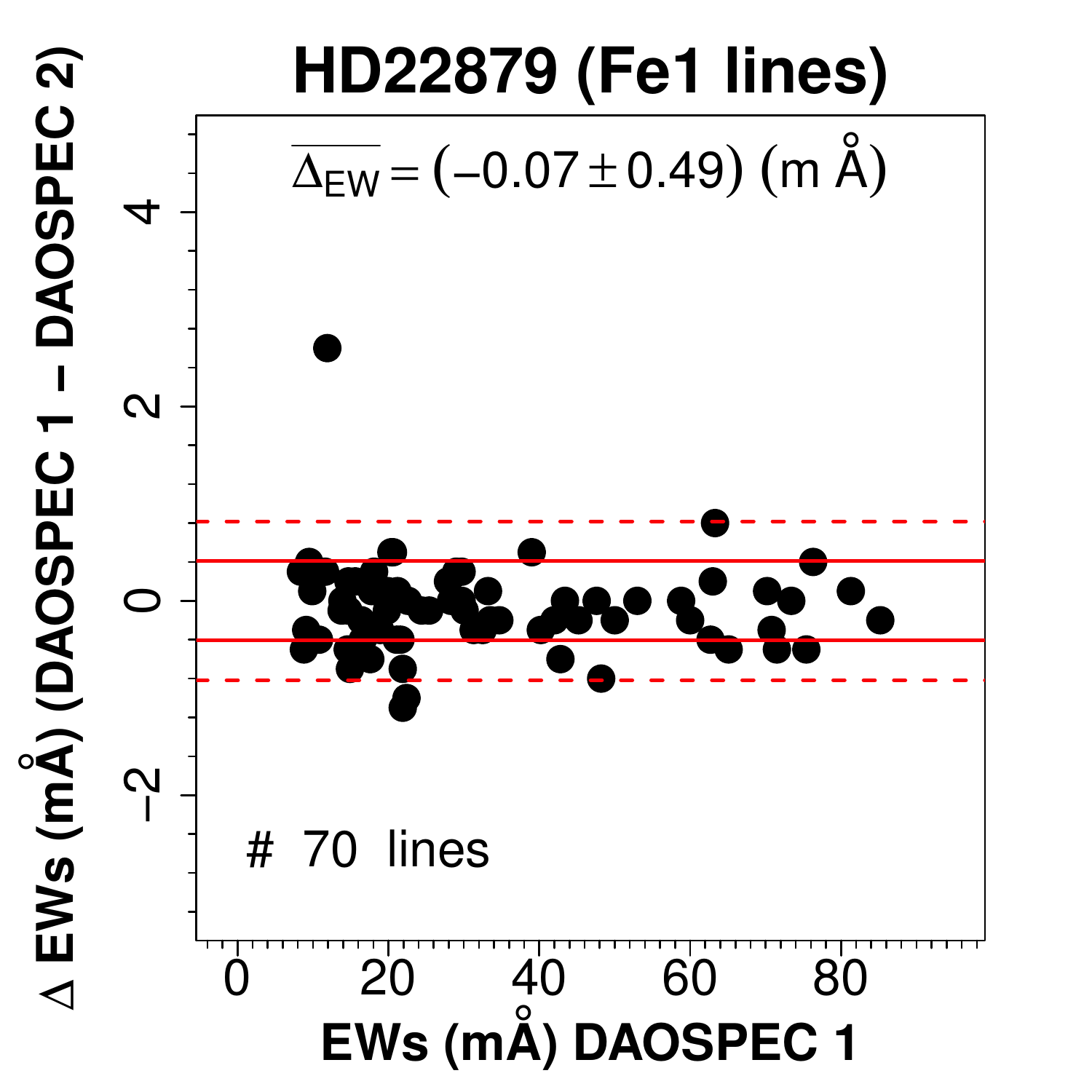}
\includegraphics[height = 4.5cm]{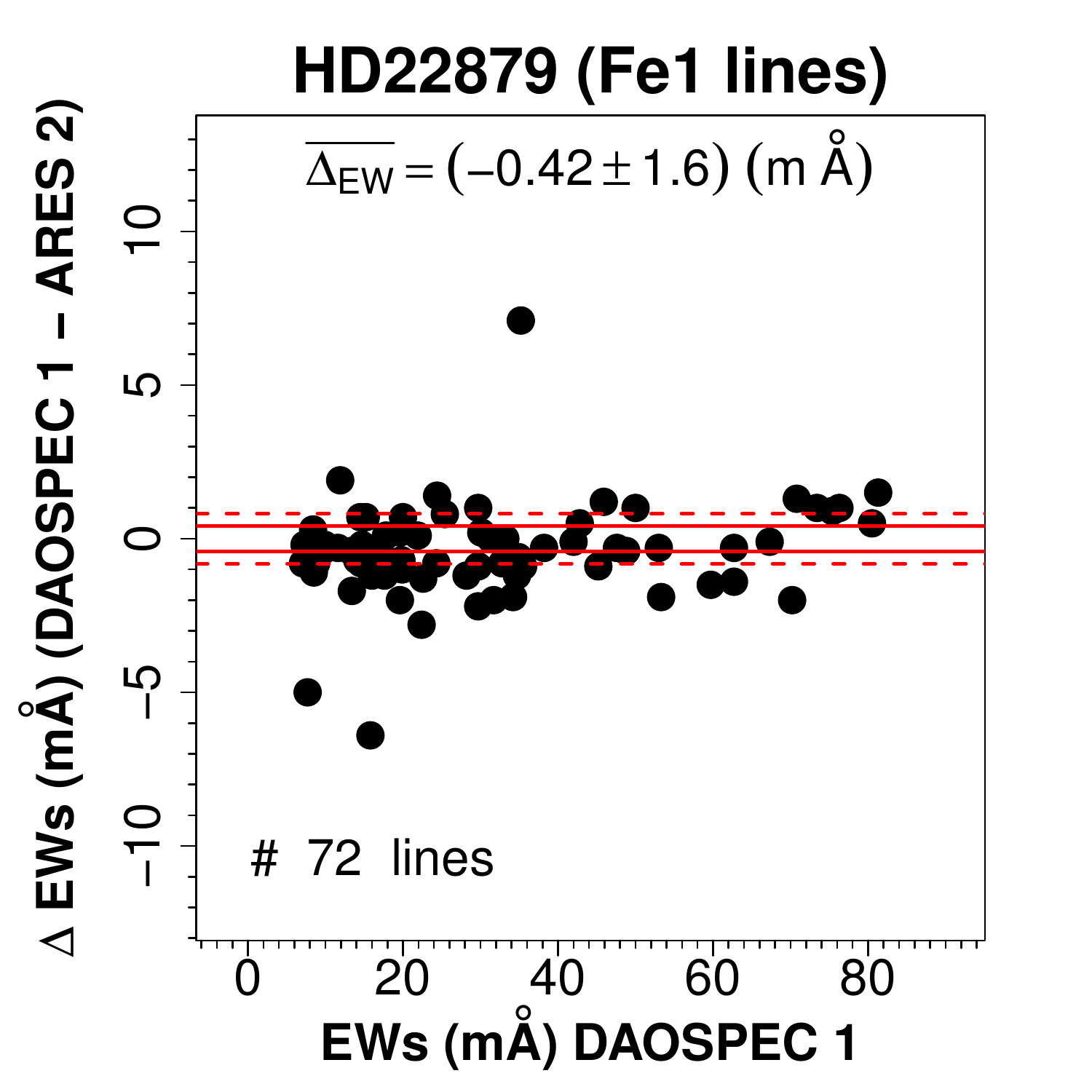}
\includegraphics[height = 4.5cm]{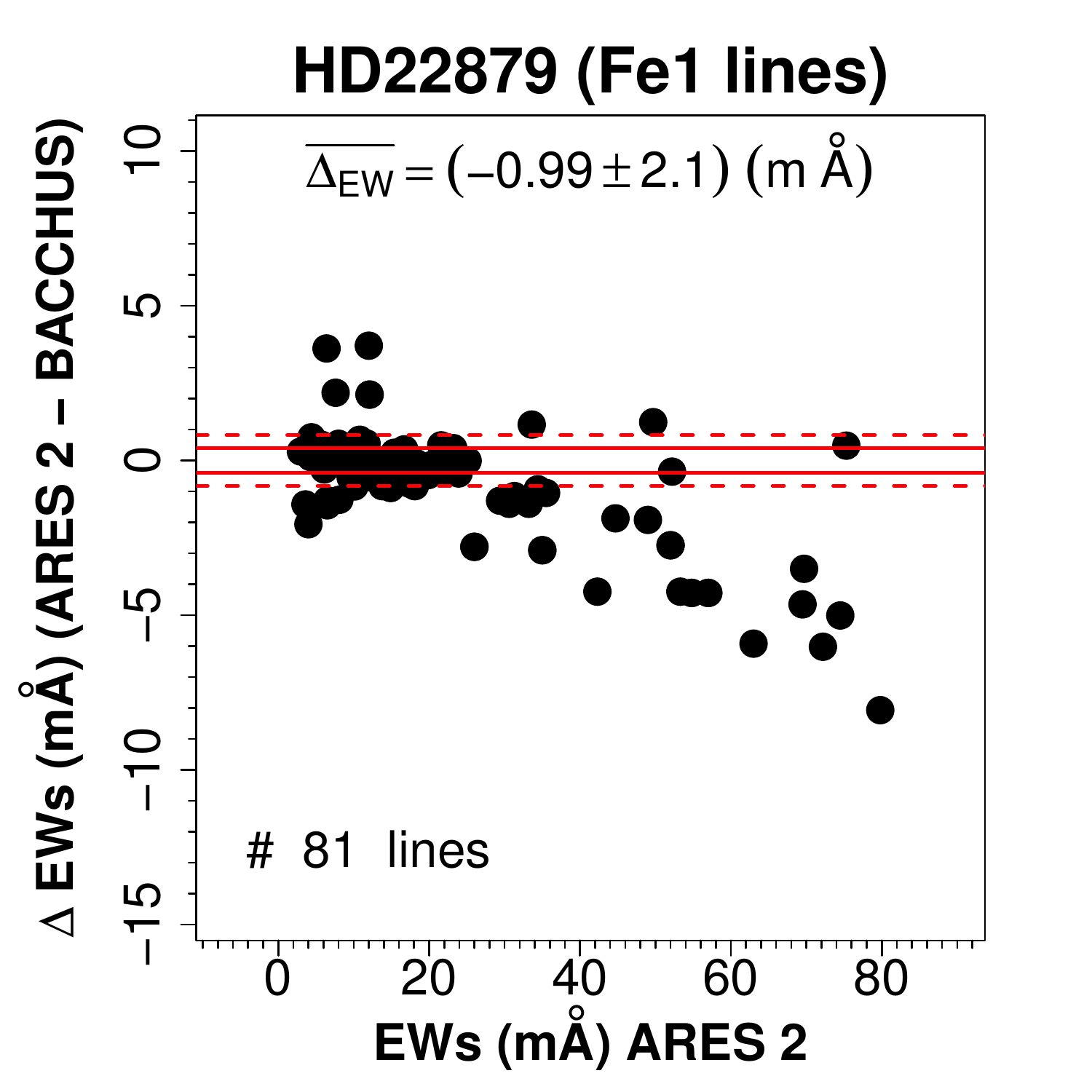}
\includegraphics[height = 4.5cm]{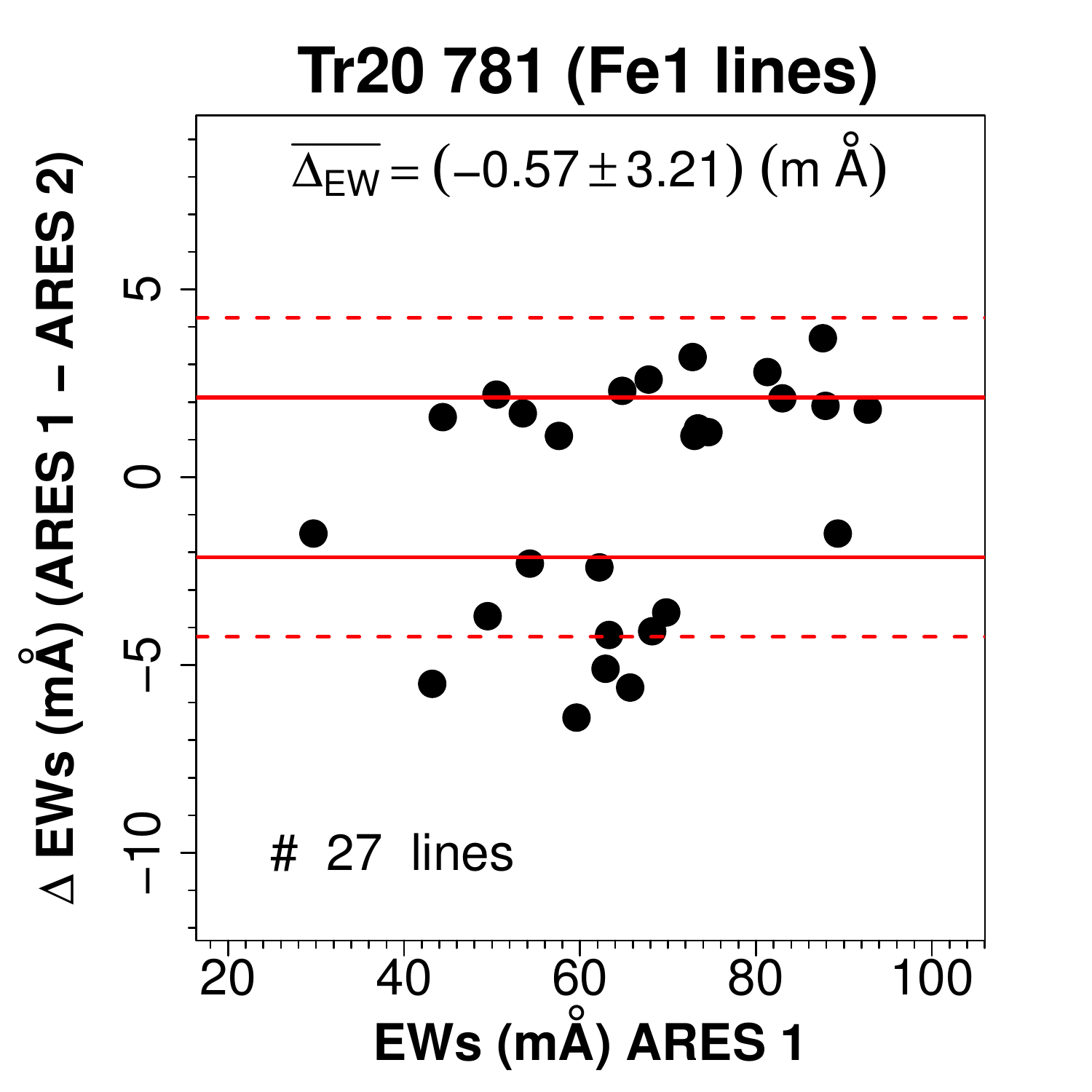}
\includegraphics[height = 4.5cm]{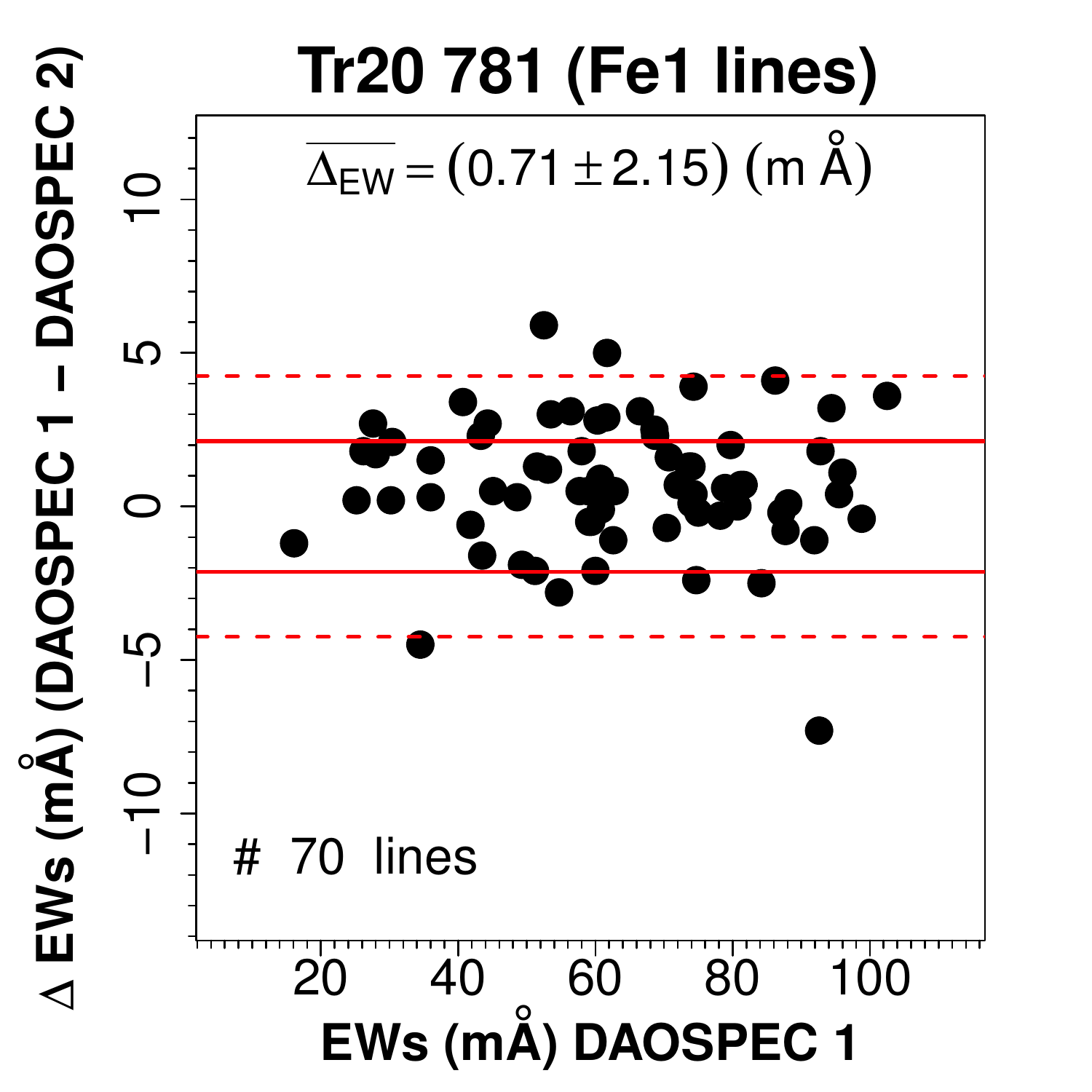}
\includegraphics[height = 4.5cm]{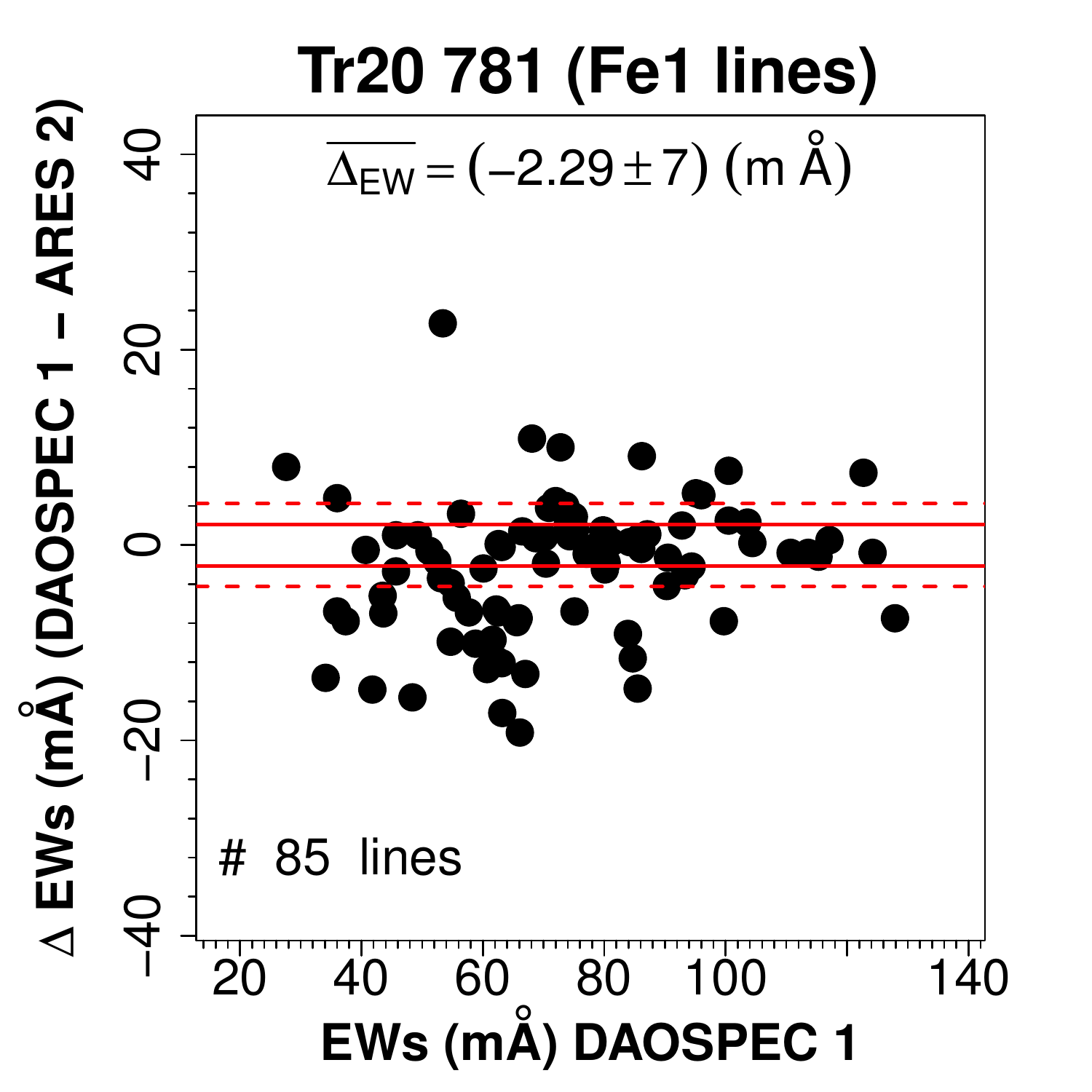}
\includegraphics[height = 4.5cm]{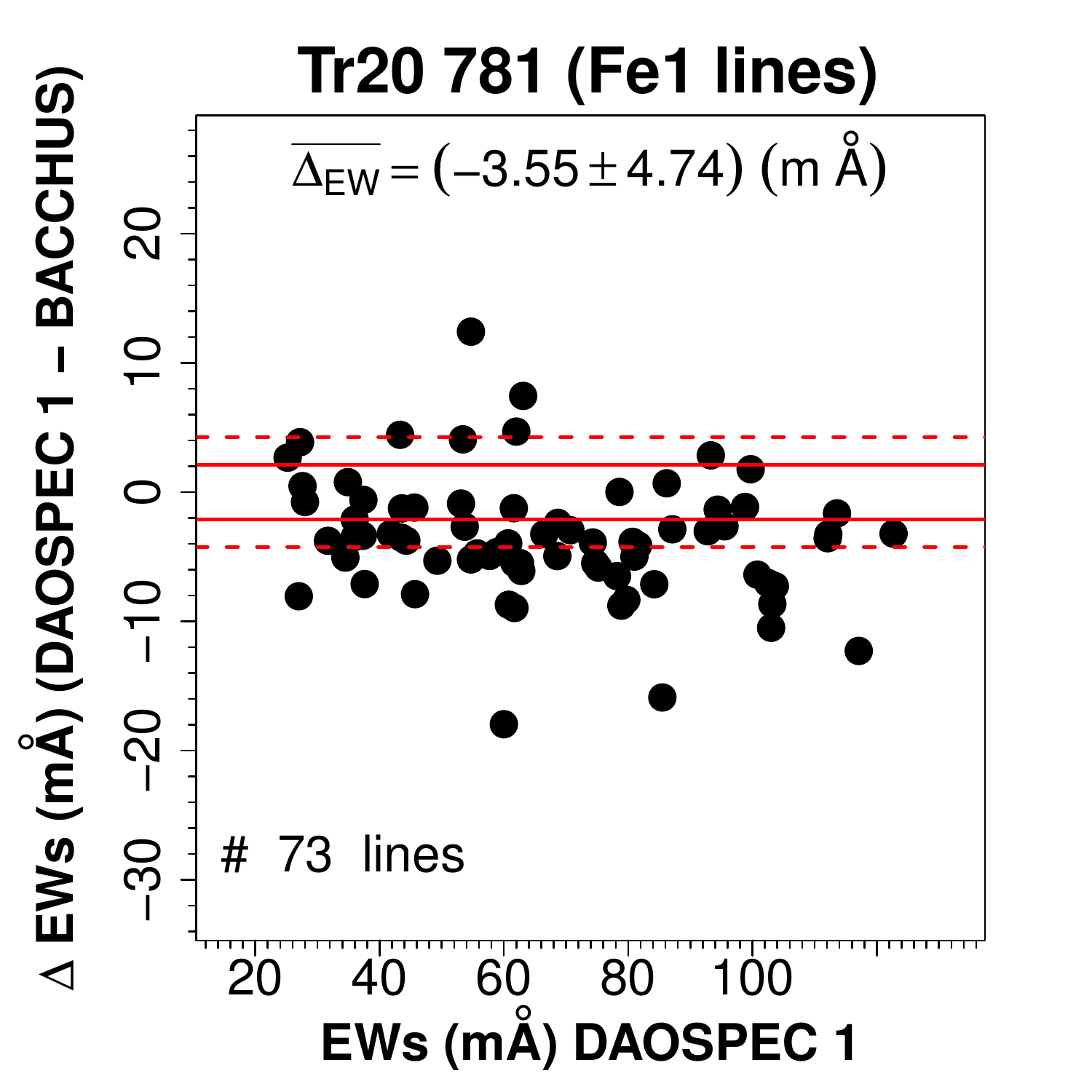}
 \caption{Comparison between \ion{Fe}{i} equivalent widths measured by different Nodes for two stars. \emph{Top row:} Star \object{HD 22879}, a benchmark star used for calibration with $T_{\rm eff}$ = 5786 K, $\log g$ =  4.23, and [Fe/H] = $-$0.90. The median S/N of its spectra are 239 and 283 for the blue and red part of the spectra, respectively. The red lines indicate the typical 1$\sigma$ (solid line) and 2$\sigma$ (dashed line) uncertainty of the EW computed with the \citetads[][]{1988IAUS..132..345C} formula, adopting FWHM = 0.190 \AA, pixel size = 0.0232 \AA, and S/N = 260.  \emph{Bottom row:} A clump giant in the open cluster \object{Trumpler 20} \citepads[\object{Trumpler 20 MG 781} in the numbering system of][]{2005ApJS..161..118M}, with $T_{\rm eff}$ = 4850 K, $\log g$ = 2.75, and [Fe/H] = +0.15. The median S/N of its spectra are 36 and 68 for the blue and red part of the spectra, respectively. The red lines indicate the typical 1$\sigma$ (solid line) and 2$\sigma$ (dashed line) uncertainty of the EW computed with the \citetads[][]{1988IAUS..132..345C} formula, adopting FWHM = 0.190 \AA, pixel size = 0.0232 \AA, and S/N = 50. In each panel, the average difference of the EWs and its dispersion are also given.}\label{fig:ewcompidr1}
\end{figure*}

Figure \ref{fig:ewcompidr1} shows the comparison between the EWs of \ion{Fe}{i} lines measured by different groups in the two stars discussed in Sect. \ref{sec:eqw} (the metal-poor dwarf HD 22879 with S/N $\sim$ 260 and the metal-rich giant Trumpler 20 MG 781 with S/N $\sim$ 50). 

The EWs measured with the same code by different Nodes (left and center-left plots in Fig. \ref{fig:ewcompidr1}) tend to agree to within 2$\sigma$, although a systematic difference is present in some case. When comparing the EWs measured with {\sf ARES} and {\sf DAOSPEC} (center-right plots in Fig. \ref{fig:ewcompidr1}), it is noticeable that the scatter increases. As discussed before, this is probably related to the different ways that the continuum is defined in each code (global vs. local continuum for {\sf DAOSPEC} and {\sf ARES} respectively). 

The comparison between {\sf BACCHUS} and the other two codes (right plots in Fig. \ref{fig:ewcompidr1}) show systematic differences that are under investigation. {\sf BACCHUS} measures the EWs not from the observed spectrum, but from a best fitting synthetic spectrum once the abundance and the parameters are fixed. It removes from the line the contribution of any known blending feature that is included in the line list. The synthetic line is computed in 1D LTE, using all the line information possible: line broadening, HFS and blends. In this sense, the {\sf BACCHUS} EWs should be the more robust measurements (assuming that the atmospheric parameters are perfectly known and that the blends are perfectly synthesized). The continuum placement might be another source of error. {\sf BACCHUS} fits the continuum relying on the synthetic spectrum, adapting it from star to star, and from wavelength region to wavelength region. However, if the continuum match is poor around the measured line, the continuum may be wrong, and so will be the final abundance and EW. The issue is complex and we are investigating the causes of the discrepancies and improving the measurements for future releases.

\begin{figure*}
\centering
\includegraphics[height = 6cm]{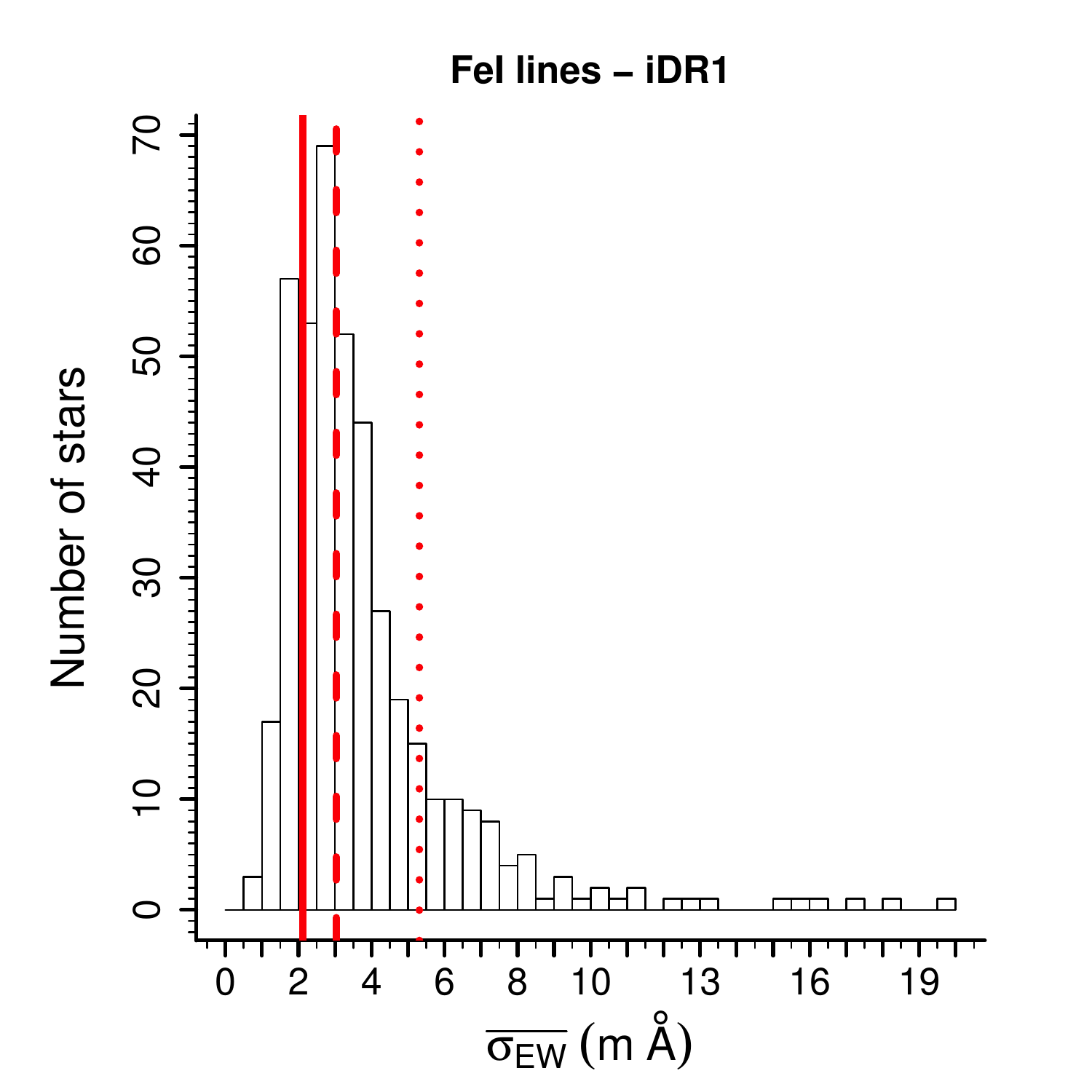}
\includegraphics[height = 6cm]{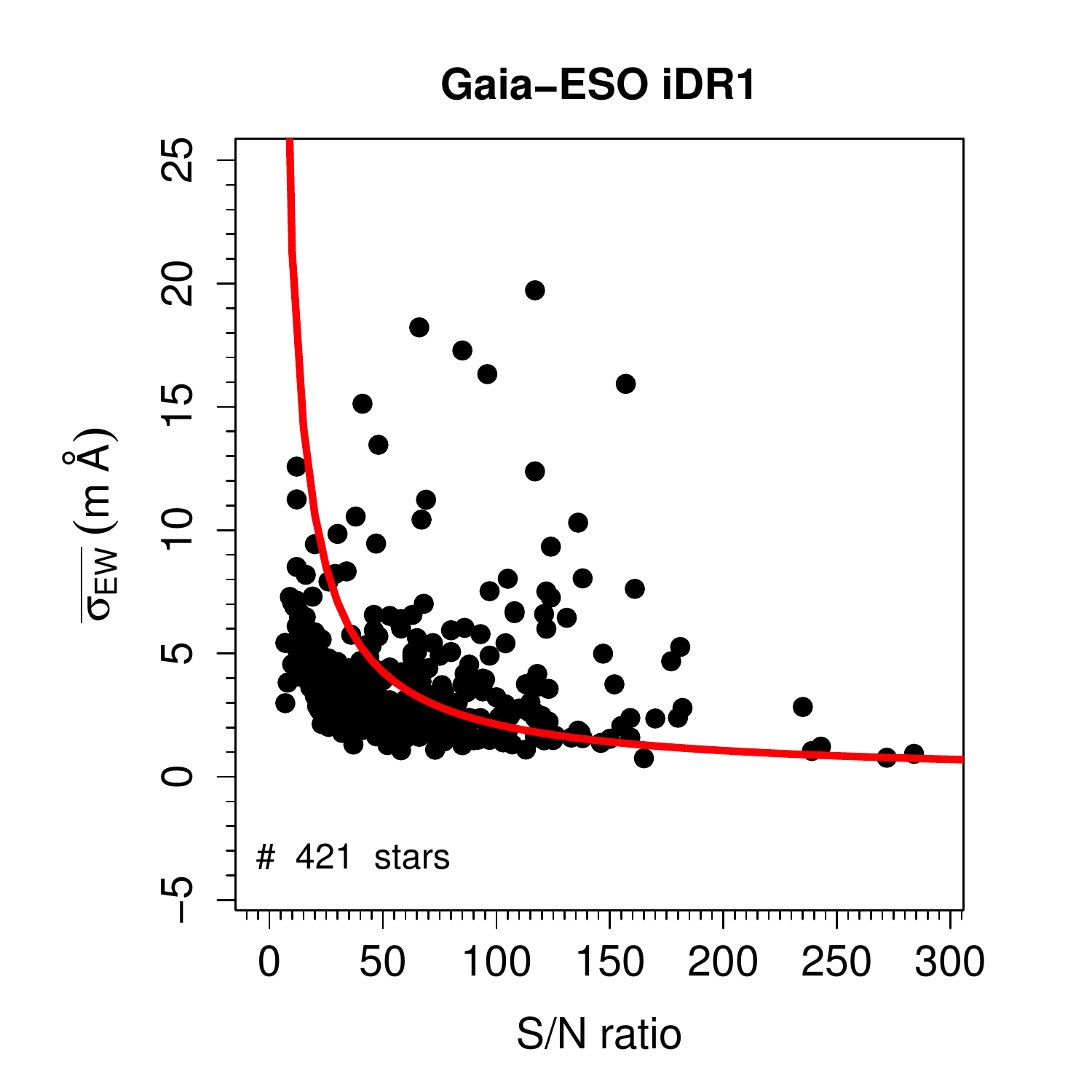}
 \caption{\emph{Left panel:} Histogram of $\overline{\sigma_{\rm EW}}$ per star, taking into account the measurements of all Nodes. Also shown are lines indicating the 2$\sigma$ uncertainty calculated with \citetads[][]{1988IAUS..132..345C} formula for S/N = 40 (dotted line at 5.31 m\AA), S/N = 70 (dashed line at 3.04 m\AA), and S/N = 100 (solid line at 2.12 m\AA). \emph{Right panel:} The dependence of $\overline{\sigma_{\rm EW}}$ with respect to S/N. Also shown is the expected 2$\sigma$ value given by the \citetads[][]{1988IAUS..132..345C} formula (as a red line).}\label{fig:ewfullcompidr1}%
\end{figure*}

Figure \ref{fig:ewfullcompidr1} depicts the behavior of $\overline{\sigma_{\rm EW}}$ measured in GESviDR1Final. For each \ion{Fe}{i} line of a star, the average value of the EW is computed, together with its standard deviation. For each star, we define $\overline{\sigma_{\rm EW}}$ as the mean of all the standard deviations of the \ion{Fe}{i} lines in that star. For most stars, the standard deviations are small ($<$ 3 m\AA), with a few cases reaching up to $\sim$ 20 m\AA. Figure \ref{fig:ewfullcompidr1} shows that for the majority of the stars, the multiple measurements of EWs tend to agree within the expected statistical uncertainty given by the S/N of the spectra.

\begin{figure*}[ht]
\centering
\includegraphics[height = 6cm]{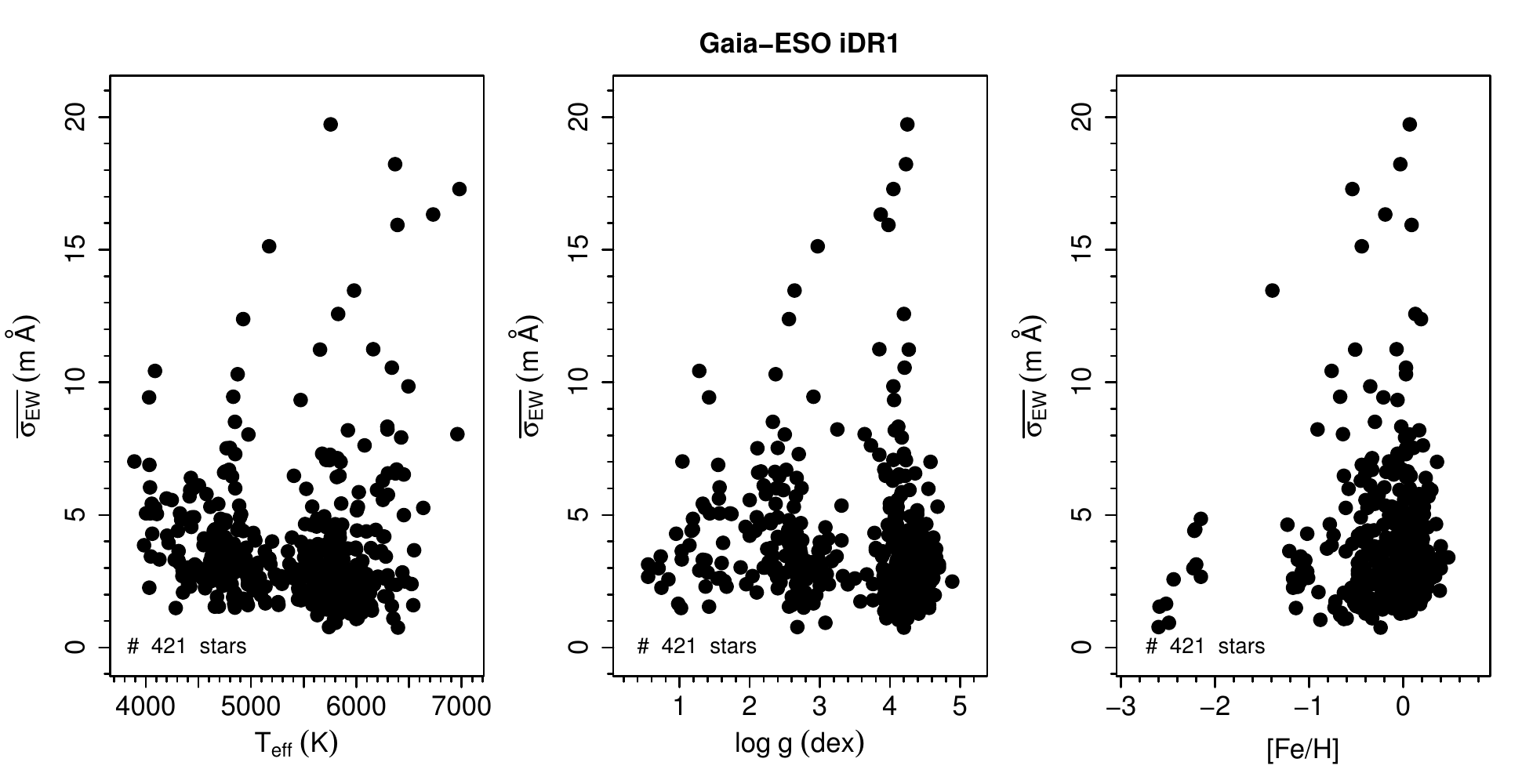}
 \caption{Mean of all the standard deviations of the \ion{Fe}{i} lines in a star, $\overline{\sigma_{\rm EW}}$, as a function of the atmospheric parameters.}\label{fig:ewparamidr1}%
\end{figure*}

\begin{figure}[h]
\centering
\includegraphics[height = 6cm]{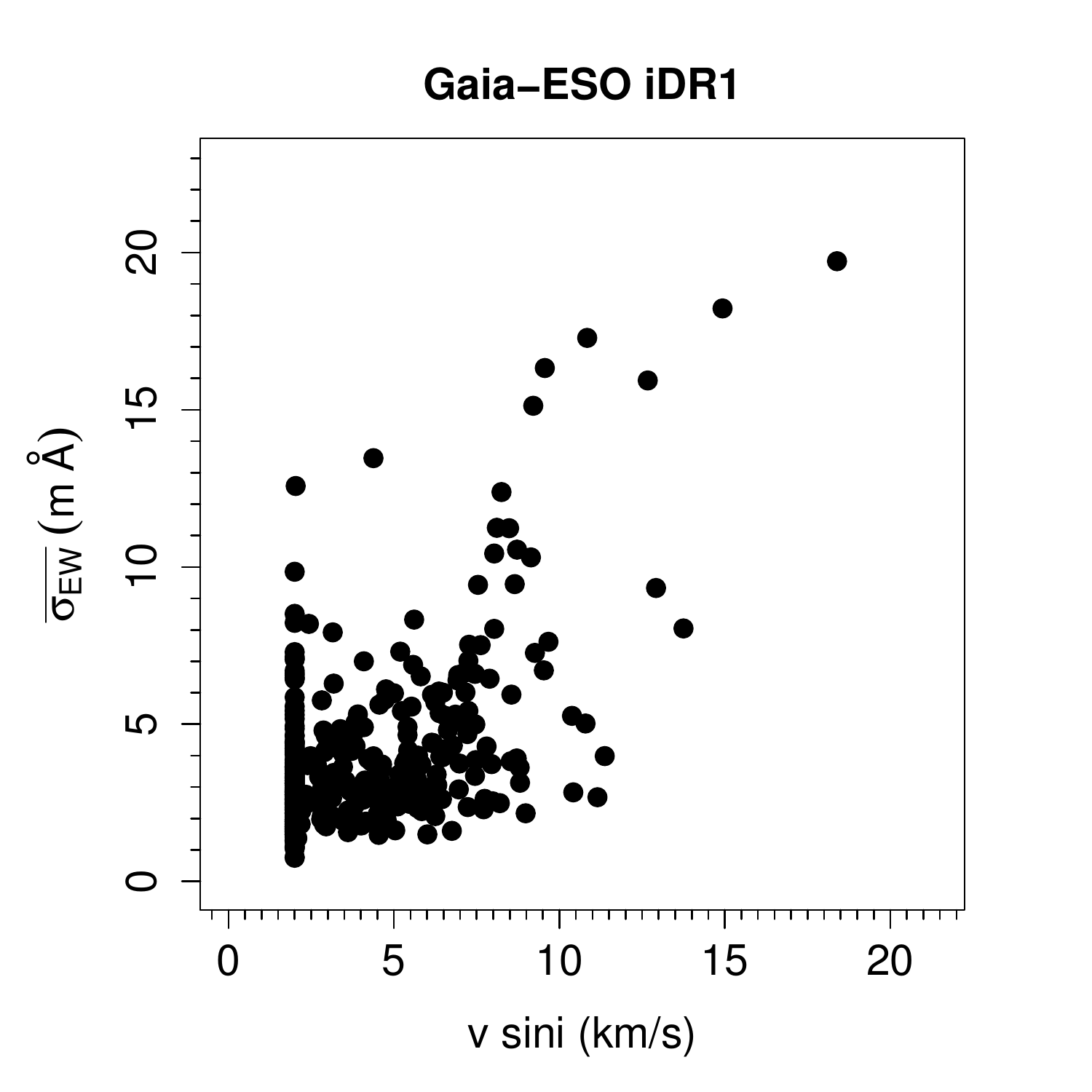}
 \caption{Mean of all the standard deviations of the \ion{Fe}{i} lines in a star, $\overline{\sigma_{\rm EW}}$, as a function of the rotational velocity of the star.}\label{fig:ewsvsiniidr1}%
\end{figure}

In Fig. \ref{fig:ewparamidr1}, $\overline{\sigma_{\rm EW}}$ is plotted against the atmospheric parameters, $T_{\rm eff}$, log $g$, and [Fe/H]. Most of the stars where $\overline{\sigma_{\rm EW}}$ $>$ 10m\AA\ tend to be warm, metal-rich subgiants or dwarfs. Many of these stars display significant rotation (Fig. \ref{fig:ewsvsiniidr1}).

We remind here that the ULB results for EWs (using the {\sf BACCHUS} code), and finally for atmospheric parameters and abundances for iDR2, were not used to compute the final recommended values that will be released, as the Node withdrew its results.

\subsection{Atmospheric Parameters in iDR1}\label{sec:atmidr1}

\subsubsection{Benchmark stars}

For the iDR1 analysis, only eight benchmark stars were available\footnote{The stars were: Bet Vir, Eta Boo, Gam Sge, Ksi Hya, HD 22879, HD 107328, HD 122563, and HD 140283.} and they did not cover the parameter space as well as the 21 stars used in iDR2. The accuracy of the Node results was judged by evaluating if the Node could reproduce $T_{\rm eff}$ and $\log~g$ of most benchmark stars to within $\pm$ 150 K and $\pm$ 0.30 dex, respectively. If yes, the Node results were considered to be accurate. If not, the Node results were disregarded. In practice, only the results of one Node were discarded. 

For iDR1 weights were not computed and the parameter space was not divided in three regions. The individual results were then combined using a simple median. The comparison with the fundamental parameters of the benchmark stars ensures that the final parameters are also in the scale defined by them, to within the accuracy level adopted above ($\pm$ 150 K for $T_{\rm eff}$ and $\pm$ 0.30 dex for $\log~g$).

\subsubsection{Calibration Clusters}

\begin{figure*}[ht]
\centering
\includegraphics[height = 4.2cm]{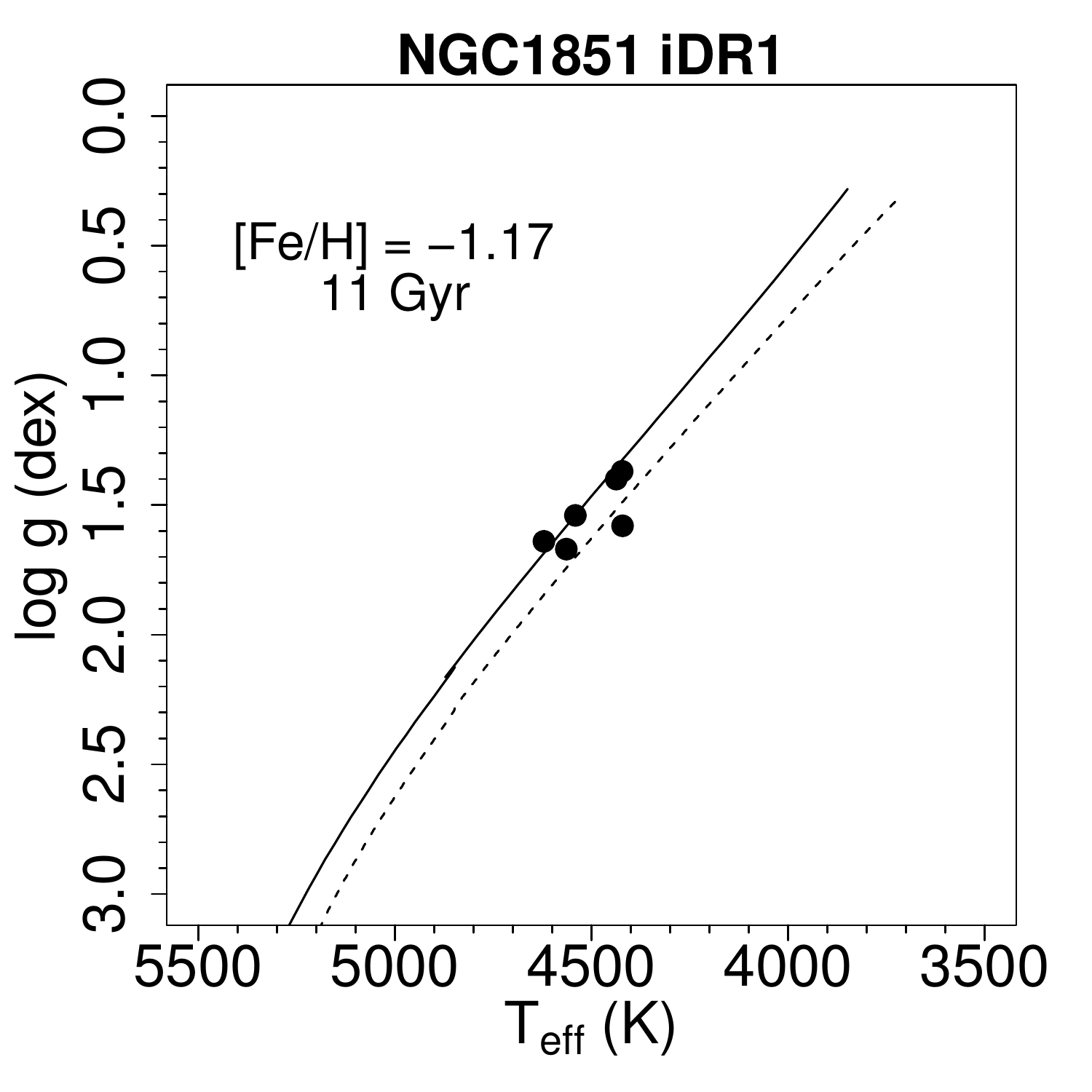}
\includegraphics[height = 4.2cm]{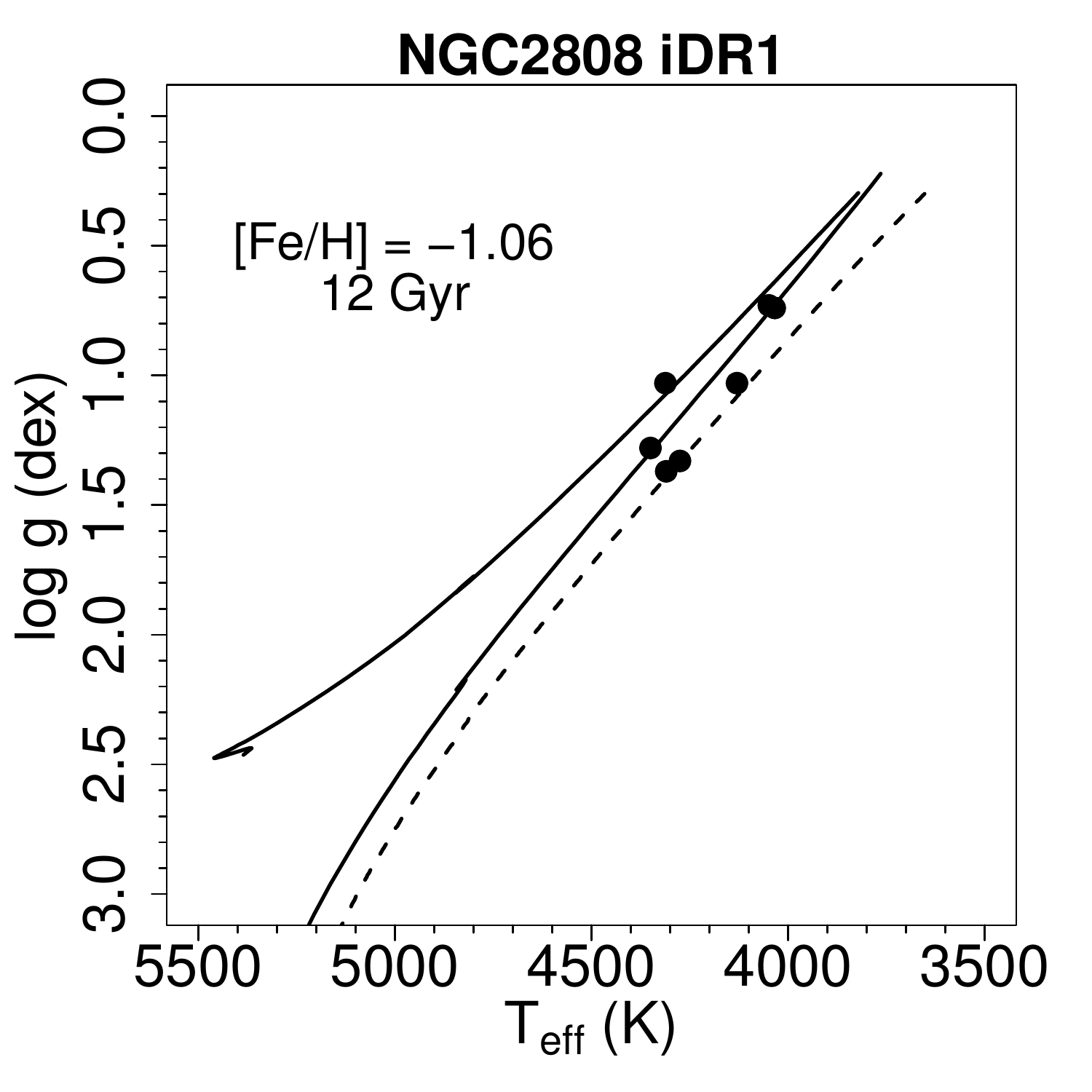}
\includegraphics[height = 4.2cm]{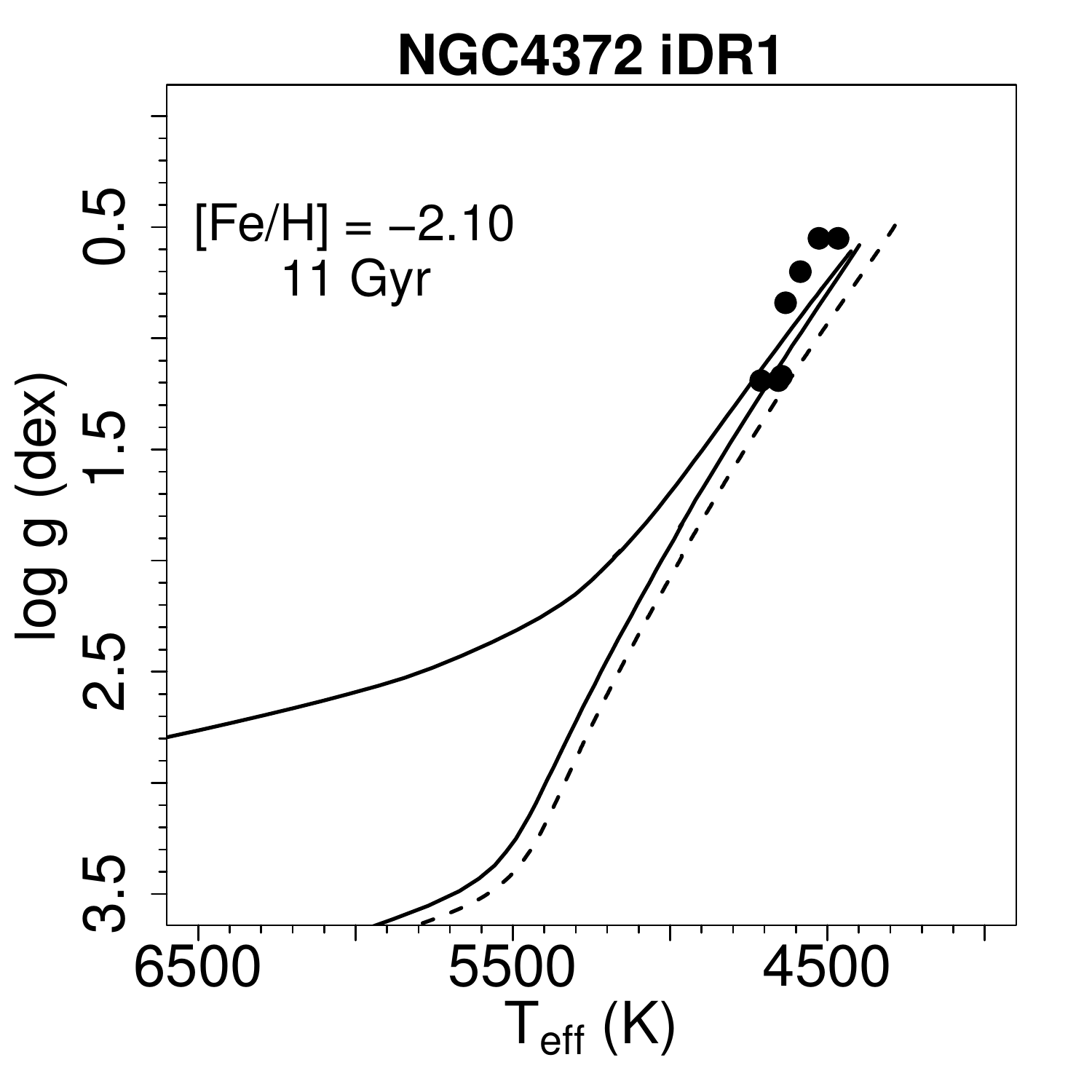}
\includegraphics[height = 4.2cm]{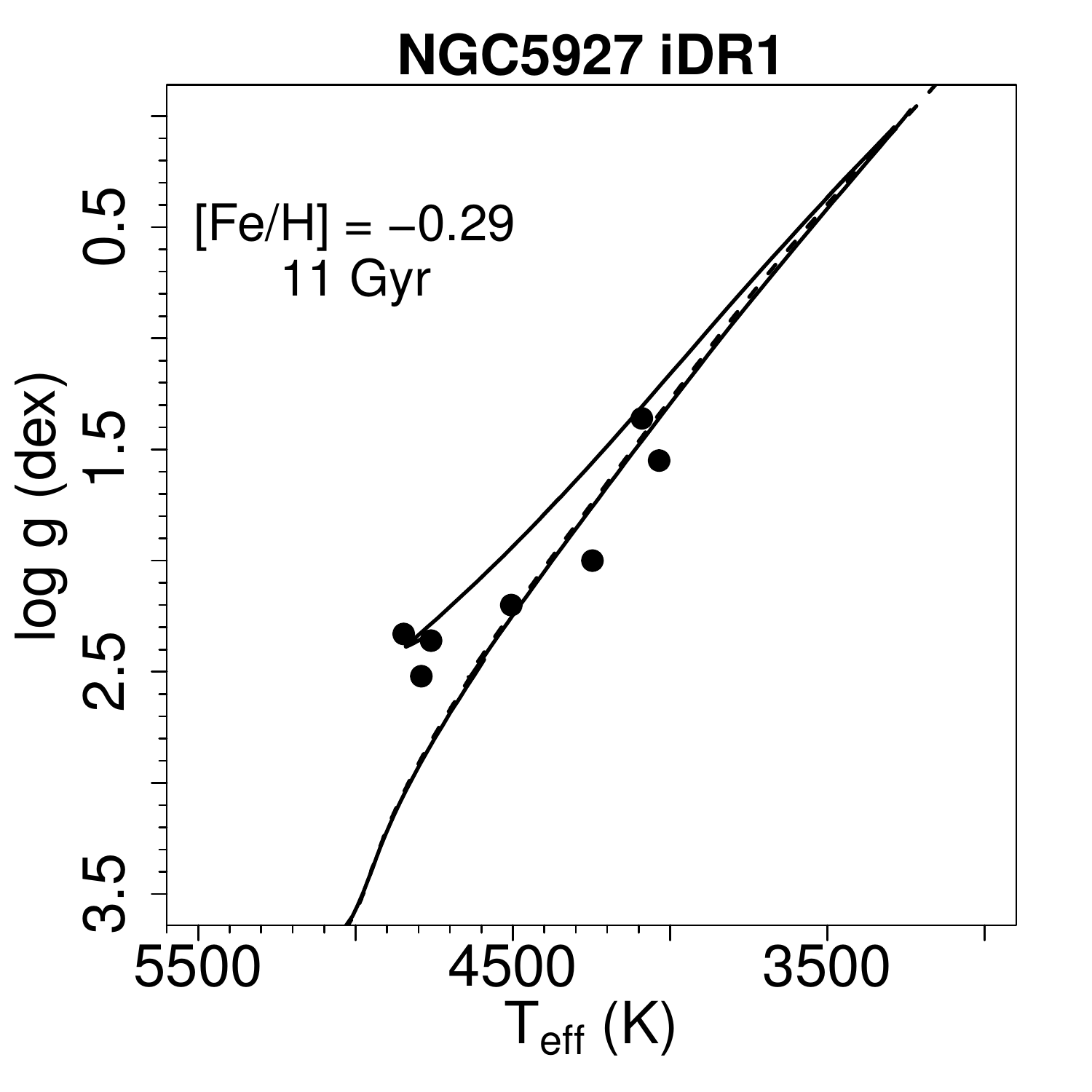}
 \caption{Recommended parameters of the stars in the calibration clusters of iDR1 in the $T_{\rm eff}$-$\log~g$ plane. No attempt was made to identify non-member stars, the plots include all stars observed in the field of the clusters. The ages, metallicities, and isochrones are the same as in Fig. \ref{fig:clusters}.}\label{fig:clustersidr1}%
\end{figure*}

The number of calibration clusters available during the iDR1 analysis was also smaller. Four calibration globular clusters (\object{NGC 1851}, \object{NGC 2808}, \object{NGC 4372}, and \object{NGC 5927}) and one calibration open cluster (NGC 6705) were analyzed. The NGC 6705 AB-type stars were mostly found to be fast rotators. The results for them were deemed uncertain and were excluded during quality control. In Fig. \ref{fig:clustersidr1} we show the final recommended parameters of the stars observed in the cluster fields in comparison with isochrones. The agreement is very good, lending confidence on the final recommended parameters of iDR1. 

\subsubsection{Method-to-method dispersion}

As done for iDR2, we compare the results of different Nodes and quantify the method-to-method dispersion of each parameter using the associated median absolute deviation (MAD). The MAD is defined as the median of the absolute deviations from the median of the data.

\begin{figure*}[ht]
\centering
\includegraphics[height = 6cm]{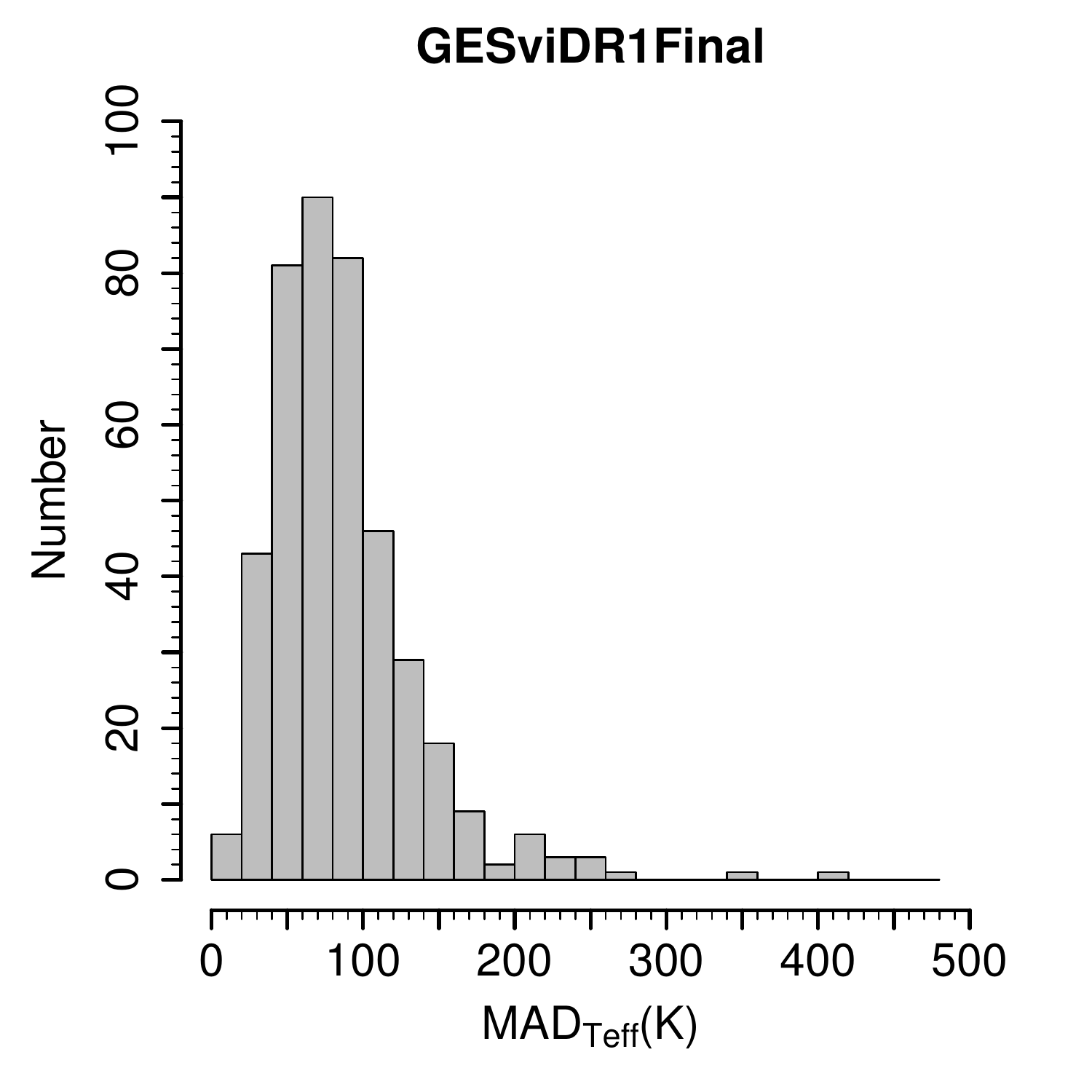}
\includegraphics[height = 6cm]{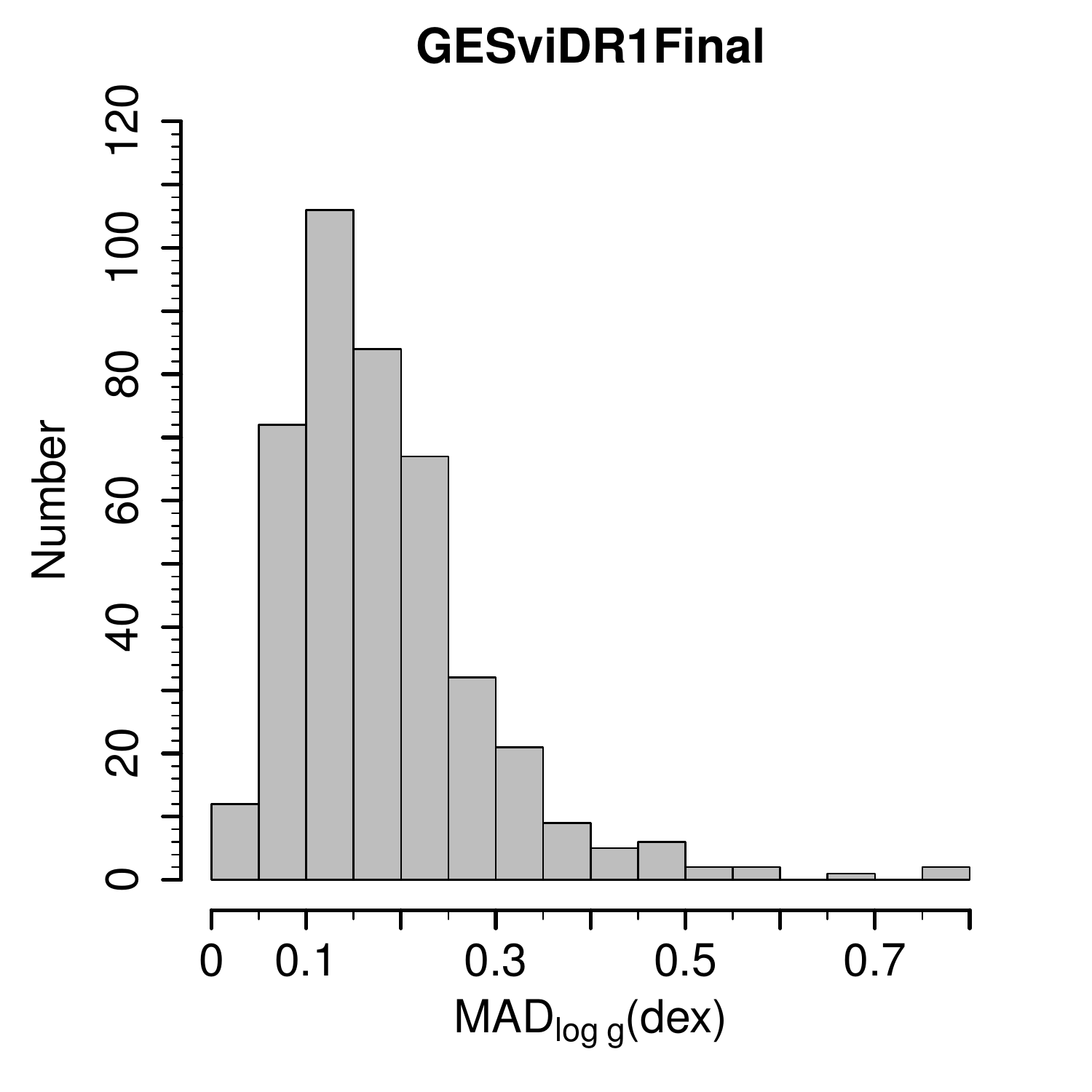}
\includegraphics[height = 6cm]{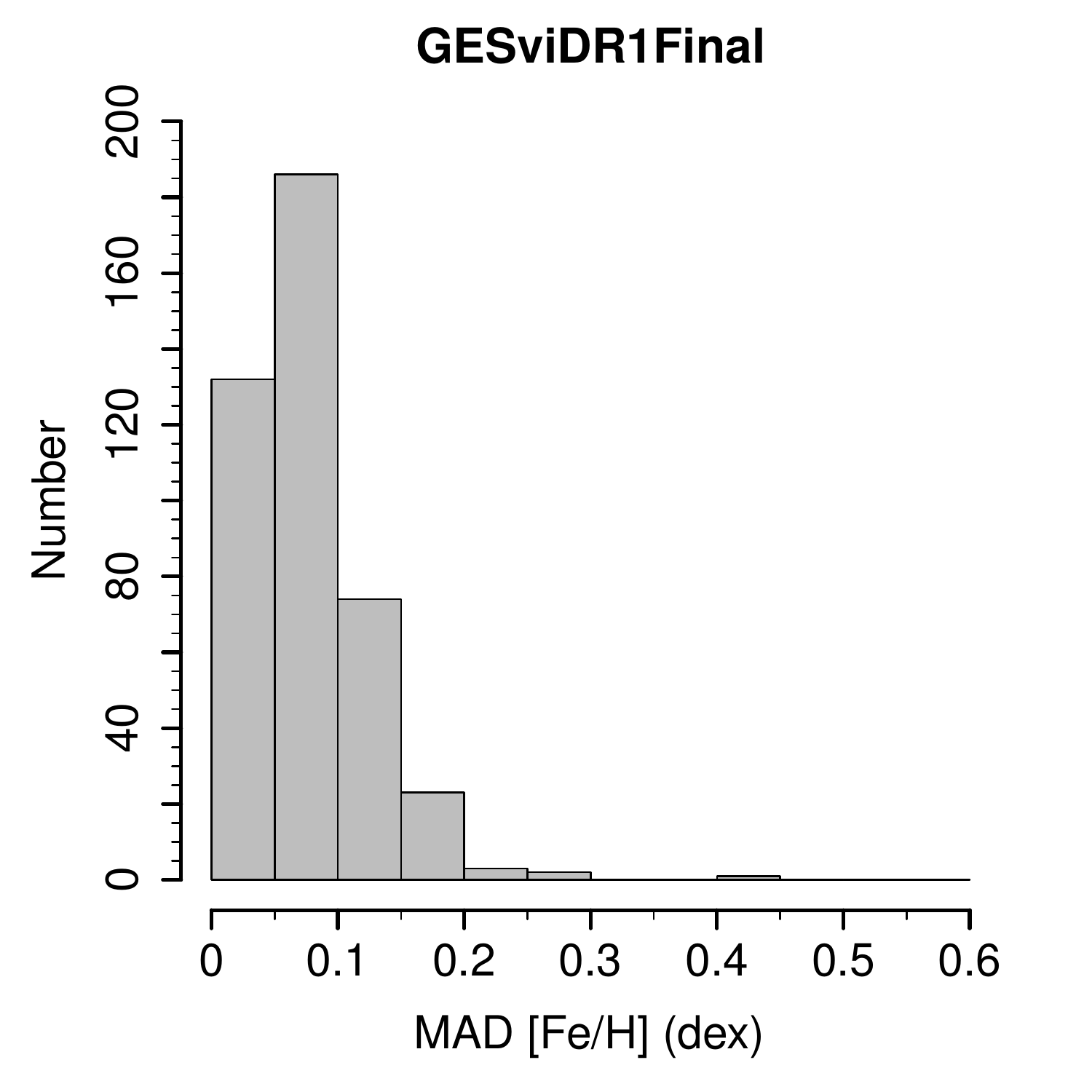}
 \caption{Histograms showing the distribution of the method-to-method dispersion of the atmospheric parameters of the 421 stars that are part of GESviDR1Final. \emph{Left:} The dispersion of $T_{\rm eff}$. \emph{Center:} The dispersion of $\log g$. \emph{Right:} The dispersion of [Fe/H].}\label{fig:histidr1}%
\end{figure*}

For the GESviDR1Final results, the median of the method-to-method dispersion is 78 K, 0.17 dex, and 0.07 dex for $T_{\rm eff}$, $\log g$, and [Fe/H], respectively. These values are slightly larger than for iDR2. The third quartile of the distribution has values of 108 K, 0.23 dex, and 0.10 dex for $T_{\rm eff}$, $\log g$, and [Fe/H], respectively. Histograms of these dispersions are shown in Fig. \ref{fig:histidr1}. For $T_{\rm eff}$, the dispersion is within reasonable expectations. For the surface gravity, the dispersion is perhaps too high. However, the surface gravity is a quantity notoriously difficult to derive for field stars with uncertain distances. For the metallicity, there is a very good agreement among the Nodes.

\subsubsection{Recommended Atmospheric Parameters}


Based on the comparisons of the individual Node results with the calibrators, as shown above, the following scheme has been adopted to calculate the recommended values of atmospheric parameters of the FGK-type stars with UVES spectra for iDR1:

\begin{enumerate}
\item The accuracy of the Node results is judged using the eight available benchmark stars as reference, with a tolerance of $\pm$150\,K and $\pm$0.30\,dex, for $T_{\rm eff}$ and $\log g$ respectively.
\item Further consistency tests of the Node results are conducted using the calibration clusters.
\item Nodes that fail to reproduce the reference atmospheric parameters of most of the benchmark stars, or that produce unreliable results for the calibration clusters are disregarded.
\item The median value of the validated results is adopted as the recommended value of that parameter. The median should minimize the effect of eventual outlier results.
\item The MAD is computed to quantify the method-to-method dispersion (analysis precision) and is adopted as an indicator of the uncertainties.
\item The number of results on which the recommended value is based is also reported.
\end{enumerate}

\begin{figure*}[ht]
\centering
\includegraphics[height = 7cm]{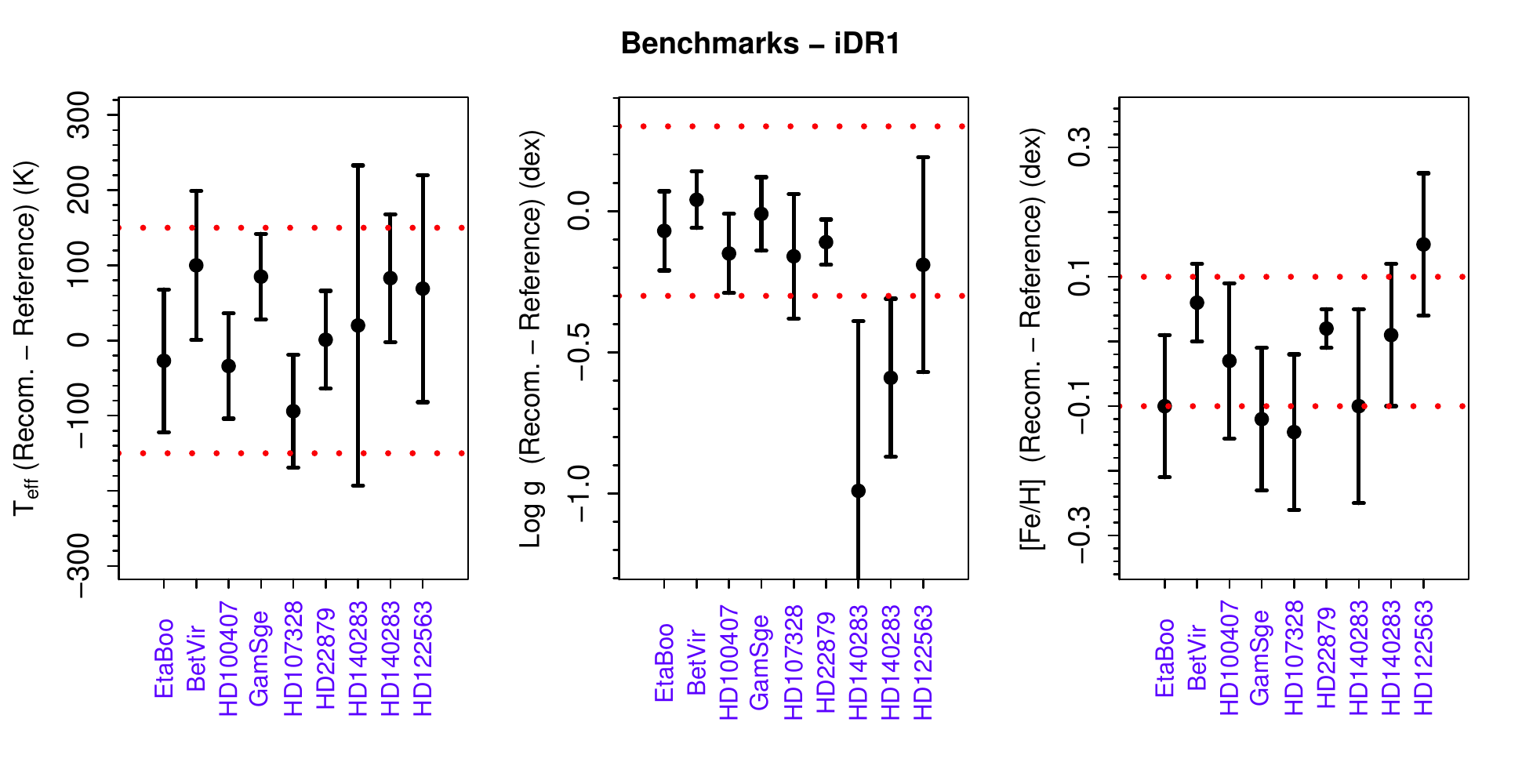}
 \caption{Difference between the recommended values of $T_{\rm eff}$, $\log g$, and [Fe/H] for the benchmark stars of GESviDR1Final and the reference values. The error bars are the method-to-method dispersions. The stars are sorted in order of decreasing [Fe/H] (left to right). The dotted red lines indicate limits of $\pm$ 150 K for $T_{\rm eff}$, of $\pm$ 0.30 dex for $\log g$, and of $\pm$ 0.10 dex for [Fe/H]. Star HD 140283 appears twice because two different spectra of this star (based on different exposures) were produced and analyzed separately.}\label{fig:benchcompidr1}%
\end{figure*}

Table \ref{tab:analysisidr1} summarizes the number of stars for which atmospheric parameters were determined during the science verification analysis and are part of the GESviDR1Final internal release. The analysis of a fraction of the stars ($\sim$ 17\%) was not completed for different reasons (e.g. high-rotation, double-lined signatures, too low S/N). 

A comparison of the recommended values of the atmospheric parameters of the benchmark stars (computed as described above) with the reference values is shown in Fig. \ref{fig:benchcompidr1}. All recommended values of $T_{\rm eff}$ are within $\pm$ 150 K of the reference ones. Good agreement is also present for $\log g$ (within $\pm$ 0.30 dex), except for HD 140283, a metal-poor subgiant (two spectra of this star were analyzed separately and thus it appears twice in the plot). Gravity values for metal-poor stars are known to be affected by NLTE effects \citepads[see e.g.][]{2012MNRAS.427...27B}, therefore it is no surprise that the results of LTE-based analyses shown here are discrepant when compared to the reference values, since the latter are independent from spectroscopy. The results included in GESviDR1Final for metal-poor stars should be used with care. The recommended [Fe/H] values agree with the reference ones to within $\pm$ 0.15 dex.

\begin{table*}[t]
\caption{Outcome of the analysis of the iDR1 data. Number of FGK-type stars observed with UVES and with atmospheric parameters in the GESviDR1Final internal release.}
\label{tab:analysisidr1} 
\centering
\begin{tabular}{lcl}
\hline\hline
Gaia-ESO type &  Number of stars & Comment \\ 
\hline 
Analyzed stars & 508 & \\
Stars with results & 421 & \\ 
GES\_MW & 271 & Milky Way fields. \\
GES\_CL & 98 & Open clusters fields. \\
GES\_SD & 52 & Calibration targets. \\
\hline
\end{tabular}
\end{table*}

\subsection{Abundances}\label{sec:abunidr1}

\begin{figure*}[t]
\centering
\includegraphics[height = 4.4cm]{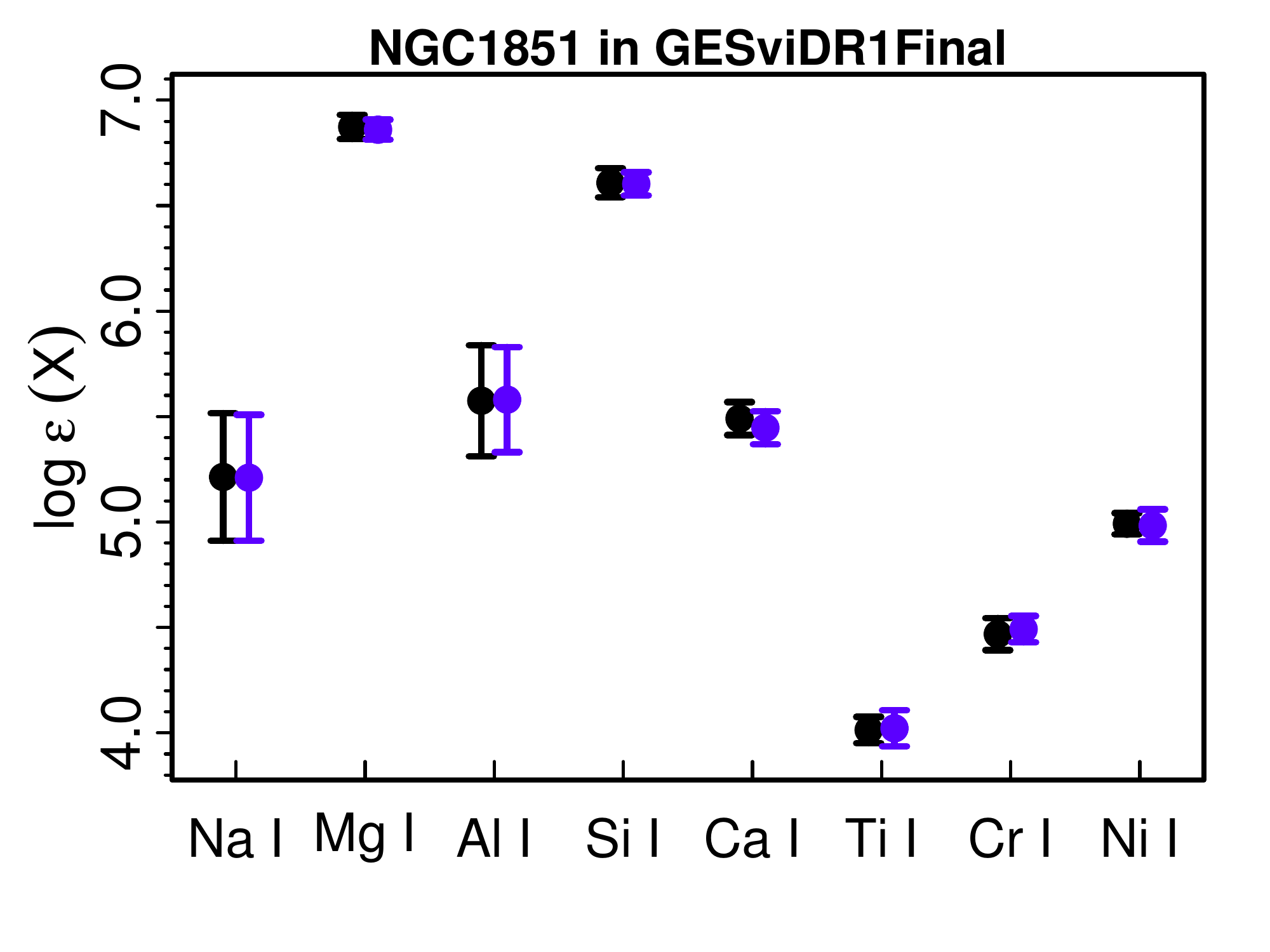}
\includegraphics[height = 4.4cm]{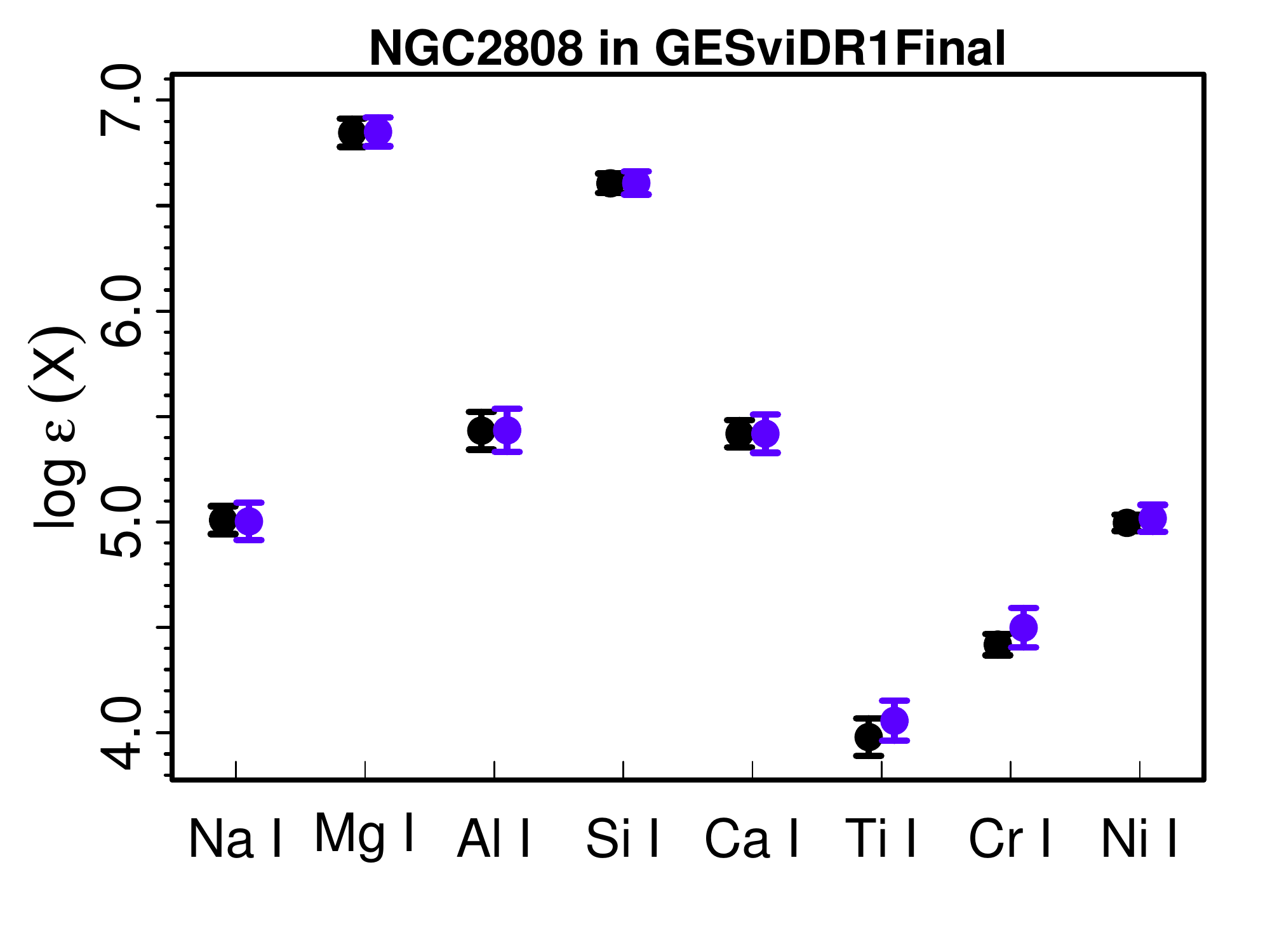}
\includegraphics[height = 4.4cm]{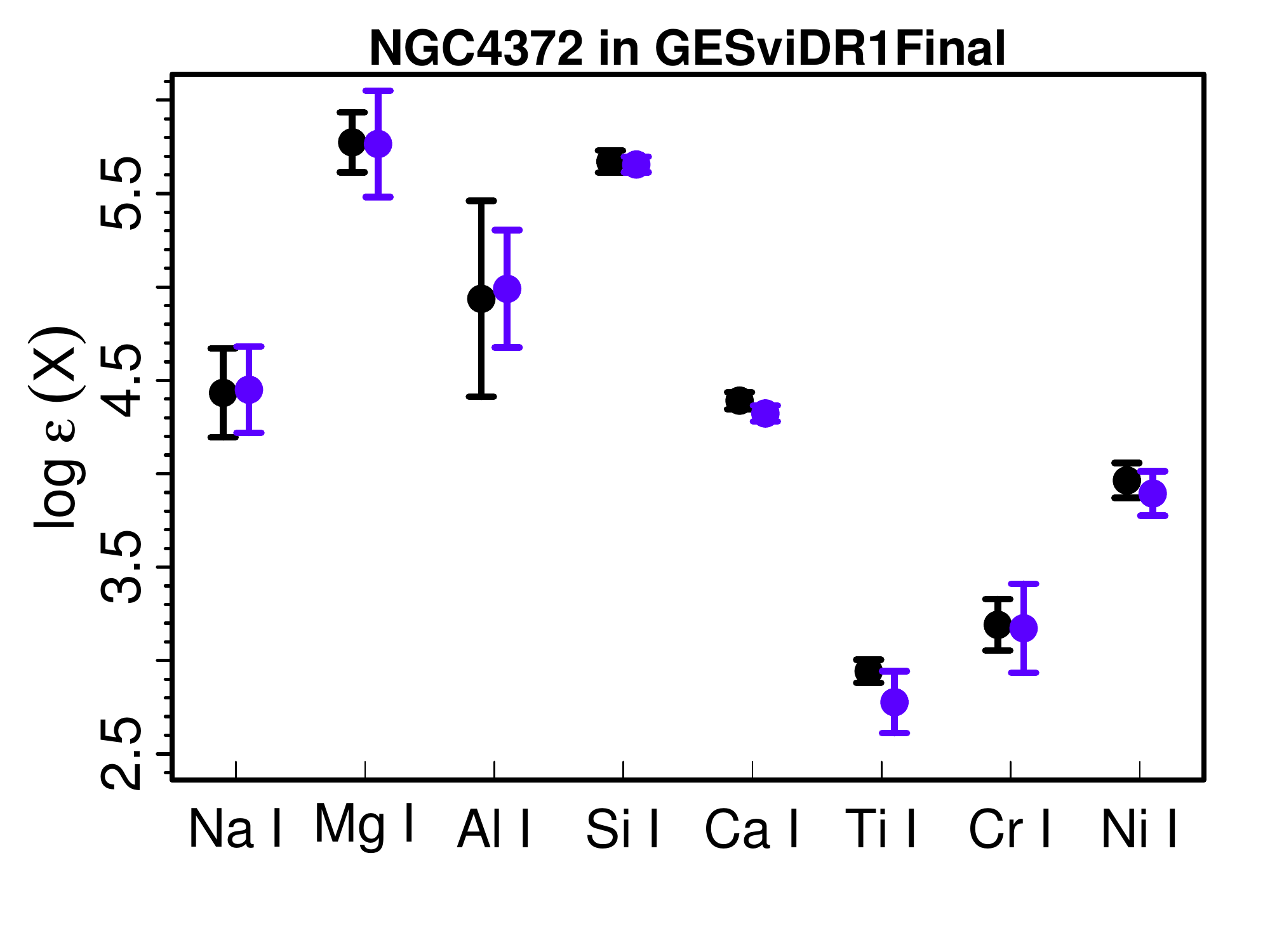}
\includegraphics[height = 4.4cm]{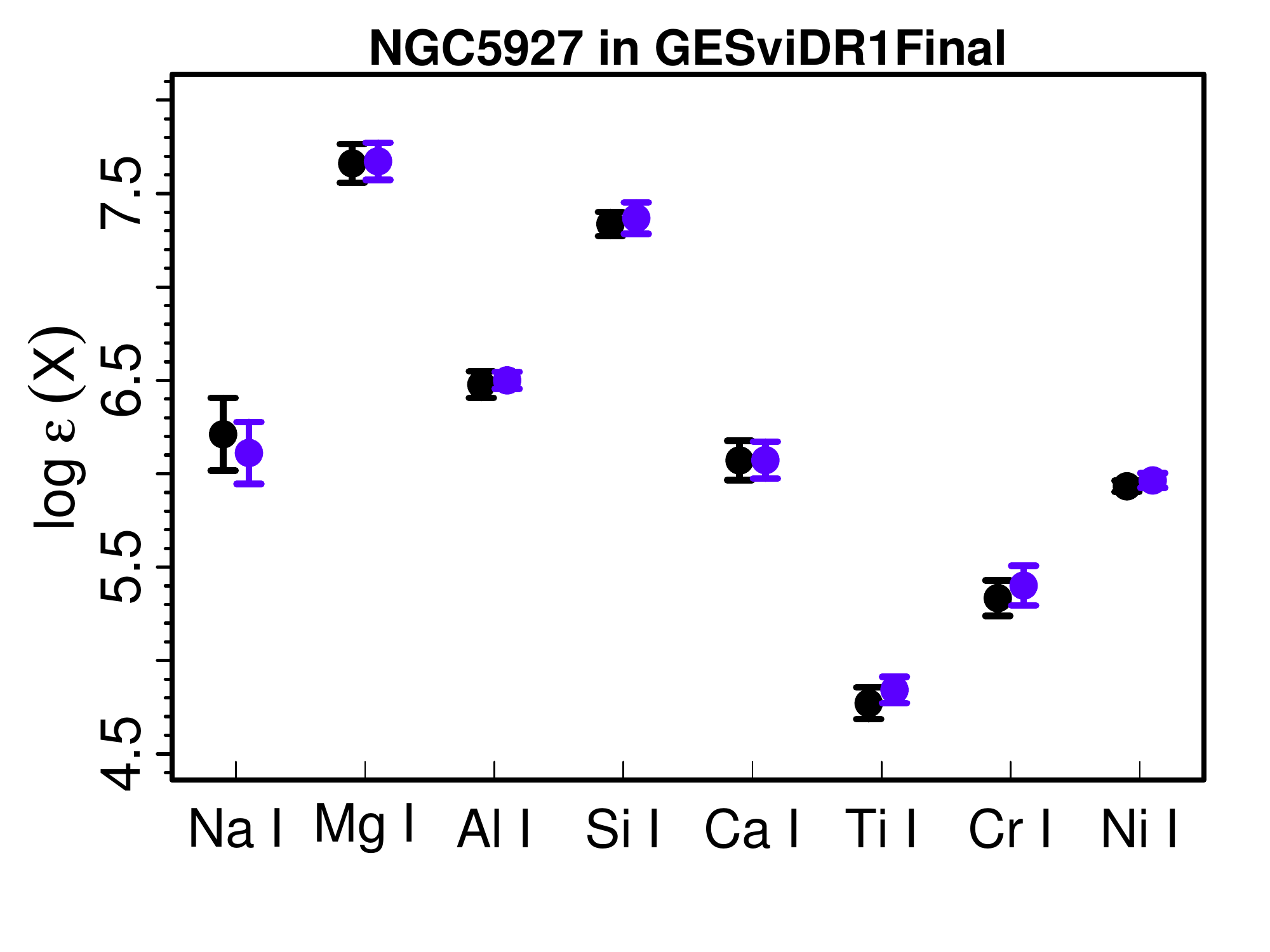}
\includegraphics[height = 4.4cm]{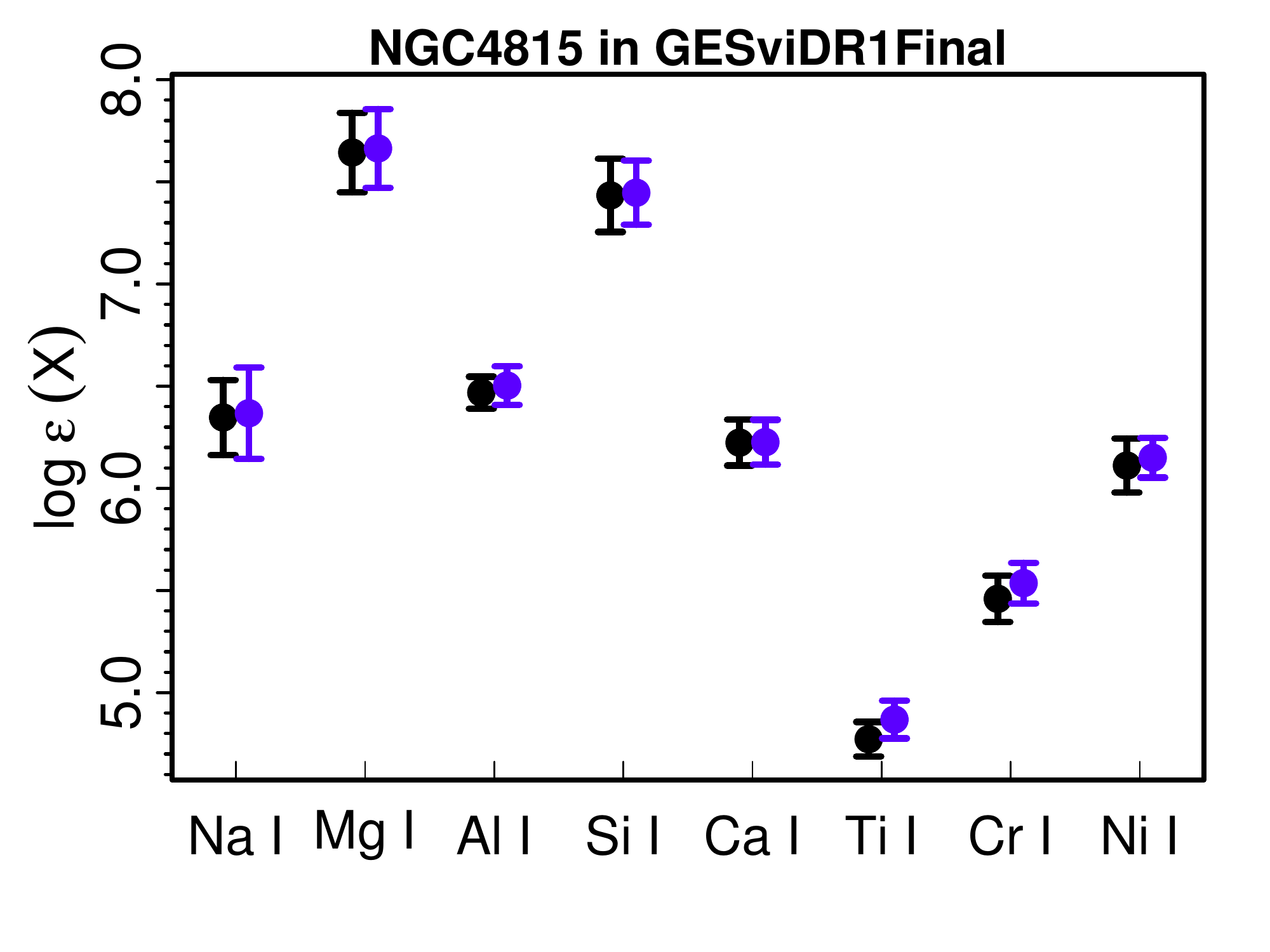}
 \caption{Comparison of two sets of abundances for the stars of calibrating globular and open clusters included in iDR1. The points are the averages for all the stars observed in the field of a given cluster. No attempt was made to identify non members. Symbols in black are the abundances computed with the recommended atmospheric parameters. Symbols in blue are the abundances computed with the atmospheric parameters of the Nodes. The error bars are the standard error of the mean. }\label{fig:abunclustersidr1}%
\end{figure*}

As for the atmospheric parameters, the elemental abundances were computed in different ways for the iDR1 and the iDR2 datasets. 

Multiple determinations of the abundances were conducted. All Nodes that have tools for abundance determinations performed the analysis in parallel for all the stars. For iDR1, the Nodes were asked to compute abundances using two sets of atmospheric parameters for each star, i.e. i) the atmospheric parameters derived by the Node itself and ii) the set of recommended atmospheric parameters, computed as described above. 

We then computed the median of the multiple determinations for each of these two cases. For iDR1 we did not homogenize the line-by-line abundances, but only the final values of each element in each star. In Fig. \ref{fig:abunclustersidr1} we compare the two sets of abundances for a few elements in stars of globular and open clusters. It is clear from this plot that there is no significant difference between the final abundances computed with the two sets of atmospheric parameters. 

In addition, it is apparent that the star-to-star scatter of the abundances does not seem to increase when using one or the other set of atmospheric parameters. This lend confidence to the approach adopted here, of having multiple abundances determined by different groups and adopting the median values as the recommended best values.

For the final set of recommended abundances included in GESviDR1Final, we decided to adopt the median of the results calculated using as input the recommended values of $T_{\rm eff}$, $\log g$, [Fe/H], and $\xi$. The MAD was again adopted as an indicator of the uncertainties (as it is a measurement of the precision with which multiple methods agree). The following 16 elements were analyzed and abundances for at least a handful of stars are included in GESviDR1Final: Li, O, Na, Mg, Al, Si, S, Ca, Ti, Cr, Fe, Ni, Zn, Y, Zr, and Ce. Except for Li, O, S, Zn, Zr, and Ce, all the abundances have been determined by at least three different Nodes. Elements that have important hyperfine structure were not included, as this kind of data were not part of the Gaia-ESO line list (version 3.0) when the abundances were calculated.

\subsubsection{Method-to-method dispersion}

\begin{figure*}[h]
\centering
\includegraphics[height = 4.5cm]{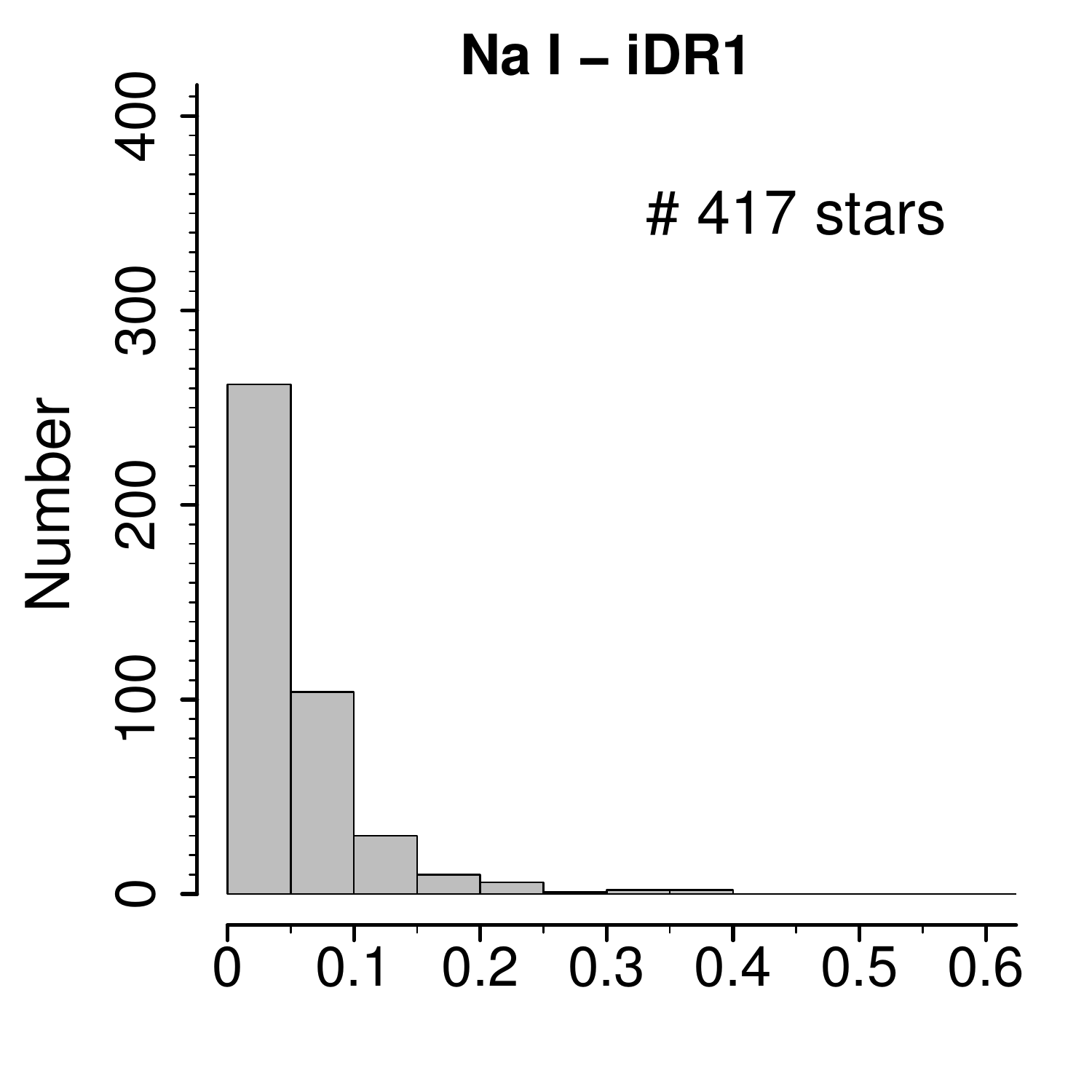}
\includegraphics[height = 4.5cm]{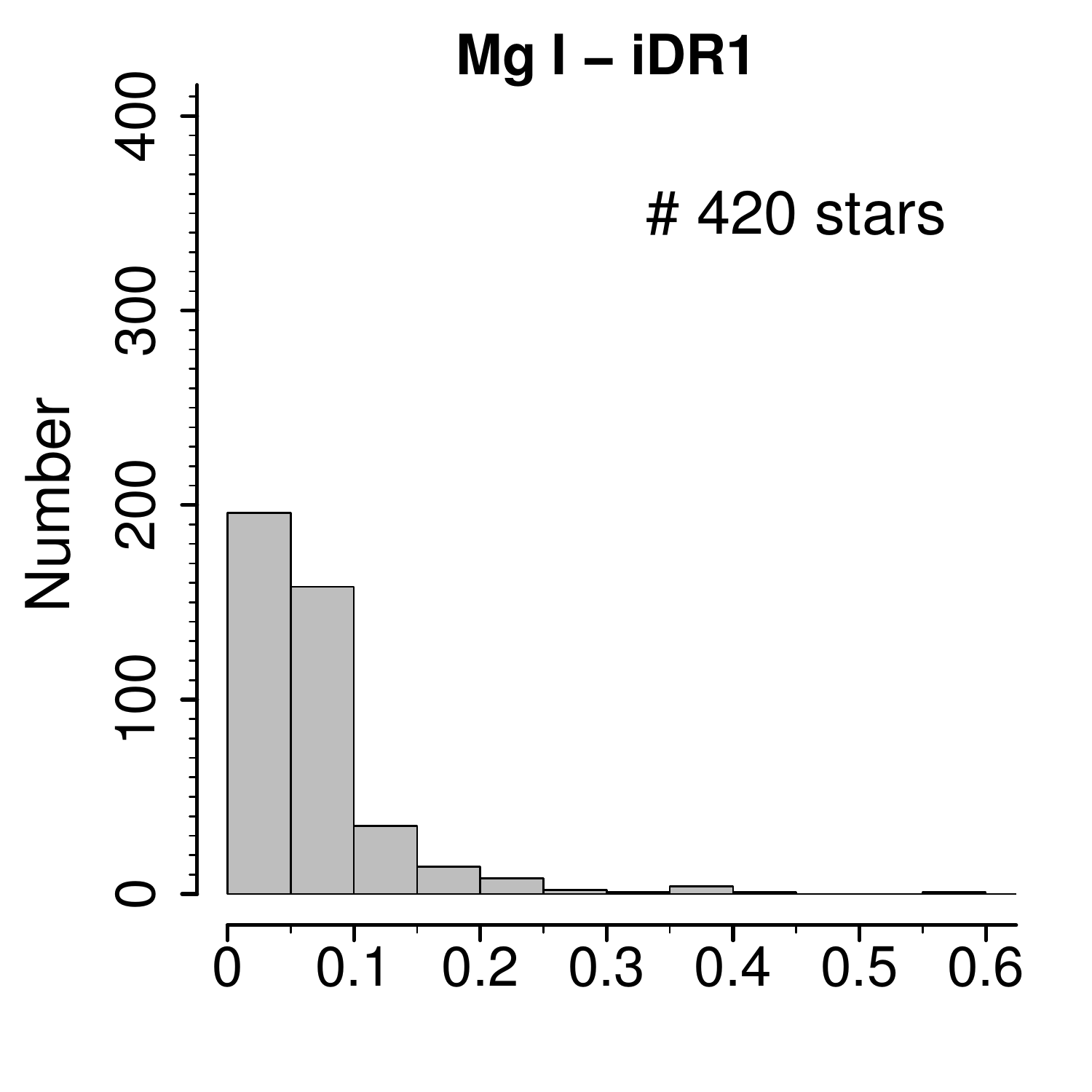}
\includegraphics[height = 4.5cm]{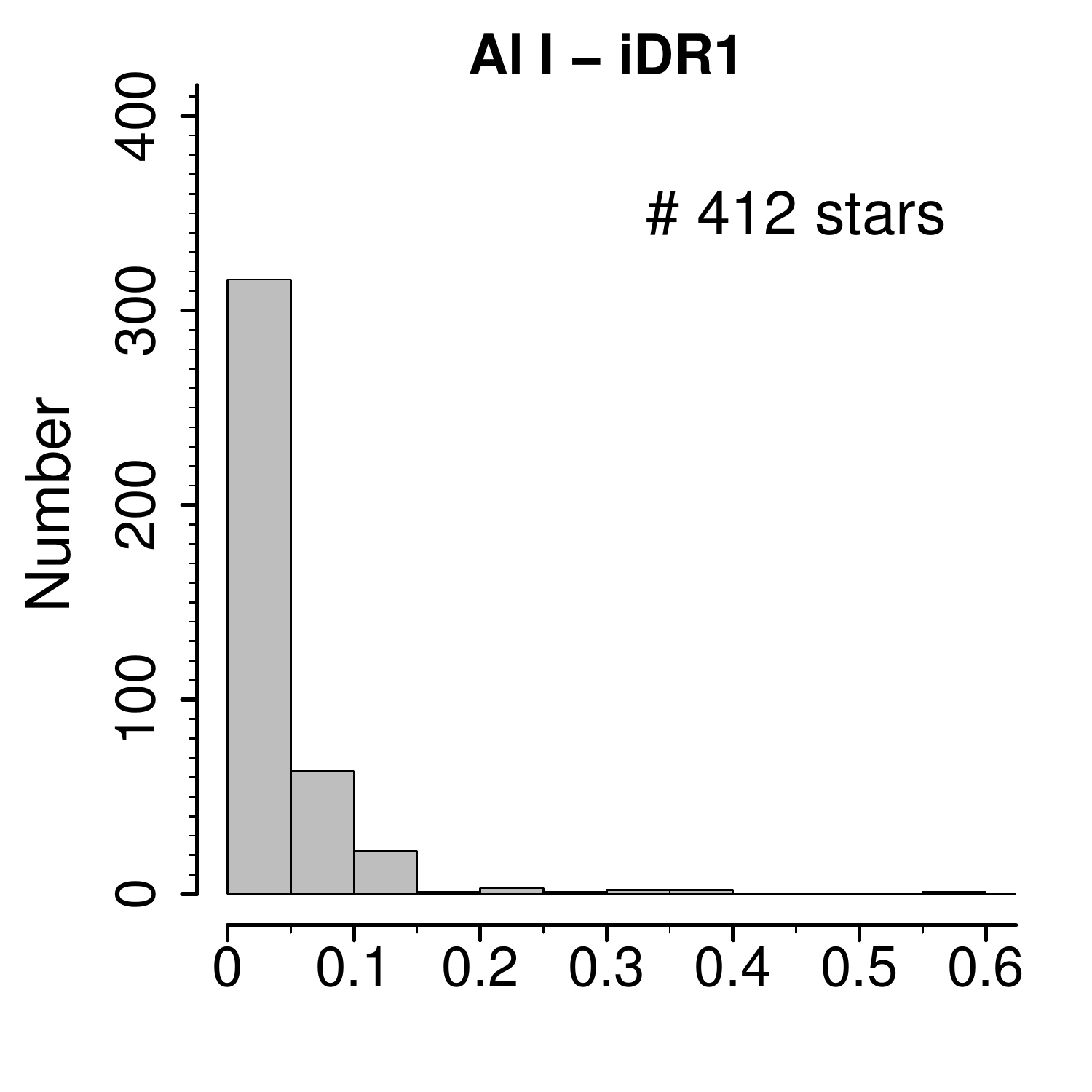}
\includegraphics[height = 4.5cm]{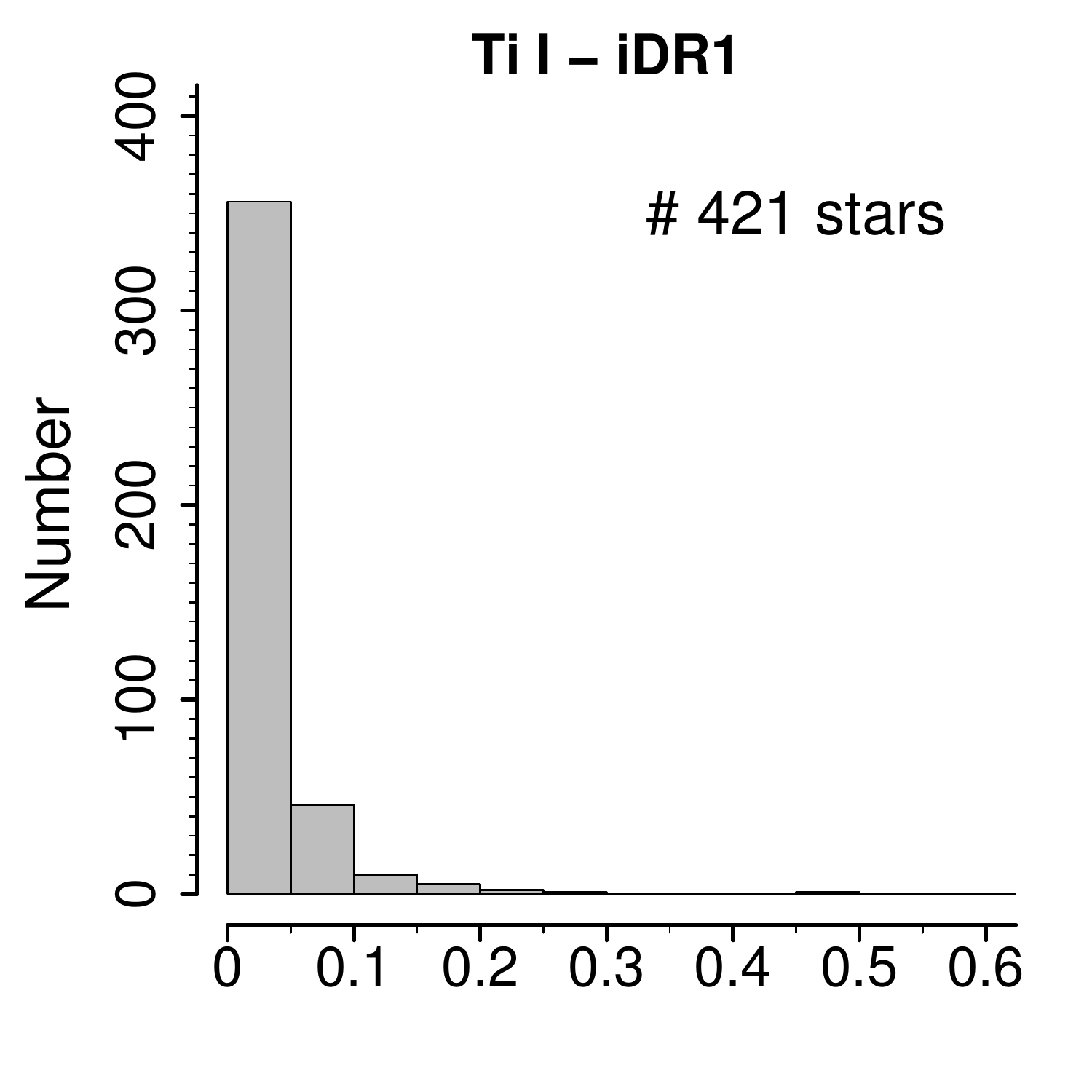}
\includegraphics[height = 4.5cm]{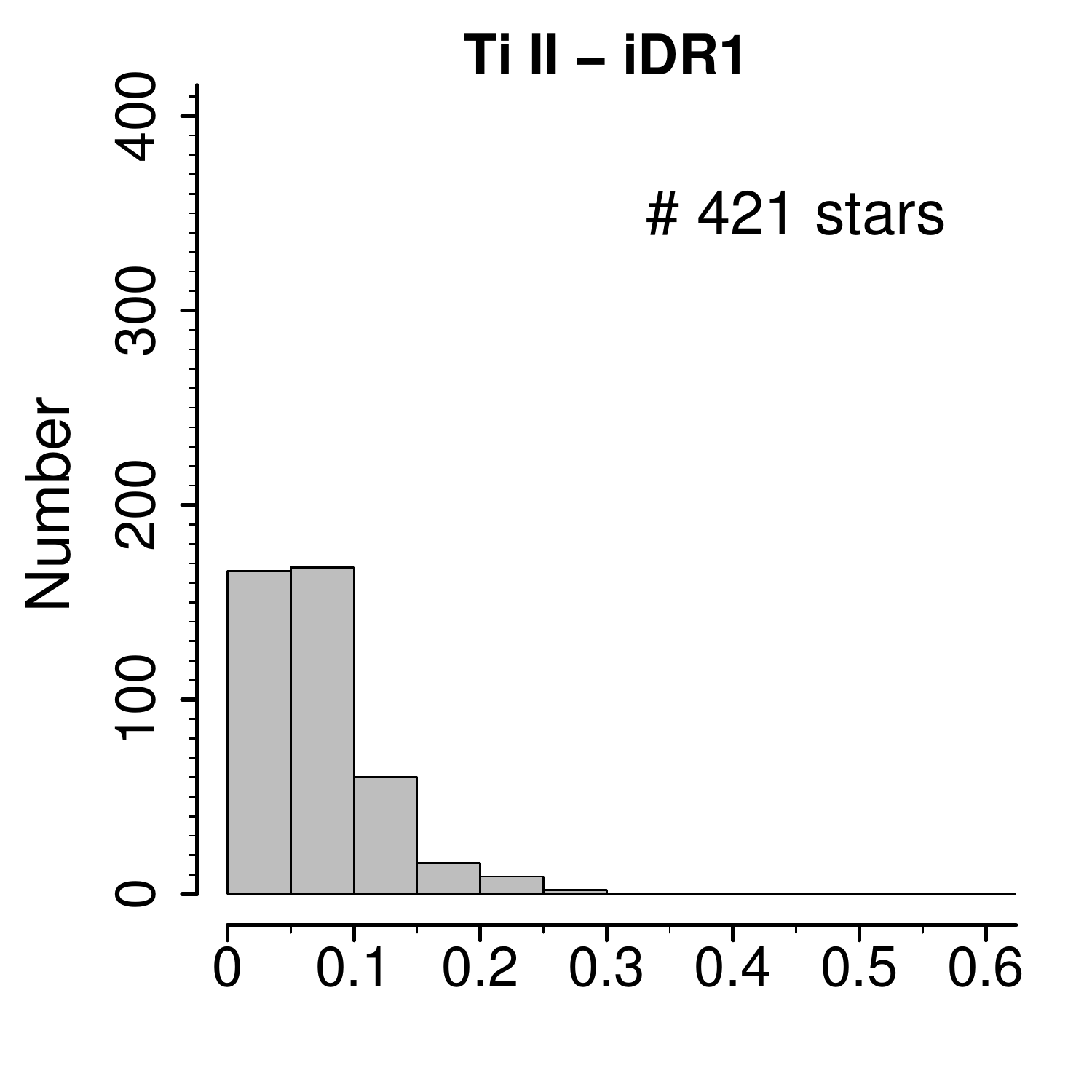}
\includegraphics[height = 4.5cm]{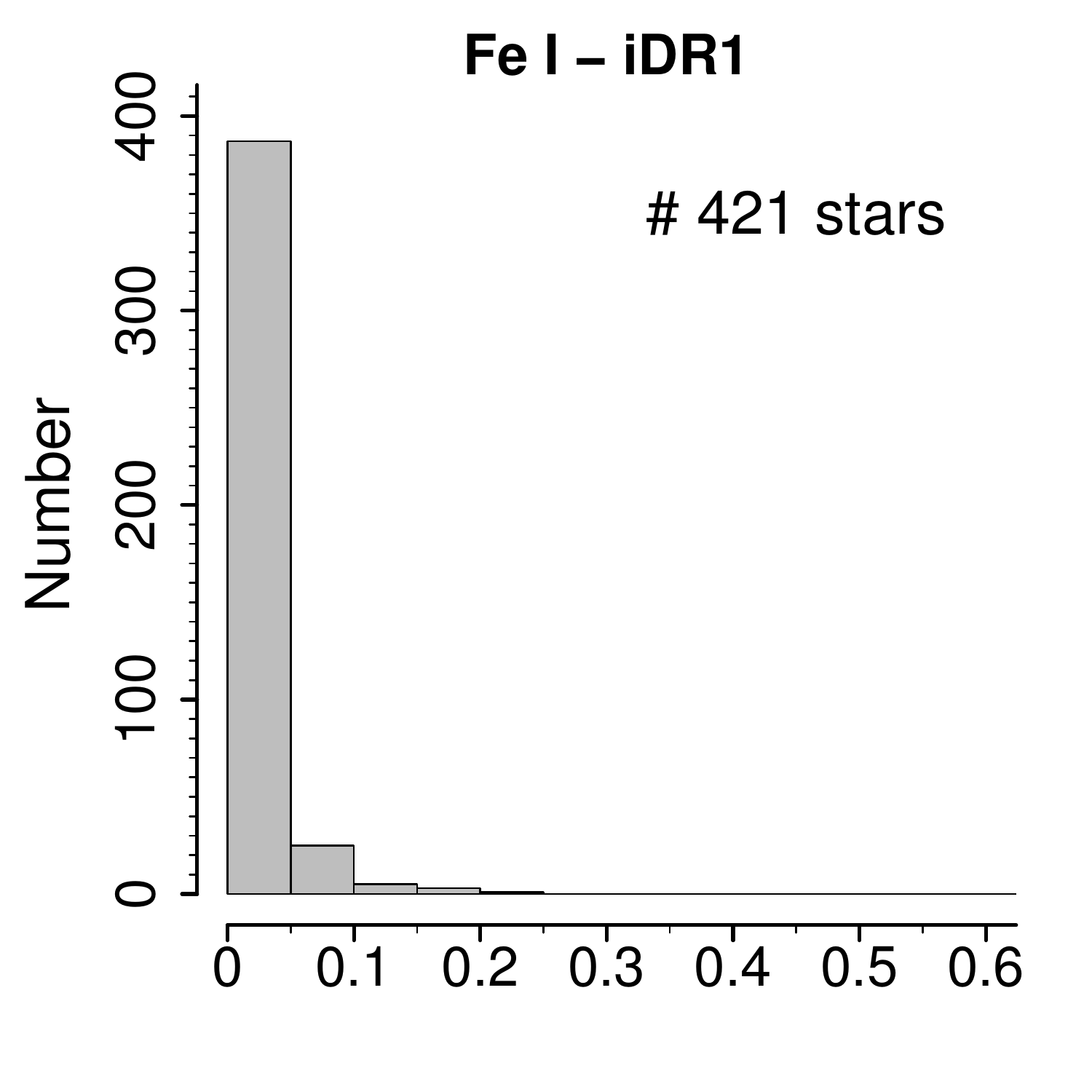}
\includegraphics[height = 4.5cm]{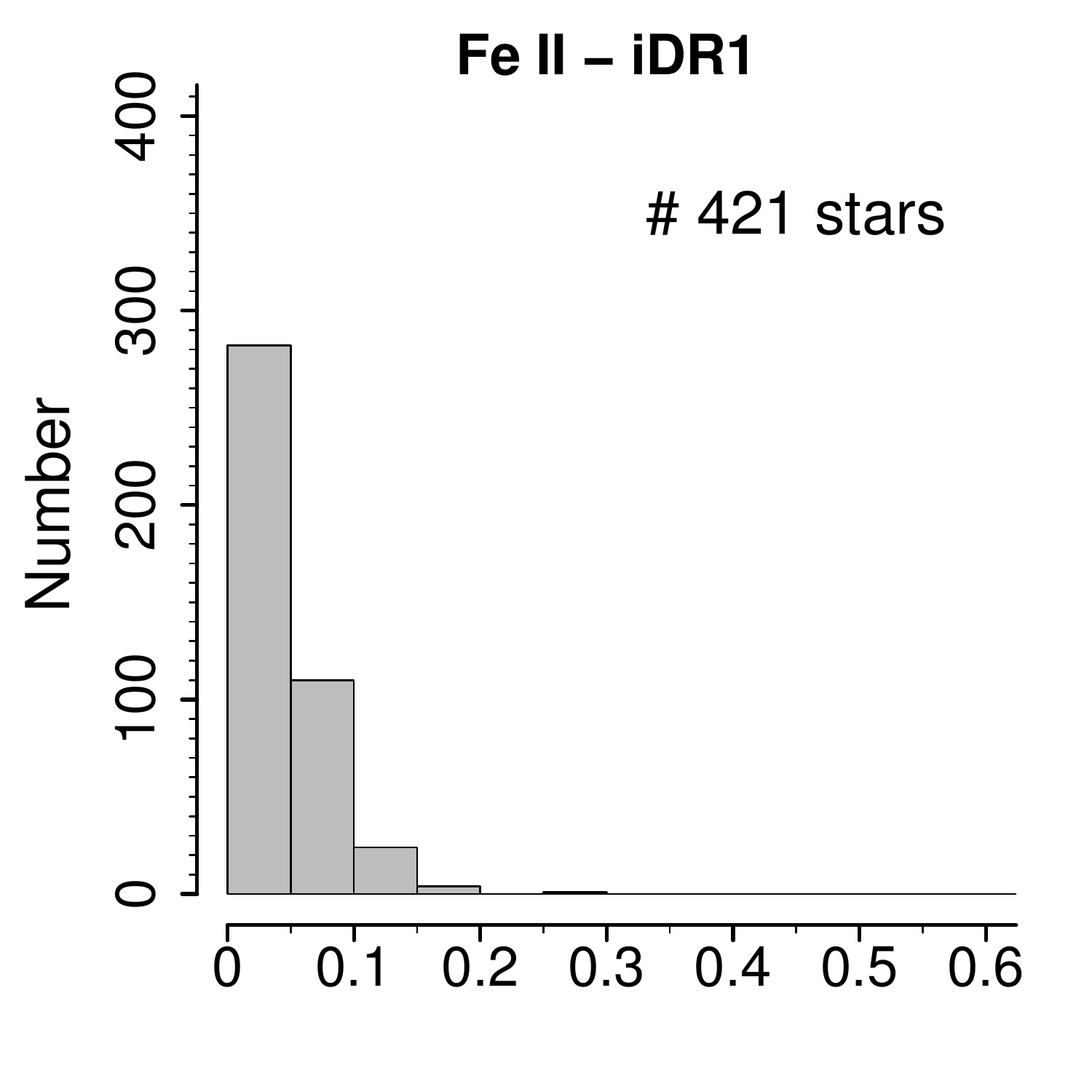}
\includegraphics[height = 4.5cm]{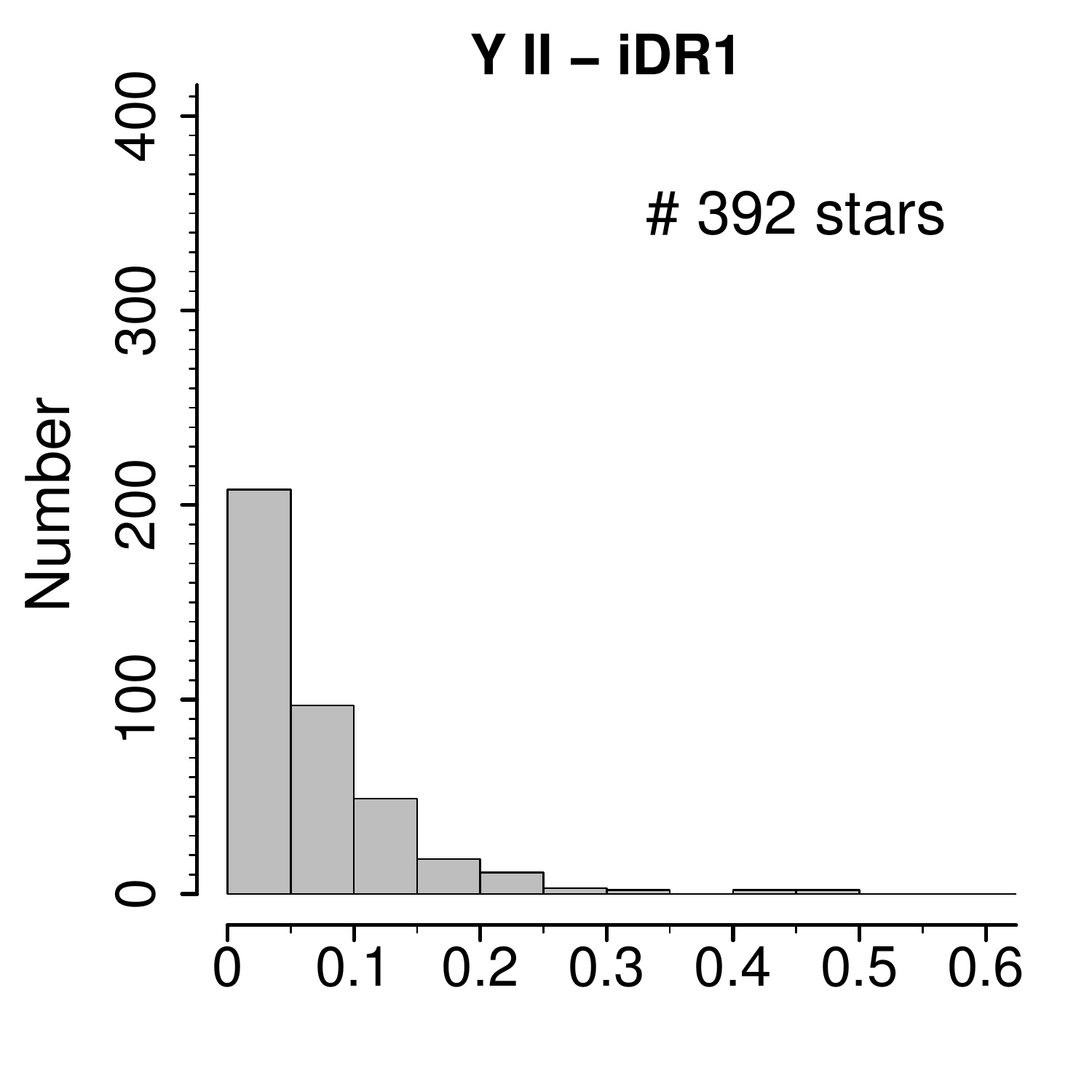}
 \caption{Histograms with the method-to-method dispersion of selected species included in the iDR1 results.}\label{fig:madelemidr1}%
\end{figure*}

The method-to-method dispersion of the abundances can be used as an indicator of the precision with which the results were derived. In Fig. \ref{fig:madelemidr1} we show the histogram of the MADs of a few selected elements. The third quartile of the method-to-method dispersion distribution is equal to or below 0.05 dex for the elements: \ion{Al}{i}, \ion{Ti}{i}, \ion{Fe}{i}, and \ion{Ni}{i}. It is between 0.06 and 0.10 dex for the other elements with multiple measurements: \ion{Na}{i}, \ion{Mg}{i}, \ion{Si}{i}, \ion{Ca}{i}, \ion{Ti}{ii}, \ion{Cr}{i}, \ion{Cr}{ii}, and \ion{Fe}{ii}. The MADs were adopted as the typical uncertainties.

%


\end{document}